\documentclass[preprint]{elsarticle}
\biboptions{sort&compress}
\usepackage{amssymb}
\usepackage{amsmath}
\usepackage{graphicx}
\usepackage{pdfpages}
\usepackage{adjustbox}
\usepackage{rotating}
\usepackage[T1]{fontenc}
\usepackage{mathrsfs}
\usepackage{multirow}
\usepackage{hhline}
\usepackage{enumerate}
\usepackage{enumitem}
\usepackage{booktabs} 
\usepackage[english]{babel}

\usepackage{float}
\usepackage[font=footnotesize,caption=false]{subfig}
\usepackage{lscape}
\usepackage{etoolbox}
\usepackage{framed}
\usepackage{xcolor}[2007/01/21]
 \usepackage{makecell}
 \renewcommand\theadfont{\bfseries}
 
 \usepackage{listings}
 
 \definecolor{mygreen}{rgb}{0,0.6,0}
 \definecolor{mygray}{rgb}{0.5,0.5,0.5}
 \definecolor{mymauve}{rgb}{0.58,0,0.82}
 \lstdefinestyle{customc}{
   belowcaptionskip=1\baselineskip,
   breakatwhitespace=false,         
   breaklines=true,                 
   captionpos=b,                    
   commentstyle=\color{mygreen},    
   deletekeywords={...},            
   escapeinside={\%*}{*)},          
   extendedchars=true,      
   frame=single,	                   
   language=C,
   showstringspaces=false,
   basicstyle=\footnotesize\ttfamily,
   keywordstyle=\bfseries\color{green!40!black},
   commentstyle=\itshape\color{purple!40!black},
   identifierstyle=\color{blue},
   stringstyle=\color{orange},
 }
 
\usepackage{hyperref}
\usepackage[
 open,
 openlevel=2,
 atend
 ]{bookmark}[2011/12/02]

 \bookmarksetup{color=blue}

\usepackage[ruled]{algorithm2e} 

\journal{Computer Science Review}

\begin{document}
\renewcommand{\thesection}{\Roman{section}}

\newtheorem{theoremm}{Theorem}
\newtheorem{eqed}{Example}
\newtheorem {lemmaa}{Lemma}
\newtheorem{proposition}{Proposition}
\newtheorem {observation}[theoremm]{Observation}
\newtheorem {defnn}{Definition}
\newtheorem {corollaryy}{Corollary}
\newtheorem {conjecturee}{Conjecture}
\newtheorem {fact}[theoremm]{Fact}
\newtheorem {procd}{Procedure}
\newtheorem {rules}{Rule}
\newenvironment{example}{\begin{eqed} \rm}{\hfill\end{eqed}}
\newenvironment{proof}{\noindent {\bf Proof:\ } }{\hfill $\Box$ }
\newenvironment{lemma}{\begin{lemmaa} \sl}{\end{lemmaa}}
\newenvironment{theorem}{\begin{theoremm}{\bf:}\sl}{\end{theoremm}}
\newenvironment{corollary}{\begin{corollaryy}{\bf:}\sl}{\end{corollaryy}}
\newenvironment{procd1}{\begin{procd} \sl}{\end{procd}}
\newenvironment{conjecture}{\begin{conjecturee} \sl}{\end{conjecturee}}
\newenvironment{definition}[1][Definition]{\begin{defnn} \sl}{\end{defnn}}

\begin{frontmatter}

\title{A Search for Good Pseudo-random Number Generators: Survey and Empirical Studies\\ (Draft Version)\tnoteref{t1}}

\author[mymainaddress]{Kamalika Bhattacharjee\corref{mycorrespondingauthor}}
\cortext[mycorrespondingauthor]{Corresponding author}
\ead{kamalika.it@gmail.com}

%
\author[sirmainaddress]{Sukanta Das}
\ead{sukanta@it.iiests.ac.in}

\address[mymainaddress]{Department of Computer Science and Engineering, National Institute of Technology, Tiruchirappalli, Tamilnadu, India 620015}
\address[sirmainaddress]{Department of Information Technology, Indian Institute of Engineering Science and Technology, Shibpur, West Bengal, India 711103}

\tnotetext[t1]{Copyright of this paper is the property of Elsevier. Please cite this paper as : Kamalika Bhattacharjee and Sukanta Das.
	A search for good pseudo-random number generators: Survey and empirical studies,
	\textit{Computer Science Review}, 45: 100471, 2022,	ISSN 1574-0137, \url{https://doi.org/10.1016/j.cosrev.2022.100471.}}
\begin{abstract}
This paper targets to search so-called \emph{good} generators by doing a brief survey over the generators developed in the history of pseudo-random number generators (PRNGs), verify their claims and rank them based on strong empirical tests in same platforms. To do this, the genre of PRNGs developed so far are explored and classified into three groups -- linear congruential generator based, linear feedback shift register based and cellular automata based. From each group, the well-known widely used generators which claimed themselves to be `\emph{good}' are chosen. Overall $30$ PRNGs are selected in this way on which two types of empirical testing are done -- blind statistical tests with Diehard battery of tests, battery \emph{rabbit} of TestU01 library and NIST statistical test-suite as well as graphical tests (lattice test and space-time diagram test). Finally, the selected PRNGs are divided into $24$ groups and are ranked according to their overall performance in all empirical tests.
\end{abstract}


\end{frontmatter}

\section{Introduction}\label{sec:Introduction}
\noindent Society, arts, culture, science and even daily life is engulfed by the concept of randomness. One of the major usage of randomness is in generating numbers for diverse fields of practices. History of human race gives evidence that, since the ancient times, people has generated random numbers for various purposes. As an example, for them, the output of rolling a dice was a sermon of God! However, in the modern times, researchers and scientists have discovered diverse applications and fields, like probability theory, game theory, information theory, statistics, gambling, computer simulation, cryptography, pattern recognition, VLSI testing etc., which require random numbers. Most of these applications entail numbers, which appear to be random, but which can be reproduced on demand. Such numbers, which are generated by a background algorithm, are called pseudo-random numbers and the implementation of the algorithms as pseudo-random number generators (PRNGs). In this work, however, by random number, we will mean pseudo-random numbers only.

Even, PRNGs have a long history of development -- its modern journey starting in late $19^{th}$ century \cite{galton1890dice} to early $20^{th}$ century \cite{tippett1927random,2980655,2983623,von195113} and evolving and getting more powerful ever since \cite{l1996maximally,Lewis:1973:GFS:321765.321777,Matsumoto:1998:MTE:272991.272995,Panneton:2005:XRN:1113316.1113319,Panneton:2006:ILG:1132973.1132974,Saito2008,Saito2009,Tausworthe,Tezuka:1987:DGP:31846.31848,wolfram86c,Marco99,Vigna:2016:EEM:2956571.2845077}.
Initially, two dominating categories of PRNGs existed -- Linear Congruential Generator (LCG) based and Linear Feedback Shift Register (LFSR) based. In $1985$, due to inherent randomness quality of some cellular automata (CAs), CAs have also been introduced as a source of randomness \cite{Wolfram85c}. Since then, it has bewitched many researchers to use it as PRNG, see as example \cite{Horte89a,Horte89c,114093,COMPAGNER1987391,Tomassini96, Marco99,122655,alonso2009elementary,tcad/DasS10,SukantaTH, Guan03, Guan04, Marco00,Guan04a,HOSSEINI2014149}.

Usually, latest PRNG claims to be superior to the previous ones. One of the reasons of this claim is based on the PRNG's performance in some statistical tests, like Diehard \cite{diehard}, TestU01 \cite{L'Ecuyer:2007:TCL:1268776.1268777}, NIST \cite{rukhin2001statistical} etc. battery of tests, which empirically detect non-randomness in the generated numbers. However, many questions arise in this regard -- \emph{How much effective are these statistical tests? Will numbers of a PRNG, which performs well in all the statistical tests, really appear random or noisy to the human eye? Is the claim of a PRNG to be superior than others really correct?} In this work, we target to address these questions. 
To do this, we have selected all uniform PRNGs developed in the history which have been widely used and are considered to be \emph{good}. Some of these PRNGs are not used now, but in the history of development of PRNG, they have played a significant role making them part of our selection. Similarly, some of them do not have very good randomness quality but are still widely used as general purpose portable generators. So, we include them in our list. 

{Our target is to test all the PRNGs on the same framework to judge their randomness quality. Like already mentioned, one of the ways of a PRNG to claim its superiority is through statistical tests. Several battery of tests exists which offer to detect non-randomness in a PRNG. However, each of these batteries has some limitations. They (e.g., Diehard) are based on some approximations about distributions (and some of them are false), and the thresholds are taken quite arbitrarily. Therefore, they are \emph{inherently incomplete} in nature. In fact, the PRNGs fools the testbed(s) by exploiting this inherent incompleteness such that the testbed cannot detect the non-randomness in it (which off course exists as it is \emph{pseudo}-random). Nevertheless, as these are the only available metric to judge the PRNGs uniformly, in our work we build a common framework by using three of the well-known existing battery of tests, namely Diehard, TestU01 and NIST. 
	
However, to address the inherent incompleteness of these \emph{blind} test-beds, we also use graphical tests that allow human intervention in the form that an user can visualize if the numbers generated by a PRNG form some patterns. More specifically, we use lattice tests and introduce \emph{space-time diagram} as a graphical randomness test in this work. In the next sections reader can find that, these graphical tests will help us to verify the blind test results. We will also see that  using human intervention, space-time diagram will guide us to decide the \emph{goodness} of the PRNG in those times when the blind test results and result from graphical tests do not tally. 

Based on the overall performance of all these PRNGs on these selected testbeds and the graphical tests on the same setting, we rank the PRNGs. Obviously this ranking is not the last word, and we agree that linear ranking is not absolute. Because, sometimes the working principles of two PRNGs differ and are not comparable. Nevertheless, we do that by collecting data about existing tests and generators in a systematic way to classify the PRNGs into some groups according to our observation in this common platform.
}

This paper is organized as follows. In Section~\ref{sec:property}, the essential properties of the PRNGs are described. Section~\ref{class} classifies the journey of the PRNGs through three technologies -- Linear Congruential Generators (LCGs), Linear Feedback Shift Registers (LFSRs) based and Cellular Automata (CAs) based. Total $30$ currently used PRNGs are selected for empirical testing. The empirical tests and test-beds are described in Section~\ref{sec:empirical}. 
For each of these PRNGs, numbers are generated using the \emph{C} programs available on the Internet. These numbers are tested uniformly using all existing statistical testbeds. Then, some visual tests are applied on these numbers. If a PRNG is really good, then the result of these statistical tests and visual tests should correlate and the numbers is to appear noisy to the human eye. Section~\ref{sec:facts} depicts the test results of these PRNGs. The result of testing for all these PRNGs are further interpreted. We have observed that, for many PRNGs, the claim and actual independent result do not tally. {A relative ranking of these PRNGs based on the empirical results is given in Section~\ref{sec:final_rank}.} For any intended application, an user may choose a PRNG according to its rank. Finally, Section~\ref{conclusion} concludes the paper.

\section{PRNGs and their Properties} \label{sec:property}
\noindent PRNGs are simple deterministic algorithms which produce deterministic sequence of numbers that appear random.
In general, a PRNG produces uniformly distributed, independent and uncorrelated real numbers in the interval $[0,1)$. However, generation of numbers in other probability distribution is also possible. Mathematically, a PRNG is defined as the following \cite{LEcuyer1990}:

\begin{definition}\label{Chap:randomness_survey:def:prng}
	A pseudo-random number generator $G$ is a structure $(\mathscr{S},\mu, f,\\ \mathscr{U}, g)$, where $\mathscr{S}$ is a finite set of states, $\mu$ is the probability distribution on $\mathscr{S}$ for the initial state called seed, $f: \mathscr{S}\rightarrow \mathscr{S}$ is the transition function, $\mathscr{U}$ is the output space and $g: \mathscr{S} \rightarrow \mathscr{U}$ is the output function. The generator $G$ generates the numbers in the following way.
	\begin{enumerate}
		\item Select the seed $s_0 \in \mathscr{S}$ based on $\mu$. The first number is $u_0 = g(s_0)$.
		\item At each step $i\geq 1$, the state of the PRNG is $s_i = f(s_{i-1})$ and output is $u_i = g(s_i)$. These outputs of the PRNG are the pseudo-random numbers, and the sequence ${(u_i)}_{i\ge 0}$ is the pseudo-random sequence.
	\end{enumerate}
\end{definition}
Since a PRNG is a finite state machine with a finite number of states, after a finite number of steps, eventually it will come back to an old state and the sequence will be repeated. This property is common to all sequences where the function $f$ transforms a finite set into itself.
This repeating cycle is known as \emph{period}. The length of a period is the smallest positive integer $\rho$, such that, $\forall n\geq k, s_{\rho+n}=s_n$, here $k\geq 0$ is an integer.
If $k=0$, the sequence is \emph{purely periodic}\label{purelyperiodic}. Preferably, $\rho\approx|\mathscr{S}|$, or, $\rho \approx 2^{b}$, if $b$ bits represent each state, that is, the period covers (almost) the whole state space.

\begin{figure}[hbtp]
	\centering
	\vspace{-1.0em}
	\subfloat[Maximum-period PRNG\label{cycle1}]{%
		\resizebox{0.40\textwidth}{!}{
			\includegraphics[width=3.5in, height = 1.3in]{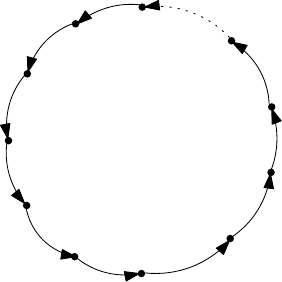}}
	}
	\hfill
	\subfloat[PRNG with non-maximum periods \label{carry}]{%
		\resizebox{0.42\textwidth}{!}{
			\includegraphics[width=3.5in, height = 1.3in]{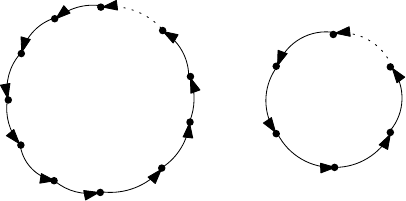}
		}
	}
	\caption{Cycle structure of PRNGs}
	\label{Chap:randomness_survey:fig:cycle} %
\end{figure}

Ideally, a PRNG has only one period, that is, all unique numbers of the output space are part of the same cycle. In that case, the PRNG is \emph{maximum-period generator}. However, many PRNGs exist which have more than one cycle. So, depending on the seeds, completely different sequence of numbers from distinct cycles may be generated. This situation is shown in Figure~\ref{Chap:randomness_survey:fig:cycle}.

Nevertheless, only LCGs (described in Section~\ref{sec:lcg}) can attain the maximum possible period. For LFSRs based generators (described in Section~\ref{sec:lfsr}), the largest achievable period is one less than the maximum period -- {such a generator is called a \emph{maximal-period} or \emph{maximal-length} generator}. Similarly, CAs are also non-maximum period generators.

Every PRNG is classified by the functions $ f$ and $g$, and the seed $s_0$.
Therefore, when a PRNG is observed for its randomness quality, it is considered that the algorithm is not known to the adversary. Following are the desirable properties, which are to be observed in a good PRNG.

\begin{description}[leftmargin=1pt]
	\item[1. Uniformity:] This property implies that, if we divide the set of possible numbers generated by the PRNG (that is, the range of the PRNG) into $K$ equal subintervals, then expected number of samples $(e_i)$ in each subinterval $i, 1\leq i \leq K$, is equal; that is, $e_i = \frac{N}{K}$, where $N$ is the range of the numbers. This ensures that, the generated numbers are equally probable in every part of the number space.
	{Let $F_i$ be the number of samples generated in subinterval $i$, $1\leq i \leq K$, then the random variable $F_i$ follows uniform distribution. Obviously, this property is not applicable for PRNGs that does not generate numbers in uniform distribution.}
	
	
	\item[2. Independence:] The generated numbers are to be independent of each other; that is, there should not be any correlation between numbers generated in succession. Let $u_t, u_{t'}$ be any two random numbers generated by the PRNG. Then, correlation between them, $Cor(u_t, u_{t'}) = 0$\footnote{Correlation of two random variables $X$ and $Y$ is related with variance of them. If $Var(X) Var(Y)$ is positive then it is defined by $$Cor(X, Y) = \dfrac{Cov(X, Y)}{\sqrt{Var(X)Var(Y)}}$$ If $Cor(X, Y) = 0$, then $X$ and $Y$ are said to be \emph{uncorrelated}.}. In that case, any subsequence of numbers have no correlation with any other subsequences. This means, given any length of previous numbers, one can not predict the next number in the sequence by observing the given numbers.
	
	\item[3. Large Period:] Every PRNG has a period after which the sequence is repeated. A PRNG is considered good if it has a very large period, {that is, $\rho\approx|\mathscr{S}|$}. Otherwise, if one can exhaust the period of a PRNG, the sequence of numbers become completely predictable.
	
	\item[4. Reproducibility:] One of the prominent reason of developing a PRNG is its property of reproducibility. This ensures that given the same seed $s_0$, the same sequence of numbers is to be generated. This is very useful in simulation, debugging and testing purposes. {Mathematically, the random sequence is $u_i = g(s_i) \text{ where } s_i = f(s_{i-1}) \text{ for all } i \ge 0$, so, when same seed $s_0$ is provided, same $u_i$s are generated.}
	
	\item[5. Consistency:] The above properties of the PRNGs are to be independent of the seed. {That is, for all $s_0$, these properties are to be maintained.}
	
	\item[6. Disjoint subsequences:] There is to be little or no correlation between subsequences generated by different seeds. {Let $u=\{\forall i~ u \text{ such that } s_0 = x\}$ and $v=\{\forall j~ u_j \text{ such that } s_0 = y\}$. Then for any $X\in u$ and $Y \in v$, $Cor(X,Y)\approx 0$.} However, this criterion is difficult to achieve in an algorithmic PRNG.
	
	
	\item[7. Portability:] A PRNG is to be portable; that is, the same algorithm can work on every system. Given the same seed, different machines with varied configuration are to give the same output sequence.
	
	\item[8. Efficiency:] The PRNG is to be very fast; which means, generation of a random number takes insignificant time. Moreover, a PRNG should not use much storage or computational overhead. This is to make certain that, the use of PRNGs in an application is not a hindrance to its efficiency.

	
	\item[9. Coverage:] {The PRNG has to cover the complete output space for a seed. This property is linked with large period. Many PRNGs have less coverage. In case the PRNG has more than one cycle, then it may happen that, only a part of its output space is covered by its seed indicating less period length.} 
	
	\item[10. Spectral Characteristics:] {In any long run of a good PRNG, the expected frequency of generation of each number should be same.}
	
	\item[11. Cryptographically Secure:] To be used in cryptographic applications, the generated numbers have to be cryptographically secure, {that is, the next output number cannot be predicted by a polynomial time algorithm (see \cite{goldreich2003foundations} for more details)}. This is a desirable property often missing in most of the algorithmic PRNGs. 
\end{description}

\noindent Many of these properties are inter-related. For example, if the numbers are not uniform, they are correlated and have identifiable patterns (poor spectral characteristics). Ideally, the numbers of a good PRNG are to satisfy all these properties. However, practically, most of the PRNGs do not possess all these properties; for example, the properties $6$ and $11$ are often missing in the existing PRNGs. Still, in terms of usage in the applications for which they are intended, many PRNGs are considered good in today's standard. 

In the next section, we tour to the existing PRNGs to classify them with respect to their underlying architecture and test their randomness quality.

\section{Classification of the PRNGs}\label{class}
\noindent Earliest PRNGs which satisfied the properties of uniformity and independences with a relatively large period were based on linear recurrences modulo a prime number, popularly called \emph{linear congruential sequence}. Introduced by Lehmer \cite{harvard1951annals}, such a PRNG is named \emph{linear congruential generator} (LCG). Most of the existing PRNGs are variants of it. 

However, another type of linear recurrences, where the modulo operator is $2$, soon became popular due to their ease of implementation and efficiency in computer's binary arithmetic. These types of recurrences work mostly based on a \emph{linear feedback shift register} (LFSR). Introduced by Tausworthe \cite{Tausworthe}, this scheme has instigated many researchers to implement their PRNGs based on its variants. For example, the celebrated PRNG \emph{Mersenne Twister} \cite{Matsumoto:1998:MTE:272991.272995} is implemented using a variation of this technology.

Another type of research on random number generators also exists, where the target is to exploit the intricate chaotic behavior originated by simple functions with local interaction to develop the random numbers. This research was initiated by Wolfram \cite{Wolfram85c}, where he used a cellular automaton (CA) as the source of psudo-randomness. Therefore, we can classify the PRNGs in three main categories -- 1) LCG based, 2) LFSR based and 3) CAs based.

\subsection{LCG based PRNGs}\label{sec:lcg}
\noindent One of the most popular random number generation technique is based on linear recursions on modular arithmetic. These generators are specialization on the linear congruential sequences, represented by 
\begin{equation}
\label{lcg-form}
x_{n+1}=(ax_n +c) \pmod{m}, ~n \geq 0
\end{equation}
Here $m>0$ is the modulus, $a$ is the multiplier, $c$ is the increment and $x_0$ is the starting value or seed;  $0 \leq a < m$, $0 \le c < m$, $0\le x_0 < m$, that is, $a,c, x_0 \in \mathbb{Z}_m$ \cite{Knuth2}. The sequence $(x_i)_{i\ge 0}$ is considered as the desired sequence, and the output is $u_i = \frac{x_i}{m}$, if anybody wants to see the numbers from $[0,1)$. However, not all choices of $m,a,c,x_0$ generate a random sequence. {For example, if $a=c=1$, the sequence becomes $x_0+1, x_0+2, x_0+3, \cdots$, which, off course, is not random}. Similarly, if $a=0$, the case is even worse. Therefore, selection of these magic numbers is crucial for getting a random sequence of numbers.

We can observe that, maximum period possible for an LCG is $m$. However, to get a maximum-period LCG, the following conditions need to be satisfied \cite{Knuth2}: 
\begin{enumerate}
	\item $c$ is relatively prime to $m$;
	\item if $m $ is multiple of $4$, $a-1$ is also multiple of $4$;
	\item for every prime divisor $p$ of $m$, $a-1$ is multiple of $p$.
\end{enumerate}

A good maximum-period LCG is Knuth's LCG \verb|MMIX| \cite{Knuth2} where $a=636413622\\3846793005$, $m = 2^{64}$ and $c=1442695040888963407$. The PRNGs used in computer programming are mainly LCGs having maximum period. Some popular examples are \verb|rand| of GNU C Library \cite{GCC} where $a=1103515245$, $c= 12345$ and $m=2^{31}$, and \verb|lrand48| of same library where $a=25214903917$, $c= 11$ and $m=2^{48}$. Another well-known LCG is \verb|drand48| of GNU C Library which is similar to \verb|lrand48|, except it produces normalized numbers. Borland LCG is also a well-liked PRNG having $a=22695477$, $c=1$ and $m=2^{32}$.

Many variations of LCGs were proposed. For example, if we take the increment $c=0$, then the generator is called \emph{multiplicative ({\em or,} mixed) congruential generator} (MCG):
\begin{equation}\label{mcg_eq}
x_{n+1}=ax_{n} \pmod{m}, ~n \geq 0
\end{equation}
Although generation of numbers is slightly faster in this case, but the maximum period length of $m$ is not achievable. Because, here $x_n=0$ can never appear unless the sequence deteriorates to zero. When $x_n$ is relatively prime to $m$ for all $n$, the length of the period is limited to the number of integers between $0$ and $m$ that are relatively prime to $m$ \cite{Knuth2}. Now, if $m=p^e$, where $p$ is a prime number and $e\in \mathbb{N}$, Equation~\ref{mcg_eq} reduces to: \[x_n = a^nx_0 \pmod{p^e}\]
Taking $a$ as relatively prime to $p$, the period of the MCG is the smallest integer $\lambda$ such that, \[x_0 = a^{\lambda}x_0 \pmod{p^e}\]
Let $p^f$ be the gcd of $x_0$ and $p^e$, then this condition turns down to \[a^{\lambda} = 1 \pmod{p^{e-f}}\]
When $a$ is relatively prime to $m$, the smallest integer $\lambda$ for which $a^{\lambda} = 1 \pmod{p^{e-f}}$ is called the \emph{order of $a \text{ modulo }{m}$}. Any value of $a$ with maximum possible order modulo $m$ is called a \emph{primitive element modulo $m$}. Therefore, the maximum achievable period for MCGs is the order of a primitive element. It can be at maximum $m-1$ \cite{Knuth2}, when 
\begin{enumerate} [topsep=1pt,itemsep=1ex]
	\item $m$ is prime;
	\item $a$ is a primitive element modulo $m$;
	\item $x_0$ is relatively prime to $m$.
\end{enumerate}

Some MCGs with large period are reported in \cite{5388354,fishman1990multiplicative,doi:10.1137/0907002}.  C++11's \verb|minstd_rand| \cite{Park:1988:RNG:63039.63042,Park:1993}, a good PRNG, is an MCG where $a=48271$ and $m=2^{31}-1$. However, the MCGs perform unsatisfactorily in spectral tests \cite{Knuth2}. So, higher order linear recurrences are proposed of the form 
\begin{equation}\label{mrg-form}
x_{n} = a_1x_{n-1}+\cdots +a_kx_{n-k} \pmod{m}
\end{equation}
where $k\geq 1$ is the order.
Here, $x_0, \cdots, x_{k-1}$ are arbitrary but not all zero. For these recurrences, the best result can be derived when $m$ is a large prime. In this case, according to the theory of finite fields, multipliers $a_1,\cdots, a_k$ exist, such that, the sequence of Equation~\ref{mrg-form} has period of length $p^k-1$, if and only if the polynomial
\begin{equation}\label{mrg-polynomial}
P(z)=z^k-a_1z^{k-1}-\cdots - a_k
\end{equation}
is a \emph{primitive polynomial modulo $p$} \cite{Knuth2}. That is, if and only if, the root of $P(z)$ is a primitive element of the Galois field with $p^k$ elements\footnote{A nonzero polynomial $P(z)$ is \emph{irreducible} if it cannot be factored into two non-constant polynomials over the same field. The straightforward criterion for a polynomial $P(z)$ of degree $k$ over Galois Field $\mathbb{F}(m)$ to be irreducible is -- (1) it divides the polynomial $z^{m^k}-z$ and (2) for all divisors $d$ of $k$, $P(z)$ and $z^{m^d}-z$ are relatively prime. The polynomial $P(z)$ is \emph{primitive}, if it is irreducible and $\min\limits_{n\in \mathbb{N}}\{n ~|~ P(z) \text{ divides } z^n-1\}=m^k-1$. In this case, $P(z)$ has a root $\alpha$ in $\mathbb{F}(m^k)$ such that, $\{0,1,\alpha, \alpha^2, \cdots, \alpha^{m^k-2}\}$ is the entire field $\mathbb{F}(m^k)$.}. A generator with such recurrence is called \emph{multiple recursive generator} (MRG) \cite{LEcuyer1990}.

A variant of MRG is the additive \emph{lagged-Fibonacci} generators \cite{brent1994periods}, which take the following form:
\[x_{n} = (\pm x_{ n-r} \pm x_{ n-s}) \pmod{2^w}\] 
general form of which is a linear recurrence \[q_0x_n+q_1x_{n+1}+\cdots +q_rx_{n+r} = 0 \pmod{2^w}\] defined by a polynomial \[Q(t)= q_0 +q_1t+\cdots + q_rt^r\] with integer coefficients and degree $r>0$. Here, $w$ is an exponent, which may be chosen according to the word length of computer. The desired random sequence is $(x_i)_{i\ge 0}$ where $x_0,\cdots,x_{r-1}$ are initially given and not all are even. However, if $Q(t)=q_0 + q_st^s+q_rt^r$ is a primitive trinomial with $r>2$, and if $q_0$ and $q_r$  are chosen as odd, the sequence $(x_i)_{i\ge 0}$ attains the maximal period of length $2^{w-1}(2^r-1)$. The PRNG, proposed in \cite{brent1994periods}, uses this type of trinomials. Some extensions of lagged-Fibonacci generators are the PRNGs named \emph{add-with-carry} (AWC) and \emph{subtract-with-borrow} (SWB) generators \cite{10.2307.2959748}, \emph{Recurring-with-carry} generators \cite{10.2307.3215210}, \emph{multiply-with-carry} (MWC) generators \cite{couture1997distribution} etc. 

Another type of generators, named as \emph{inversive congruential generators} (ICGs) are proposed in \cite{Eichenauer1986,EICHENAUERHERRMANN1992345,eichenauer1993statistical}. These generators are defined by the recursion \[x_{n+1}=a{x_{n}}^{-1} + c \pmod{p},~~ n \geq 0\]
where $p$ is a large prime, $x_n$ ranges over the set $\{0,1,\cdots,p-1,\infty\}$ and the $x_n^{-1}$ is the inverse of $x_n$, defined as:  $0^{-1}=\infty$, $\infty^{-1}=0$, otherwise $x^{-1}x \equiv 1 \text{ (modulo }p)$. For the purpose of implementation, one can consider $0^{-1}=0$, as 0 is always followed by $\infty$ and then by $c$ in the sequence. Here, for many choices of $a$ and $c$, maximum period length $p+1$ is attainable \cite{Knuth2}. 

To improve the randomness of an LCG, several techniques have been proposed. One important class of PRNGs exists which deals it by combining more than one LCG, see for example \cite{L'Ecuyer:1988:EPC:62959.62969,10.2307.2347988,doi:10.1287.opre.44.5.816,doi:10.1287.opre.47.1.159}. Several combining techniques are suggested in the literature, like addition using integer arithmetic \cite{10.2307.2347988,L'Ecuyer:1988:EPC:62959.62969}, shuffling \cite{nance1978some}, bitwise addition modulo $2$ \cite{bratleyguide} etc. 

In \cite{doi:10.1287.opre.44.5.816}, it is shown that, we can get an MRG equivalent (or approximately equivalent) to the combined generator of two or more component MRGs, where the equivalent MRG has modulus equal to the product of the individual moduli of the component MRGs.
Let us consider $J\geq 2$ component MRGs where each MRG satisfies the recurrence
\begin{equation}\label{cmrg-form}
x_{j,n} = a_{j,1}x_{j,n-1}+\cdots + a_{j,k}x_{j,n-k} \pmod{m_j} \text{,~~~~    $1\leq j \leq J$,}
\end{equation}
having order $k_j$ and coefficients $a_{j,i}$ ($1\le i \le k$). Here, the moduli $m_j$s are pairwise relatively prime and each MRG has period $\rho_j={m_j^{k_j}}-1$.
Now, two combined generators can be defined \cite{doi:10.1287.opre.44.5.816}:
\begin{equation}\label{cmrg_2}
w_{n} = (\sum_{j=1}^{J} \frac{\delta_jx_{j,n}}{m_j}) \pmod{1} 
\end{equation}
\begin{equation}\label{cmrg_1}
z_{n} = (\sum_{j=1}^{J} \delta_jx_{j,n}) \pmod{m_1} \text{;~~~~  $\tilde{u}_n = \frac{z_n}{m_1}$,}
\end{equation}
Here, $\delta_1, \delta_2, \cdots$ are arbitrary integers such that each $\delta_j$ is relatively prime to $m_j$. The MRGs of Equation~\ref{cmrg_2} and \ref{cmrg_1} are approximately equivalent to each other. Further, it can be shown that, the MRGs of equations \ref{cmrg-form} and \ref{cmrg_2} are equivalent to an MRG of Equation~\ref{mrg-form} with $m=\prod\limits_{j=1}^{J} m_j$ and period length = $lcm(\rho_1,\cdots, \rho_J)$.
A well-known example of combined MRG is \verb|MRG31k3p| \cite{L'Ecuyer:2000:FCM:510378.510476} where two component MRGs of order $3$ are used having the following parameters:  
\[m_1=2^{31}- 1,~ a_{1,1}=0,~ a_{1,2}=2^{22},~ a_{1,3}=2^7+1\]
\[m_2=2^{31}-21069,~ a_{2,1}=2^{15},~ a_{2,2}=0,~ a_{2,3}=2^{15}+1\]
Here, for ease of implementation, each component MRG has two non-zero coefficients of the form $2^q$ and $2^q+1$. The combined MRG follows Equation~\ref{cmrg_1} and its period length is approximately $2^{185}$.

Another technique of improving randomness quality of a generator is to use a randomized algorithm over the outputs of a single LCG. This randomized algorithm is an efficient permutation function or hash function in \cite{o1988pcg}, which introduces the family of generators as \emph{permuted congruential generator} or \emph{PCG}. Here, several operations are performed on the outputs of a fast LCG, like random shifts to drop bits, random rotation of bits, bitwise exclusive-or(XOR)-shift and modular multiplication to perturb the lattice structure inherent to LCGs and improve its randomness quality. A good implementation is \verb|PCG-32|, which produces $32$-bit output and has a period length $2^{64}$. Here, the multiplier is $6364136223846793005$ and increment is taken as $1$.


Sometimes, LCG can be written in a matrix form as 
\begin{equation}\label{LCG_matrix}
\mathbf{X_n}=\mathbf{AX_{n-1}+C}\pmod{m}
\end{equation}
Here, $S=\{\mathbf{X}=(x_1,\cdots,x_k)^T ~|~ 0 \leq x_0,\cdots,x_k < m\}$ is the set of $k$-dimensional vectors with elements in $F=\{0,1,\cdots,m-1\}$, $\mathbf{A}=(a_{ij})$ is a $k \times k$ matrix with elements in $F $, $\mathbf{C} \in S$ is a constant vector and $\mathbf{X_0}$ is the seed \cite{LEcuyer1990}. If $k=1$,  the recurrence of Equation~\ref{LCG_matrix} reduces to Equation~\ref{lcg-form}. When $\mathbf{C}=0$, the generator is an MCG:
\begin{equation}\label{MCG_matrix}
\mathbf{X_n}=\mathbf{AX_{n-1}}\pmod{m}
\end{equation}
This form is useful because of its jumping-ahead property. Even for a large $v$, $\mathbf{X}_{i+v}$ can be reached from  $\mathbf{X}_{i}$, by first computing $\mathbf{A}^v \pmod{m}$ in $\mathscr{O}(\log{ v})$ time and applying a matrix-vector multiplication $\mathbf{X}_{i+v} = (\mathbf{A}^v \pmod{m})\mathbf{X}_i \pmod{ m}$ \cite{LEcuyer1990}.
Moreover, using this matrix, any LCG of order $k$ can be expressed by an MCG of oder $k+1$: modify $\mathbf{A}$ to add $\mathbf{C}$ as its $(k+1)^{th}$ column and a $(k+1)^{th}$ row containing all $0$s except $1$ in $(k+1)^{th}$ position; modify $\mathbf{X_n}$ to add $1$ as its $(k+1)^{th}$ component. When $m$ is prime and $\mathbf{C}=0$, $F$ and $S$ are equivalent to $\mathbb{F}(m)$ and $\mathbb{F}(m^k)$, where $\mathbb{F}(m^k)$ is the Galois field with $m^k$ elements. In this case, the MCGs have maximal possible period = $m^k-1$ if and only if the characteristic polynomial of $\mathbf{A}$,   
\begin{equation}\label{charac_poly}
f(x)=|xI-\mathbf{A}|\pmod{m} = (x^k - \sum_{i=1}^{k}a_ix^{k-i})\pmod{m}
\end{equation}
with coefficients $a_i$ in $\mathbb{F}(m)$ is a primitive modulo $m$. For attaining this period, $\mathbf{A}$ must be nonsingular in modulo $m$ arithmetic. Nevertheless, a polynomial of Equation~\ref{charac_poly} has a companion matrix $\mathbf{A}$:
\begin{equation}\label{a_matrix}
\mathbf{A}={\begin{bmatrix}
	0 & 1 &  \cdots & 0\\
	\vdots  & \vdots & \ddots & \vdots \\
	0 & 0  & \cdots & 1 \\
	a_k & a_{k-1} & \cdots & a_1
	\end{bmatrix}}
\end{equation}
In this case, by taking $\mathbf{X_n}=(x_n,\cdots,x_{n-k+1})^T$, MCG of Equation~\ref{MCG_matrix} is converted to recurrence of MRG (Equation~\ref{mrg-form}), where $\mathbf{X_n}$ obeys the recursion: \[\mathbf{X}_{n} = a_1\mathbf{X}_{n-1}+\cdots a_k\mathbf{X}_{n-k} \pmod{m}\]

 \subsection{LFSR based PRNGs}\label{sec:lfsr} 
\begin{figure}[hbtp]
\centering
\includegraphics[width=3.5in, height = 0.5in]{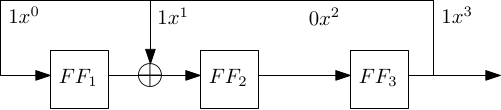}
\caption{A schematic diagram of $3$-bit LFSR with characteristic polynomial $P(x)=1+x+x^3$}
\label{fig:lfsr_schema}
\vspace{-1.0em}
\end{figure}

\noindent If modulus of the linear recurrence (Equation~\ref{lcg-form}) $m=2$ and $c=0$, the linear recurrence is based on the Galois field $\mathbb{F}(2)$. These recurrences can be implemented on a linear feedback shift register (LFSR). A LFSR is a shift register where the output of some bit positions are XOR-ed and feed as input to the register. This feedback connection ensures that the register cycles endlessly through repetitive sequences of values. To implement an LFSR in hardware, $k$ number of memory elements (flip-flops) are connected via XOR gates (see Figure~\ref{fig:lfsr_schema}). 
The positions of XOR in LFSR determines the characteristic polynomial of the LFSR, whereas the number of flip-flops ($k$) determines the degree of the polynomial. If flip-flop (FF) $i$ is associated with a feedback connection, coefficient of $x^i$ is $1$ for the characteristic polynomial $P(x)$. If this characteristic polynomial is primitive over $\mathbb{F}_2$, a $k$-bit LFSR can generate a maximal length sequence of period $2^k-1$, where $k$ is the degree of the polynomial. {The initial states of all the flip-flops together is the seed of the LFSR.} Likewise MCGs, seed of a LFSR, should always be a non-zero value, otherwise, the sequence degrades to zeros. 

As the generated numbers are binary, elementary bit string operations like rotation, shift, mask, exclusive-or etc. can be applied on them efficiently on a computer. The advantage of using this scheme is, LFSRs can be implemented on hardware; therefore, the generated circuits can be fast, cost-effective and efficient in terms of computational overhead. For many applications demanding PRNG, like VLSI testing, pattern recognition, computer simulation etc., efficient hardware implementation of the PRNG with very low overhead is a basic requirement. For this reason, most of today's research on PRNG is directed towards these genre of PRNGs. Many variations of this scheme are proposed, like Tausworthe generator, generalized linear feedback shift register (GFSR), twisted GFSR (TGFSR), Mersenne Twister, xorshift generators, WELL etc.
 
Tausworthe generator \cite{Tausworthe} is a linear recurrence of order $k>1$ like Equation~\ref{mrg-form} where $m=2$, defined by the recurrence
\begin{equation}\label{lfsr_form}
x_n = (a_1x_{n-1} + \cdots + a_kx_{n-k}) \pmod{2}
\end{equation}
Here, $a_k = 1$ and $a_1,a_2,\cdots, a_{k-1} \in \mathbb{F}_2$. This recurrence can be implemented on a linear feedback shift register (LFSR). However, the random number is represented by 
\begin{equation}
u_n = \sum_{l=1}^{L}x_{ns+l-1}2^{-l}
\end{equation}
which is a number with $L$ consecutive bit sequence of Recurrence \ref{lfsr_form}; with successive $u_n$s spaced $s$ bits apart \cite{Tausworthe}. Here $s$ and $L$ are positive integers.
This PRNG can have a maximal period $\rho = 2^k -1$, if and only if, the characteristic polynomial
\begin{equation}\label{poly}
P(z)= 1+a_1z + a_2z^2+\cdots + z^k
\end{equation}
is primitive over $\mathbb{F}(2)$ and $s$ is relatively prime to $2^k-1$. Then the generated sequence is called \emph{maximal-length linearly recurring sequence modulo $2$}. A popular example of this type of PRNG is \verb|random| of GNU C Library, which returns numbers between $0$ to $2147483647$ having period $ \rho \approx 16\times(2^{31}-1)$

Initially LFSR-based Tausworthe generators used primitive trinomials \cite{Tausworthe,Tootill_1973}. In \cite{Tootill:1971:RUP:321650.321655}, it is shown that, any Tausworthe generator that uses primitive trinomials of form 
\begin{equation}\label{trinomial}
P(z)=z^p+z^q+1 ~~(1 \leq q \leq (p-1)/2)
\end{equation}
as the characteristic polynomial can be represented by a simple linear recurrence in $\mathbb{F}(2^p)$. Moreover, likewise combined MRGs, \emph{combined} Tauseworthe generators have also been proposed \cite{l1996maximally}. 
It consists of $J\geq 2$ Tausworthe generators with primitive characteristic polynomials $P_j(z)$ of degree $k_j$ where $s_j$ is mutually prime to $2^{k_j}-1$, $1\leq j \leq J$. The sequence is denoted by $x_{j,n}$ (see Equation~\ref{cmrg-form} with modulus $2$) and random number by $u_{j,n} = \sum\limits_{l=1}^{L}x_{j,ns_j+l-1}2^{-l}$. $L$ is usually the word size of the computer. The output of the combined generator is 
\[u_n = (u_{1,n}\oplus u_{2,n}\oplus\cdots\oplus u_{J,n})\]
where $\oplus$ is the bitwise XOR operation. As discussed before in Page~\pageref{cmrg_1}, this generator has period $\rho = lcm(2^{k_1}-1,2^{k_2}-1,\cdots, 2^{k_J}-1)$, if the polynomials $P_j(z)$ are pairwise relatively prime, that is, every pair of polynomials have no common factor. \verb|Tauss88| \cite{l1996maximally} is such a generator where three component PRNGs are used with order $k_1 = 31$, $k_2 = 29$, and $k_3 = 28$ respectively. This PRNG has period length $\rho = (2^{31} -1)(2^{29}-1)(2^{28}-1) \approx 2^{88}$ and 
returns either $32$-bit unsigned integer or its normalized version. The $C$ code for this PRNG is accessible from $https://github.com/LuaDist/gsl/blob/master/rng/taus.c$. There are two other good combined Tausworthe generators, named \verb|LFSR113| and \verb|LFSR258.| For \verb|LFSR113|, number of component PRNGs $J = 4$ with period length $\rho \approx 2^{113}$. However, for \verb|LFSR258|, $J = 5$ and period length $\rho \approx 2^{258}$. Both these PRNGs return $64$ bit normalized numbers; for \verb|LFSR113| the numbers are normalized by multiplying the unsigned long integer output of the LFSR with $2.3283064365387\times 10^{-10}$ and for \verb|LFSR258|, the same is done by multiplying the unsigned $64$-bit output of the LFSR with $5.421010862427522170037264\times 10^{-20}$. $C$ codes for these two PRNGs are available at \cite{RNG}.

A generalization of LFSR (GFSR) has been proposed in \cite{Lewis:1973:GFS:321765.321777} to improve the quality of the PRNGs.
A GFSR sequence can be represented in binary as 
\begin{equation}
{X}_n = x_{j_1+n-1}x_{j_2+n-1}\cdots x_{j_k+n-1}
\end{equation}
where $X_n$ is a sequence of $k$-bit integers and $x_i$ is a LFSR sequence of Equation~\ref{lfsr_form}. 
%
However, this generator fails to reach its theoretical upper bound on period (equal to number of possible states) and has large memory requirement. So, another variation, named twisted GFSR (TGFSR), was proposed in \cite{matsumoto1992twisted,Matsumoto:1994:TGG:189443.189445}. This generator is same as GFSR, but, its linear recurrence is 
\begin{equation}
\mathbf{X}_{l+n} = \mathbf{X}_{l+m} \oplus \mathbf{X}_l \mathbf{A},~~ (l=0,1,\cdots)
\end{equation}
where $\mathbf{A}$ is a $w \times w$ matrix over $\mathbb{F}(2)$, $n,m,w$ are positive integers with $n>m$ and $\mathbf{X}_i$s are vectors in $\mathbb{F}(2^w)$. Usually, matrix $\mathbf{A}$ is chosen as Equation~\ref{a_matrix}. The seed is the tuple $(\mathbf{X}_0,\mathbf{X}_1,\cdots,\mathbf{X}_{n-1})$ with at least one non-zero value.

Likewise LCGs, all LFSR based generators can be represented in the following matrix form: 
\begin{equation}\label{lfsr1}
\mathbf{X_n} = \mathbf{AX}_{n-1}
\end{equation}
\begin{equation}\label{lfsr2}
\mathbf{Y_n}=\mathbf{BX}_n
\end{equation}
\begin{equation}
u_n = \sum_{l=1}^{w}y_{n,l-1}2^{-l}
\end{equation}
Here, $k,w > 0$, $\mathbf{A}$ is a $k \times k$ matrix, called transition matrix, $\mathbf{B}$ is a $w \times k$ matrix, called output transformation matrix and elements of $\mathbf{A},\mathbf{B}$ are in $\mathbb{F}_2$. The $k$-bit state vector at step $n$ is $\mathbf{X}_n = (x_{n,0} , \cdots, x_{n,k-1} )^T$,  the $w$-bit output vector is $\mathbf{Y}_n = (y_{n,0} , \cdots, y_{n,k-1} )^T$ and output at step $i$ is $u_n \in [0, 1)$. All the operations in equations \ref{lfsr1} and \ref{lfsr2} are modulo $2$ operations. The characteristic polynomial of matrix $\mathbf{A}$ is same as Equation~\ref{charac_poly} with modulus $2$:
\begin{equation}
P(z)=det(z\mathbf{I}-\mathbf{A})=(z^k - \sum_{i=1}^{k}a_iz^{k-i})
\end{equation}
where $a_j \in \mathbb{F}_2$ and $\mathbf{I}$ is the identity matrix. If $a_k = 1$, this recurrence is purely periodic (see Page~\pageref{purelyperiodic}) with order $k$. The period of $\mathbf{X}_n$ is maximal, that is, $2^k - 1$, if and only if, $P (z)$ is a primitive polynomial in $\mathbb{F}_2$. In this way, these PRNGs can be portrayed as LCGs in polynomials over $\mathbb{F}_2$.

\begin{figure}[hbtp]
	\centering
	\includegraphics[width=3.5in, height = 2.5in]{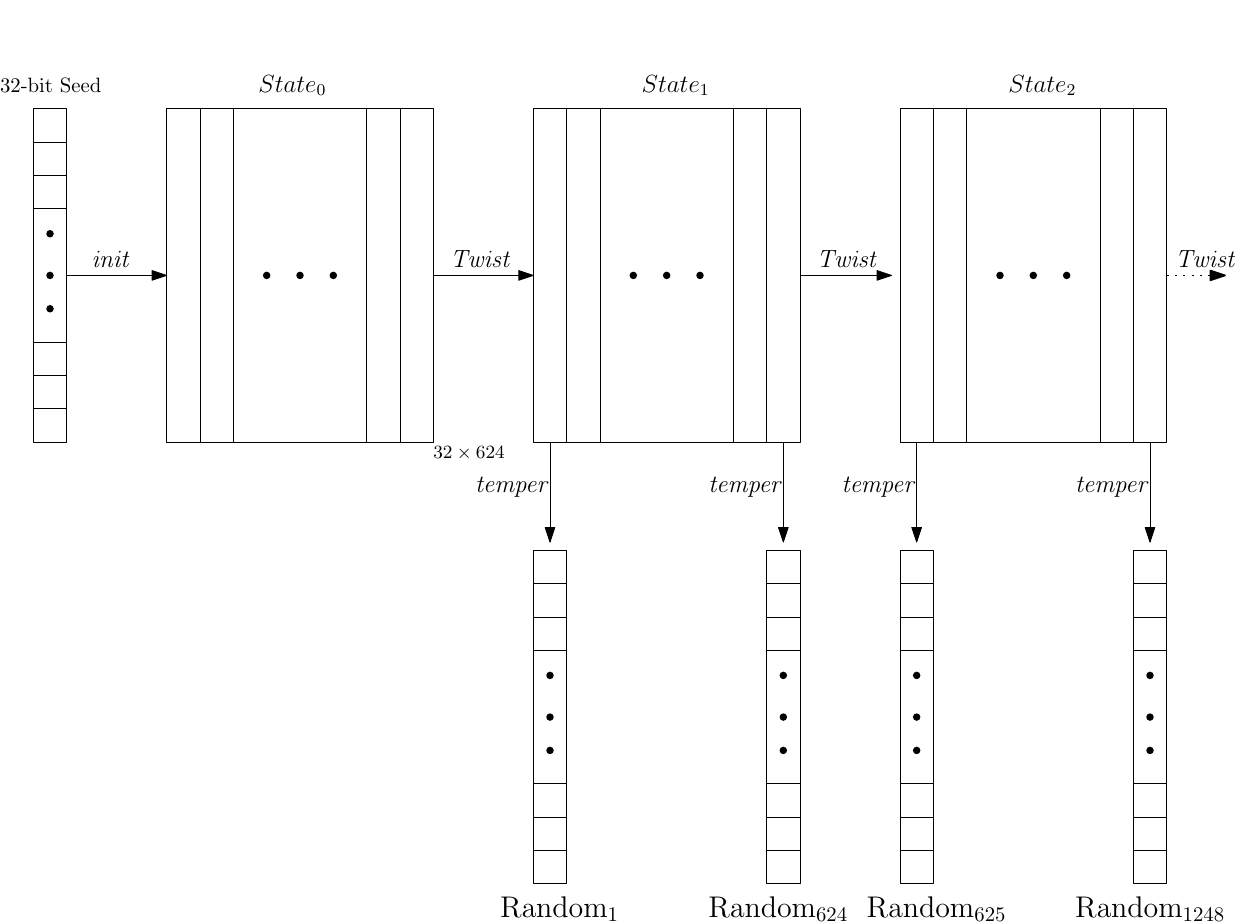}
	\caption{Working Principle of Mersenne Twister}
	\label{fig:MT}
	\vspace{-1.0em}
\end{figure}

The matrix $\mathbf{B}$ is usually used for tempering to improve \emph{equidistribution}\footnote{A sequence of real numbers $\{x_n\}$ is equidistributed on an interval $[a,b]$ if the probability of finding $x_n$ in any subinterval is proportional to the subinterval length. Equidistribution of $t$-dimensional unit hypercube into $2^{kl}$ cubic cells is defined as the set of all cubic cells in $[0,1)^k$ with side length $2^{-l}$ having coordinates of the corners as multiples of $2^{-l}$ \cite{Tausworthe}.} property of the PRNG by elementary bitwise transformation operations, like XOR, AND and shift \cite{Matsumoto:1994:TGG:189443.189445}. A TGFSR with tempering operations is called \emph{tempered} TGFSR. The well-known PRNG \emph{Mersenne Twister} (MT) is a variation of TGFSR where the linear recurrence is \cite{Matsumoto:1998:MTE:272991.272995}:
\begin{equation}\label{MT_rec}
\mathbf{X}_{k+n} = \mathbf{X}_{k+m} \oplus (\mathbf{X}^u_{k} | \mathbf{X}^l_{k+1})\mathbf{A}
\end{equation}
Here, $r,m,w$ are positive integers with $0\le r \le w-1$, $m$ ($1 \le m \le n$) is middle term and $r $ is separation point of one word. $\mathbf{A}$ is a $w \times w$ matrix (like Equation~\ref{a_matrix}) with entries in $\mathbb{F}(2)$, $|$ denotes bit vector wise concatenation operation, $\mathbf{X}_{k}^{u}$ is the upper $ w-r$ bits of $\mathbf{X}_{k}$, and $\mathbf{X}_{k+1}^{l}$ is the lower $r$ bits of $\mathbf{X}_{k+1}$. 
If $r=0$, this recurrence reduces to TGFSR and if $r=0$ and $\mathbf{A}=\mathbf{I}$, it reduces to GFSR \cite{Matsumoto:1998:MTE:272991.272995}.
Tampering is done by the following transformations in succession:
\begin{equation}\nonumber
\mathbf{y} = \mathbf{x} \oplus (\mathbf{x} >> u)
\end{equation}
\begin{equation}\nonumber
\mathbf{y} = \mathbf{y} \oplus ((\mathbf{y} << s) \text{ AND } \mathbf{b})
\end{equation}
\begin{equation}\nonumber
\mathbf{y} = \mathbf{y} \oplus ((\mathbf{y} << t) \text{ AND } \mathbf{c})
\end{equation}
\begin{equation}\nonumber
\mathbf{z} = \mathbf{y} \oplus (\mathbf{y} >> l)
\end{equation}
where $l,s,t,u$ are integers called tempering parameters, $\mathbf{b}$ and $\mathbf{c}$ are suitable bitmasks of size $w$ and $\mathbf{z}$ is the returned vector. 
The state transition is directed by a linear transformation 
\begin{equation}\nonumber
\mathbf{B} = 
{\begin{bmatrix}
	0 & \mathbf{I}_w & 0 & 0 & \cdots&\cdots &\cdots \\
	0 & 0 & \mathbf{I}_w & 0 &\cdots & \cdots & \cdots\\
	\vdots&  \ddots&  & &\\
	0 &  &\ddots & & \\
	\mathbf{I}_w &  & &\ddots & \\
	0 & &  & & \ddots\\
	\vdots&  &  & &  & \ddots\\
	0 & \cdots &\cdots &\cdots & 0 & \mathbf{I}_w & 0 \\
	0 & \cdots& \cdots& \cdots& 0 & 0 & \mathbf{I}_{w-r}\\
	\mathbf{S} & \cdots &\cdots & \cdots& 0 & 0 & 0\\
	\end{bmatrix}}
\end{equation}
on an array of size $p=nw-r$ (or an $(n\times w-r)$ array with $r$ bits missing at the upper right corner).
Here, $\mathbf{I}_j$ is a $j\times j$ identity matrix, $\mathbf{0}$ is the zero matrix and 
\begin{equation}
\mathbf{S} = 
{\begin{bmatrix}
	\mathbf{0} & \mathbf{I}_r\\
	\mathbf{I}_{w-r} & \mathbf{0}\\
	\end{bmatrix}}\mathbf{A}
\end{equation}
The generated numbers are integers between $0$ and $2^w-1$ provided $p$ is chosen as a Mersenne exponent such that, the characteristic polynomial of $\mathbf{B}$ is primitive and period is a Mersenne prime $2^p-1 = 2^{nw-r}-1$.
A commonly used example of Mersenne Twister is \verb|MT19937| \cite{Matsumoto:1998:MTE:272991.272995} having a period $\rho = 2^{19937}-1$. There are $32$-bit and $64$-bit word size variations of it. The working principle of Mersenne Twister is represented in Figure~\ref{fig:MT}. Here, a $32$-bit integer is taken as seed which is used to initialize the starting state of the $19937$ length TGFSR. Every time \emph{twist} function is used to generate the next state of the PRNG (Note that, $32\times 624 = 19936$). But, instead of directly producing each state as output, each of the internal states of Mersenne Twister is divided into 624 $32$-bit words where each word (integer) is tempered using the \emph{temper} function before getting produced as output. So, after each application of \emph{twist}, the current state of the TGFSR is used to generate 624 $32$-bit integers consecutively. Once these 624 numbers are exhausted, then only again the state of the TGFSR is updated using the twist function. 

In case of \verb|MT19937-32| ($32$-bit), the associated parameters are $(w,n,m,r)=(32,624,397,31)$, $\mathbf{a}$=0x9908B0DF, $u=11$, $s=7$, $\mathbf{b}$=0x9D2C5680, $t=15$, $\mathbf{c}$=0xEFC60000, $l=18$ and number of terms in the characteristics polynomial is $135$. However, for \verb|MT19937-64| ($64$-bit), $(w, n, m, r) = (64, 312, 156, 31)$, $\mathbf{a}$ = 0xB5026F5AA96619E916, $u=29$, $s = 17$, $\mathbf{b}$ = 0x71D67FFFEDA6000016, $t=37$, $\mathbf{c}$ = 0xFFF7EEE00000000016 and $l = 43$. These parameters ensure that the period is the Mersenne prime $2^{19937}-1$ which is longer than any other random number generator proposed before and is one of the reasons for the popularity of Mersenne Twister.

In \cite{Saito2008}, a variation of Mersenne Twister named single instruction multiple data (SIMD)-oriented Fast Mersenne Twister (SFMT) is proposed. It uses all features of MT along with multi-stage pipelines and Single Instruction Multiple Data (SIMD) (like 128-bit integer) operations of today's computer system. A popular implementation is \verb|SFMT19937| which has same period as \verb|MT19937|. It can generate both $32$-bit and $64$-bit unsigned integer numbers.
Further, there is a variation of SFMT specialized in producing double precision floating point numbers in IEEE 754 format. This PRNG is, in fact, named as double precision floating point SFMT (dSFMT) \cite{Saito2009}. Two versions of it are available - \verb|dSFMT-32| and \verb|dSFMT-52| \cite{sfmt}. For \verb|dSFMT-52|, the output of the PRNG is a sequence of $52$-bit pseudo-random patterns along with $12$ MSBs (sign and exponent) as constant. Here, instead of linear transition in $\mathbb{F}_2$, an affine transition function is adopted which keeps the constant part as 0x3FF.

Another PRNG, named \emph{well-equidistributed long-period linear} generator or WELL is also based on tampered TGFSR \cite{Panneton:2006:ILG:1132973.1132974}. For this PRNG, the characteristic polynomial of matrix $\mathbf{A}$ has degree $k = rw - j$, where $r,j$ are unique integers such that $r > 0$ and $0 \le j < w$, and it is primitive over $\mathbb{F}_2$. Two such good generators are \verb|WELL512a| and \verb|WELL1024a| \cite{RNG}. In case of \verb|WELL512a|, the parameters are $k = 512$, $w = 32$, $n = 16$ and $r = 0$; so expected period is $\rho = 2^{512}-1$. However, for \verb|WELL1024a|, the parameters are $k = 1024, w = 32, n = 32$ and $r = 0$ with period length $\rho = 2^{1024}-1$. The return values for these PRNGs are $32$-bit numbers normalized by multiplying with $2.32830643653869628906 \times 10^{-10}$. WELL generators follows the general equations \ref{lfsr1} and \ref{lfsr2}.

In \cite{marsaglia2003xorshift}, Marsagila has proposed a very fast PRNG, named \emph{xorshift} generator. The basic concept of such generators is $-$ to get a random number, first shift $a$ positions of a block of bits and then apply XOR on the original block with this shifted block. In general, a xorshift generator has the following recurrence relation \cite{brent2004note,Panneton:2005:XRN:1113316.1113319}:
\begin{equation}\label{xorshift_form}
\mathbf{v}_n = \sum_{j=1}^{t}\mathbf{\tilde{A}_j}\mathbf{v}_{n-m_j} \pmod{2} 
\end{equation}
where $t,m_j >0$, for each $n$, $\mathbf{v}_n$ is a $w$-bit vector and $\mathbf{\tilde{A}}_j$ is either $\mathbf{I}$ or product of $v_j$ xorshift matrices for $v_j\ge 0$. At step $n$, the state of the PRNG is $\mathbf{x}_n = (\mathbf{v}_{n-r+1}^T,\cdots, \mathbf{v}_{n}^T)^T$ where $\mathbf{v}_n = (v_{n,0},\cdots, v_{n,w-1})^T$ and output is $u_n=\sum_{l=1}^{w}v_{n,l-1}2^{-l}$ where $r = \max\lim_{1\le j \le t} m_j$. This generator converts into the general LFSR PRNG of equations \ref{lfsr1} and \ref{lfsr2}, if 
\begin{equation}
\mathbf{A} = 
{\begin{bmatrix}
	0 & \mathbf{I} & \cdots & 0\\
	\vdots &  & \ddots & \vdots \\
	0 & 0 & \cdots & \mathbf{I} \\
	\mathbf{A}_r & \mathbf{A}_{r-1} &  \cdots & \mathbf{A}_1
	\end{bmatrix}}
\end{equation}
where $k=rw$, $\mathbf{y}_n = \mathbf{v}_n$ and $\mathbf{B}$ matrix has $\mathbf{I}$ matrix of size $w \times w$ in upper left corner with zeros elsewhere. Matrix $\mathbf{A}$ has characteristic polynomial of the form \[P(z)=det(z^r\mathbf{I} + \sum_{j=1}^{r}z^{r-j}\mathbf{A}_j)\]
Therefore, the generator has maximal period length of $2^{rw}-1$, if and only if, this polynomial $P(z)$ is primitive. Marsagila's \verb|xorshift32| generator \cite{marsaglia2003xorshift} uses $3$ xorshift operations $-$ first XOR with left shift of $13$ bits, then with right shift of $17$ bits and finally again XOR with left shift of $15$bits. Here the returned number is a $32$ bit unsigned integer. There are three other good xorshift generators -- \verb|xorshift64*|, \verb|xorshift1024*M_8| and \verb|xorshift128+.| In \verb|xorshift64*| generator, the returned number is the current state perturbed by a non-linear operation, which is multiplication by $2685821657736338717$ \cite{Vigna:2016:EEM:2956571.2845077}. Here also $3$ xorshifts are performed -- left with $12$ bits, right with $25$ bits and again left with $27$ bits. However, in \verb|xorshift1024*M_8|, the multiplier is $1181783497276652981$ and shift parameters are $31, 11$, and $30$. In both cases, the generated numbers are $64$-bit unsigned integers. In 
\verb|xorshift128+|, the outputs are $64$-bits and two previous output states are added to get the result \cite{VIGNA2017175}. 

Although most of the above mentioned generators are linear, many researchers have developed LFSR based PRNGs by combining these with some non-linear operations \cite{l2003combined,Vigna:2016:EEM:2956571.2845077,L'Ecuyer:2007:TCL:1268776.1268777} to scramble the regularity of linear recurrence. For example, in \cite{l2003combined}, two component combined generators are proposed, where the major component is linear (LFSR or LCG), but the second component is distinct (nonlinear or linear). Whereas, in \cite{Vigna:2016:EEM:2956571.2845077}, to remove the flaws of xorshift generators, a non-linear operation is applied to scramble the results. The \emph{xorshift*} PRNGs, discussed above, are a well-known example.

\subsection{Cellular Automata based PRNGs}
\label{sec:cas}
\noindent A cellular automaton (CA) is a discrete dynamical system comprising of a regular network of cells, where each cell is a finite state automaton\footnote{A CA is identified by a quadruple ($\mathscr{L},\mathcal{S},\mathcal{N},\mathcal{R}$), where $\mathscr{L} \subseteq \mathbb{Z}^{D}$ is the $D$-dimensional cellular space, $\mathcal{S}$ is the finite set of states which a cell can take, $\mathcal{N} = (\overrightarrow{v_1}, \overrightarrow{v_2}, \cdots, \overrightarrow{v_m})$ identifies $m$ neighbors of each cell and $\mathcal{R}:\mathcal{S}^{m}\rightarrow \mathcal{S}$ is the rule of the automaton. For more details on CAs, interested reader may see \cite{Kari05, bhattacharjee2020survey}}. During evolution, a cell of a CA updates its state depending on the present states of its \emph{neighbors} following a {\em next state function}, also known as {\em local rule}, or simply \emph{rule}, whose arguments are the present states of the cell's neighbors. Each  of these combination of neighborhoods to $R$ is also named as \emph{Rule Min Term} or \emph{RMT}, usually represented by its decimal equivalent. For example, Table~\ref{rule} shows a rule for $2$-state $3$-neighborhood $1$-D CAs, where each of these neighborhood combinations $000$ to $111$ is an RMT. Hence, any rule can be represented by the string corresponding to the next state values of the RMTs.
Collection of the states of all cells at a given time is called {\em configuration} of the CA. So, during evolution a CA hops from one configuration to another. 


In \cite{Wolfram85c}, CAs were introduced as a source of pseudo-randomness. Here, an infinite array of cells is considered, where each cell can take either of the states $0$ or $1$ depending on the present states of its left neighbor, itself and its right neighbor. Such a CA is an example of $1$-dimensional $3$-neighborhood $2$-state CAs, popularly known as, \emph{elementary CAs} or \emph{ECAs}. For instance, the ECA rule chosen as source of randomness in \cite{Wolfram85c} is rule $30$\footnote{ Historically, the notion of representing these rules by the decimal equivalent of the binary string corresponding to the neighborhood combinations $\mathcal{R}(x,y,z)$, $x,y,z \in \{0,1\}$ was introduced in \cite{Wolfr83}, where $\mathcal{R}:\{0,1\}^3\rightarrow \{0,1\}$ is a rule.} (decimal of the string $00011110$) of Table~\ref{rule}. 
\begin{table}[h]
	\begin{center}
		\vspace{-1.0em}	
		\caption{Some rules of $1$-dimensional $3$-neighborhood $2$-state CAs.}
		\label{rule}
		\resizebox{0.85\textwidth}{!}{
			\begin{tabular}{cccccccccc}
				\toprule
				\thead{Neighborhood Combination} & \thead{111} & \thead{110} & \thead{101} & \thead{100} & \thead{011} & \thead{010} & \thead{001} & \thead{000} & \theadfont{Rule Number}\\ 
				(RMT) & (7) & (6) & (5) & (4) & (3) & (2) & (1) & (0) & \\
				\midrule
				& 0 & 0 & 0 & 1 & 1 & 1 & 1 & 0 & 30\\
				\thead{Next State}  & 0 & 0 & 1 & 0 & 1 & 1 & 0 & 1 & 45\\
				& 0& 1& 0& 1& 1& 0& 1 & 0 & 90\\
				& 1& 0& 0& 1& 0& 1& 1& 0 & 150\\
				\bottomrule
			\end{tabular}
		}
	\end{center}
	\vspace{-1.5em}
\end{table}
In this case, a random sequence is generated using the next state values of the single cell with initial state $1$ among all cells, initiated with state $0$.

\begin{figure}[hbtp]
	\vspace{-1.0em}
	\subfloat[Null boundary\label{Fig:NullB}]{%
		\includegraphics[scale = 0.35]{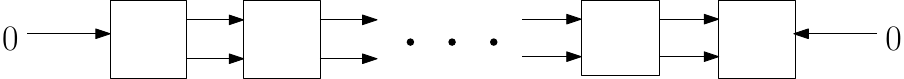}
	}
	\hfill
	\subfloat[Periodic boundary\label{Fig:PeriodicB}]{%
		\includegraphics[scale = 0.35]{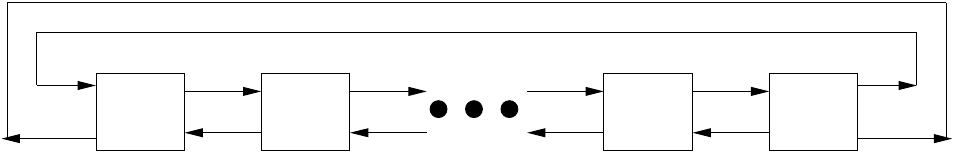}
	}
	\caption{Boundary conditions for 1-D finite CAs. Arrows pointing to a cell indicate the dependencies of the cell. Here all the CAs use 3-neighborhood dependency}
	\label{Fig:BoundaryC}
	\vspace{-0.5em}
\end{figure}

According to the original proposal, only a single bit is collected from each configuration of the CA where number of cells is infinite. However, to implement as PRNG, one needs to consider only a finite number of cells having boundaries. Two boundary conditions are usually considered -- (1) null boundary, where the boundary cells are connected to null or $0$ state (see Figure~\ref{Fig:NullB}), (2) periodic boundary, where the boundary cells are neighbors of each other (see Figure~\ref{Fig:PeriodicB}).
So, to use as a PRNG, one may abuse \cite{Wolfram85c} slightly by running the CA under periodic boundary condition, and collecting $32$ bits from $32$ consecutive configurations to report a single $32$-bit integer \footnote{We have taken the number of cells as $101$ and next states of the middle cells are collected to generate $32$-bit numbers.}.

\begin{figure}[hbtp]
	\centering
	\includegraphics[width=0.85\textwidth, height = 2.5cm]{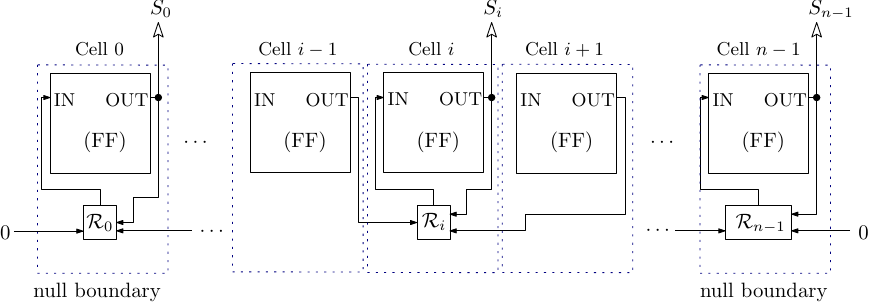}
	\caption{Hardware implementation of an n-cell ECA under null boundary condition}
	\label{fig:CA_hardware}
\end{figure}

The most exciting aspect of CAs is their complex global behavior, which is resulted from simple local interaction and computation and massive parallelism. Another important property of CAs is, like LFSR, CAs can be easily implemented in hardware -- each cell consists of a memory element to store its state and a combinational logic circuit to find the next state of the cell. Figure~\ref{fig:CA_hardware} represents hardware implementation of an ECA with $n$ cells under null boundary condition. These properties of CAs along with ease of scalability have made CAs, especially ECAs, an area of extensive research for applications like VLSI circuit testing \cite{Horte89a,Makato98,ppc1,tcad/DasS10,SukantaTH,Horte89c,card89}, Monte-Carlo simulations \cite{870571}, Field Programmable Gate Arrays (FPGAs) \cite{comer2012random,6195137}, cryptography \cite{Wolfr86b,machicao2012chaotic,ESLAMI20102889} etc. However, most of these works considers $1$-dimensional CAs where the cells follow $3$-neighborhood dependency. So, in this paper we have taken only such CAs based PRNGs for comparison with the other class of generators. 

\begin{figure}[hbtp]
	\centering
 \includegraphics[width=0.65\textwidth, height = 2.5cm]{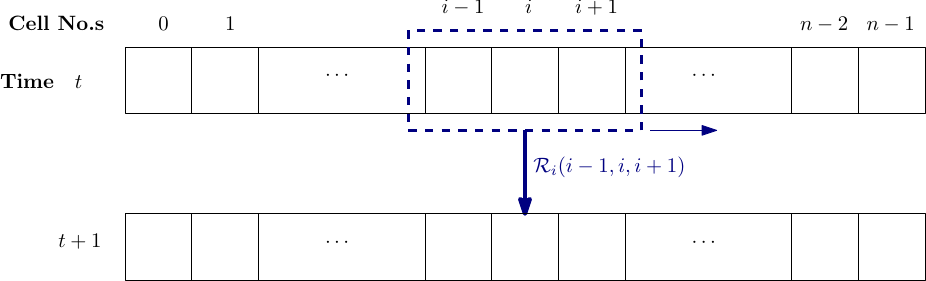}
 \caption{Next state calculation for an $n$-cell hybrid CA with $3$ neighborhood dependency}
 \label{fig:ca_evolution}
 \vspace{-1.0em}
\end{figure}

Randomness of a CA-based PRNG is, in general, effected by its transition rule, cell size, initial configuration (seed) and boundary condition. So, research on CA-based PRNGs have been to find the best possible result by varying these structures of the CAs. {The rule of the CA is chosen as \emph{autoplectic}\footnote{This term, first coined by Stephen Wolfram indicates intrinsic randomness where without any outside input, randomness can be generated inside the system by the system itself \cite{Wolframbook1}. In this case, even simple initial conditions can derive randomness.} such that feature extraction require more sophisticated computation than the simple run of the CA. For instance, ECA rule $30$ used in \cite{Wolfram85c} is a autoplectic rule.} However, to improve the cycle length of the CA and introduce more complexity in the system, non-uniformity in the local rule is instigated \cite{Pries86}. In this case, instead of using a single rule $\mathcal{R}$, a vector $\mathscr{R}=\langle \mathcal{R}_0, \mathcal{R}_1, \cdots, \mathcal{R}_{n-1}\rangle$, called \emph{rule vector}, is used, where $n$ is the number of cells and the $i^{th}$ cell uses rule $\mathcal{R}_i$. 
The next state calculation in this case, is governed by $\mathcal{R}_i$ for each $i$ (see, for example,  Figure~\ref{fig:ca_evolution}). 
Such a CA where different cells may take different rules is called a \emph{hybrid} or \emph{non-uniform} CA.  For instance, in \cite{Horte89a}, a rule vector $\mathscr{R}=\{30,45\}^{16}$ is used under periodic boundary condition to generate $32$-bit numbers from the whole configuration of the CA. Moreover, if $\mathcal{R}_0 = \mathcal{R}_1= \cdots= \mathcal{R}_{n-1}$, the CA is uniform CA. In Figure~\ref{fig:CA_hardware}, if combinational logic circuits for each cells are different, the ECA is a hybrid ECA.

\begin{figure}[hbtp]
	\vspace{-1.0em}
	\subfloat[\label{space0}]{%
		\includegraphics[width=0.45\textwidth, height = 1.2cm]{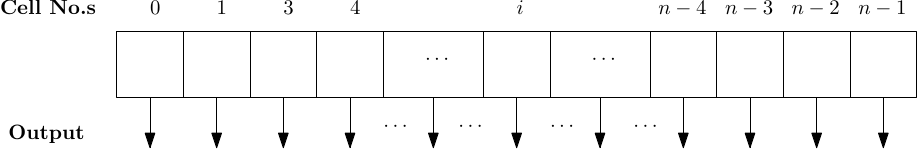}
	}
	\hfil
	\subfloat[\label{space1}]{%
		\includegraphics[width=0.45\textwidth, height = 1.2cm]{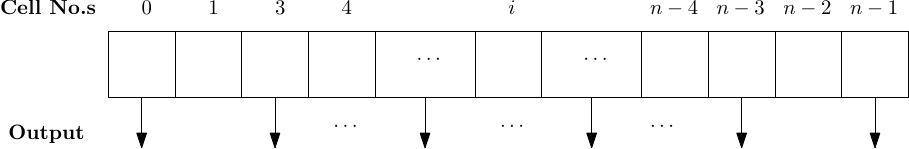}
	}
	\caption{Site Spacing for even cell length $n$. Here, $\gamma=0$ for Figure~\ref{space0} and $\gamma=1$ for Figure~\ref{space1}. The random numbers are collected from the cells with arrows.}
	\label{fig:sitespacing}
	\vspace{-0.5em}
\end{figure}

\begin{figure}[hbtp]
	\centering
	\includegraphics[width=3.0in, height = 1.5in]{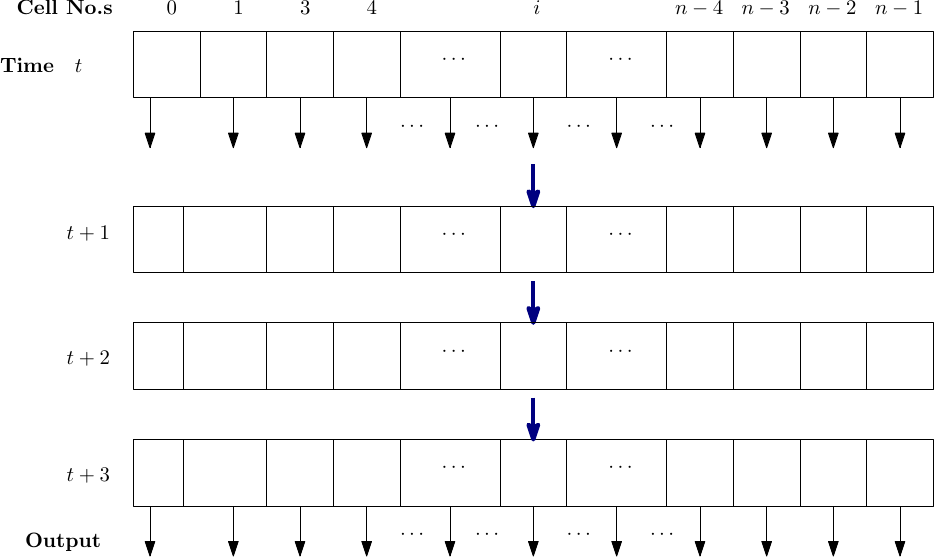}
	\caption{Time spacing with no site spacing. Here, $\alpha$ is taken as $2$.}        
	\label{fig:timespacing}
\end{figure}

Many other ways are also proposed to generate random numbers from the configuration of a CA. In \cite{Horte89a}, concept of site spacing (output number is collected from cells spaced by $\gamma$ distance) and time spacing (output numbers are taken $\alpha$ time steps apart) are introduced (see Figure~\ref{fig:sitespacing} and Figure~\ref{fig:timespacing}). If $\gamma=0$, the whole configuration of the CA is treated as a number.

However, to further improve the period of the PRNGs, a special type of non-uniform CAs are introduced where the rules are \emph{linear}\footnote{If all rules of the CA can be expressed by a linear function, that is, 
\[\forall i,~ \mathcal{R}_i (a_1, a_2, \cdots, a_N) = \sum\limits_{j=1}^N c_j.a_j\]
where $c_j$ $\in \mathcal{S}$ is a constant and $a_j$ is the state of the $j^{th}$ neighbor of cell $i$, 
then the CA is called a \emph{linear} CA. In this case, the set $\mathcal{S}$ forms a commutative ring with identity.}.
For example, the ECAs $90$ and $150$ are linear CAs as they can be represented as: 
\[90:  S_i(t+1)~=~S_{i-1}(t) \oplus S_{i+1}(t)\]
 \[150: S_i(t+1)~=~S_{i-1}(t) \oplus S_{i}(t) \oplus S_{i+1}(t)\]
where $S_i(t)$ is the state of $i^{th}$ cell at time $t$. Linear CAs can be easily characterized by using standard algebraic tools \cite{ppc1}. For instance, a binary $n$-cell linear CA can be represented by an $n \times n$ characteristics matrix ($T$) operating on $\mathbb{F}(2)$. In this matrix, the $i^{th}$ row represents the dependency of the $i^{th}$ cell to its neighbors. The characteristics matrix ($T$), in this case, is formed as:
\begin{small}
	\begin{equation}
  \begin{array}{l}
   \mbox{$T\left[ i, j\right] $ =}
    \left\{
	\begin{array}{ll}
	    \mbox{1  if the next state of the $i^{th}$ cell depends on the present state of the $j^{th}$ cell} \\
	    \mbox{0  otherwise}
	    \end{array}\right.
    \end{array}
 \end{equation}  
\end{small}
For example, the characteristics matrix of a $4$-cell hybrid CA with rule vector $\mathcal{R}=\langle 150, 150, 90, 150 \rangle$ under null boundary condition is: 
\[
   T=
\begin{bmatrix}
    1 & 1 & 0 & 0 \\
    1 & 1 & 1 & 0 \\
    0 & 1 & 0 & 1 \\
    0 & 0 & 1 & 1
\end{bmatrix}
\]
Here, as boundary condition is null, left neighbor of first cell and right neighbor of last cell are $0$. For any one-dimensional linear CA using $3$-neighborhood condition, this matrix is tridiagonal \cite{Serra90c}. Now, if the characteristic polynomial of this $T$ matrix is primitive over $\mathbb{F}_2$, the CA can generate maximal cycle length $2^n-1$. Such CAs are called \emph{maximal-length CAs} which has to be defined over null-boundary condition.

Several researches have been conducted to establish the isomorphism of a $1$-dimensional linear hybrid CA with its corresponding LFSR, where both have the same primitive characteristic polynomial \cite{Serra90c,114093}. It is shown that, for every irreducible polynomial, there are exactly two hybrid CAs with rules $90,150$ under null boundary condition; the construction process of such a CA is shown in \cite{cattell1996synthesis}. In 
\cite{Jetta95}, a list of maximal-length CAs for each degree from $1$ to $500$ is synthesized for the corresponding primitive polynomials given in \cite{Bardell,Bardell1992}. Needless to say, only specific combinations of the local rules $90$ and $150$ over null boundary condition can generate a maximal-length CA. Due to this maximal cycle length property, many researchers have used these CAs as their generators \cite{Horte89a,Horte89c,114093,COMPAGNER1987391,10.1007/11861201_25,122655}.
 For example, in \cite{Horte89a}, the maximal length CA with rule vector $\mathcal{R}=(90,150,90,90,90,150,150,90,90,90,90,90,150,90,90$,$150,150,90,150,150,150$,\\$90,150,150,150,150,90,150,90,150,90,150)$ is used. To implement this PRNG, we have considered two types of site spacing -- $1$ site spacing (that is, $\gamma=1$) and no site spacing ($\gamma=0$). For $1$ site spacing, two consecutive output sequences are concatenated to get one $32$-bit number, whereas, for no site spacing, the whole configuration of the CA at each time instant is treated as a $32$-bit number.

However, due to difficulty in finding the primitive polynomial required for a maximal-length CA, non-linear CAs were introduced as PRNGs \cite{tcad/DasS10,SukantaTH}. For example, in \cite{tcad/DasS10}, an algorithm is given to select a ($2$-state $3$-neighborhood) non-uniform non-linear CA as the random number generator. 
One such CA has rule vector $\mathscr{R}=\{5,105,90,90,165,150,90,105,150,105,90,165,150,150,165,90,165,\\90,165,150,150,90,165,105,90,165,150,90,105,150,165,90,105,105,90,150,90,\\90,165,150,150,105,90,165,20\}$. Each of its $45$-bit configuration is considered as a number.
Some other works of using hybrid CAs are \cite{Horte89a,Horte89c,alonso2009elementary}. In \cite{alonso2009elementary}, cells of the CAs were allowed to hold memory of their last two state values. Here, numbers were taken from overlapping window of size $50$ and the CAs are with rules $30$, $90$ or $150$.

\begin{figure}[hbtp]
 	\centering
 	\includegraphics[width=3.5in, height = 0.5in]{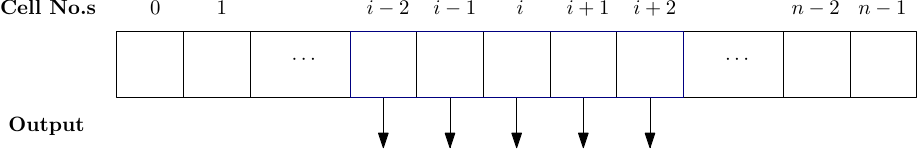}
 	\caption{Window of size $5$ taken from the middle cells.}        
 	\label{fig:window}
 \end{figure}

In some recent works, the numbers are generated by a small window ($w<n$) out of the $n$ cells using $3$-neighborhood $1$-D CAs having more than $2$-states per cell. For example, in \cite{IJMC2017}, a $3$-state CA with local rule $\mathcal{R}=1200211200210211200\\21021210$\footnote{Here, the rule signifies the string corresponding to all possible RMTs starting from $222$ up to $000$.} under periodic boundary condition is used. The base-$3$ numbers, observed through the window, are considered as pseudo-random numbers (Figure~\ref{fig:window}).
To generate ternary strings of length $20$ and $40$, the CA sizes can be taken as $51$ and $101$ with window length $w=20$ and $40$ respectively. When converted, these ternary numbers are equivalent to $32$-bit and $64$-bit numbers respectively.

In \cite{BHATTACHARJEE2019104878}, $1$-dimensional $3$-neighborhood decimal CAs are exploited as source of randomness. Here, it is argued that the CAs which are \emph{chaotic}\footnote{Chaos is defined over infinite lattice. However, while constructing a PRNG, we need a finite system.} while defined over infinite lattice, but have no \emph{self-similar} and \emph{self-organizing} patterns can be used as good source of randomness. Further, this paper also identifies a list of desirable properties for these finite CAs to be good candidates as PRNGs, such as, \emph{balancedness}, \emph{information flow}, \emph{asymmetric configuration space}, large cycle length etc. (see Ref~\cite{BHATTACHARJEE2019104878} for more details).
A heuristic synthesis algorithm is reported to generate candidate CAs following these properties. One can observe that, a rule of decimal CAs has $1000$ RMTs which may be cumbersome to represent using, say, format like Table~\ref{rule}. To simply the representation, the RMTs of a rule is divided into $100$ non-overlapping strings where each string contains $10$ consecutive RMTs. These strings are named as $Sibl_0$ to $Sibl_{99}$. For example, the RMTs $0 (000),1 (001),2 (002),3 (003),4 (004),5 (005),6 (006),7 (007),8 (008),9 (009)$ in the  order $9876543210$ forms $Sibl_0$. Similarly $Sibl_i$ contains the RMTs $\langle 10j+9, 10i+8, \cdots, 10j\rangle$. Therefore, a rule is a concatenation of the strings $Sibl_{99}~Sibl_{98}\\\cdots Sibl_0$.
A simple synthesis algorithm is also provided in \cite{BHATTACHARJEE2019104878} which generates the complete rule from a small fragment of 10 RMTs given as a permutation of $`0123456789$'; for the sake of completeness, the algorithm is reproduced below:

 \begin{algorithm}[H]
 	\scriptsize
 	\BlankLine
 	\SetKwInOut{Input}{Input}
 	\SetKwInOut{Output}{Output}
 	
 	\Input{A permutation of $`0123456789$' ($initPerm$)}
 	\Output{A CA rule}
 	\BlankLine
 	\rule[4pt]{0.95\textwidth}{0.99pt}\\
 	\BlankLine
 	\hspace{0.04\textwidth} \nlset{Step 1}\label{st1}  Set $Sibl_0 \leftarrow initPerm$\; 
 	\hspace{0.04\textwidth}	\nlset{Step 2}\label{st2}
 	\For {$i = 1$ to $9$}{ 
 		Set	$Sibl_i \leftarrow Sibl_{0} >> i$ ; \tcp{Set $Sibl_i$ with $i$ times circular right shift of $Sibl_0$} 
 	}
 	\hspace{0.04\textwidth}	\nlset{Step 3}\label{st3}
 	\For {$j = 10$ to $99$}{ 
 		Set	$Sibl_j \leftarrow Sibl_{j \pmod{10}} >> (j/10)$ ; \tcp{Set $Sibl_j$ with $j/10$ times circular right shift of $Sibl_k$ where $j \pmod{10}$} 
 	}
 	\hspace{0.04\textwidth}	\nlset{Step 4}\label{st4} 
 	\For {$i = 0$ to $8$}{ 
 		\For {$j=i+1$ to $9$}{
 			Set $Sibl_{j*10+i} \leftarrow Sibl_{i*10+j}$ ; \tcp{The RMTs of $Sibl_{j*10+i}$ and $Sibl_{i*10+j}$ have same values}
 		}
 	}	
 	\BlankLine
 	\caption{\emph{GenerateRuleFromPermutation}}
 	\label{algo:CA_generation}
 \end{algorithm}
For example, when $1632405789$ is given as input to Algorithm~\ref{algo:CA_generation}, first RMTs of $Sibl_0$ are assigned, that is, RMT 9 gets 1, RMT $8$ gets 6, and so on. Similarly, by cyclic permutation of this string, the other strings are generated. Finally, we can get the actual rule of the decimal CA by concatenating those strings. It is shown that, rules synthesized in this way also follow the desirable properties mentioned earlier to be a good source of randomness. Further, to extract random numbers from the configurations of the CAs, two window-based schemes are proposed. A specialty of this generator is it can create random decimal numbers of any arbitrary length as well as random binary numbers of any length as multiple of $32$. For our purpose, we have taken the binary output generator version of this PRNG with rule $1632405789$ and generated $32$ as well as $64$ bits depending on the size of the seed. 
 
Sometimes, optimization techniques are applied to the CAs to improve their randomness qualities.
In \cite{Tomassini96, Marco99}, genetic algorithms are applied to co-evolve hybrid CAs for generating random numbers. For example, in \cite{Tomassini96}, cellular programming is used over a CA of size $50$, where the first $22$ cells have rule $165$, next $22$ cells have rule $90$ and last $6$ cells have rule $150$ to develop a PRNG. In \cite{Guan03}, numbers are generated using evolutionary multi-objective optimization techniques on controllable CA, whereas, in \cite{Guan04}, self-programmable CA is used. In a controllable CA, the update of some cells is controlled via some control signals, while in programmable CA, spatial and temporal variations are allowed in the CA rules using some external control scheme. Another example of self-programmable CAs used for implementation of PRNG on FPGA is \cite{PETRICA2018251}. In \cite{PURKAYASTHA201632}, quantum-dot cellular automata is explored in developing a pseudo random number generator based cryptographic architecture. 
In \cite{Marco00,Guan04a,HOSSEINI2014149,COMPAGNER1987391,TOMASSINI2001151,10.1007/3-540-45356-3_71, chowdhury1994class,Torres-Huitzil2012,4459620,shackleford2002fpga}, $2$-dimensional CAs are used as the PRNGs. For example, in \cite{HOSSEINI2014149}, $2$-state periodic boundary CA with the rules $165, 105, 90, 150, 153, 101, 30, 86$ is combined with Langton's ants to generate the numbers. Here, Langton's ant is a simple $2$-dimensional Turing machine with complex behavior \cite{langton86}.

\subsection{Special Purpose Generators}
\noindent Recently, a new class of generators are proposed where chaotic systems are utilized to introduce non-linearity in developing pseudo-random number generators. Mostly, these generators are developed for some special applications, like cryptographic purposes etc. For example, in \cite{DONG201940} high dimensional Hamiltonian conservative chaotic systems are exploited to design pseudo-random number generator using FPGA technique. Another PRNG for video encryption is proposed in \cite{XU20169305}, which uses the output of a 16-cell LFSR perturbed by two chaotic maps, namely a bit-reorganizer, and a nonlinear function. Its output is 32 bit word. In \cite{REZK2019174}, the Lorenz and Lü chaotic systems based reconfigurable PRNG is designed on FPGA. Some other similar generators are \cite{MERANZACASTILLON2019239,TUTUEVA2020109615,wang2016pseudorandom,li2019novel,lv2018novel,ozturk2015novel,sahari2018pseudo,HAMZA2017119,8531753,AYUBI2020102472,cang2021pseudo}.
In \cite{CHEN201857}, SRAM Physical Unclonable Functions (PUFs) are taken as the entropy source to provide seed to a hash based Deterministic Random Bit Generator (DRBG). Sometimes, chaotic source are added with the existing PRNGs to produce new hybrid PRNGs; see, for example \cite{avarouglu2015hybrid,NEUGEBAUER2018316,ULLAH2021102619}. In \cite{SAVVIDY2015161} a family of matrix random number generators called MIXMAX random number generator is defined that uses linear matrix recursions modulo a prime number $p$. Its output is 53-bit double precision number with the default $p = 2^{61}-1$. This class of generators is a generalization of LCGs that uses the ergodicity of chaotic dynamical systems. A recent survey on FPGA implemented PRNGs is reported in \cite{BAKIRI2018135}. With the advancement of quantum computing, some researchers are using a quantum computation model, called quantum random walks (QRWs) for constructing PRNGs \cite{yang2016novel,ELLATIF2020123869}. On the other hand, with the introduction of GPUs, a new research direction is opened for generating uniform random numbers for parallel computing environment \cite{LECUYER20173}. However, in this work, we have not considered these special class of generators for testing and comparison.

\subsection{Remark and Discussion}
\hspace{0.1em}$\bullet$ In most of the PRNGs, the underlying backbone is existence of a primitive polynomial of large degree. This polynomial ensures that the PRNG has a large period. All celebrated PRNGs today depend on this theory. A primitive polynomial belongs to the class of irreducible polynomials. There exists algorithms to determine whether a polynomial $P$ is reducible or not \cite{McEliece:1987:FFS:22839}. However, testing primitivity of an irreducible polynomial requires prime factorization which is difficult to handle. To avoid this, researchers use known tricks (like, use of Mersenne primes) that guaranty that the characteristic polynomial of the PRNG is primitive and the period is maximal. Therefore, the main problem is synthesizing a primitive polynomial. If there was efficient ways to synthesize a primitive polynomial, it would have been possible to develop PRNGs with any desirable period.

\hspace{0.1em}$\bullet$ The reason of development of LFSR and CA based PRNGs is mainly ease of cost effective hardware implementation. However, for PRNGs like Mersenne Twister, which takes a primitive polynomial of large degree, this hardware implementation is so costly that, it is infeasible. Moreover, for applications like VLSI circuit testing, efficiency and portability (see Section~\ref{sec:property}) of a PRNG is more essential than intricate randomness. Nevertheless, for CA-based PRNGs, as feedback connections are from neighboring memory elements (cells), cost of interconnection on hardware implementation can be lesser than LFSR. Due to this reason, and option of parallelism, CA-based, more specifically ECA-based PRNGs are attractive as VLSI test pattern generators.

%

To conclude this short survey, we list out the good PRNGs in Table~\ref{tab:PRNG_list}, which we have already discussed in this section. 
\begin{table}[!h]
	\setlength{\tabcolsep}{1.5pt}
	\scriptsize
	\renewcommand{\arraystretch}{1.20}
	\centering
	\caption{List of $30$ good PRNGs}\label{tab:PRNG_list}
	\begin{tabular}{|c|p{6.0cm}|}
		\hline
		\theadfont{Class of PRNGs} & \theadfont{Name of the PRNGs}\\
		\hline
		\multirow{2}{*}{LCGs} & \verb|minstd_rand|, Borland LCG, Knuth's LCG \verb|MMIX|, \\
		& \verb|rand|, \verb|lrand48|, \verb|MRG31k3p|, \verb|PCG-32| \\
		\hline
		\multirow{7}{*}{LFSRs} & \verb|random|, \verb|Tauss88|, \verb|LFSR113|, \verb|LFSR258|, \\
		& \verb|WELL512a|, \verb|WELL1024a|, \\
		& \verb|MT19937-32|, \verb|MT19937-64|, \\
		& \verb|SFMT19937-32|, \verb|SFMT19937-64|, \\
		& \verb|dSFMT19937-32|, \verb|dSFMT19937-52|, \\
		& \verb|xorshift32|, \verb|xorshift64*|, \verb|xorshift1024*|, \\ & \verb|xorshift128+| \\
		\hline
		\multirow{4}{*}{CAs} & Rule $30$ with CA size $101$,\\
		& Hybrid CA with Rules $30$ \& $45$, Maximal Length CA with $\gamma=0$ and $\gamma=1$, Non-linear $2$-state CA,\\
		&$3$-state CA, Decimal CA\\
		\hline
	\end{tabular}
\end{table}  
Our target is to measure the randomness quality of these PRNG uniformly on the same platform. To do this we use some well-known empirical testbeds which we briefly discuss next.

\section{Empirical Tests}\label{sec:empirical}
\noindent Empirical tests target to find some pattern in the generated numbers of a PRNG to prove its non-randomness. These tests aim to check the local randomness property, that is, randomness of the numbers are approximated over a minimum sequence length, rather than the whole period \cite{Knuth2}. Note that, for empirical tests, numbers of a complete period are not necessary. Innumerable such tests can be developed which aim to find any violation of the desirable properties (described in Section~\ref{sec:property}), if exists, in a PRNG. If a PRNG passes all \emph{relevant} empirical tests, it is declared as a good PRNG. However, usually, there is no known method to find which tests are pertinent for a PRNG to certify its randomness quality. Therefore, the common practice is to use empirical testbeds to identify non-randomness in the generated numbers.

In general, empirical tests can be classified into two groups -- blind tests and graphical tests. In case of blind tests, the tests are based on statistics and computation, so, no human intervention is required in taking a decision. On the other hand, in case of graphical tests, the performance is measured by finding visible patterns in the generated image; so here decision is taken by the coordinating person(s). In the next subsections, the tests used for our purpose are described in more details.

\subsection{Blind (Statistical) tests}\label{sec:BTest}
\noindent 
The target of these tests is to find evidence against a specific null hypothesis $(\mathscr{H}_0)$. Usually, this $\mathscr{H}_0$ is, ``the sequence to be tested is random''. For each test, based on the random sequence produced by a PRNG, a decision is taken either to reject or not to reject the null hypothesis $\mathscr{H}_0$. To do this, a suitable randomness statistic, having a distribution of possible values, has to choose which determines the rejection of $\mathscr{H}_0$. Under $\mathscr{H}_0$, the reference theoretical distribution (usually standard normal or chi-square distribution) of this statistic is calculated. A \emph{critical value} $(t)$ is computed for this reference distribution. During a statistical test, the relevant statistic is calculated on the generated random sequence and compared to the critical value. If the test statistic value is greater than the critical value, $\mathscr{H}_0$ is rejected, otherwise it is not rejected. The probable conclusions for any situation are shown in Table~\ref{stat_error}. 


\begin{table}
	\caption{Conclusions and Errors in statistical test}
	\label{stat_error}
	\begin{tabular}{|c|c|c|}
		\hline
		\multirow{2}{*}{Real situation} & \multicolumn{2}{c|}{Conclusion}\\
		\hhline{~--} 
		& $\mathscr{H}_0$ is rejected & $\mathscr{H}_0$ is not rejected\\
		\hline
		Data is random ($\mathscr{H}_0$ is true) & Type I error & Correct decision \\
		\hline
		Data is not random ($\mathscr{H}_0$ is not true) & Correct decision & Type II error \\
		\hline
	\end{tabular}
\end{table}

When a conclusion is made to reject the null hypothesis, while in truth, the data is random, is called a \emph{Type I} error. However, if the data is not random, but in conclusion, $\mathscr{H}_0$ is not rejected, it results in generating \emph{Type II} error. In other cases, the conclusion is correct. 
The \emph{level of significance} ($\alpha$) of a test is defined as the probability of generating a Type I error. Usually, it is set prior to the test as a number between $0.0001$ and $0.01$. {Whereas, the probability of generating a Type II error is denoted by $\beta$.}

However, the $p$-value of a test measures the strength of evidence against the null hypothesis. It is defined as 
\[p=\Pr[X\geq t~|~\mathscr{H}_0 ~~{ is~ true }]\]
where $X$ denotes the test statistics and $t$ the critical value \cite{L'Ecuyer:2007:TCL:1268776.1268777}. If $p$-value is very close to $0$ or $1$, it indicates that the sequence generated by the PRNG is not random. Normally, if $p \geq \alpha$, then the tested sequence is considered random, and $\mathscr{H}_0$ is not rejected, otherwise it is rejected.
However, if the test statistic has a discrete distribution, the $p$-value is redefined as:
\begin{equation*}
p={\begin{cases}
	p_R,  &\text{if $p_R > p_L$ }\\
	1-p_L, &\text{if $p_R \geq p_L$ and $p_L < 0.5$}\\
	0.5, &\text{otherwise}
	\end{cases}}
\end{equation*} 
where $p_R= \Pr(X \geq t ~|~ \mathscr{H}_0 \text{ is true })$ and $p_L = \Pr(X \leq t ~|~ \mathscr{H}_0 \text{ is true })$.

There are many statistical battery of tests available which target to find non-randomness based on a collection of statistical tests. The first known statistical battery of tests was offered by Donald Knuth in $1969$ in his book ``The Art of Computer Programming, Vol. $2$''\cite{Knuth2}. Later, other batteries have been developed improving the testing procedures of Knuth, {such as Diehard, NIST, TestU01, PractRand, Dieharder, etc.} In this paper, we have selected three well-known test suites, namely \emph{Diehard}, \emph{TestU01} and \emph{NIST}. A PRNG is supposed to be a good PRNG, if it passes all tests of every testbed for any seed. 


\subsubsection{Diehard battery of Tests}\label{sec:diehard}
\noindent George Marsaglia in $1996$ provided this battery of tests \cite{diehard}, which is the basic testbed for PRNGs. It consists of $15$ different tests - 
\begin{small}
	\begin{framed}
		\vspace{-1.0em}
		\label{diehardtests}
		\begin{center}
			\underline{\textbf{Diehard Battery of Tests}}
		\end{center}
		1. Birthday spacings, 2. Overlapping permutations, 3. Ranks of $31\times31$ and $32\times32$ matrices, 4. Ranks of $6\times8$ matrices, 5. Monkey tests on $20$-bit Words, 6. Monkey tests: OPSO (Overlapping-Pairs-Sparse-Occupancy), OQSO (Overlapping-Quadruples-Sparse-Occupancy) and DNA tests, 7.  Count the $1$'s in a stream of bytes, 8. Count the $1$'s in specific bytes, 9. Parking lot test, 10. Minimum distance test, 11. Random spheres test, 12. The squeeze test, 13. Overlapping sums test, 14. Runs up and runs down test, and 15. The craps test (number of wins and throws/game). 
	\end{framed}
\end{small}
To test a PRNG on Diehard for a particular seed, a binary file of size $10-12$ MB is created using the generated numbers of the PRNG with that seed. In our case, we have taken the file size as $11.5$ MB. For each test, one or multiple $p$-values are derived. A test is called \emph{passed}, if every $p$-value of the test is within $0.025$ to $0.975$ \cite{diehard}. 

\subsubsection{TestU01 library of Tests}\label{sec:testu01}
\noindent This library offers implementations of many stringent tests -- the classical ones as well as many recent ones. It was developed by Pierre L'Ecuyer and Richard Simard \cite{L'Ecuyer:2007:TCL:1268776.1268777} to remove the limitations of existing testbeds -- like inability to modify the test parameters (such as, the input file type, $p$-values etc.) as well as to encompass new updated tests. It comprises of several battery of tests, including most of the tests in Diehard and many more with more flexibility
to select the test parameters than in Diehard. However, we have selected the battery \emph{rabbit} to test the PRNGs. The reason for choosing this test-suite is -- this battery is specifically designed to test a sequence of random bits produced by a generator. 
It contains the following $26$ tests from different modules (mentioned in parenthesis for each test):
\begin{small}
	\begin{framed}
		\vspace{-1.0em}
		\label{rabbittests}
		\begin{center}
			\underline{\textbf{Battery Rabbit of TestU01}}
		\end{center}
		1. MultinomialBitsOver test (smultin), 2. ClosePairsBitMatch in $t = 2$ dimensions (snpair) and 3. ClosePairsBitMatch in $t = 4$ dimensions (snpair), 4. AppearanceSpacings test (svaria), 5. LinearComplexity test (scomp), 6. LempelZiv test (scomp), 7. Spectral test of Fourier1 (sspectral), and 8. Spectral test of Fourier3 (sspectral), 9. LongestHeadRun test (sstring), 10. PeriodsInStrings test (sstring), 11. HammingWeight with blocks of $L = 32$ bits test (sstring), 12. HammingCorrelation test with blocks of $L = 32$ bits (sstring), 13. HammingCorrelation test with blocks of $L = 64$ bits (sstring) and 14. HammingCorrelation test with blocks of $L = 128$ bits (sstring), 15. HammingIndependence with blocks of $L = 16$ bits (sstring), 16. HammingIndependence with blocks of $L = 32$ bits (sstring) and 17. HammingIndependence with blocks of $L = 64$ bits (sstring), 18.AutoCorrelation test with a lag $d = 1$ (sstring)and 19. AutoCorrelation test with a lag $d = 2$ (sstring), 20. Run test (sstring), 21. MatrixRank test with $32 \times 32$ matrices (smarsa) and 22. MatrixRank test with $320 \times 320$ matrices (smarsa), 23. RandomWalk1 test with walks of length $L = 128$ (swalk), 24. RandomWalk1 test with walks of length $L = 1024$ (swalk), and 25. RandomWalk1 test with walks of length $L = 10016$ (swalk).
	\end{framed}
\end{small}

Here $smultin$ is a module of tests based on the multinomial distribution \cite{S1064827598349033} which tests uniformity in the $t$-dimensional unit hypercube. The module $snpair$ implements tests based on the distances between the closest points in a sample of $n$ uniformly distributed points in the unit torus in $t$-dimensions \cite{opre.48.2.308.12385}. $svaria$ is a module that implements different uniformity tests, mainly based on some simple statistics. The module $scomp$ contains tests based on linear complexity of bit sequence as well as on the compressibility of it, measured by the Lempel-Ziv complexity \cite{Lempel}. The statistical tests developed by George Marsaglia and his collaborators in \cite{marsaglia1985current} are implemented in $smarsa$ module. In case these tests are spacial cases of the tests of module $smultin$, the function smultin\_MultinomialOver is called. The module $sspectral$ contains tests based on spectral methods, which computes the discrete Fourier transform of a bit string of size $n$ and looks for deviations in the spectrum inconsistent with $\mathscr{H}_0$. $sstring$ module implements tests on strings of random bits made by concatenating blocks of $s$ bits from each. In module $swalk$, statistical tests based on discrete random walks over $\mathbb{Z}$ is implemented \cite{SHCHUR1997579}. Among these tests, spectral tests are the most difficult ones to pass.

The battery \emph{rabbit} takes two arguments -- a filename and number of bits $(nb)$. The first $nb$ bits of the binary file, filled by the random numbers generated by the PRNG, is tested. For each test, the parameters are a function of $nb$, to make it dynamic. Here, to test a PRNG using \emph{rabbit}, we have set $nb=10^7$ and the file size is taken as $10.4$ MB. A test is declared to be passed for a seed, if each of the $p$-values of the test is within $0.001$ to $0.999$ \cite{L'Ecuyer:2007:TCL:1268776.1268777}.

\noindent\textbf{Remark:} Between Diehard and TestU01's rabbit battery of tests, we have observed that, for the selected PRNGs, some tests of Diehard are more stringent to pass than that of battery rabbit of TestU01. For example, for a specific seed, even if the PRNG passes all tests of rabbit, but it may fail to pass overlapping permutations test of Diehard. 

\subsubsection{NIST Statistical Test suite}\label{sec:nist}
\noindent The NIST Statistical Test Suite is developed to test a PRNG for cryptographic properties \cite{rukhin2001statistical}. It has mainly three tasks -- (1) investigate the distribution of $0$s and $1$s, (2) using spectral methods analyze the harmonics of bit stream and (3) detect patterns based on information theory and probability theory. This test suite consists of $15$ tests --
\begin{small}
	\begin{framed}
		\vspace{-1.0em}
		\label{nisttests}
		\begin{center}
			\underline{\textbf{NIST Test Suite}}
		\end{center}
		1. The Frequency (Monobit) Test,
		2. Frequency Test within a Block,
		3. The Runs Test,
		4. Tests for the Longest-Run-of-Ones in a Block,
		5. The Binary Matrix Rank Test,
		6. The Discrete Fourier Transform (Spectral) Test,
		7. The Non-overlapping Template Matching Test,
		8. The Overlapping Template Matching Test,
		9. Maurer's ``Universal Statistical'' Test,
		10. The Linear Complexity Test,
		11. The Serial Test,
		12. The Approximate Entropy Test,
		13. The Cumulative Sums (Cusums) Test,
		14. The Random Excursions Test, and
		15. The Random Excursions Variant Test.
	\end{framed}
\end{small}

For this test suite, the level of significance $\alpha = 0.01$. So, for a sample size $m$ generated by a PRNG with a particular seed, if $x$ is the minimum pass rate, then for the PRNG to pass the test, minimum number of sequences with $p$-values $\geq 0.01$ has to be $x$.
This minimum pass rate is approximately $615$ for a sample size of $629$ for the random excursion (variant) test and $980$ for sample size $1000$ for each of the other tests. In general, the range of acceptable proportions for $x$ is calculated as $(1-\alpha)\pm 3 \sqrt{\frac{\alpha(1-\alpha)}{m}}$, for a sample size $\alpha$.


To test a PRNG using NIST test suite, a binary or ASCII file containing the random numbers is given as input. We have taken sample size as $10^3$ with sequence length = $10^6$ and generated binary file of size $125$MB as the input. The default parameters are not updated, that is, block length ($M$) for block frequency test is $128$ and for linear complexity test is $500$. Similarly, block length ($m$) for both non-overlapping template test and overlapping template test is $9$, for approximate entropy test is $10$ and for serial test is $16$. For each run, a file named \emph{finalAnalysisReport.txt} is generated that summarizes the results of all the tests for that run. In this file, the first ten columns note the frequency of $p$-values in each of the $10$ equal sub-intervals between $0$ to $1$ and column $11$ is the $p$-value derived by applying chi-square test on these columns. For the corresponding statistical test noted in column $13$, column $12$ records the passed proportions of samples. This file also indicates the tests (or parts of a test) which are not passed, by marking it with $`*$'. If all parts of a test are passed, the PRNG is said to have passed that test.

\noindent\textbf{Remark:} In general, simple non-cryptographic PRNGs fail to pass NIST tests. However, a good PRNG, which passes all or most of the tests of TestU01 and Diehard, also perform well in NIST test-suite. Therefore, a good non-cryptographically secure PRNG may pass all tests of NIST for some seeds.

\subsection{Graphical Test}\label{sec:Gtest}
\noindent As discussed already, goal of every empirical test is to detect a pattern in the numbers generated by a PRNG to prove its non-randomness. In statistical tests, this is done by generating the $p$-value. However, it may happen that, a Type I or Type II error has occurred, and the wrong conclusion is reached. Therefore, it is useful to actually see how the numbers look like in a $2$-dimensional or $3$-dimensional plot. 

In graphical tests, the numbers are plotted in a graph to see whether any visible pattern exists or not. As period length of a PRNG is expected to be very large, so, for every graphical test also, all numbers of a period can not be used; rather a set of numbers needs to be generated based on some seed. We have mainly used two graphical tests -- (1) Lattice tests, (2) Space-time diagram.
Let us now explain these two tests.

\subsubsection{Lattice Test}
\noindent This test identifies whether the random numbers form some patterns.
To test this, the consecutive random numbers (in normalized form), generated from a seed, are paired and plotted. Two types of lattice tests are executed on these normalized numbers, namely $2$-D lattice test (takes two consecutive numbers as a point) and $3$-D lattice tests (three consecutive numbers form a point). If the random numbers are correlated, the plots show patterns. Otherwise, the PRNG is considered to be good. 

\subsubsection{Space-time Diagram}
\noindent Space-time diagram is an important theoretical tool that has long been used to observe and predict the behavior and evolution of a CA  \cite{Wolframbook1}. For CAs, it is a graphical representation of the configurations (on $x$-axis) at each time $t$ (on $y$-axis). Each of the CA states are depicted by some color. So, the evolution of the CA can be visible from the patterns generated in the state-space diagram.

In this work, we propose this tool as an useful measure of randomness of a PRNG. The $x$-axis of a diagram represents a number generated at any time instant and $y$-axis depicts time. To test a PRNG with space-time diagram, the numbers need to be non-normalized. If numbers are in base $b$, then $b$ different colors are required to represent a number where each color signifies a particular digit of that base. For example, if numbers are binary, then two colors, usually black for $1$ and white for $0$, are required to represent any number. So, each binary number is then a combination of black and white. For $b>2$, more colors are required for each number.

Starting with a seed, a set of numbers are generated over time $t$ and each number is plot against $t$. If there is a pattern among any consecutive numbers, or in any part of a number, then it can be seen from this diagram, as colors make this pattern more prominent. If there is no pattern, and the numbers appear noisy in color, the PRNG has good randomness quality. Therefore, using this diagram, clear idea about the randomness properties of a PRNG can be developed.

In the next section, result of these empirical tests applied on the PRNG are recorded. For each seed $1000$ and $250$ numbers are generated for testing using the lattice test and space-time diagram test respectively. In case of space-time diagrams, these numbers are directly represented, whereas, for lattice tests, pairs (resp. triplets) of consecutive numbers are plot as each point in the 2-D (resp. 3-D) plane.

\section{Empirical Facts}\label{sec:facts}
\noindent A usual claim of a PRNG is, it is better than others! Mostly, this claim is based on the performance of the PRNGs in the well-known test-suites. In this section, we verify the claim of the existing PRNGs to find which are really better than others with respect to randomness quality. To do so, we choose the PRNGs of Table~\ref{tab:PRNG_list} and apply empirical tests of Section~\ref{sec:empirical}. Although failure in any test of a testbed means the PRNG is non-random, but for the sake of uniformity, we take the count of the number of tests for which the null hypothesis is not rejected for a PRNG as its merit for randomness. However, we do not trust the Blind tests completely and give a preference to the results of graphical tests whenever there is discrepancy.

For all these PRNGs, we have collected the standard implementation and used them to generate numbers.
To maintain uniformity in testing, we have tested the stream of binary numbers generated in sequence by these standard implementations of the PRNGs. For each PRNG, binary (\emph{.bin}) files are produced which contain sequence of numbers (in binary form) without any gap between two consecutive numbers in the sequence. 
If the generated numbers are normalized, then, first we convert fractional part of each number into its binary equivalent and then add these bits to the .bin file. 


\subsection{Choice of Seeds}
\noindent Although a good PRNG should be independent of seeds, but to run a PRNG we need to choose the seeds. This seed can be any number from the period of a PRNG. However, it is not possible to test every PRNG for all possible seeds. So, we have taken the following greedy approach:
\begin{enumerate}
\item As we have collected the $C$ programs of the PRNGs from their respective websites, each of them has an available seed for a test run. For example, for \verb|MT19937|, the seed was $19650218$. Nevertheless, in most of the cases, this seed is $1234$ or $12345$. We, therefore, have collected all the seeds hard-coded in the $C$ programs of all PRNGs, and used these as the set of seeds for each PRNG. These seeds are $7$, $1234$, $12345$, $19650218$ and $123456789123456789$. We have tested each PRNG with all these seeds.

\item Apart from studying the behavior of PRNGs for fixed seeds, we also want to observe the rough estimate of the general behavior of the PRNGs. For this reason, we have chosen a simple LCG, \verb|rand| to generate seeds for all other PRNGs. This \verb|rand| is initialized with \verb|srand(0).| The next $1000$ numbers of \verb|rand|  are supplied as seeds to each PRNG to test it $1000$ times with these random seeds.
\end{enumerate}

All PRNGs are tested empirically for each of these seeds and the results are compared impartially. Some PRNGs, such as \verb|LFSR113|, \verb|LFSR258|, etc. require more than one (non-zero) seed to initialize its components. However, we supply here the same seed to all of its components.

\subsection{Results of Empirical Tests}
\noindent The selected PRNGs (LCG-based, LFSR-based and CA-based) are tested with statistical tests as well as with the graphical tests. Here, summary of the results of these tests is documented. A library of codes and packages used for this purpose is publicly available at Ref~\cite{githubLibraryPRNG}.

\subsubsection{Results of Statistical Tests}
\noindent Table~\ref{tab:blind_test} shows the summary of results of Diehard battery of tests, battery \emph{rabbit} of TestU01 library and NIST statistical test suite for the fixed seeds. In this table, for each PRNG, result (in terms of numbers of tests passed) of the testbeds per each seed is recorded. 

\begin{table}[!h]
	\setlength{\tabcolsep}{1.3pt}
	\renewcommand{\arraystretch}{1.30}
	\centering
	\scriptsize
	\caption{Summary of statistical test results for different fixed seeds}
	\label{tab:blind_test}
	\resizebox{1.00\textwidth}{4.5cm}{
		\begin{tabular}{|c|c|c|c|c|c|c|c|c|c|c|c|c|c|c|c|c|c|}
			\hline
			\multicolumn{2}{|r|}{\theadfont{Seeds $\longrightarrow$}} & \multicolumn{3}{c|}{\thead{$7$}} &  \multicolumn{3}{c|}{\thead{$1234$}} &   \multicolumn{3}{c|}{\thead{$12345$}} &  \multicolumn{3}{c|}{\thead{$19650218$}} &  \multicolumn{3}{c|}{\thead{$123456789123456789$}} & \thead{Ranking}\\
			\cline{1-17}
			\multicolumn{2}{|c|}{\theadfont{ Name of the PRNGs}} &  Diehard & TestU01 & NIST & Diehard & TestU01 & NIST & Diehard & TestU01 & NIST & Diehard & TestU01 & NIST & Diehard & TestU01 & NIST & (First Level)\\
			\hline
			\multirow{7}{*}{\rotatebox{90}{LCGs}}& MMIX & 6 & 19 & 7 & 5 & 18 & 7 & 6 & 17 & 8 & 4 & 16 & 8 & 5 & 18 & 8 & 8\\
			\cline{2-18}
			& minstd\_rand & 0 & 1 & 1 & 0 & 1 & 1 & 0 & 1 & 2  & 0 & 1 & 1 & 0 & 2 & 1 & 12\\
			\cline{2-18}
			& Borland LCG & 1 & 3 &5 & 0 & 3 & 5 & 1 & 3 & 5 & 1 & 3 & 4 & 1 & 3 & 5 & 11\\
			\cline{2-18}
			& rand & 1 &1 &2 & 1 & 1& 2& 1 & 3 & 2& 1 & 2 & 2& 1 & 2 & 2 & 11\\
			\cline{2-18}
			& lrand48 & 1 & 3 & 2& 1 & 2 & 2 & 1 & 3 & 2 & 1 & 3 & 2 & 1 & 2 & 2 & 11\\
			\cline{2-18}
			& MRG31k3p & 1 & 2 & 1 & 1 & 1 & 1 & 1 & 2 & 1 & 1 & 1 & 1 & 0 & 0 & 2 & 12\\
			\cline{2-18}
			& PCG-32 & 9 & 25 & 15& 9 & 25 &14 & 11 & 25  &14 & 10 & 24 &15 & 9 & 25 & 15 & 2\\
			\cline{2-18}
			\hline
			\multirow{16}{*}{\rotatebox{90}{LFSRs}}& random & 1 & 1 & 1 & 1& 1 & 1 & 1 & 3 & 1 & 1 & 2 & 1 & 1 & 2 & 1 & 11\\
			\cline{2-18}
			& Tauss88 & 11 & 21 & 15 & 9 & 23 & 15 & 11 & 23 & 15 & 11 & 23 & 14 & 10 & 23 & 15 & 4\\
			\cline{2-18}
			& LFSR113 & 5 & 6 & 1& 11 & 23 & 14& 9 & 23 & 15 & 7 & 23 & 14 & 9 & 23 & 15 & 7\\
			\cline{2-18}
			& LFSR258 & 0 & 0 & 1& 0 & 5 & 2 & 1 & 5 & 2 & 1 & 5 & 2& 1 & 5 & 0 & 12\\
			\cline{2-18}
			& WELL512a & 9 & 23 &15 & 10 & 23 & 14 & 10 & 23 & 15 & 8 & 23 & 15 & 7 & 23 & 15 & 5\\
			\cline{2-18}
			& WELL1024a & 9 & 25 & 15 & 10 & 24 & 15 & 9 & 24& 14& 9 & 25 & 15 & 9 & 25 &15 & 3\\
			\cline{2-18}
			& MT19937-32 & 10 & 25 & 13& 9 & 25 & 13& 9 & 25 &14 & 9 & 25 &15 & 9 & 25 &15 & 3\\
			\cline{2-18}
			& MT19937-64 & 10 & 25 &15 & 10 & 24 & 15& 8 & 24 & 15& 11 & 25 &15 & 10 & 25 & 15 & 2\\
			\cline{2-18}
			& SFMT19937-32 & 10 & 25 & 15& 9 & 25 & 15& 10 & 25 &15 & 9 & 25 & 15& 10 & 25 & 15 & 1\\
			\cline{2-18}
			& SFMT19937-64 & 11 & 25 & 15 & 10 & 25 &15 & 10 & 25 & 15& 9 & 25 &15 & 10 & 25 & 15 & 1\\
			\cline{2-18}
			& dSFMT-32 & 7 & 25 &15 & 8 & 25 & 15& 11 & 24 & 13& 11 & 25 &15 & 10 & 24 & 15 & 5\\
			\cline{2-18}
			& dSFMT-52 & 5 & 11 & 3& 5 & 10 & 3& 7 & 11 & 3 & 6 & 10 & 3 & 7 & 9 & 3 & 9\\
			\cline{2-18}
			&  xorshift32 & 4 & 17 & 4& 4 & 17 & 4& 4 & 17 &2 & 0 & 17 &13 & 4 & 17 & 13 & 9\\
			\cline{2-18}
			&  xorshift64* & 10 & 25 & 15 & 10 & 25 & 15& 8 & 25 & 15& 7 & 25 & 15 & 8 & 25 & 14 & 5\\
			\cline{2-18}
			&  xorshift1024* & 7 & 20 & 6& 9 & 21 & 15& 7 & 20 &15 & 8 & 20 &15 & 6 & 21 & 15 & 6\\
			\cline{2-18}
			&  xorshift128+ & 9 & 25 & 14& 9 & 25 & 14& 10 & 24 &15 & 10 & 25 &15 & 8 & 24 & 15 & 4\\
			\hline
			\multirow{7}{*}{\rotatebox{90}{CAs}}& Rule $30$ & 11& 25& 15& 10&25 & 15& 9 & 25& 15 & 8 & 25 & 15 & 11 & 24 & 15 & 2\\
			\cline{2-18}
			& Hybrid CA with Rules $30$ \& $45$ & 3 & 8 & 3& 0& 1 & 0 & 2  & 8 & 1 & 1 & 7 & 2 & 1 & 8& 2 & 11\\
			\cline{2-18}
			& Maximal Length CA with $\gamma=0$ & 2 & 12 & 10& 0& 12 & 11 & 1 & 12& 11& 1& 12& 11 & 2& 12 & 10 & 10\\
			\cline{2-18}
			& Maximal Length CA with $\gamma=1$ & 4& 17& 14 & 3 & 16 & 14 & 3  & 17 & 14 & 4 & 15 & 14 & 3 & 16 & 14 & 8\\
			\cline{2-18}
			& Non-linear $2$-state CA & 6 & 11& 4& 7& 11 & 2& 5 & 11 & 3 & 6& 11& 4& 5& 12& 4 & 9\\
			\cline{2-18}
			& $3$-state CA & ${3}$& ${12}$& ${6}$ & ${3}$ & ${12}$ & ${6}$ & ${3}$ & ${11}$ & ${5}$ & ${2}$ & ${11}$ & ${4}$ & ${3}$ & ${11}$ & ${4}$ & ${10}$\\
			\cline{2-18}
			& Decimal CA & {9}& {25}& {15} &{10} &{25} &{15} & {11} & {25} & {15} & {11} & {25} & {15} & {10} & {25} & {15} & 1\\
			\hline
	\end{tabular}}
	\vspace{-0.5em}
\end{table}  

It is observed that, none of the PRNGs can pass all tests of these blind testbeds. We also notice the following:
\begin{itemize}[leftmargin=1pt]
	\item Among all, the LCGs \verb|minstd_rand|, Borland's LCG, \verb|rand|, \verb|lrand48|, \verb|MRG31k3p| \\~and the LFSRs \verb|LFSR258| and \verb|random| perform very poorly. In case of diehard tests, these PRNGs can pass at most the runs test, except \verb|LFSR258|, which passes the rank test of $31 \times 31$ and $31 \times 32$ matrices and runs down test.
	
	\item The remaining two LCGs based PRNGs, namely Knuth's \verb|MMIX| and \verb|PCG-32| behave well in comparison to the other LCGs as well as many of the LFSRs. For example, Knuth's \verb|MMIX| is better than \verb|LFSR258| and \verb|xorshift32|, whereas \verb|PCG-32| is better than \verb|MMIX| as well as \verb|LFSR113|, Xorshift PRNGs, WELL PRNGs and dSFMTs. In fact, performance of \verb|PCG-32| is comparable to MTs and SFMTs.
	
	\item Performance of \verb|LFSR113| is dependent on seed; whereas, \verb|WELL512a| and \verb|WELL1024a| are invariant of seeds.
	
	\item Among the Mersenne Twister and its variants, performance of dSFMTs (especially, \verb|dSFMT-52)|, are unexpectedly poor in terms of the blind empirical tests. 
	
	\item Among the CA-based PRNGs, Decimal CA and ECA $30$ can compete with the elite group of PRNGs like Mersenne Twisters, WELL and PCG. However, other CA-based PRNGs are not so good. For all CAs, except ECA rule $30$, the complete (or, a block from a) configuration of the CA is taken as a number. Among these CAs based PRNGs, performance of the max-length CA ($\gamma=1$) is better than the non-linear $2$-state CA, which is better than the max-length CA ($\gamma=0$) and hybrid CA rule $30-45$.
	
\end{itemize}

Therefore, we can define first level ranking of the PRNGs from these test results for fixed seeds. Last column of Table~\ref{tab:blind_test} shows this ranking. For this ranking, we have considered the overall tests passed in each test-suite. If two PRNGs give similar results, they have the same rank. It can be observed that, among the LCG-based PRNGs, \verb|PCG-32| gives the best result and among the LFSR-based PRNGs, performances of SFMTs ($32$ and $64$ bits) are the best. Moreover, among the CA-based PRNGs, Decimal CA gives the best result, whereas result of Wolfram's rule $30$ is also comparable to the results of SFMTs and MTs.
\begin{itemize}[leftmargin=1pt]
	\item As SFMTs and Decimal CA are best performers, these are ranked as $1$. Similarly, performance of \verb|MT19937-64|, \verb|PCG-32| and rule $30$ are comparable (ranked $2$), performance of \verb|MT19937-32| and \verb|WELL1024a| are comparable (ranked $3$), whereas performance of \verb|xorshift128+| and \verb|Tauss88| are comparable (ranked $4$). These are the elite group of best performing PRNGs.
	
	\item \verb|xorshift64*| performs well, but it has more dependency on seed than the MTs and SFMTs. For this reason we rank it lower than MT and SFMT. Similarly, \verb|LFSR113| can perform better than WELLs and non-linear $2$-state CA-based PRNG for some seeds, but as performance of WELL is more invariant of seed, it has higher rank than \verb|LFSR113.   | 
	
	\item As performance of \verb|WELL512a|, \verb|Xorshift64*| and \verb|dSFMT-32| bit are comparable, they are ranked as $5$. In terms of performance in NIST test-suite, {\verb|xorshift1024*|} (rank $6$) is better than \verb|LFSR113| (rank $7$), but worse than \verb|dSFMT-32.| 
	
	\item Among the other PRNGs, maximal-length CA with $\gamma=1$ and \verb|MMIX| are ranked $8$, the non-linear $2$-state CA based PRNG, \verb|dSFMT-52| and \verb|xorshift32| are ranked $9$ and maximal-length CA with $\gamma=0$ and $3$-state CA are ranked $10$.
	
	\item Rest of the PRNGs form two groups -- \verb|rand|, \verb|lrand48|, Borland's LCG, \verb|random| and rule $30-45$ as rank $11$ and \verb|minstd_rand|, \verb|MRG31k3p| and \verb|LFSR258| as rank $12$. 
\end{itemize}

\begin{figure}[hbtp] 
	\centering
	\vspace{-1.0em}
		\subfloat[dSFMT19937-52\label{dsfmt64_avg}]{%
		\includegraphics[width=0.3\linewidth, height=3.0cm]{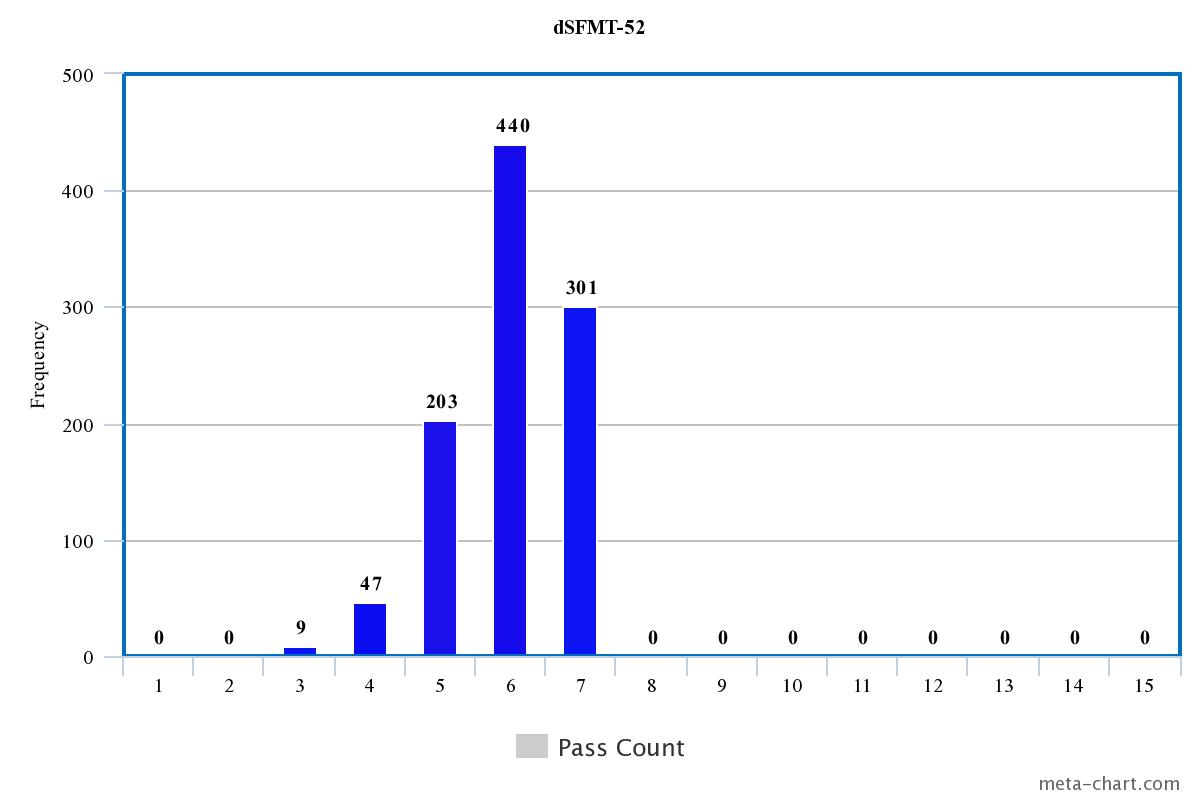}}
	\hfill
	\subfloat[ xorshift32\label{xor32_avg}]{%
		\includegraphics[width=0.3\linewidth, height=3.0cm]{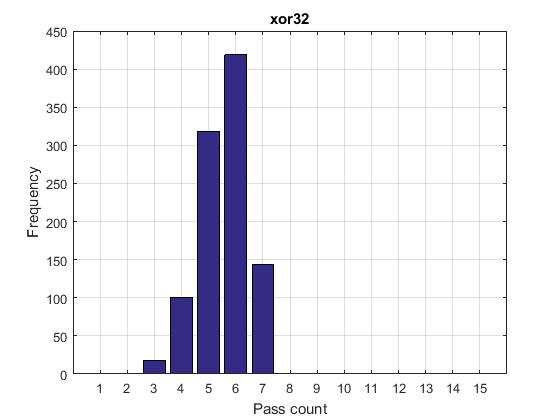}}
	\hfill
	\subfloat[ xorshift64*\label{xor64_avg}]{%
		\includegraphics[width=0.3\linewidth, height=3.0cm]{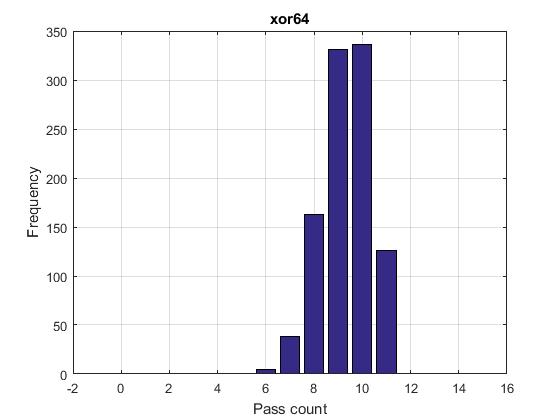}}
	\hfill
	\subfloat[ xorshift1024*\label{xor1024_avg}]{%
		\includegraphics[width=0.3\linewidth, height=3.0cm]{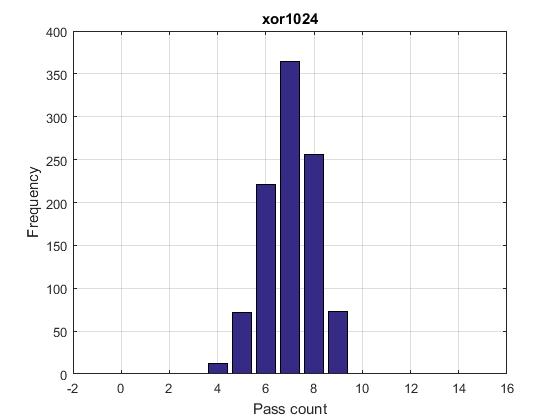}}
	\hfill
	\subfloat[ xorshift128+\label{xor128}]{%
		\includegraphics[width=0.3\linewidth, height=3.0cm]{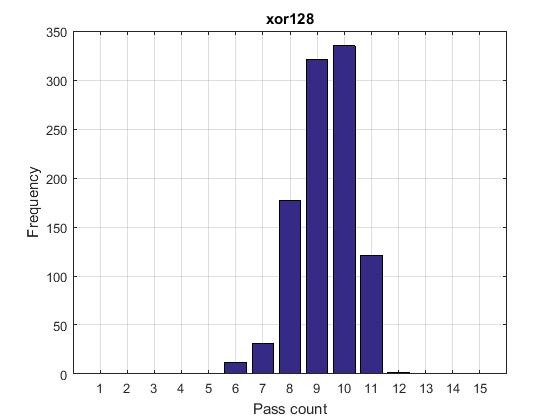}}
	\hfill
	\subfloat[Rule30\label{rule30_avg}]{%
		\includegraphics[width=0.3\linewidth, height=3.0cm]{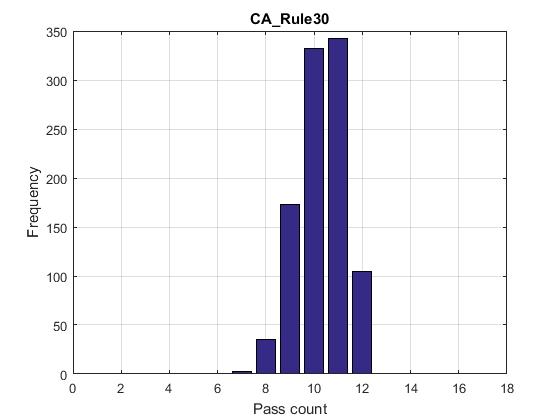}}
	\hfill
	\subfloat[Rule30-45\label{rule30-45_avg}]{%
		\includegraphics[width=0.3\linewidth, height=3.0cm]{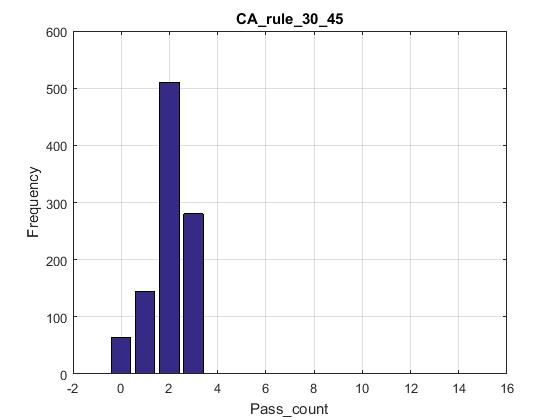}
	}\hfill
	\subfloat[max-length CA with $\gamma =0$\label{maxlength_y=0_avg}]{%
		\includegraphics[width=0.3\linewidth, height=3.0cm]{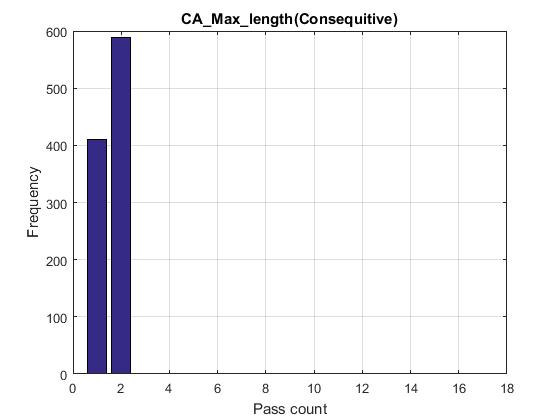}}
	\hfill
	\subfloat[max-length CA with $\gamma =1$\label{maxlength_y=1_avg}]{%
		\includegraphics[width=0.3\linewidth, height=3.0cm]{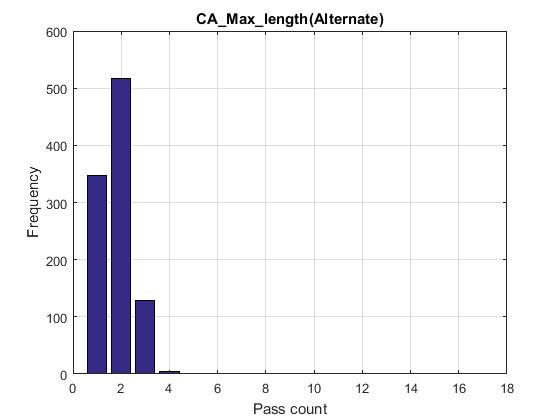}}
	\hfill
	\subfloat[non-linear $2$-state CA\label{nonlinear_avg}]{%
		\includegraphics[width=0.3\linewidth, height=3.0cm]{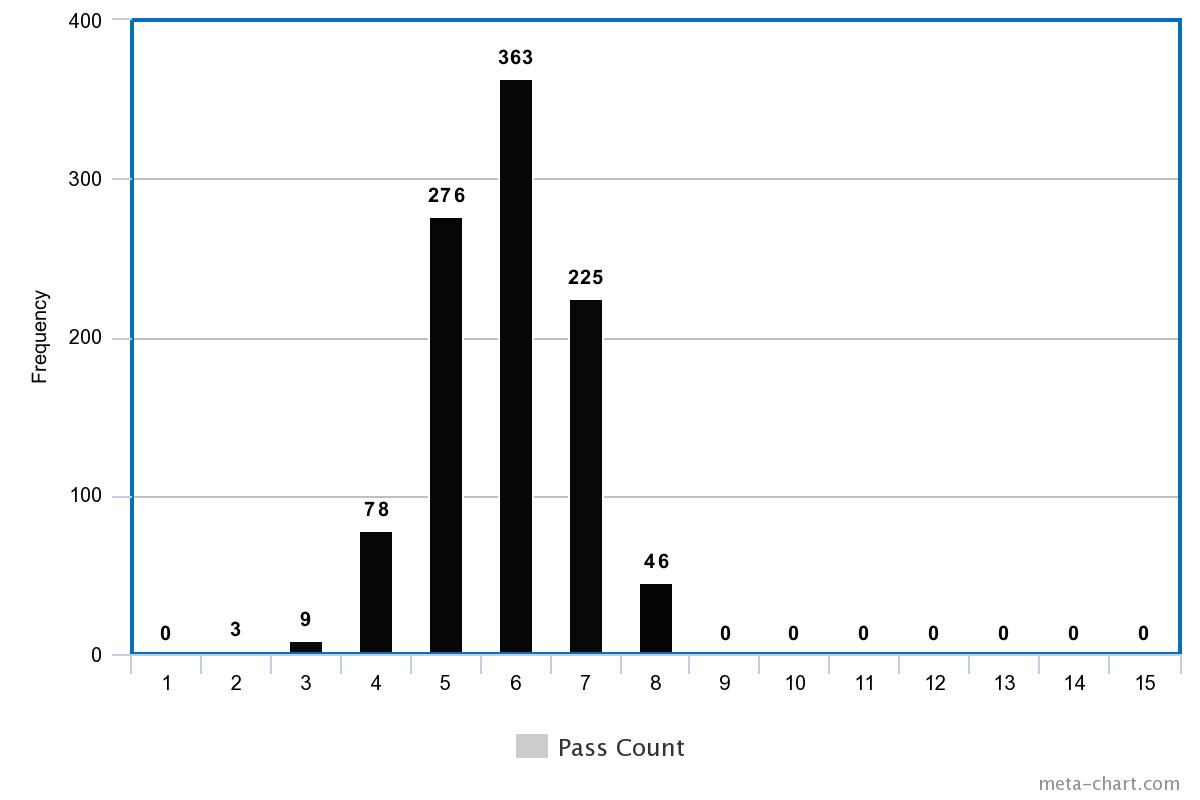}}
	              	\hfill
	       \subfloat[$3$-state CA \label{3-state_avg}]{%
	             \includegraphics[width=0.3\linewidth, height=3.0cm]{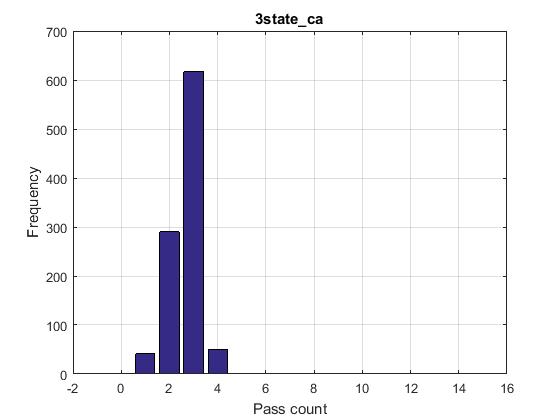}}
                      	\hfill
\subfloat[SFMT19937-32\label{sfmt32_avg}]{%
	\includegraphics[width=0.3\linewidth, height=3.0cm]{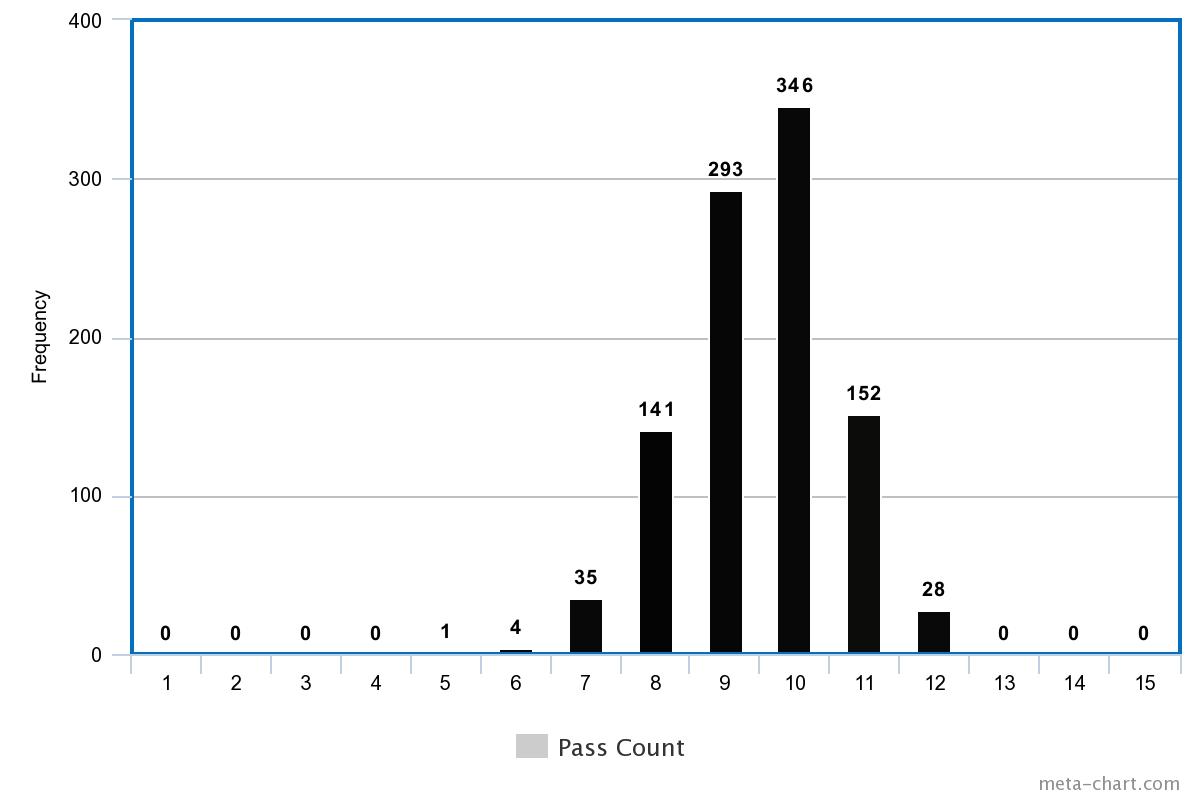}}
\hfill
\subfloat[SFMT19937-64\label{sfmt64_avg}]{%
	\includegraphics[width=0.4\linewidth, height=3.0cm]{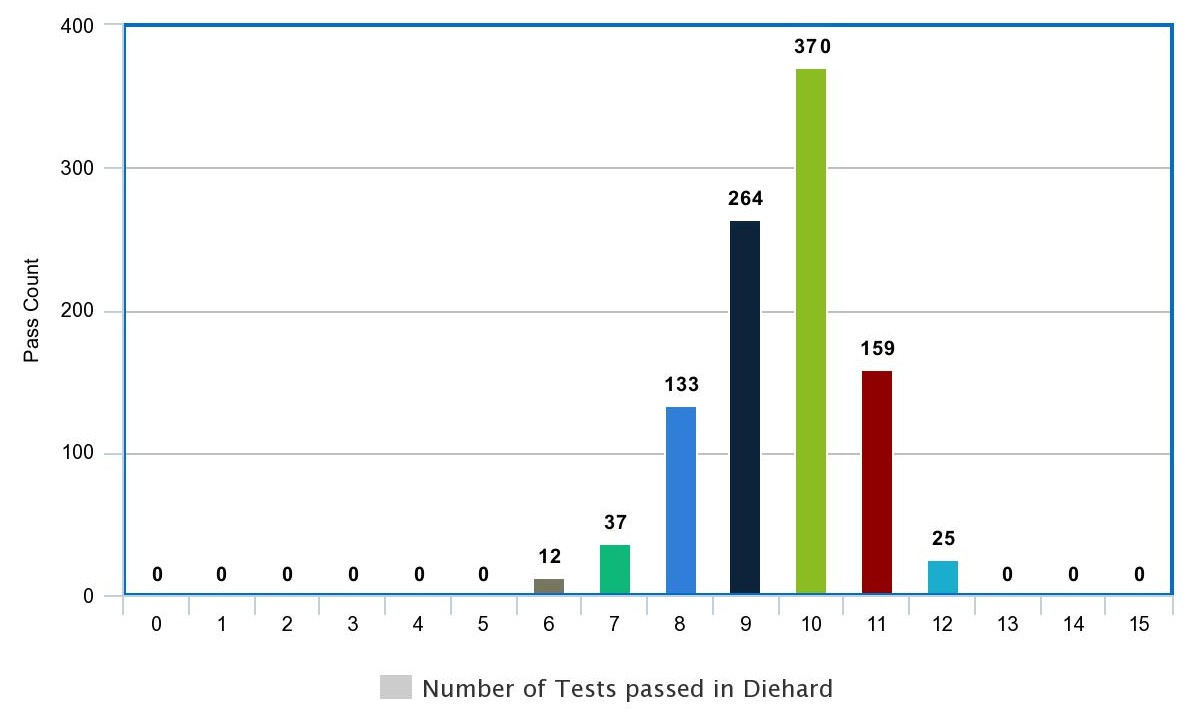}}
\hfill
               	\subfloat[Decimal CA $1632405789$\label{diehard_proposedPRNG}]{%
             		\includegraphics[width=0.4\linewidth, height=3.0cm]{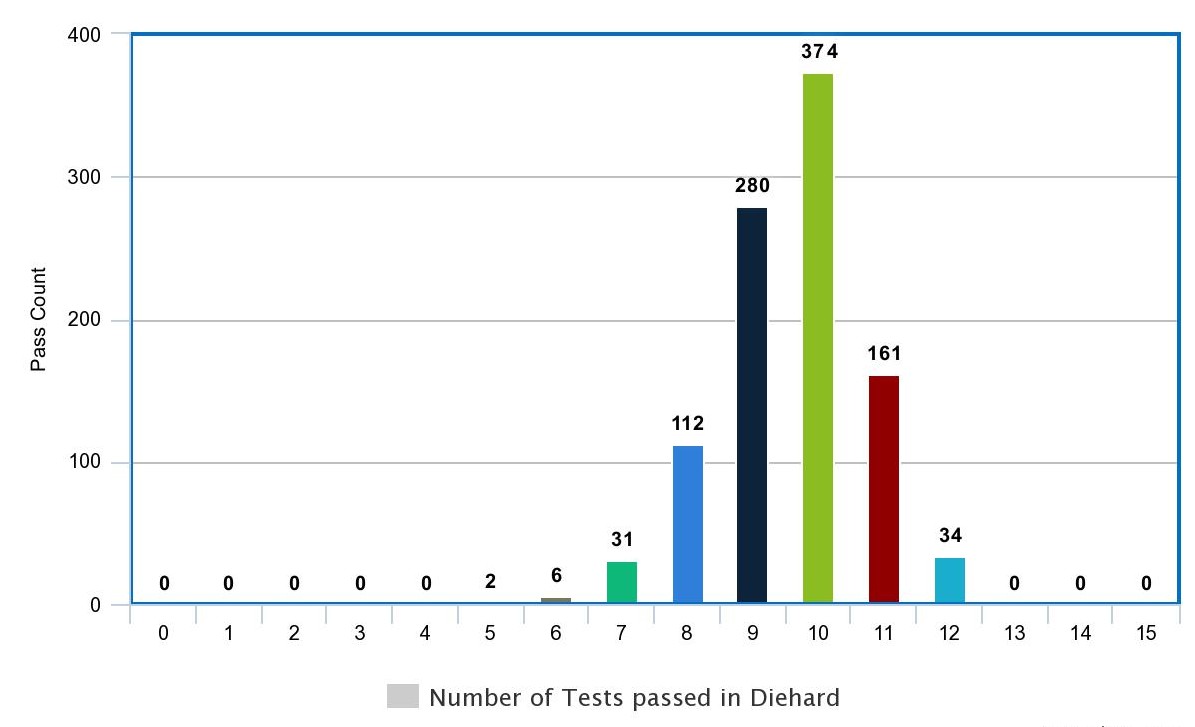}
             	}
         	\caption{Test results of PRNGs for $1000$ seeds with Diehard battery of tests to get a rough estimate (continued)}
	\label{fig:avg_plot3}
\end{figure}

In this ranking, however, there are many ties. To improve the ranking, we next use $1000$ seeds, as mentioned before, and observe the rough estimate of the performance of PRNGs in Diehard battery.
\begin{figure}[hbtp]
	\centering
	\vspace{-2.0em}
	\subfloat[MMIX\label{mmix_avg}]{%
		\includegraphics[width=0.3\linewidth, height=3.0cm]{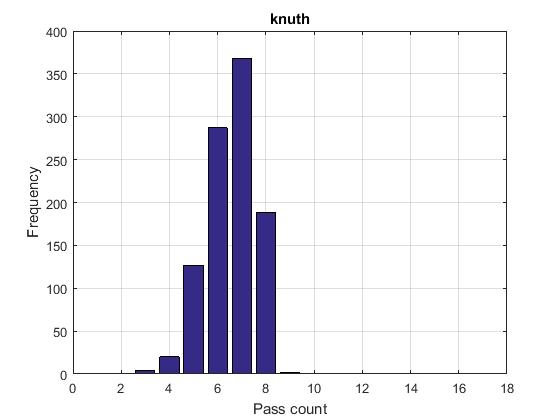}}
	\hfill
	\subfloat[minstd\_rand\label{minstd_avg}]{%
		\includegraphics[width=0.3\linewidth, height=3.0cm]{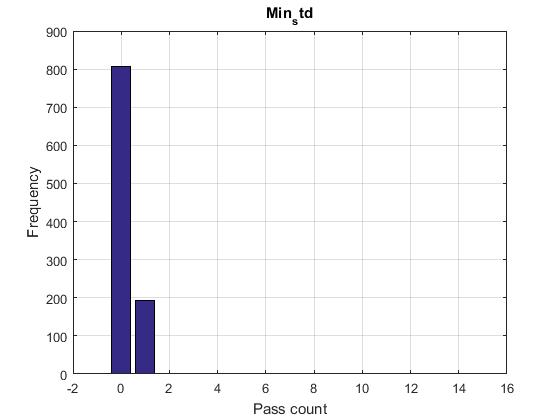}}
	\hfill
	\subfloat[Borland's LCG\label{borland_avg}]{%
		\includegraphics[width=0.3\linewidth, height=3.0cm]{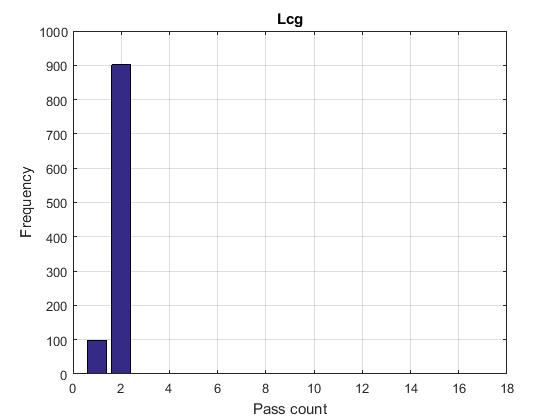}}
	\hfill  
	\subfloat[lrand48()\label{lrand_avg}]{%
		\includegraphics[width=0.3\linewidth, height=3.0cm]{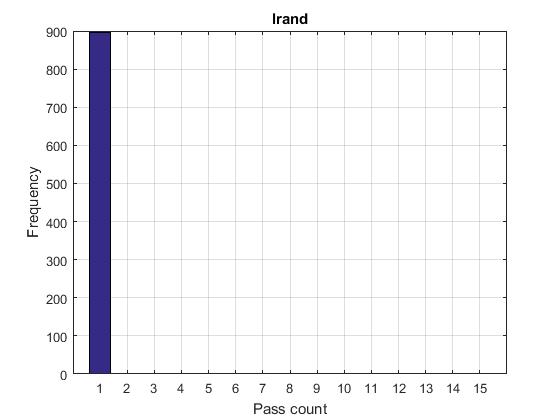}}
	\hfill
	\subfloat[MRG31k3p\label{mrg_avg}]{%
		\includegraphics[width=0.3\linewidth, height=3.0cm]{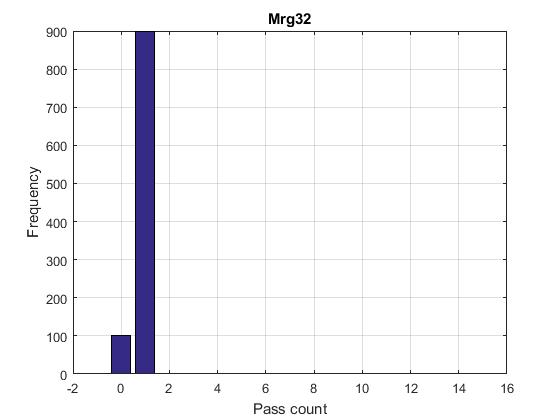}
	}
	\hfill
	\subfloat[PCG-32\label{pcg_avg}]{%
		\includegraphics[width=0.3\linewidth, height=3.0cm]{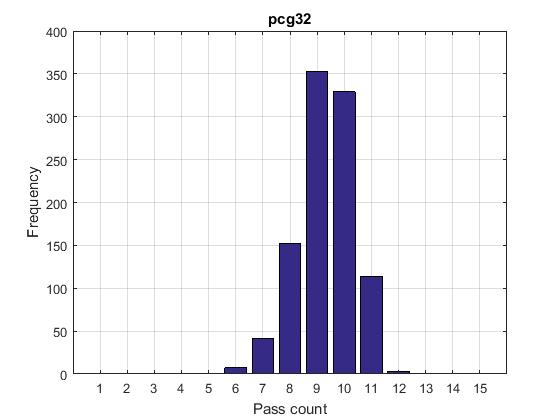}}
	\hfill
	\subfloat[random\label{random_avg}]{%
		\includegraphics[width=0.3\linewidth, height=3.0cm]{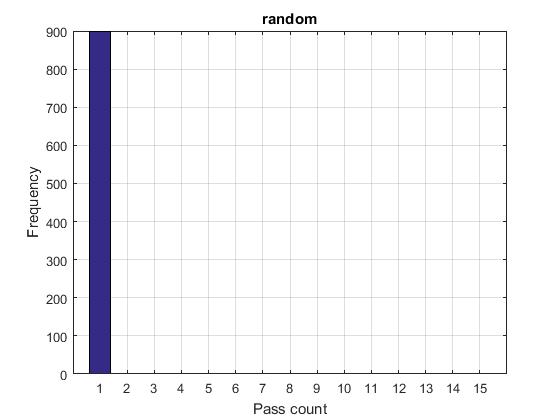}}
	\hfill
	\subfloat[Tauss88\label{tauss_avg}]{%
		\includegraphics[width=0.3\linewidth, height=3.0cm]{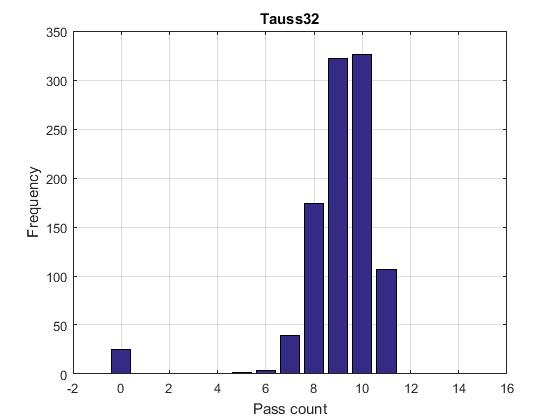}
	}
	\hfill  
	\subfloat[LFSR113\label{lfsr113_avg}]{%
		\includegraphics[width=0.3\linewidth, height=3.0cm]{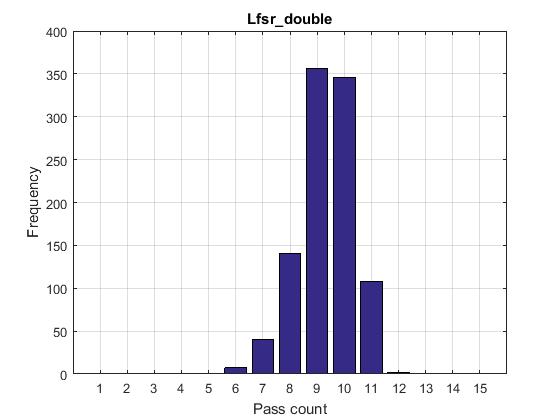}}
	\hfill
	\subfloat[LFSR258\label{lfsr258_avg}]{%
		\includegraphics[width=0.3\linewidth, height=3.0cm]{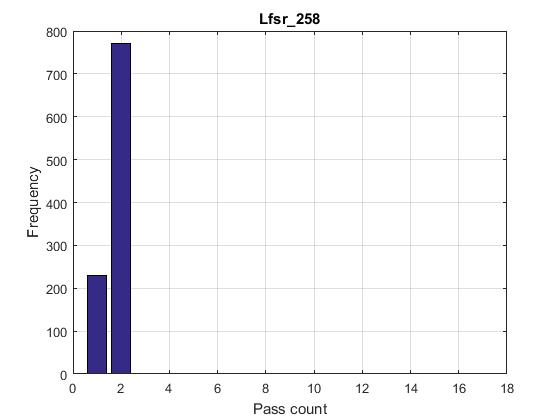}}
	\hfill  
	\subfloat[WELL512a\label{well512_avg}]{%
		\includegraphics[width=0.3\linewidth, height=3.0cm]{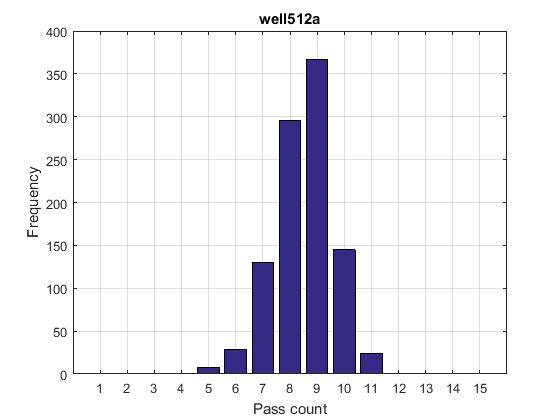}}
	\hfill
	\subfloat[WELL1024a\label{well1024_avg}]{%
		\includegraphics[width=0.3\linewidth, height=3.0cm]{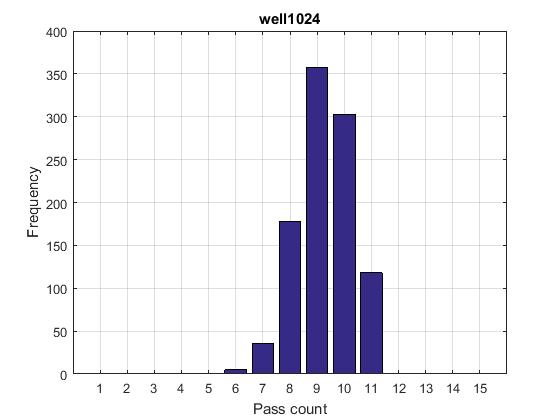}}
	\hfill
	\subfloat[MT19937-32\label{mt32_avg}]{%
		\includegraphics[width=0.3\linewidth, height=3.0cm]{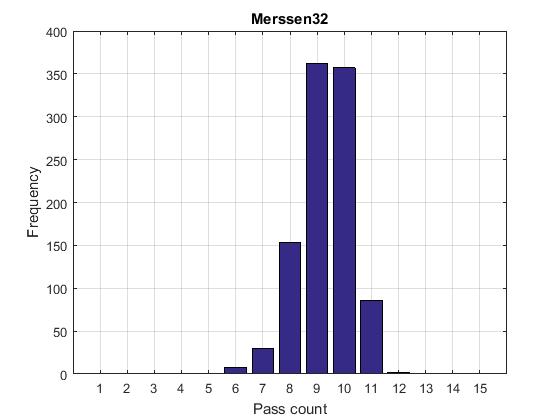}}
	\hfill
	\subfloat[MT19937-64\label{mt64_avg}]{%
		\includegraphics[width=0.3\linewidth, height=3.0cm]{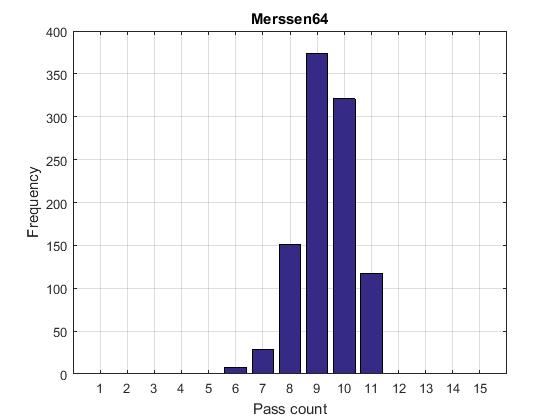}}
	\hfill
	\subfloat[dSFMT19937-32\label{dsfmt32_avg}]{%
		\includegraphics[width=0.3\linewidth, height=3.0cm]{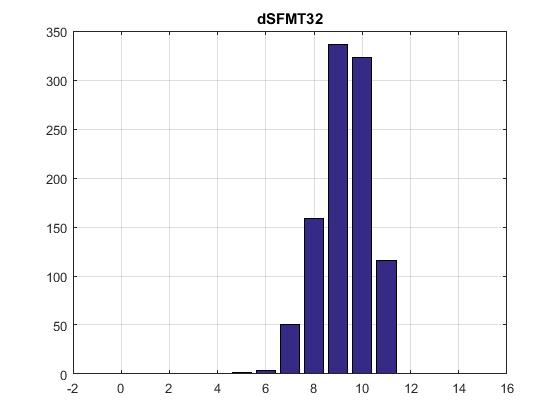}}
	
	\caption{Test results of PRNGs for $1000$ seeds with Diehard battery of tests to get a rough estimate}
	\label{fig:avg_plot2}
\end{figure}
Figure~\ref{fig:avg_plot2} and  Figure~\ref{fig:avg_plot3} show the plots for all the PRNGs (except \verb|rand|, as \verb|rand| is used to generate the seeds). For each of the figures, $x$ axis represents the number of tests passed by a PRNG and $y$ axis denotes the frequency of passing these tests. By using these plots, we can get a second level of ranking of the PRNGs, as shown in Table~\ref{tab:blind_test_avg}.   
We find rough estimate of performance by calculating $\frac{\sum\limits_i f_i \times t_i}{1000}$, where $f_i$ is the frequency of a PRNG to pass $t_i$ number of tests for $i^{th}$ seed. The column `Range' in Table~\ref{tab:blind_test_avg} indicates minimum and maximum number of tests passed in the whole experiment for a particular PRNG. The new ranks are discussed below.
\begin{table*}[hbtp]
	\vspace{-0.8em}
	\setlength{\tabcolsep}{1.5pt}
	\scriptsize
	\renewcommand{\arraystretch}{1.20}
	\centering
	\small
	\caption{Summary of Statistical test results for different seeds}
	\label{tab:blind_test_avg}
	\resizebox{0.90\textwidth}{4.5cm}{
		\begin{tabular}{|c|c|c|c|c|c|c|c|c|}
			\hline
			\multicolumn{2}{|c|}{\multirow{2}{*}{\theadfont{Name of the PRNGs}}} & \multicolumn{3}{c|}{\theadfont{Fixed Seeds}} &  \multicolumn{2}{c|}{\theadfont{Random Seeds}} & \multirow{2}{*}{\theadfont{Previous Rank}} & \multirow{2}{*}{\theadfont{$2^{nd}$ level Rank}} \\
			\cline{3-7}
			\multicolumn{2}{|c|}{ } & Diehard & TestU01 & NIST & Estimate & Range &  & \\
			\hline
			\multirow{7}{*}{\rotatebox{90}{LCGs}}& MMIX & 4-6 & 16-19 & 7-8 & 6.5 & 2-9  & 8 & 9\\
			\cline{2-9}
			& minstd\_rand & 0 & 1 & 1-2 &0.38 & 0-1 & 12 & 14\\
			\cline{2-9}
			& Borland LCG & 1 & 3 & 4-5 & 1.9 & 1-2 & 11 & 12\\
			\cline{2-9}
			& rand & 1 & 1-3 & 2-3 &  &  & 11 & 13\\
			\cline{2-9}
			& lrand48 & 1 & 2-3 & 2 & 1 & 1 & 11 & 13\\
			\cline{2-9}
			& MRG31k3p & 0-1 & 1-2 & 1-2 & 0.9 & 0-1 & 12 & 14\\
			\cline{2-9}
			& PCG-32 & 9-11 & 24-25 & 14-15 & 9.3 & 6-12 & 2 & 4\\
			\cline{2-9}
			\hline
			\multirow{16}{*}{\rotatebox{90}{LFSRs}}& random & 1 & 1-3 & 1 & 1 & 1 & 11 & 13\\
			\cline{2-9}
			& Tauss88 & 9-11 & 21-23 & 14-15 & 9.0 & 0-12 & 4 & 7\\
			\cline{2-9}
			& LFSR113 & 5-11 & 6-23 & 1-15 & 9.3 & 6-12 & 7 & 7\\
			\cline{2-9}
			& LFSR258 & 0-1 & 0-5 & 0-2 &  1.8 & 1-2 & 12 & 14\\
			\cline{2-9}
			& WELL512a & 7-10 & 23 & 14-15 & 8.5 & 5-11 & 5 & 6\\
			\cline{2-9}
			& WELL1024a & 9-10 & 24-25 & 14-15 & 9.2 & 6-11 & 3 & 4\\
			\cline{2-9}
			& MT19937-32 & 9-10 & 25 & 13-15 & 9.3 & 6-12 & 3 & 4\\
			\cline{2-9}
			& MT19937-64 & 8-11 & 24-25 & 15 & 9.4 & 6-11 & 2 & 3\\
			\cline{2-9}
			& SFMT19937-32 & 9-10 & 25 & 15 & 9.5 & 5-12 & 1 & 1\\
			\cline{2-9}
			& SFMT19937-64 & 9-11 & 25 & 15 & 9.52 & 6-12 & 1 & 1\\
			\cline{2-9}
			& dSFMT-32 & 7-11 & 24-25 & 13-15 & 9.3 & 5-11 & 5 & 5\\
			\cline{2-9}
			& dSFMT-52 & 5-7 & 9-11 & 3 & 5.98 & 3-7 & 9 & 10\\
			\cline{2-9}
			&  xorshift32 & 2-4 & 17 & 2-13 & 5.5 & 3-7 & 9 & 10\\
			\cline{2-9}
			&  xorshift64* & 7-10 & 25 & 14-15 & 8.0 & 6-11 & 5 & 6\\
			\cline{2-9}
			&  xorshift1024* & 6-9 & 20-21 & 6-15 & 7.0 & 4-9 & 6 & 8\\
			\cline{2-9}
			&  xorshift128+ & 8-10 & 24-25 & 14-15 & 9.4 & 6-12 & 4 & 4\\
			\hline
			\multirow{7}{*}{\rotatebox{90}{CAs}}& Rule $30$ & 8-11 & 24-25 & 15 & 10.2 & 7-12 & 2 & 2\\
			\cline{2-9}
			& Hybrid CA with Rules $30$ \& $45$ & 0-3 & 1-8 & 0-3 & 2.0 & 0-3 & 11 & 12\\
			\cline{2-9}
			& Maximal Length CA with $\gamma=0$ & 0-2 & 12 & 10-11 & 1.6 & 1-2 & 10 & 11\\
			\cline{2-9}
			& Maximal Length CA with $\gamma=1$ & 3-4 & 15-17 & 14 & 1.8 & 1-4 & 8 & 11\\
			\cline{2-9}
			& Non-linear $2$-state CA & 5-7 & 11 & 3-4 & 5.85 & 2-8 & 9 & 9\\
			\cline{2-9}
			& $3$-state CA  & ${2-3}$ & ${11-12}$ & ${4-6}$ & ${2.7}$ & ${1-4}$ & ${10}$ & ${11}$\\
			\cline{2-9}
			& Decimal CA  & ${9-11}$ & ${25}$ & ${15}$ & ${9.59}$ & ${6-12}$ & ${1}$ & ${1}$\\
			\hline
	\end{tabular}}
	\vspace{-1.0em}
\end{table*}

\begin{itemize}[leftmargin=1pt]
	\item In terms of (rough) estimate of tests passed and range, rule $30$ beats all other PRNGs. However, considering results of NIST and TestU01, Decimal CA and SFMTs still hold the first rank, while rule $30$ is ranked $2$. As estimated tests passed by \verb|MT19937-64| is better than \verb|PCG-32|, it is ranked $3$.
	
	\item Estimated tests passed by \verb|PCG-32|, \verb|MT19937-32|, \verb|xorshift128+| and \verb|WELL1024a| are similar and in terms of performance in NIST and TestU01 battery of tests, they are alike. So, all these PRNGs are ranked $4$.
	
	\item The next rank holders are \verb|dSFMT-32| (rank $5$), \verb|WELL512a| and \verb|xorshift64*| (rank $6$). We have observed that, \verb|Tauss88| sometimes fails to pass any tests of Diehard. So, it is put in the same group (rank $7$) with \verb|LFSR113|, which has a very good estimate of pass count despite having a not-so-good performance for the fixed seeds.
	
	\item {\verb|xorshift1024*|} has better rank (rank $8$) than non-linear $2$-state CA and \verb|MMIX| (rank $9$), because of performance in NIST and TestU01 library. 
	
	\item The next rank holders are \verb|xorshift32| and \verb|dSFMT-52| (rank $10$). 
	
	\item As estimated pass count of maximal-length CA with $\gamma=1$ is low, its rank is degraded. It is put in the same group as the $3$-state CA and maximal-length CA with $\gamma=0$ (rank $11$).
	
	\item Rule $30-45$ and Borland's LCG are ranked $12$, whereas other PRNGs previously on the same group, like \verb|rand|, \verb|lrand48| and \verb|random| are ranked $13$. 
	
	\item Like previous ranking, \verb|minstd_rand|, \verb|MRG31k3p| and \verb|LFSR258| are the last rank holders based on their overall performance and the fact that for many seeds, these PRNGs fails to pass any test.
\end{itemize}

{As mentioned in the introduction, blind tests have inherent incompleteness. Therefore, to deal with these shortcomings, in the next section, graphical tests are further incorporated on these PRNGs to verify whether this ranking is justified by visualization (human intervention) as well as to break the tie between intra-class PRNGs whenever possible.}

\subsubsection{Results of Graphical Tests}
\noindent As mentioned, we use two types of graphical tests -- lattice tests (2-D and 3-D) and space-time diagram test. Here, each of the PRNGs is tested using only the five fixed seeds, which are renamed as following for ease of presentation:
seed $7$ as $s_1$, $1234$ as $s_2$, $12345$ as $s_3$, $19650218$ as $s_4$ and seed $123456789123456789$ as $s_5$.
The motivation behind these graphical tests are -- 
$(a)$ to understand why some PRNGs perform very poorly, $(b)$ to differentiate the behavior of the PRNGs which perform similarly in the blind empirical tests and $(c)$ to visualize the randomness of the PRNGs and co-relate with the blind test results. The result of these tests are shown as following.

\begin{enumerate}[leftmargin=0pt]
	\item \textbf{Result of Lattice Tests:} For each of the seeds, the $2$-dimensional and $3$-dimensional lattice tests are performed on every PRNG. As expected, for \verb|rand|, \\~\verb|lrand48|, \verb|minstd_rand|, Borland's LCG, and \verb|random|, the points are either scattered or concentrated on a specific part of the $2$-D and $3$-D planes. However, for the good PRNGs like Decimal CA, MTs, SFMTs and WELL, the plots are relatively filled. For example, see Figure~\ref{fig:lattice} for output of \verb|MMIX|, \verb|Tauss88|, \verb|WELL1024a| \\
	and rule $30$.
	\begin{figure}[hbtp]
		\centering
		\vspace{-2.0em}
		\subfloat[MMIX ($2$-D)\label{2d_knuth}]{%
			\includegraphics[width=0.25\linewidth, height=2.0cm]{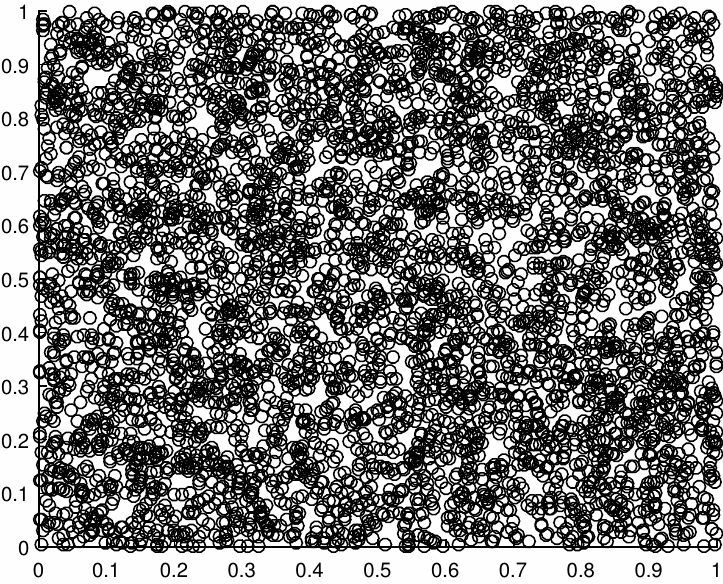}}
		\hfill
		\subfloat[MMIX ($3$-D)\label{3d_knuth}]{%
			\includegraphics[width=0.25\linewidth, height=2.0cm]{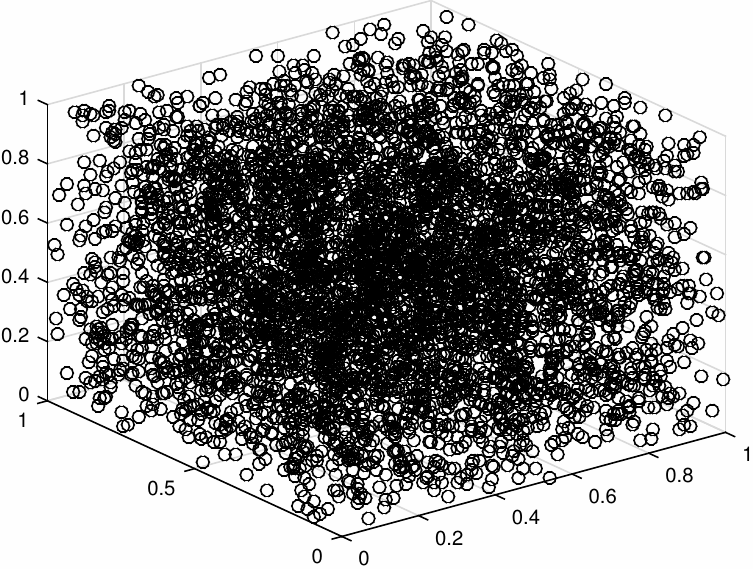}}
		\hfill
		\subfloat[Tauss88 ($2$-D)\label{2d_tauss_7-eps-converted-to}]{%
			\includegraphics[width=0.25\linewidth, height=2.0cm]{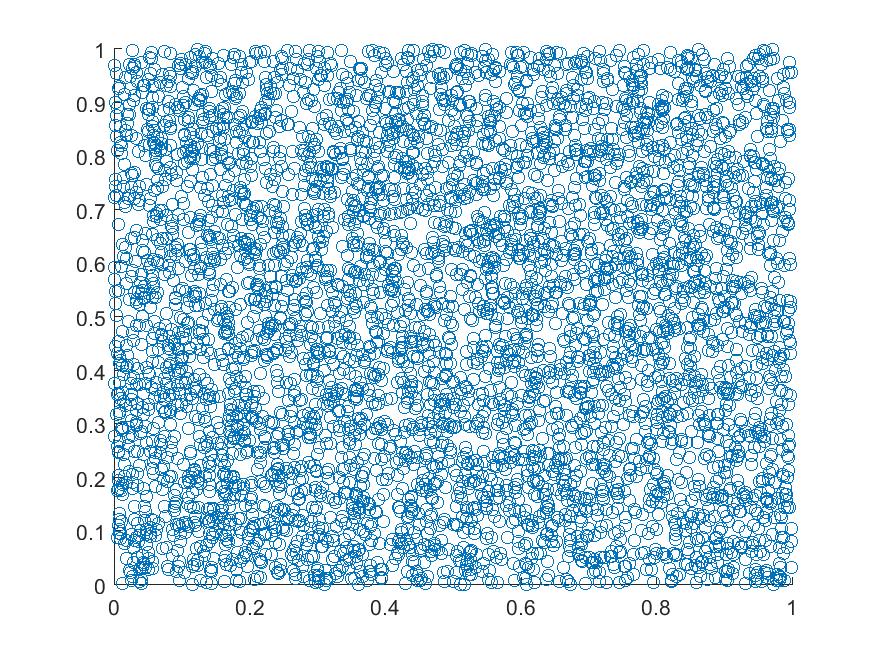}}
		\hfill
		\subfloat[Tauss88 ($3$-D)\label{3d_tauss_7-eps-converted-to}]{%
			\includegraphics[width=0.25\linewidth, height=2.0cm]{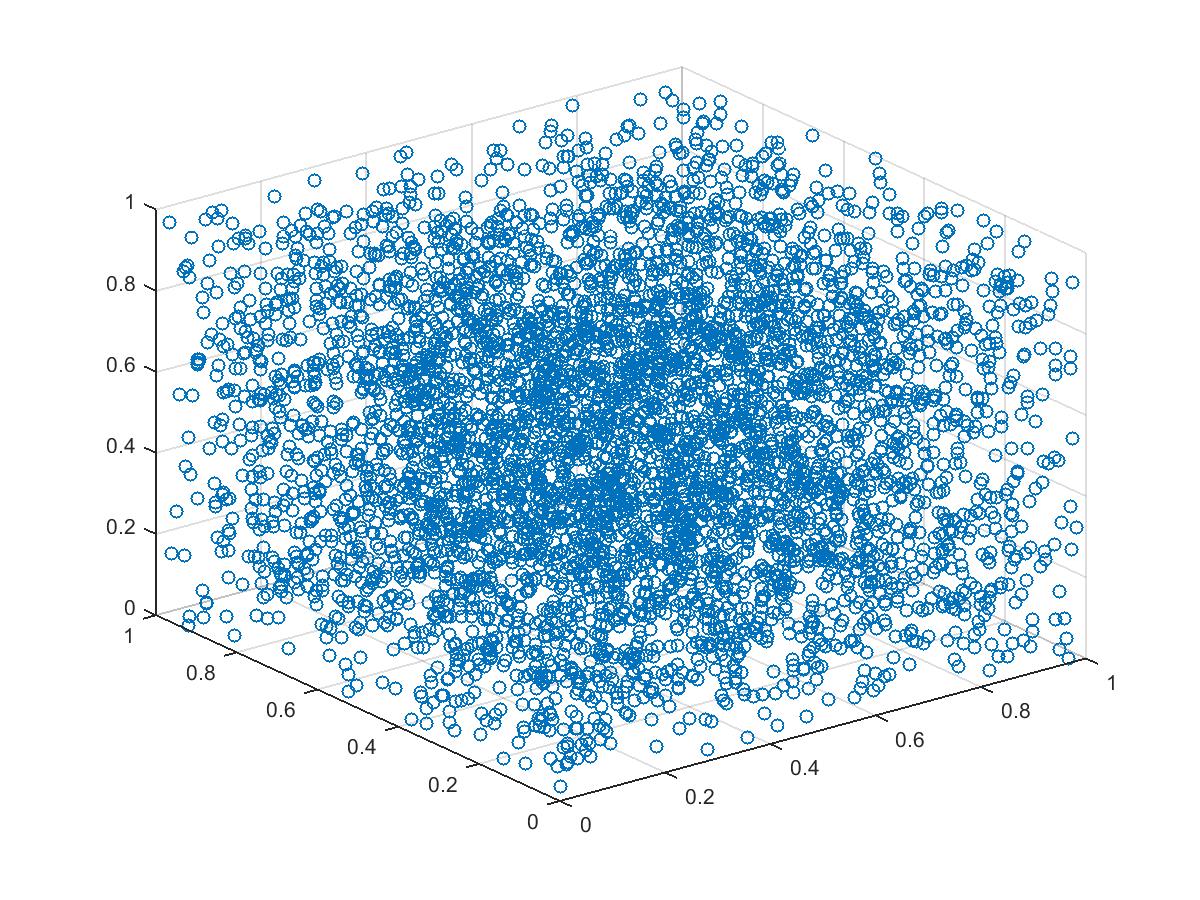}}
		\hfill\\
		\subfloat[WELL1024a ($2$-D)\label{2d_well1024_7-eps-converted-to}]{%
			\includegraphics[width=0.25\linewidth, height=2.0cm]{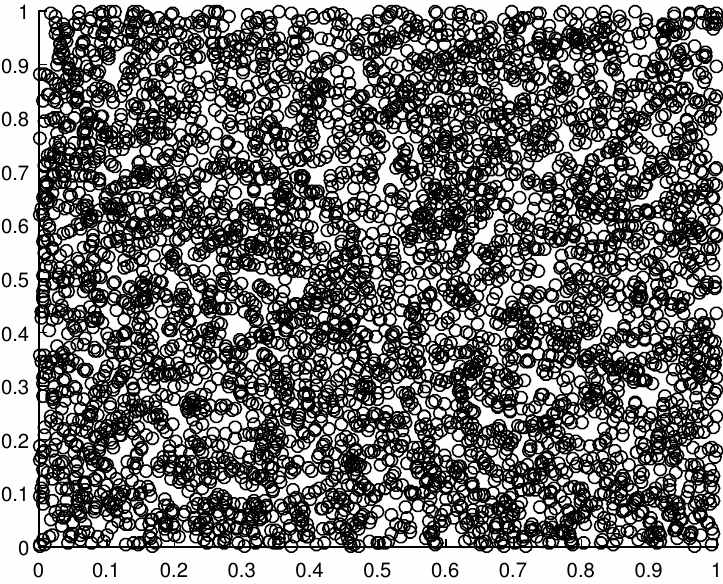}}
		\hfill
		\subfloat[WELL1024a ($3$-D)\label{3d_well1024_7}]{%
			\includegraphics[width=0.25\linewidth, height=2.0cm]{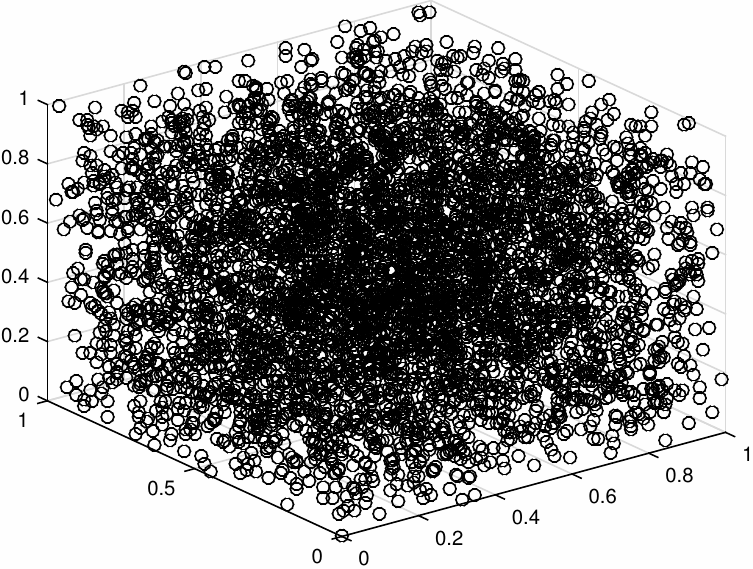}}
		\hfill
		\subfloat[Rule $30$ ($2$-D)\label{2d_rule30_7}]{%
			\includegraphics[width=0.25\linewidth, height=2.0cm]{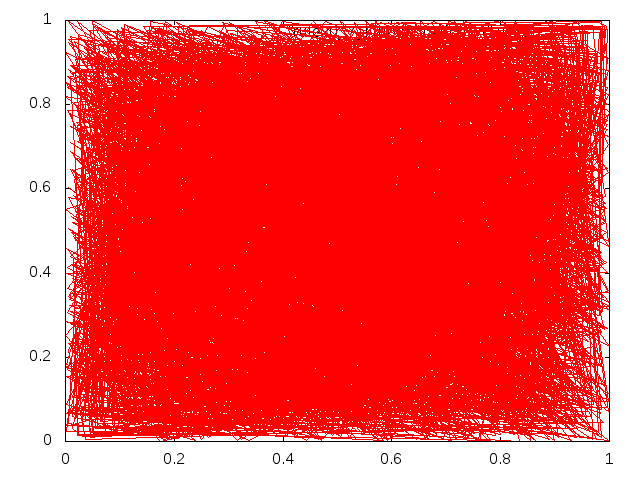}}
		\hfill
		\subfloat[Rule $30$ ($3$-D)\label{3d_rule30_7}]{%
			\includegraphics[width=0.25\linewidth, height=2.0cm]{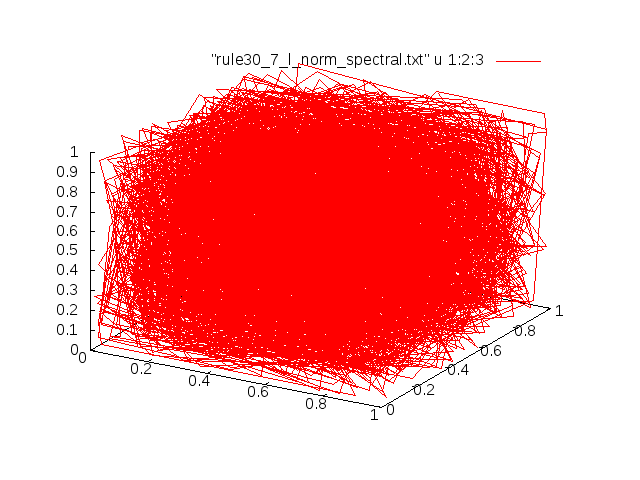}}
		\caption{Lattice Test results for rand, MMIX, Tauss88, WELL1024a 
			and Rule $30$ with $s_1$
		}
		\label{fig:lattice}
	\end{figure}
	We have observed that, if a PRNG performs badly, then in the corresponding lattice test diagrams, there is correlation between the numbers. Hence, these plots of lattice test verify our previous ranking based on blind tests.  
	However, this test fails to further enhance or modify the ranking shown in Table~\ref{tab:blind_test_avg}.
	So, we avoid supplying all the images of lattice test but move to space-time diagram.

	\item \textbf{Result of Space-time Diagram Test:} For space-time diagram, a set of $250$ numbers are generated from each seed and printed on $X-Y$ plane. The space-time diagrams of $4$ seeds $s_1,s_3,s_4,s_5$ for each PRNG, are shown in Figure~\ref{fig:ca2_space-time}
	to
	\ref{fig:ca_space-time}. From these figures, we can observe the following:
	
		\begin{figure}[H]
	\centering
	\subfloat[$s_1$\label{3state_7_space}]{%
		\includegraphics[width=0.1\linewidth, height=5.0cm]{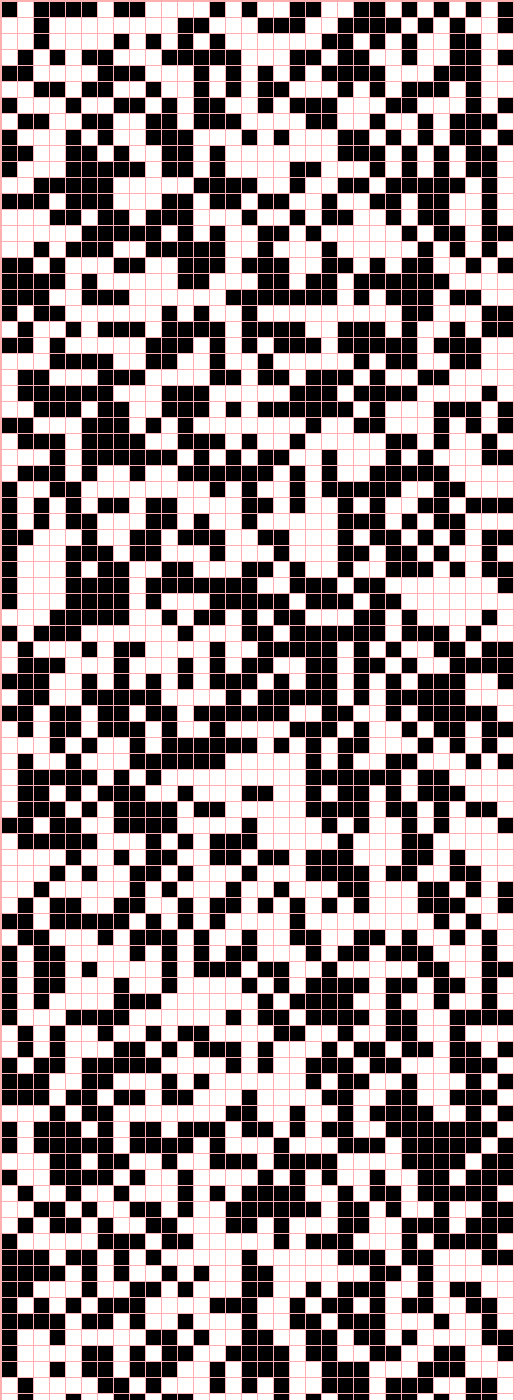}}
	\hfill
	\subfloat[$s_3$\label{3state_12345_space}]{%
		\includegraphics[width=0.1\linewidth, height=5.0cm]{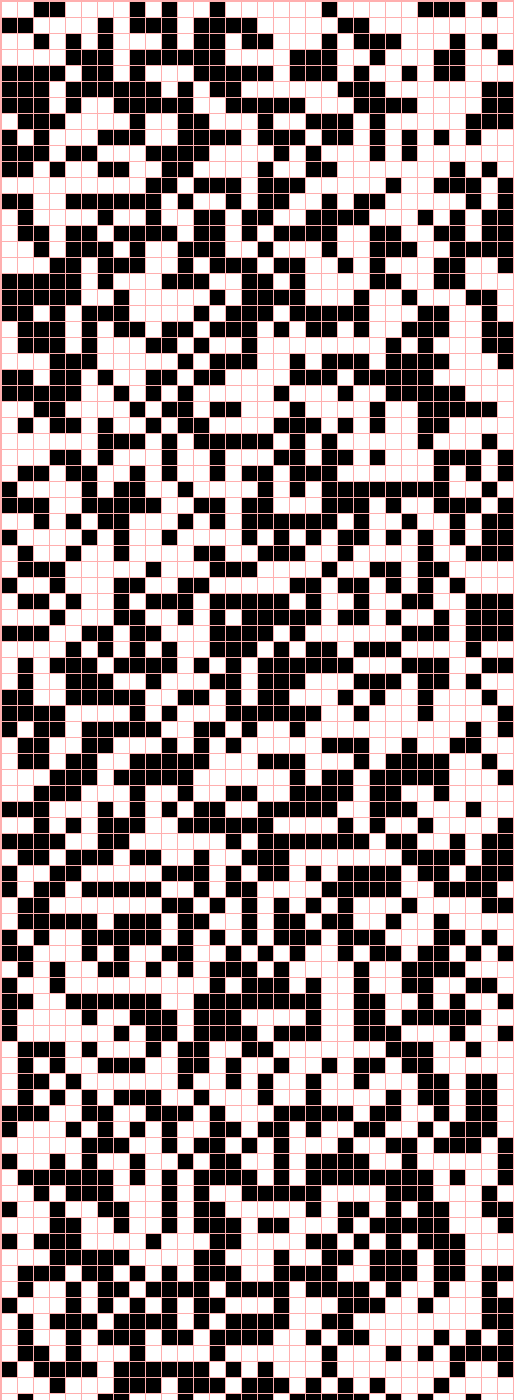}}
	\hfill
	\subfloat[$s_4$\label{3state_9650218_space}]{%
		\includegraphics[width=0.1\linewidth, height=5.0cm]{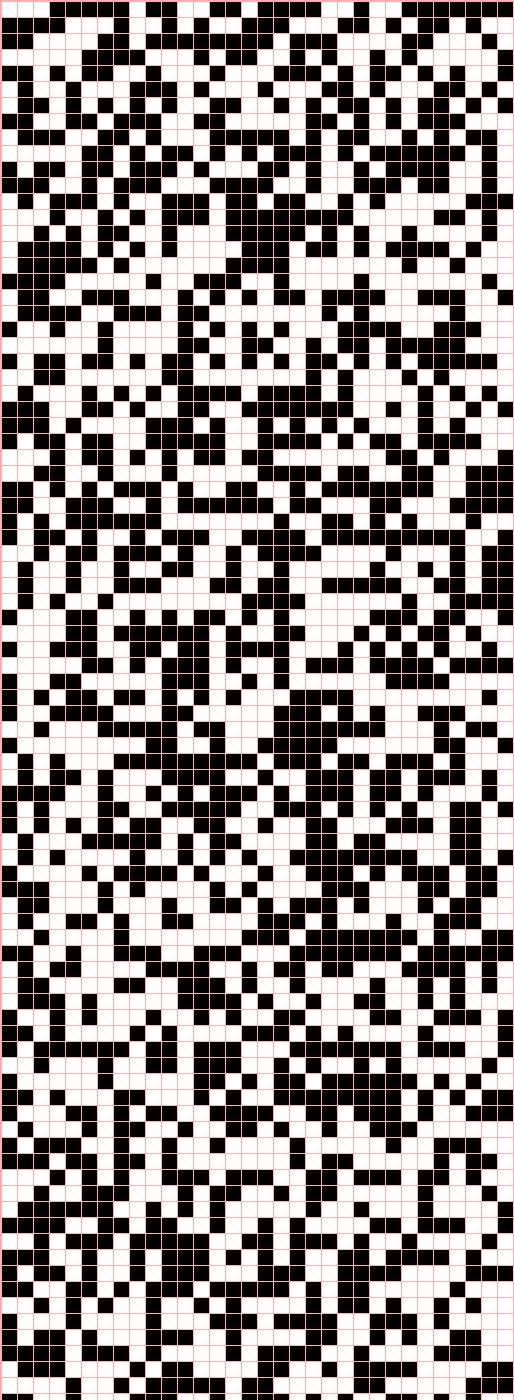}}
	\hfill
	\subfloat[$s_5$\label{3state_123456789123456789_space}]{%
		\includegraphics[width=0.1\linewidth, height=5.0cm]{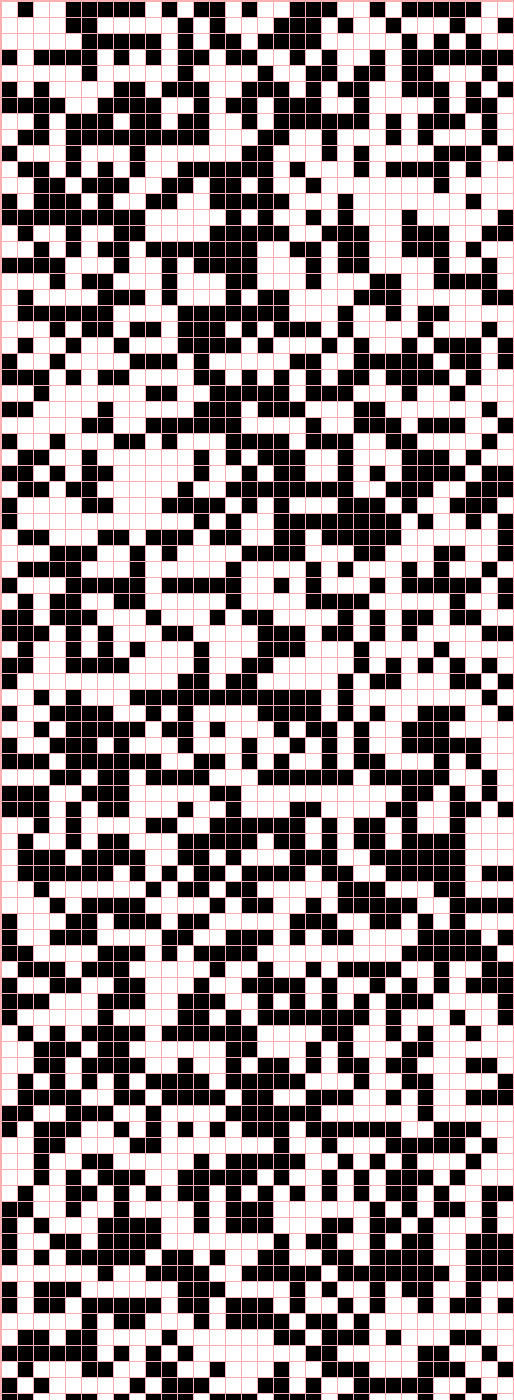}}
	\hfill
	\subfloat[$s_1$\label{10state_7_space}]{%
		\includegraphics[width=0.1\linewidth, height=5.0cm]{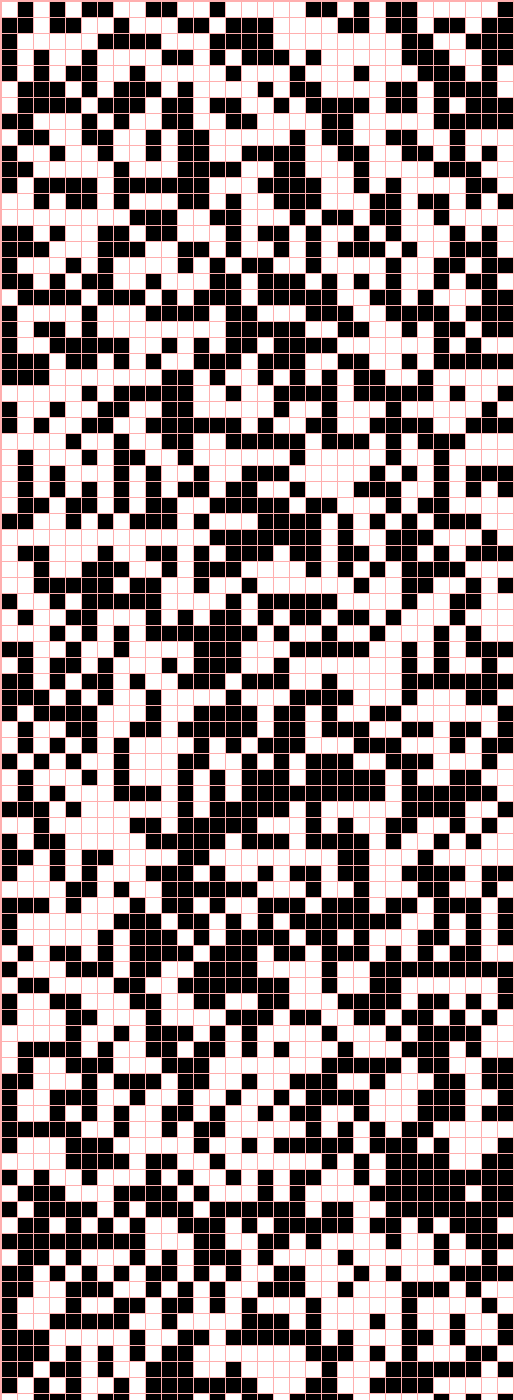}}
	\hfill
	\subfloat[$s_3$\label{10state_12345_space}]{%
		\includegraphics[width=0.1\linewidth, height=5.0cm]{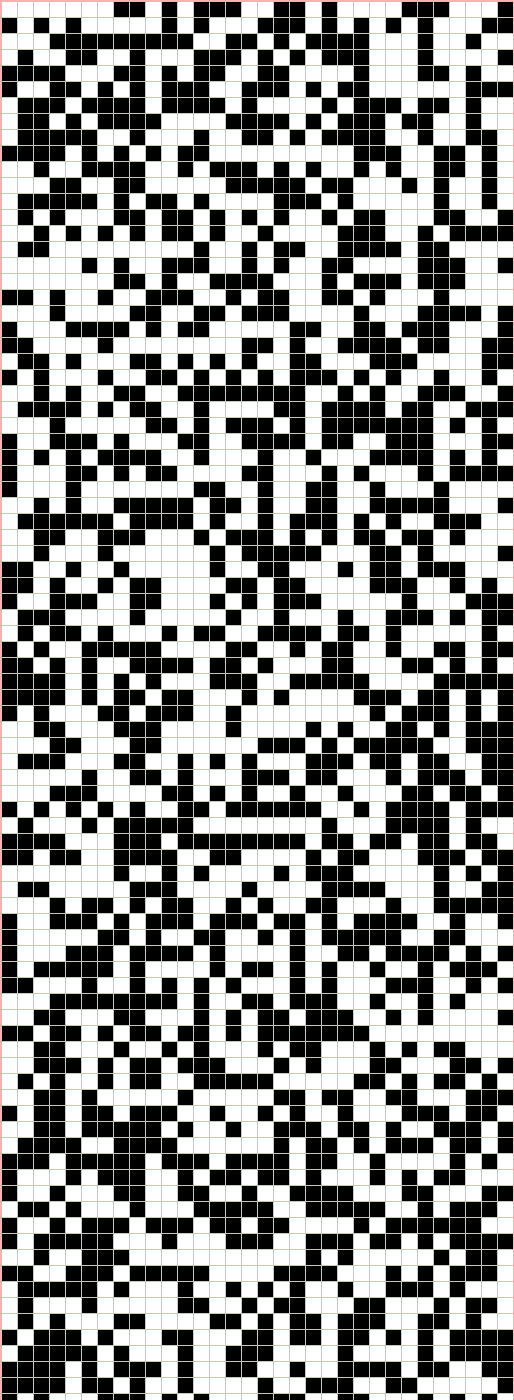}}
	\hfill
	\subfloat[$s_4$\label{10state_9650218_space}]{%
		\includegraphics[width=0.1\linewidth, height=5.0cm]{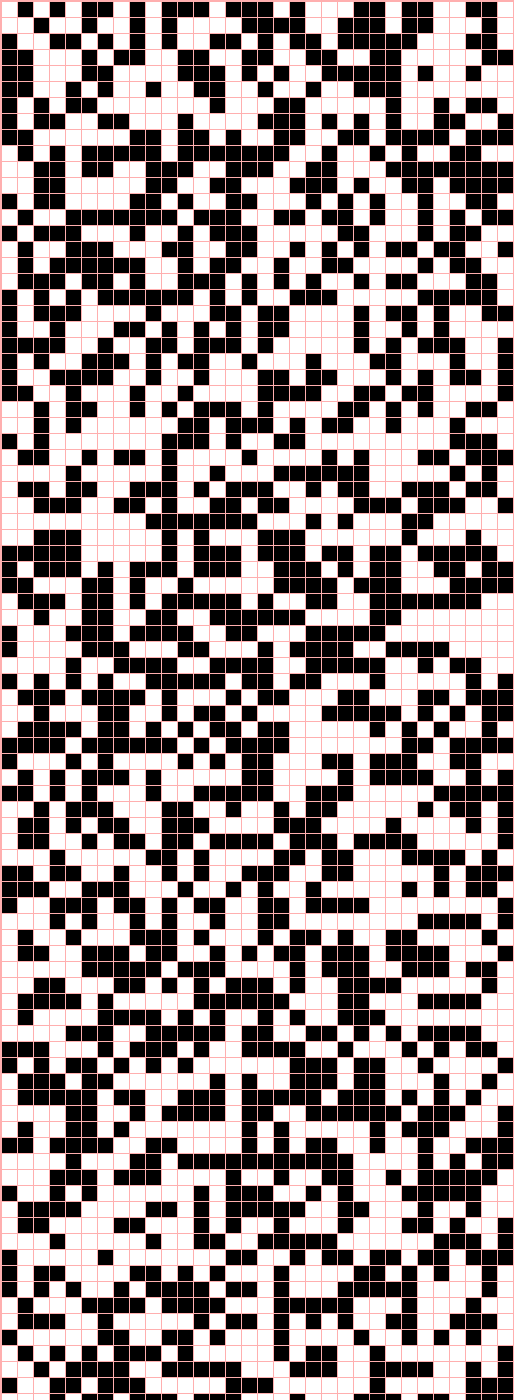}}
	\hfill
	\subfloat[$s_5$\label{10state_123456789123456789_space}]{%
		\includegraphics[width=0.2\linewidth, height=5.0cm]{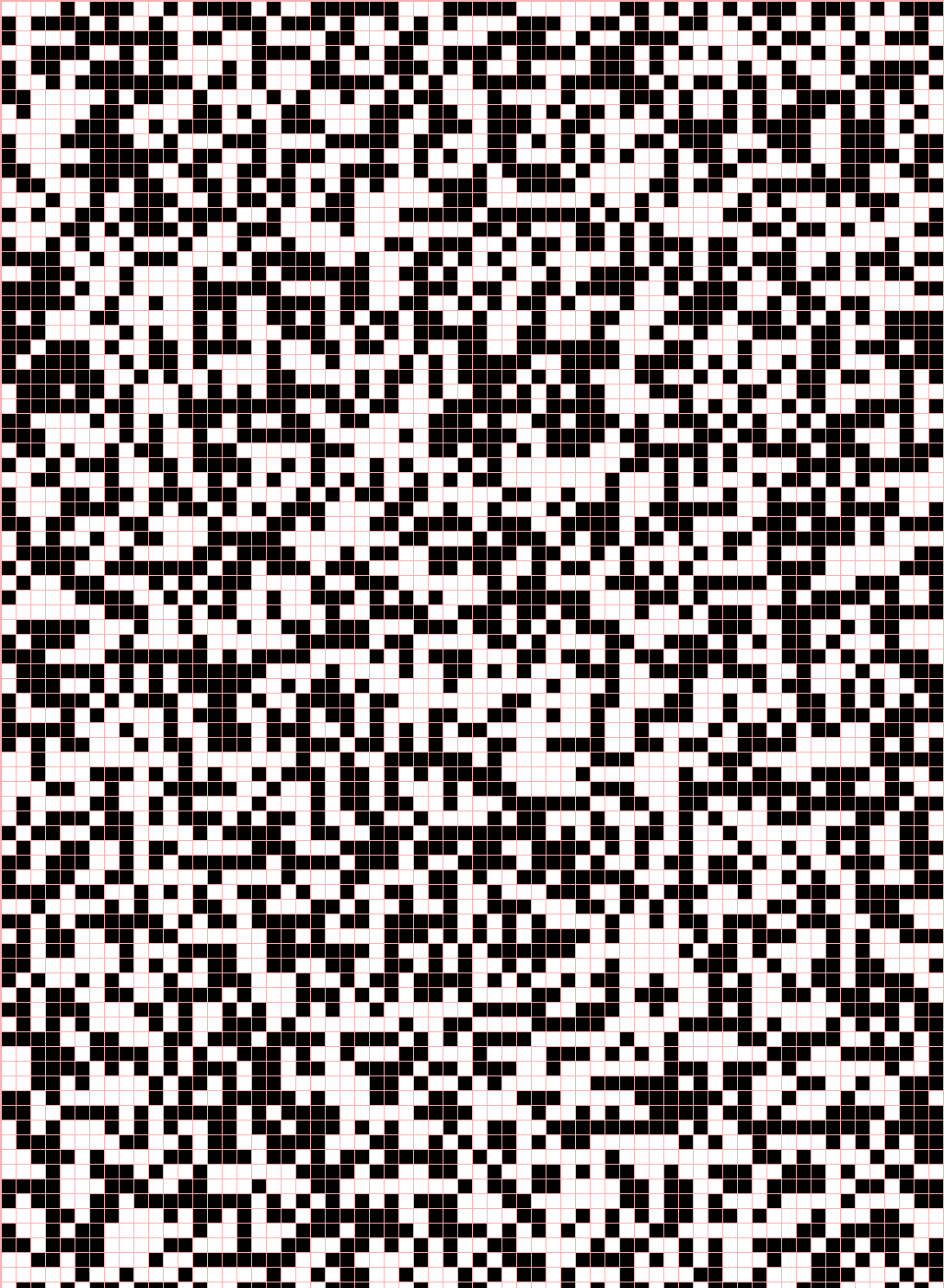}}
%
	\caption{Space-time diagram for $3$-state CA (\ref{3state_7_space}) to \ref{3state_123456789123456789_space}) and Decimal CA $1632405789$ (\ref{10state_7_space}) to \ref{10state_123456789123456789_space}). For seed $s_5$, the output of Decimal CA is taken as 64 bits.}
	\label{fig:ca2_space-time}
\end{figure}

	\begin{figure}[!h]
		\centering
		\vspace{-2.0em}
		\subfloat[$s_1$\label{knuth_lcg_7_space}]{%
			\includegraphics[width=0.2\linewidth, height=5.0cm]{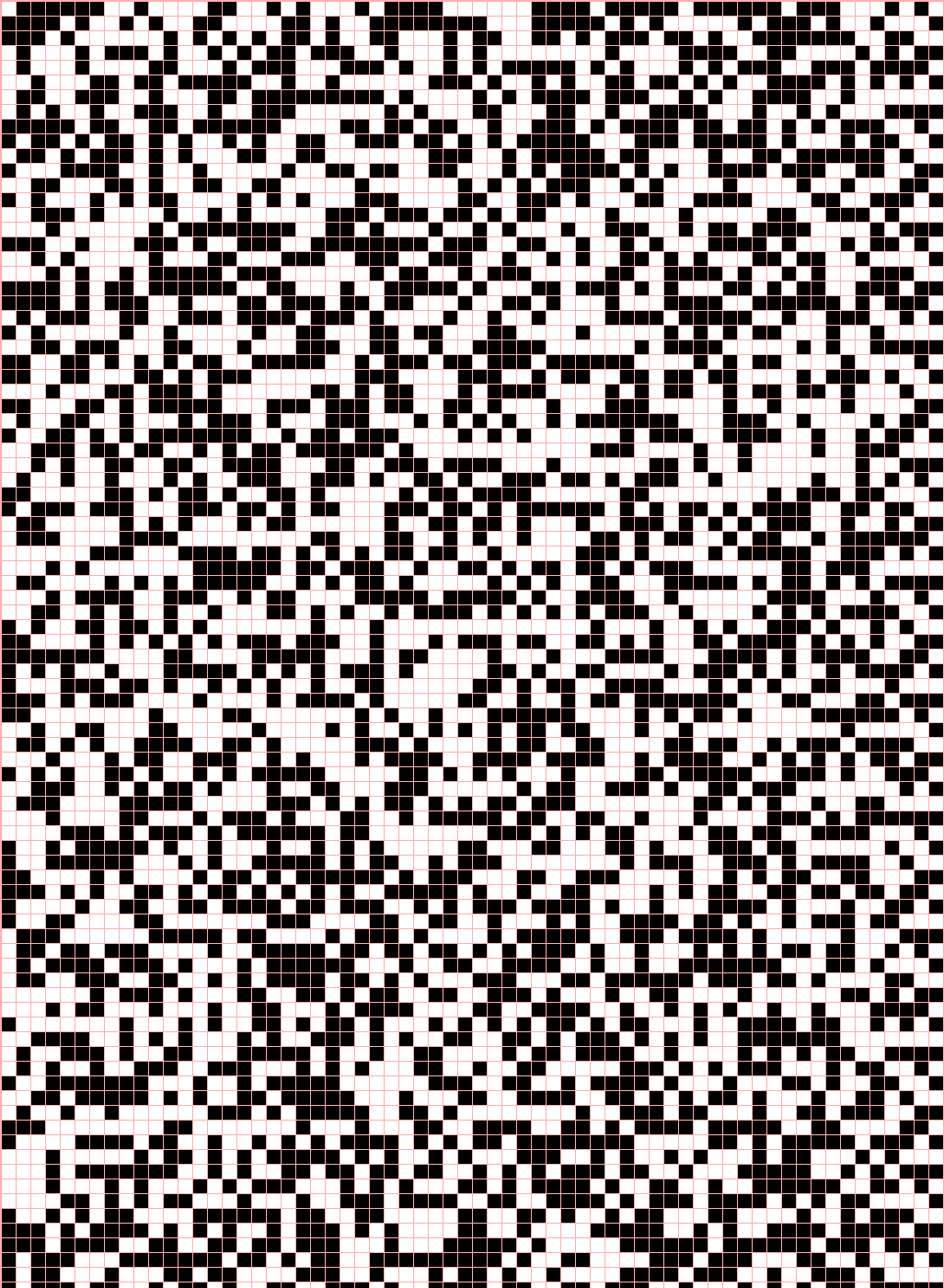}}
		\hfill
		\subfloat[$s_3$\label{knuth_lcg_12345_space}]{%
			\includegraphics[width=0.2\linewidth, height=5.0cm]{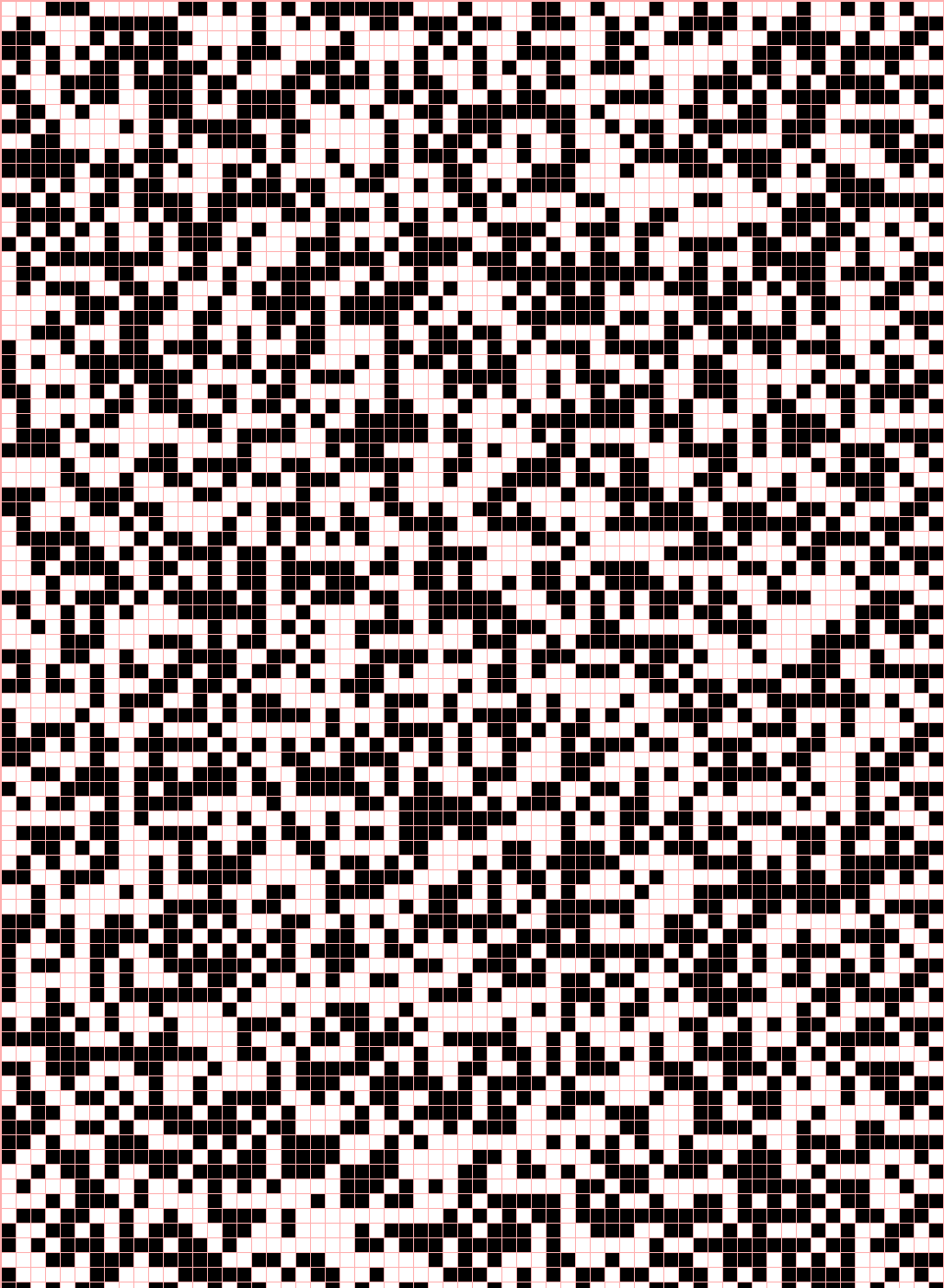}}
		\hfill
		\subfloat[$s_4$\label{knuth_lcg_9650218_space}]{%
			\includegraphics[width=0.2\linewidth, height=5.0cm]{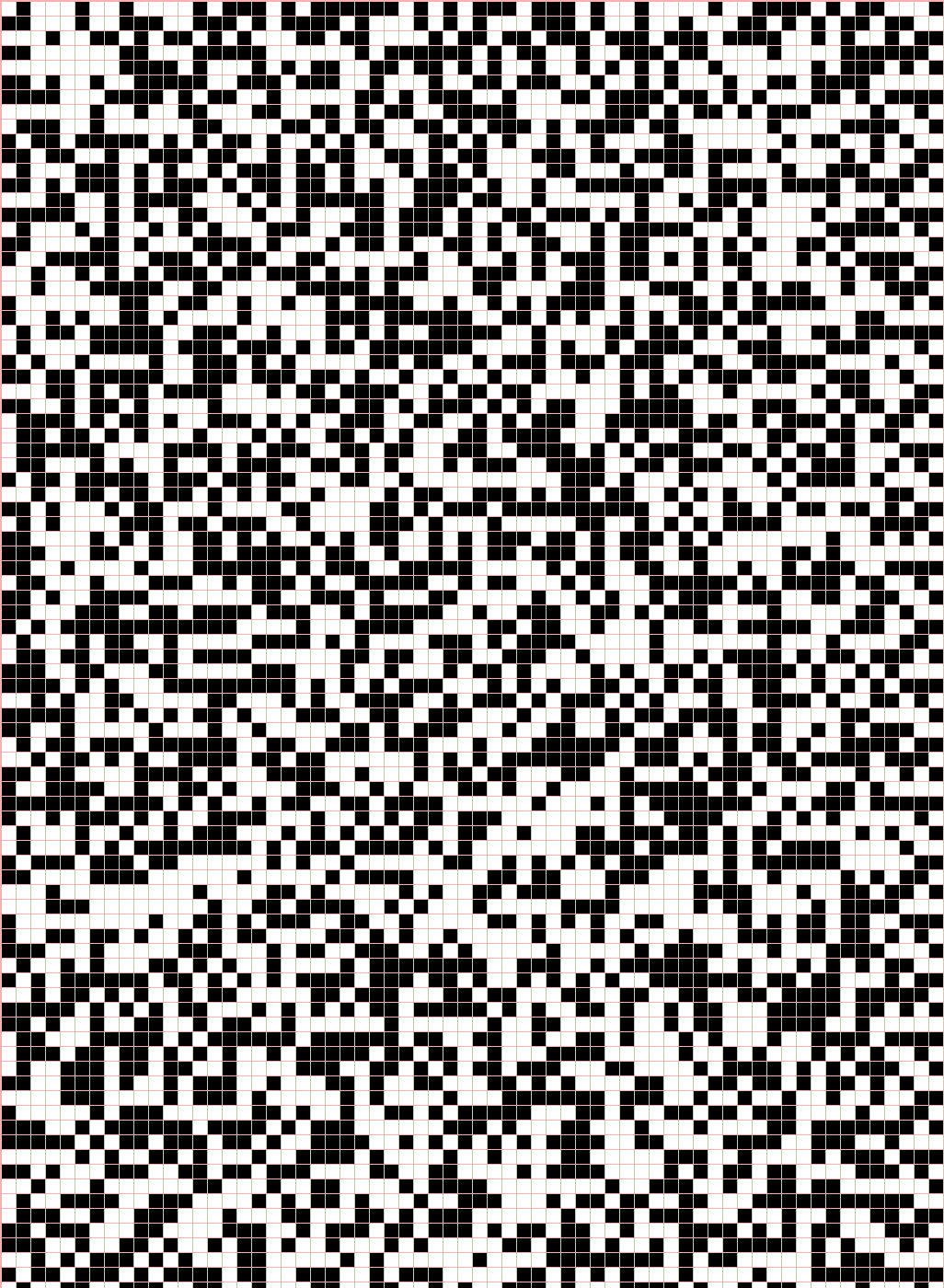}}
		\hfill
		\subfloat[$s_5$\label{knuth_lcg_123456789123456789_space}]{%
			\includegraphics[width=0.2\linewidth, height=5.0cm]{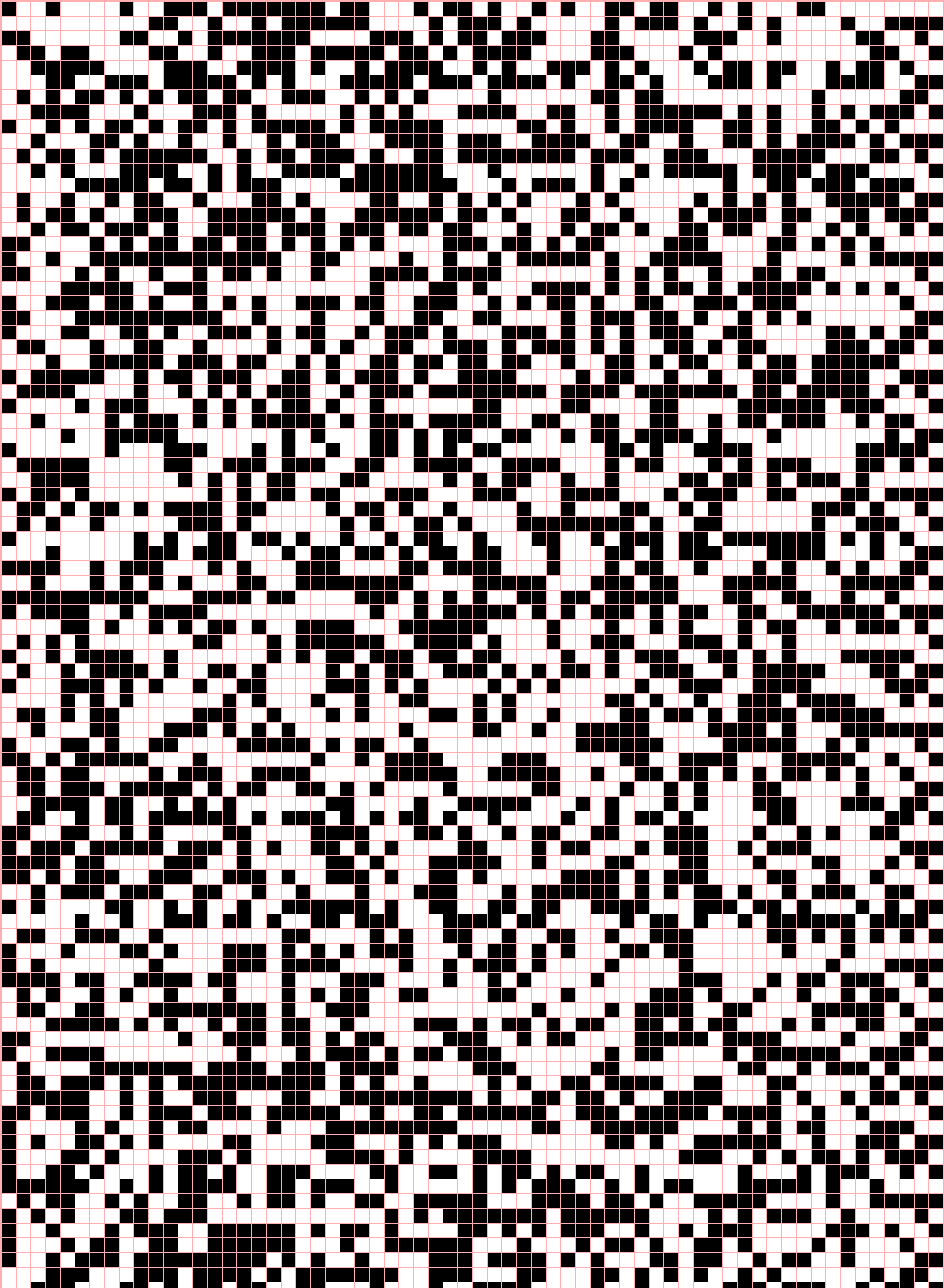}}
		%
		%
		\hfill\\
		\subfloat[$s_1$\label{borland_lcg_7_space}]{%
			\includegraphics[width=0.1\linewidth, height=5.0cm]{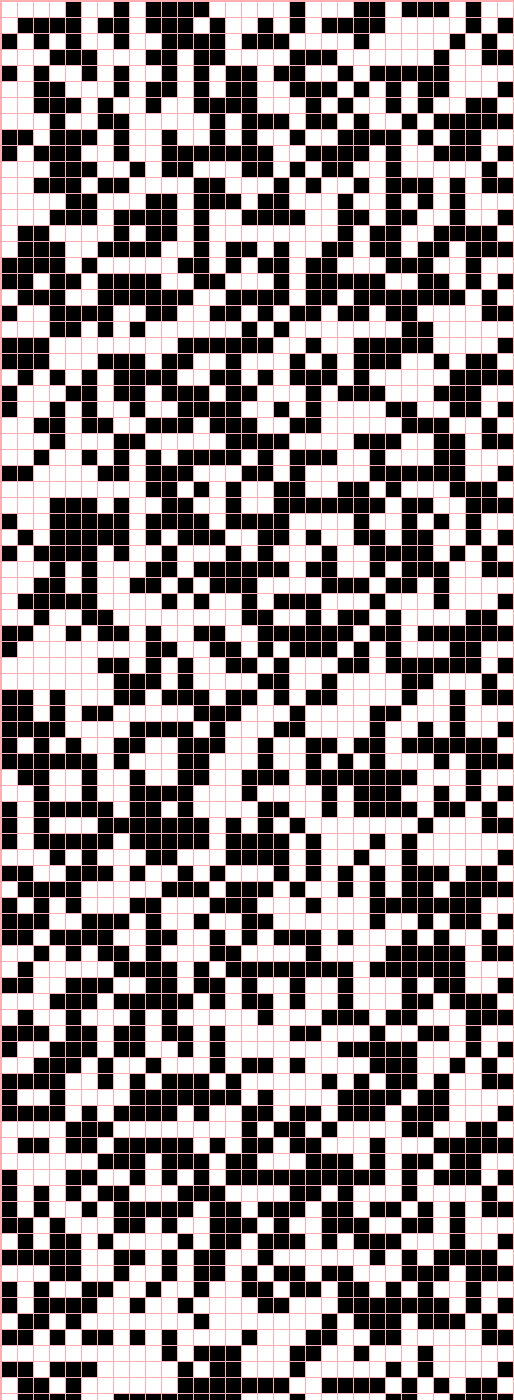}}
		\hfill
		\subfloat[$s_3$\label{borland_lcg_12345_space}]{%
			\includegraphics[width=0.1\linewidth, height=5.0cm]{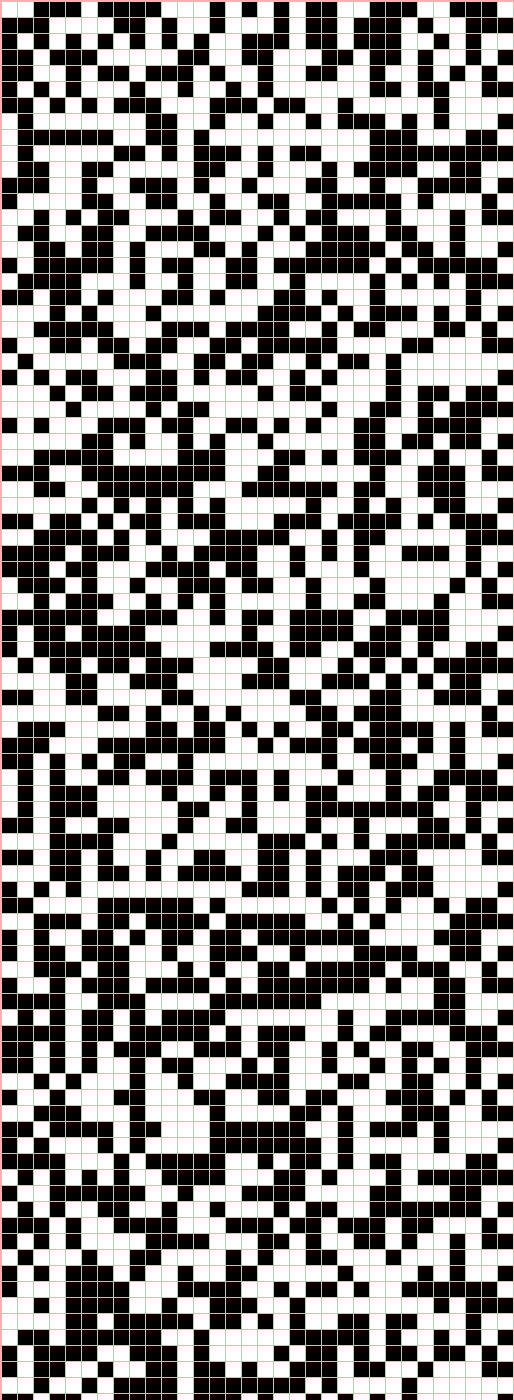}}
		\hfill
		\subfloat[$s_4$\label{borland_lcg_9650218_space}]{%
			\includegraphics[width=0.1\linewidth, height=5.0cm]{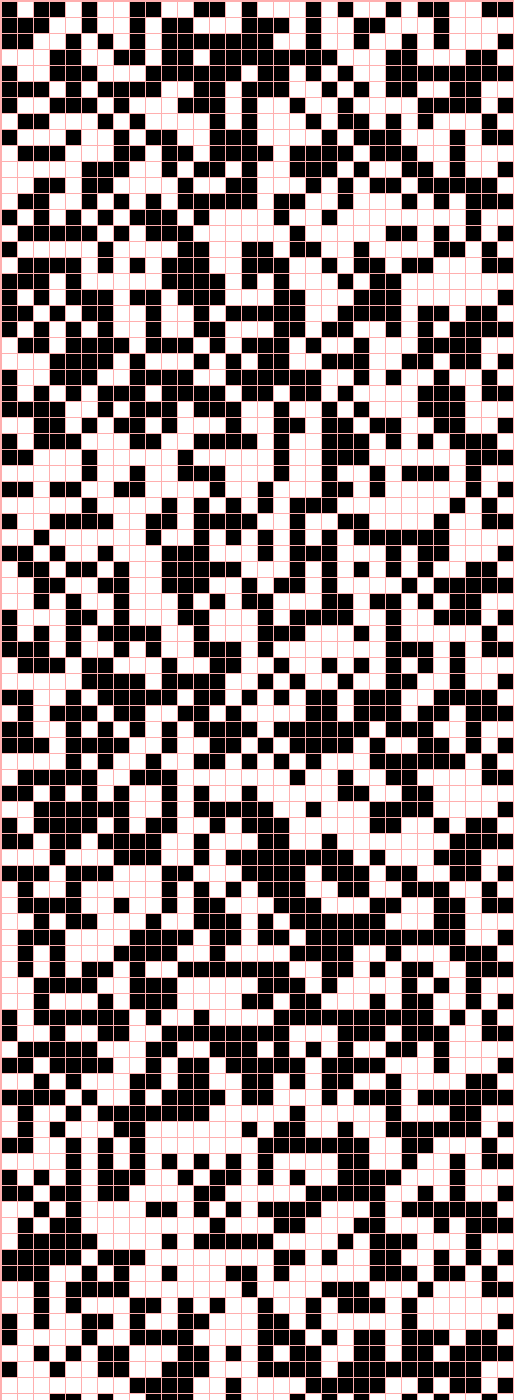}}
		\hfill
		\subfloat[$s_5$\label{borland_lcg_123456789123456789_spaceo}]{%
			\includegraphics[width=0.1\linewidth, height=5.0cm]{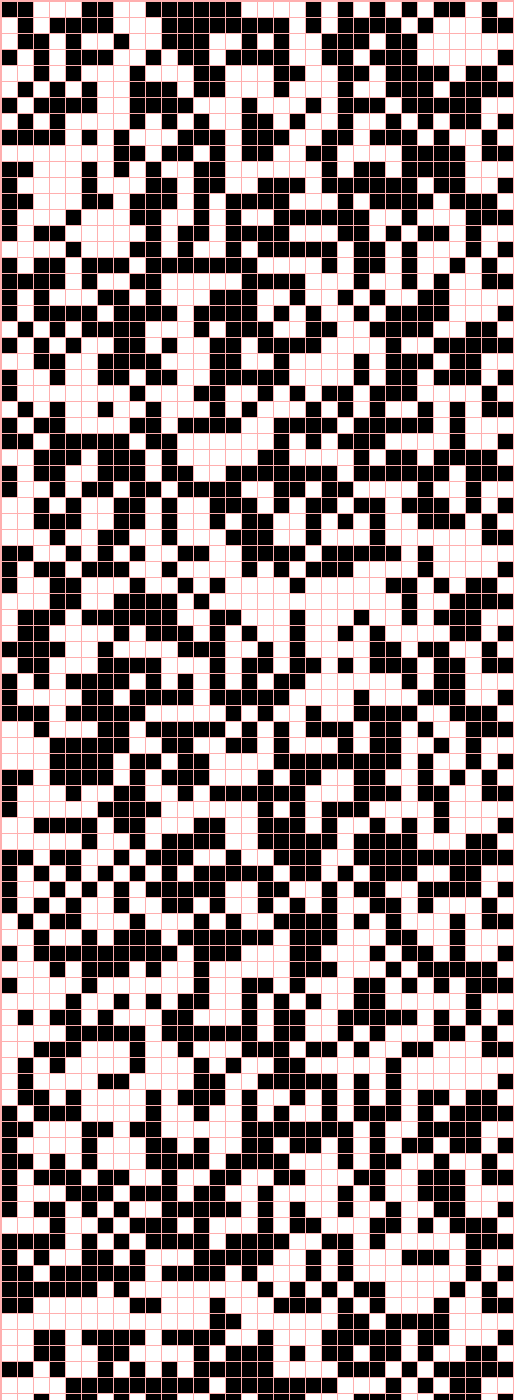}}
		\hfill
		\subfloat[$s_1$\label{min_std_7_space}]{%
			\includegraphics[width=0.1\linewidth, height=5.0cm]{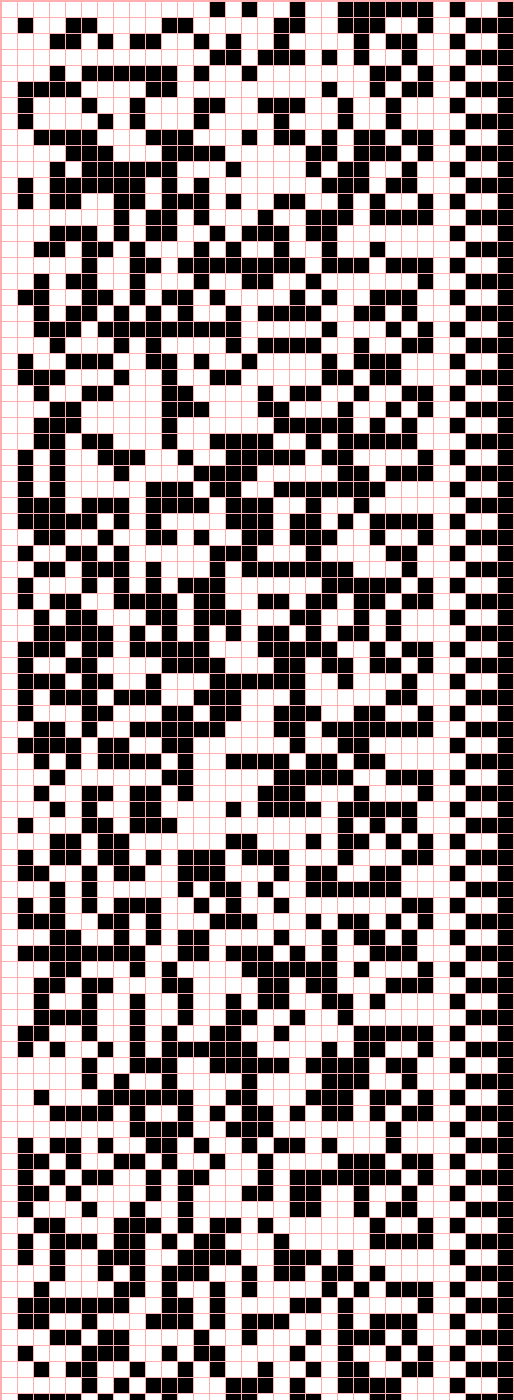}}
		\hfill
		\subfloat[$s_3$\label{min_std_12345_space}]{%
			\includegraphics[width=0.1\linewidth, height=5.0cm]{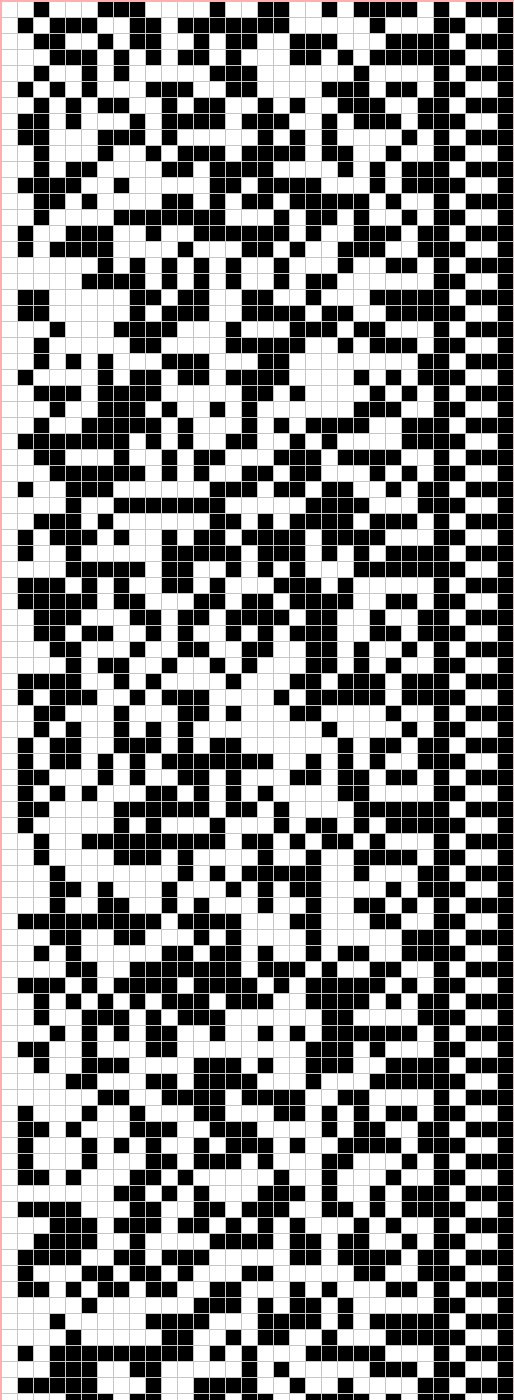}}
		\hfill
		\subfloat[$s_4$\label{min_std_9650218_space}]{%
			\includegraphics[width=0.1\linewidth, height=5.0cm]{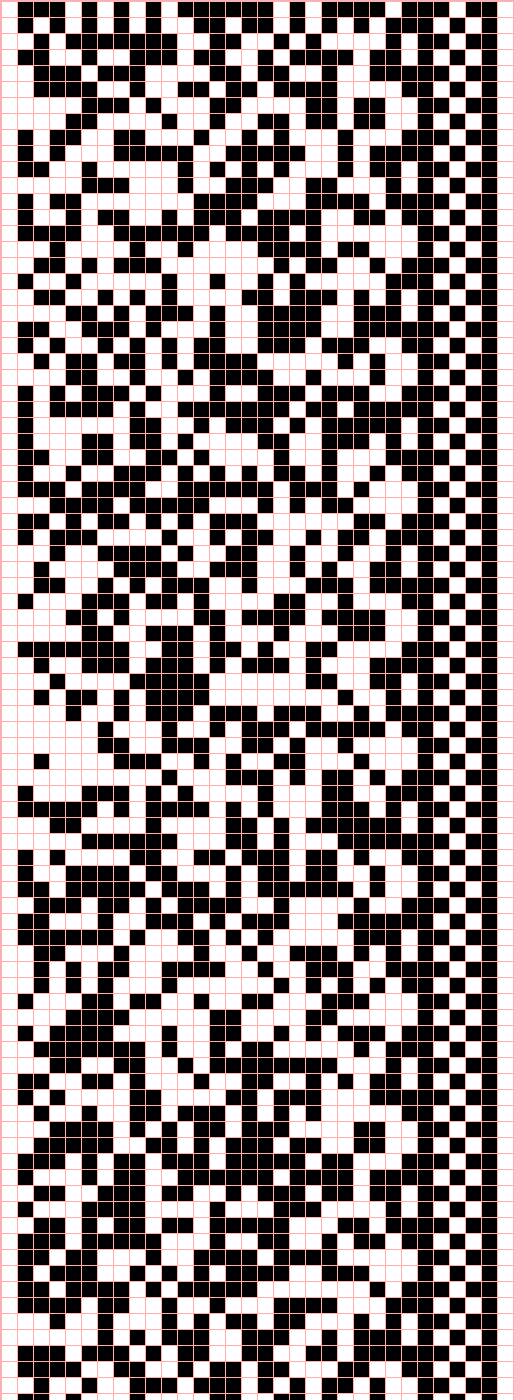}}
		\hfill
		\subfloat[$s_5$\label{min_std_123456789123456789_spaceo}]{%
			\includegraphics[width=0.1\linewidth, height=5.0cm]{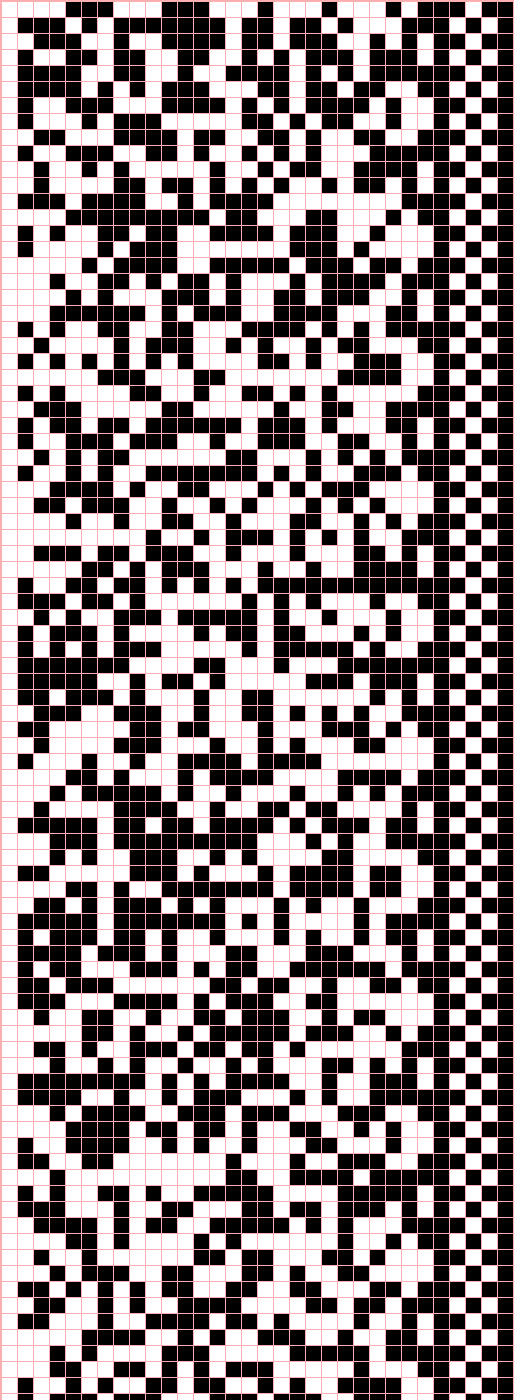}}
		\hfill\\
		%
		%
		\subfloat[$s_1$\label{rand_32_7_space}]{%
			\includegraphics[width=0.1\linewidth, height=5.0cm]{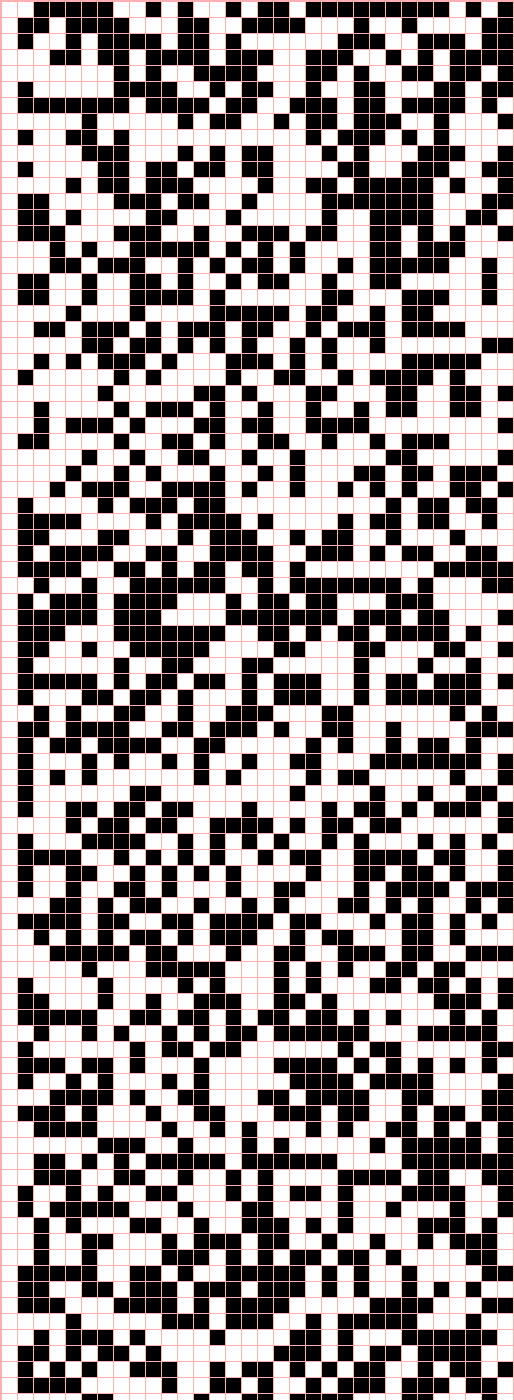}}
		\hfill
		\subfloat[$s_3$\label{rand_32_12345_space}]{%
			\includegraphics[width=0.1\linewidth, height=5.0cm]{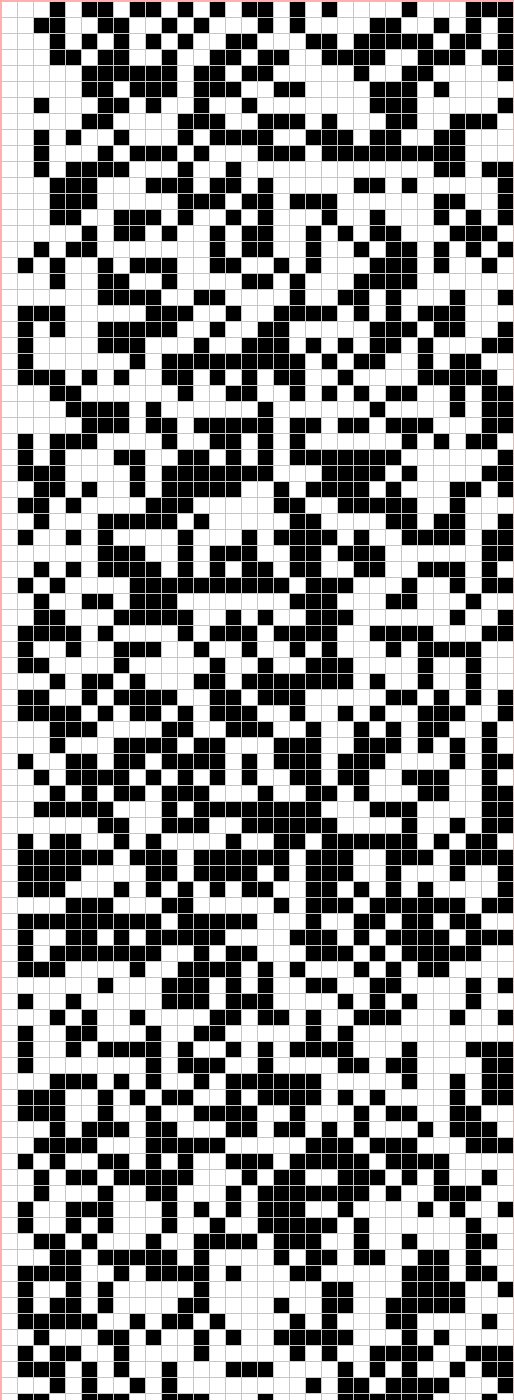}}
		\hfill
		\subfloat[$s_4$\label{rand_32_9650218_space}]{%
			\includegraphics[width=0.1\linewidth, height=5.0cm]{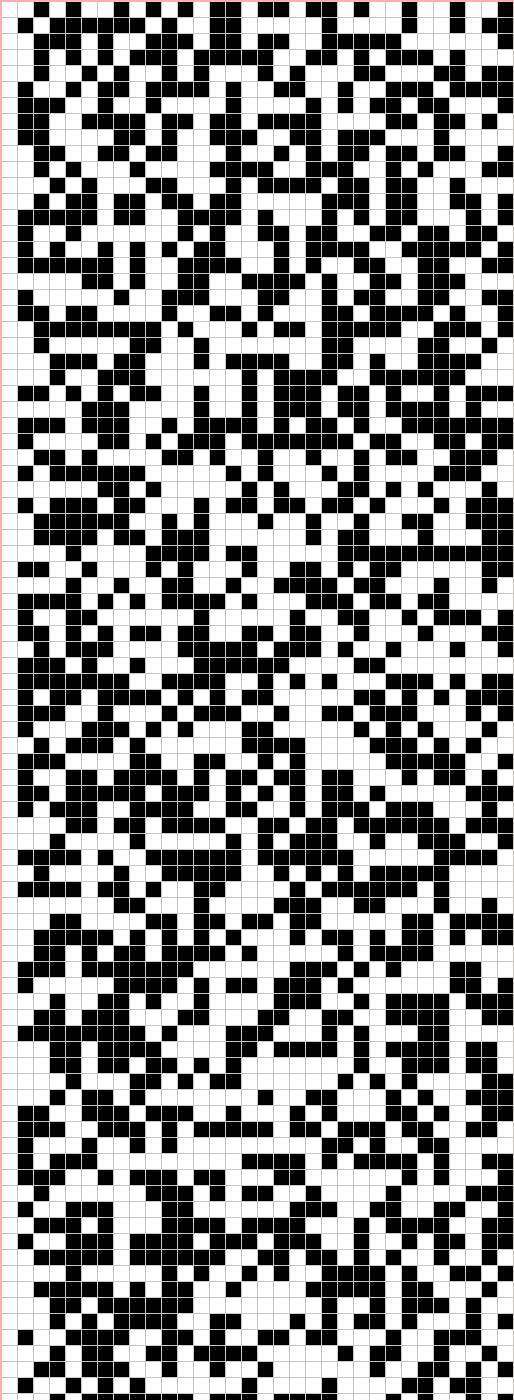}}
		\hfill
		\subfloat[$s_5$\label{rand_32_123456789123456789_spaceo}]{%
			\includegraphics[width=0.1\linewidth, height=5.0cm]{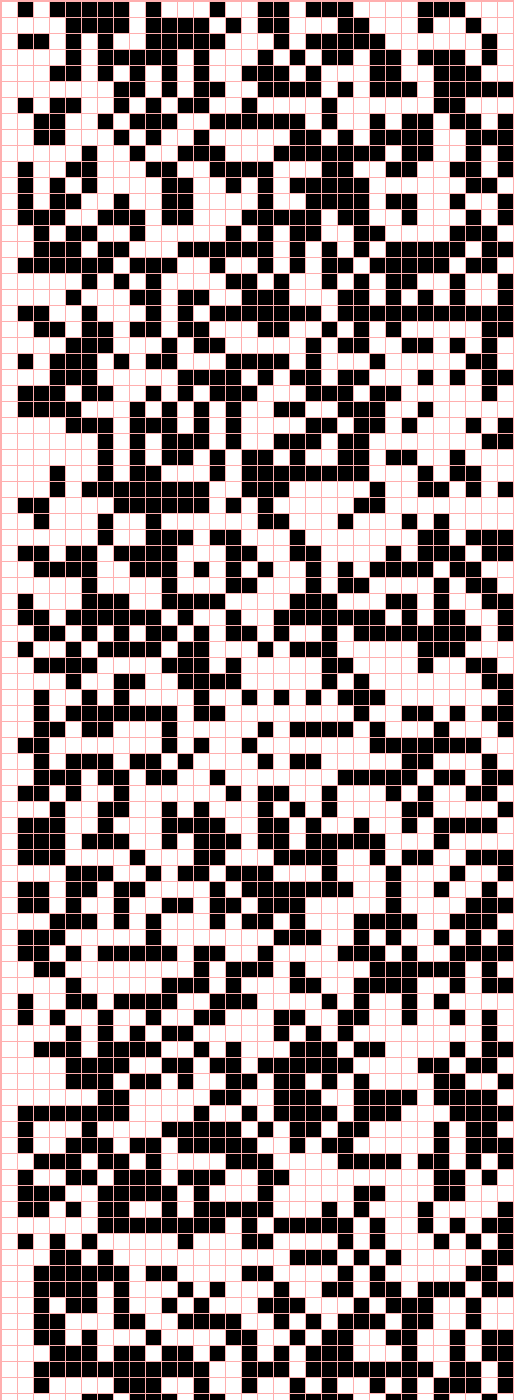}}
		\hfill
		\subfloat[$s_1$\label{lrand_32_7_space}]{%
			\includegraphics[width=0.1\linewidth, height=5.0cm]{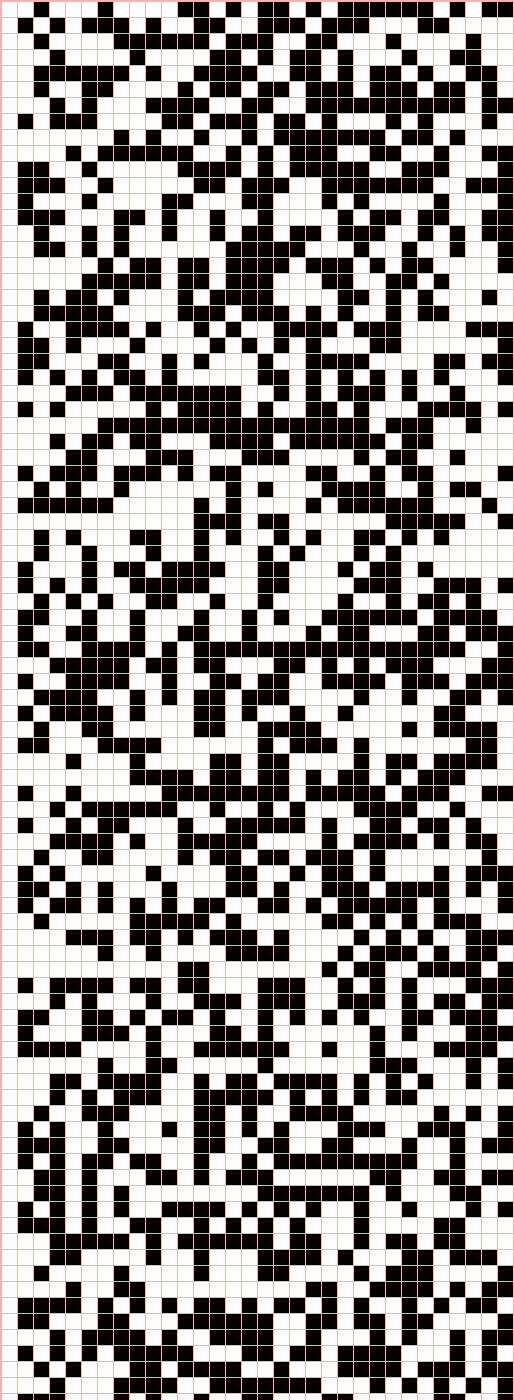}}
		\hfill
		\subfloat[$s_3$\label{lrand_32_12345_space}]{%
			\includegraphics[width=0.1\linewidth, height=5.0cm]{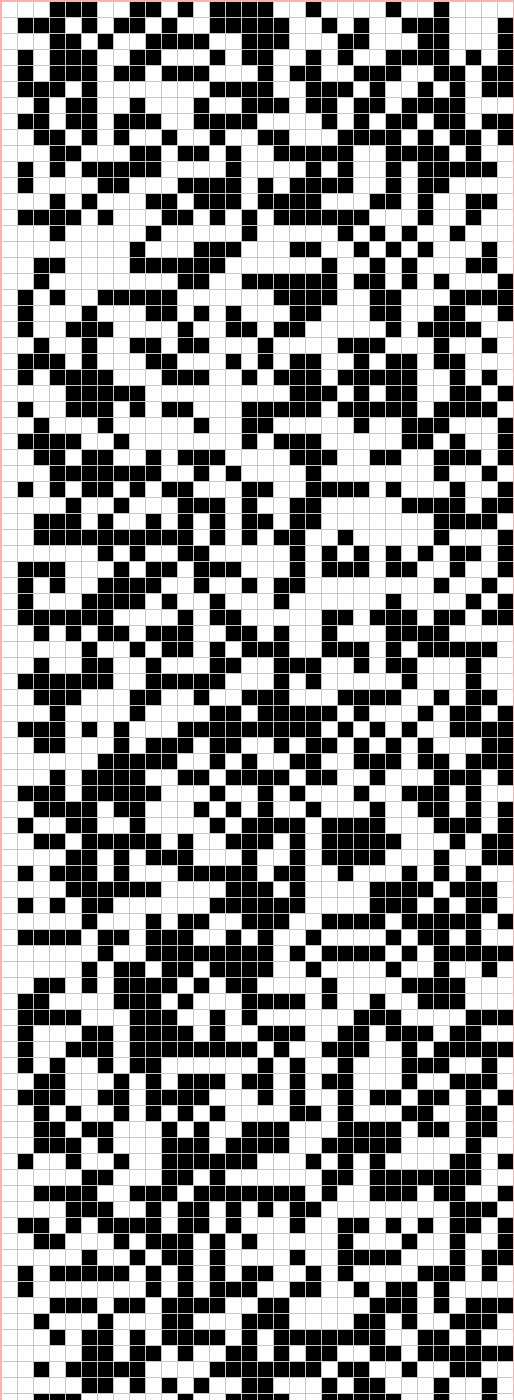}}
		\hfill
		\subfloat[$s_4$\label{lrand_32_9650218_space}]{%
			\includegraphics[width=0.1\linewidth, height=5.0cm]{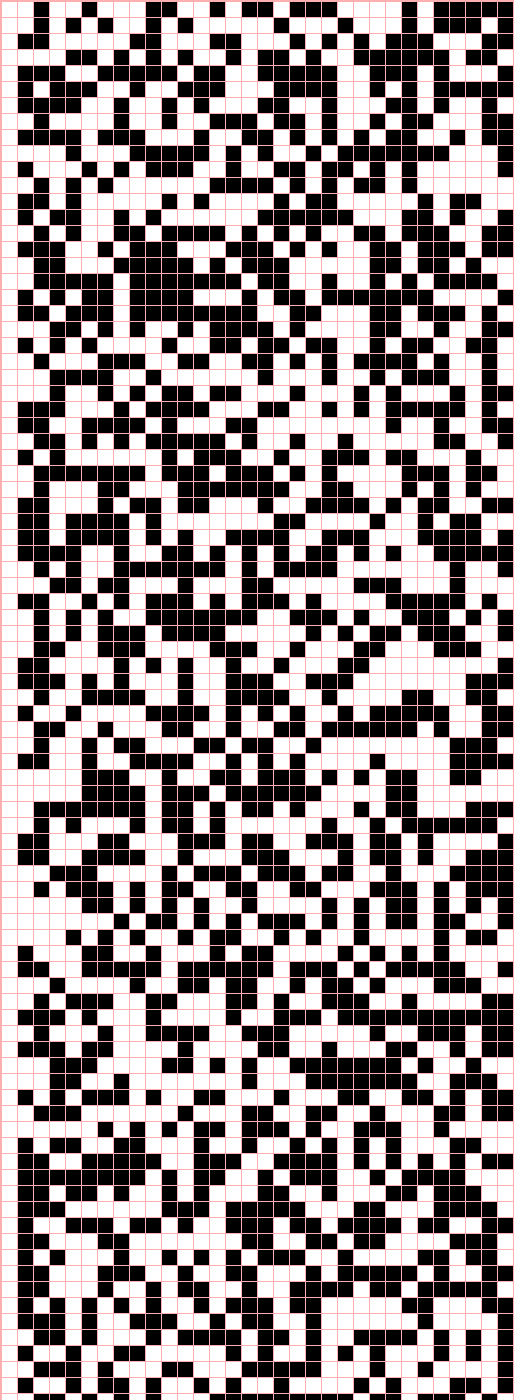}}
		\hfill
		\subfloat[$s_5$\label{lrand_32_123456789123456789_space}]{%
			\includegraphics[width=0.1\linewidth, height=5.0cm]{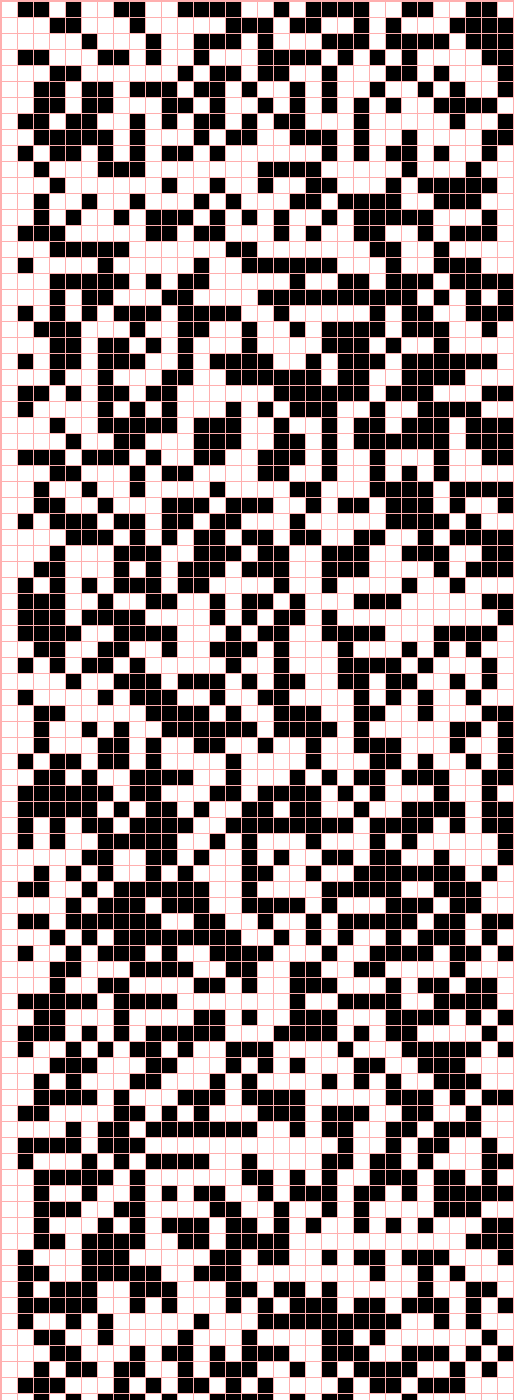}}
		\caption{Space-time diagram for Knuth's MMIX (\ref{knuth_lcg_7_space} to \ref{knuth_lcg_123456789123456789_space}), Borland's LCG (\ref{borland_lcg_7_space} to \ref{borland_lcg_123456789123456789_spaceo}) and minstd\_rand (\ref{min_std_7_space} to \ref{min_std_123456789123456789_spaceo}), rand (\ref{rand_32_7_space} to \ref{rand_32_123456789123456789_spaceo}) and lrand (\ref{lrand_32_7_space} to \ref{lrand_32_123456789123456789_space}) of UNIX}
		\label{fig:rand_space-time}
	\end{figure}          
	
	\begin{figure}[!h]
		\centering
		\vspace{-2.0em}
		\subfloat[$s_1$\label{mrg_7_space}]{%
			\includegraphics[width=0.1\linewidth, height=5.0cm]{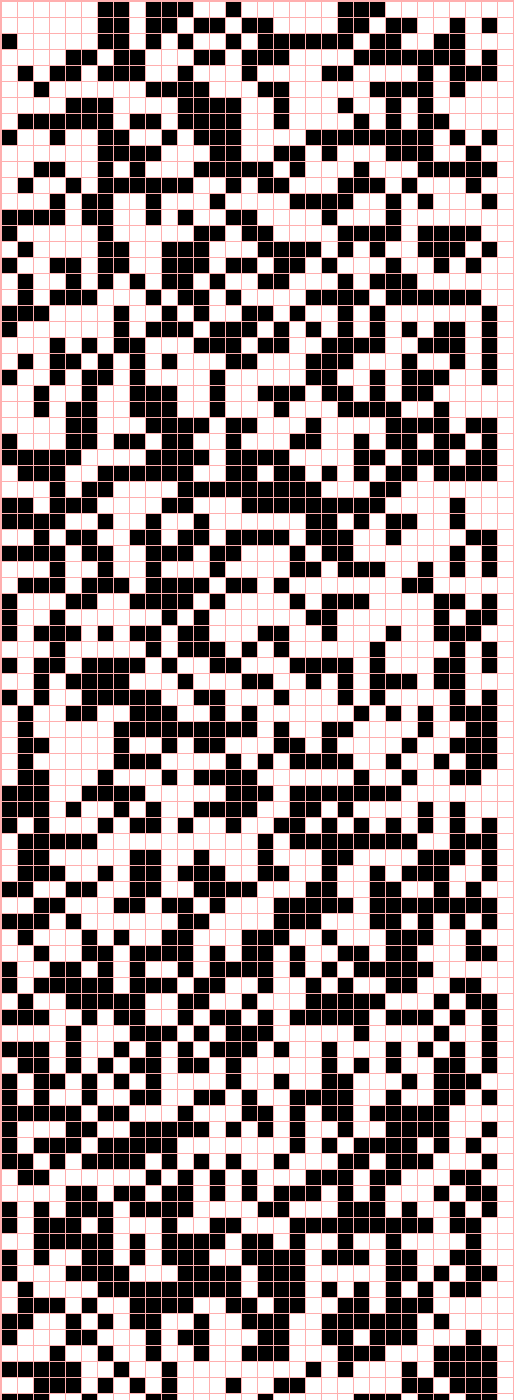}}
		\hfill
		\subfloat[$s_3$\label{mrg_12345_space}]{%
			\includegraphics[width=0.1\linewidth, height=5.0cm]{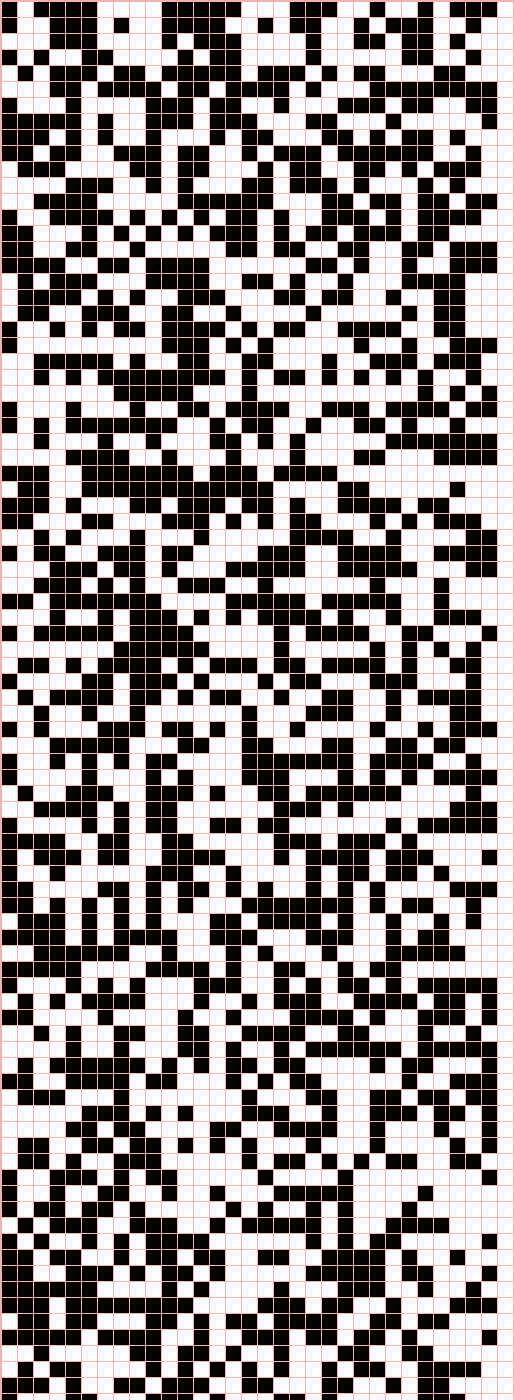}}
		\hfill
		\subfloat[$s_4$\label{mrg_9650218_space}]{%
			\includegraphics[width=0.1\linewidth, height=5.0cm]{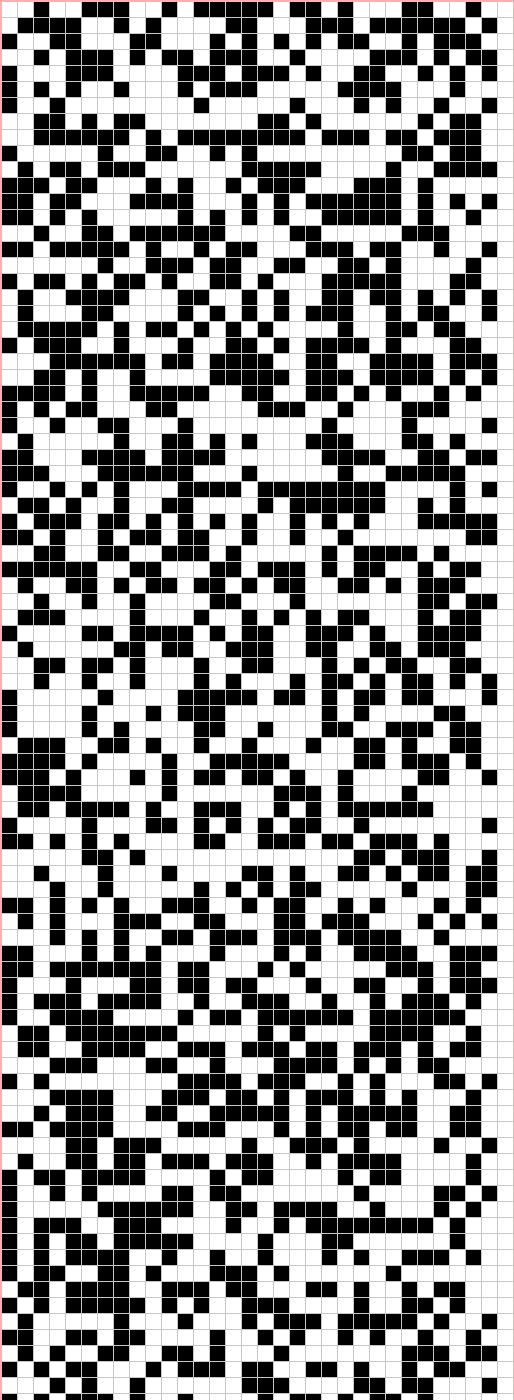}}
		\hfill
		\subfloat[$s_5$\label{mrg_123456789123456789_spaceo}]{%
			\includegraphics[width=0.1\linewidth, height=5.0cm]{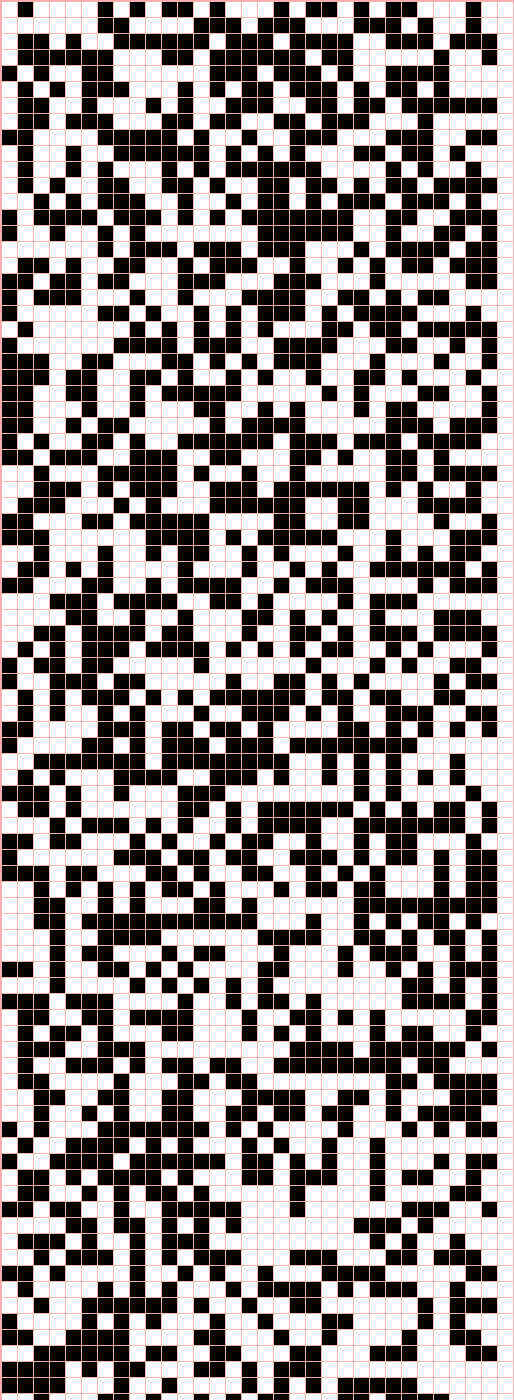}}
		\hfill
		\subfloat[$s_1$\label{pcg_32_7_space}]{%
			\includegraphics[width=0.1\linewidth, height=5.0cm]{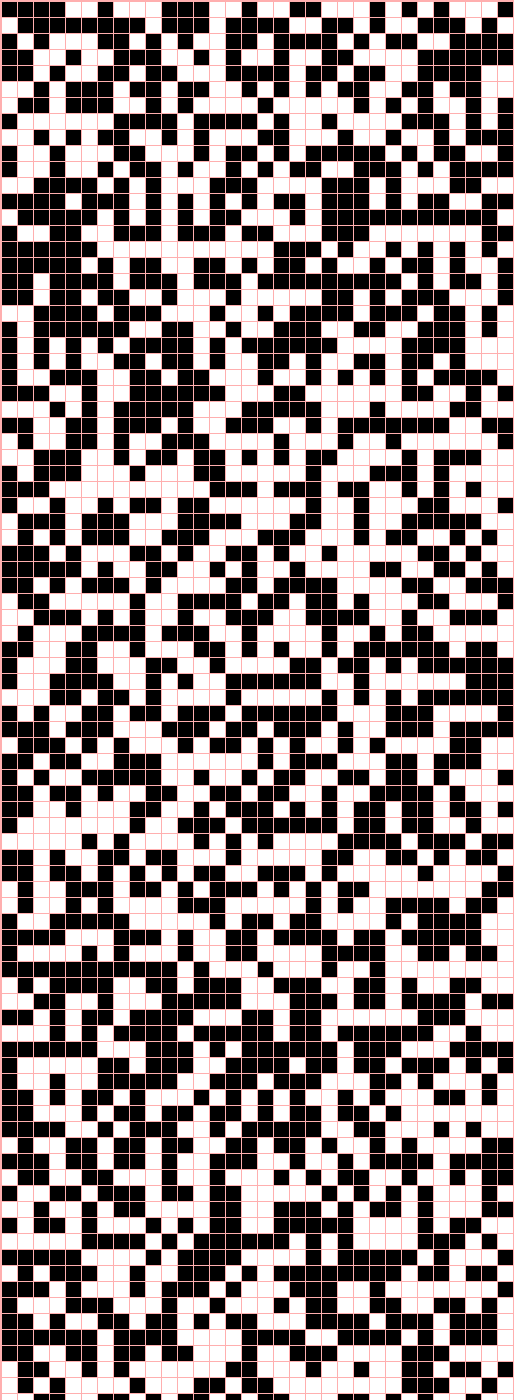}}
		\hfill
		\subfloat[$s_3$\label{pcg_32_12345_space}]{%
			\includegraphics[width=0.1\linewidth, height=5.0cm]{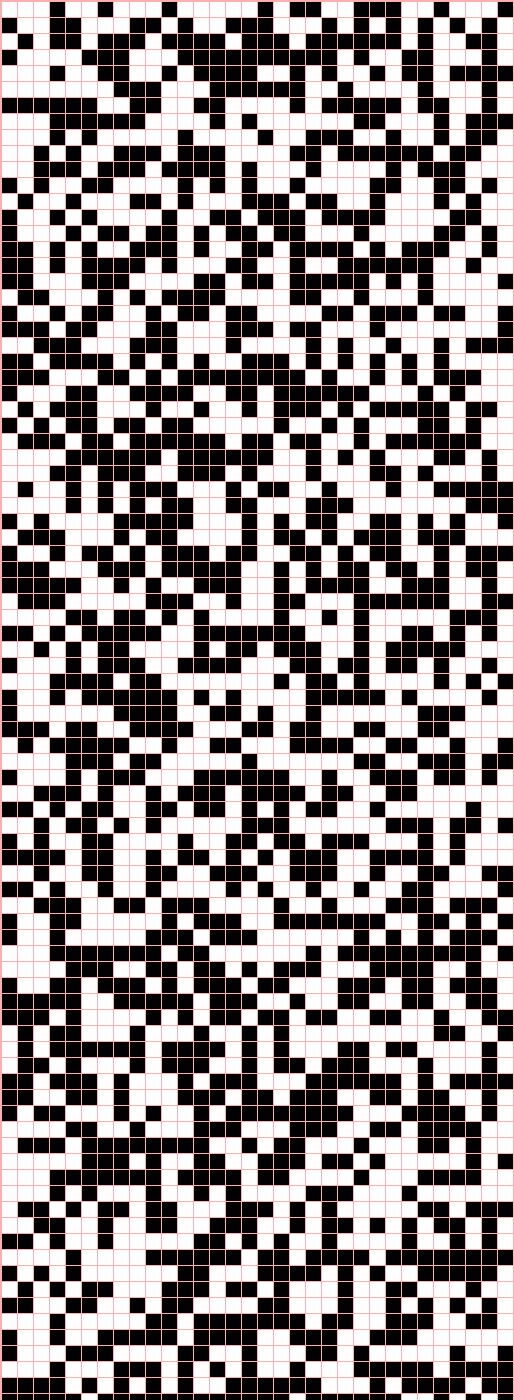}}
		\hfill
		\subfloat[$s_4$\label{pcg_32_9650218_space}]{%
			\includegraphics[width=0.1\linewidth, height=5.0cm]{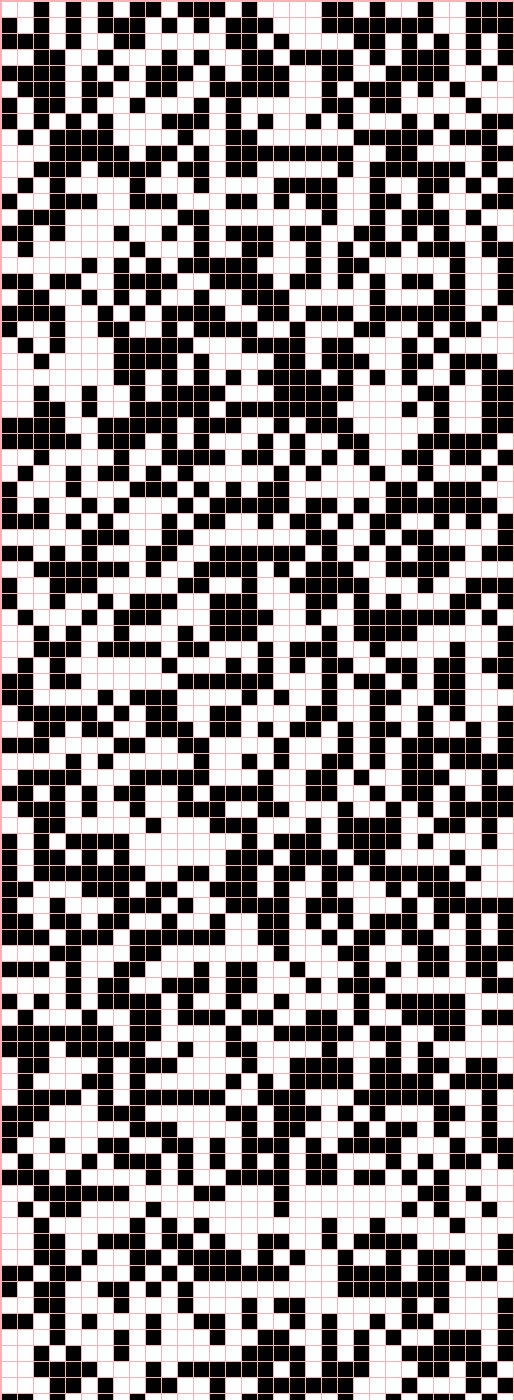}}
		\hfill
		\subfloat[$s_5$\label{pcg_32_123456789123456789_space}]{%
			\includegraphics[width=0.1\linewidth, height=5.0cm]{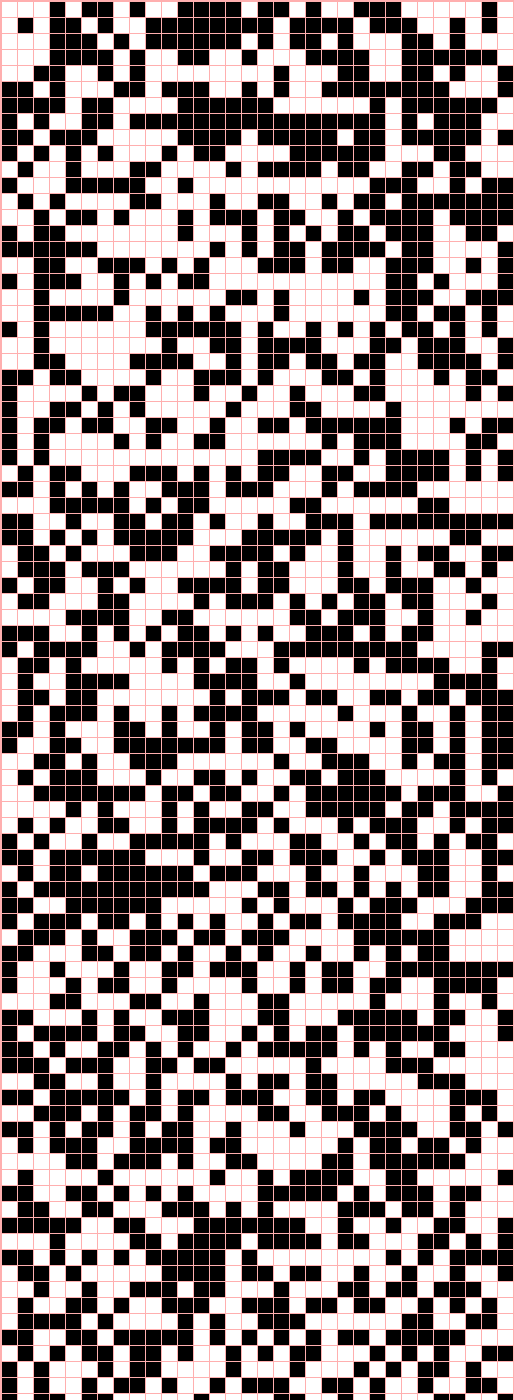}}
		%
		%
		\hfill\\
		\subfloat[$s_1$\label{random_32_7_space}]{%
			\includegraphics[width=0.1\linewidth, height=5.0cm]{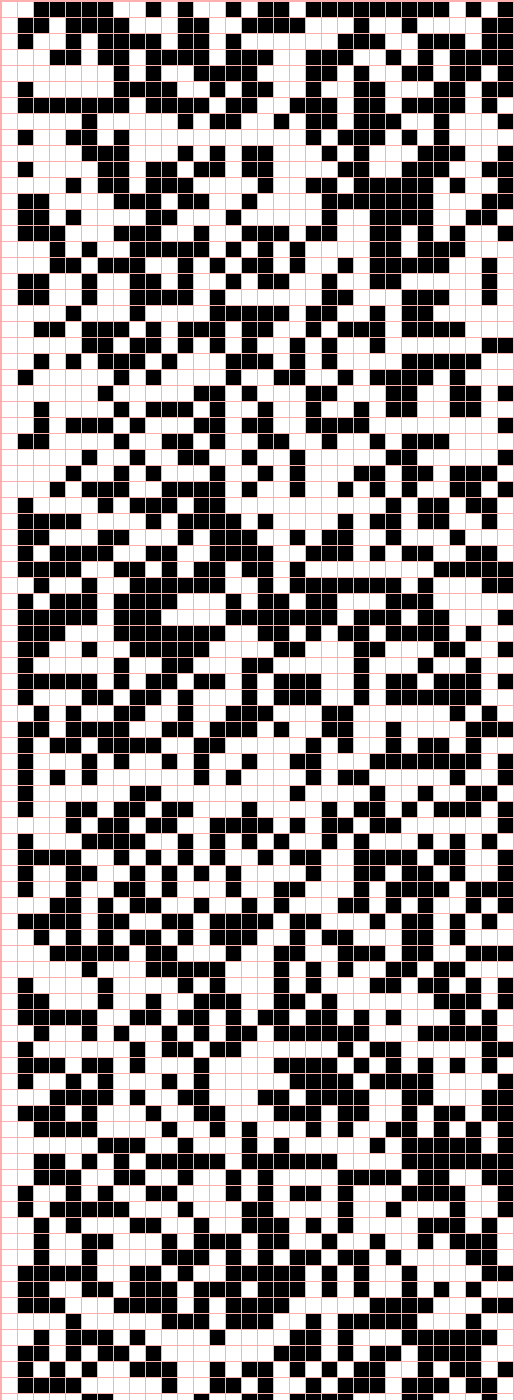}}
		\hfill
		\subfloat[$s_3$\label{random_32_12345_space}]{%
			\includegraphics[width=0.1\linewidth, height=5.0cm]{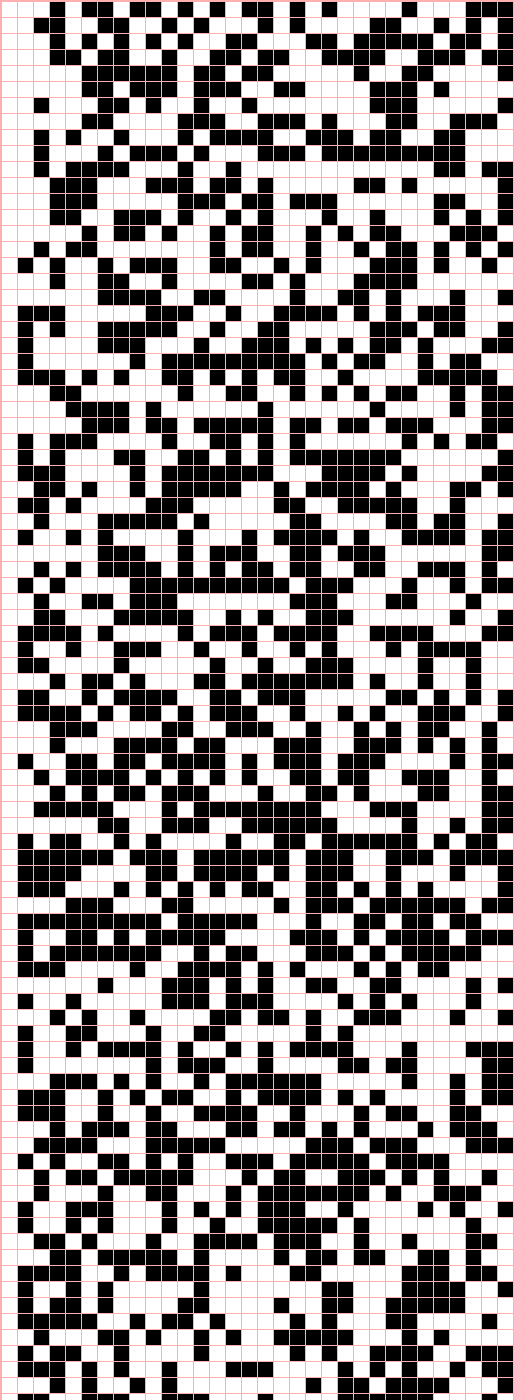}}
		\hfill
		\subfloat[$s_4$\label{random_32_9650218_space}]{%
			\includegraphics[width=0.1\linewidth, height=5.0cm]{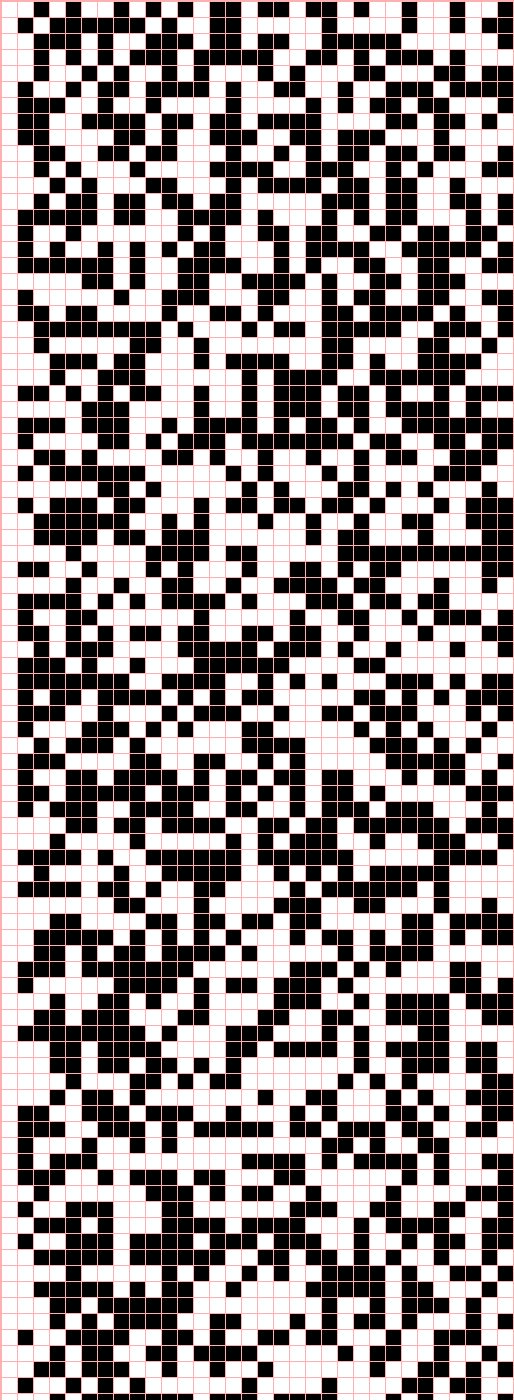}}
		\hfill
		\subfloat[$s_5$\label{random_32_123456789123456789_spaceo}]{%
			\includegraphics[width=0.1\linewidth, height=5.0cm]{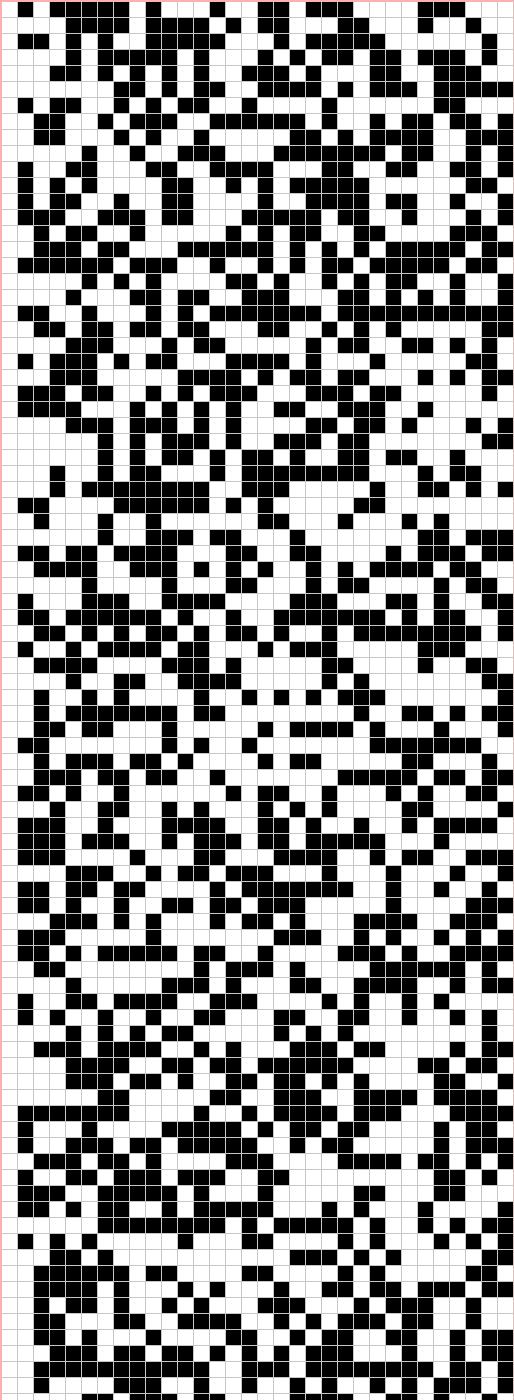}}
		\hfill
		\subfloat[$s_1$\label{taus_32_7_space}]{%
			\includegraphics[width=0.1\linewidth, height=5.0cm]{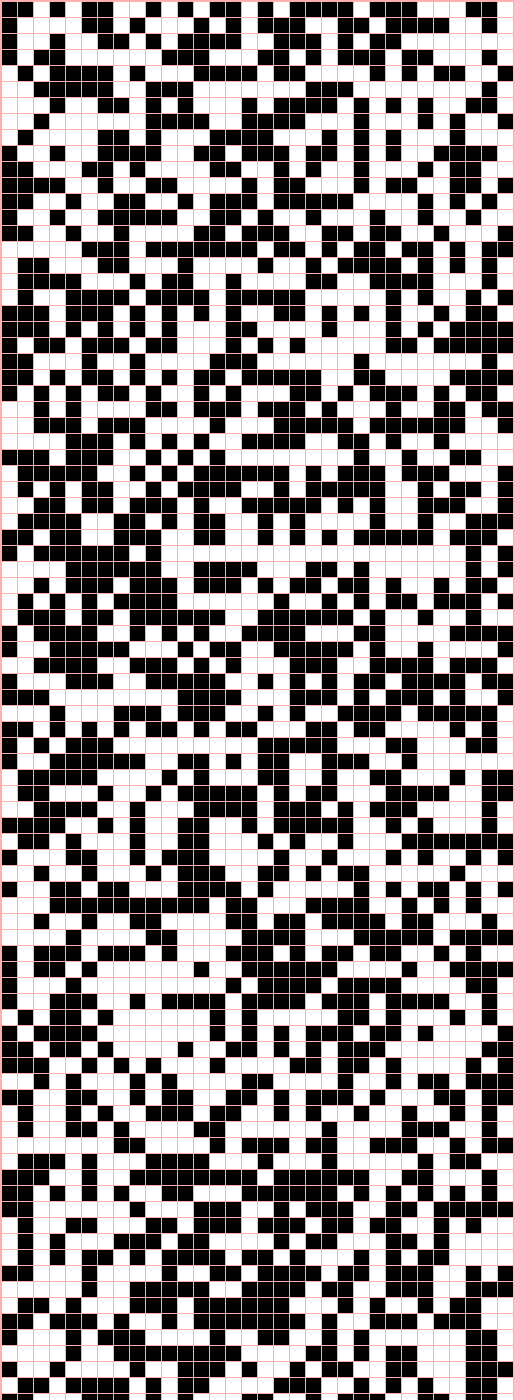}}
		\hfill
		\subfloat[$s_3$\label{taus_32_12345_space}]{%
			\includegraphics[width=0.1\linewidth, height=5.0cm]{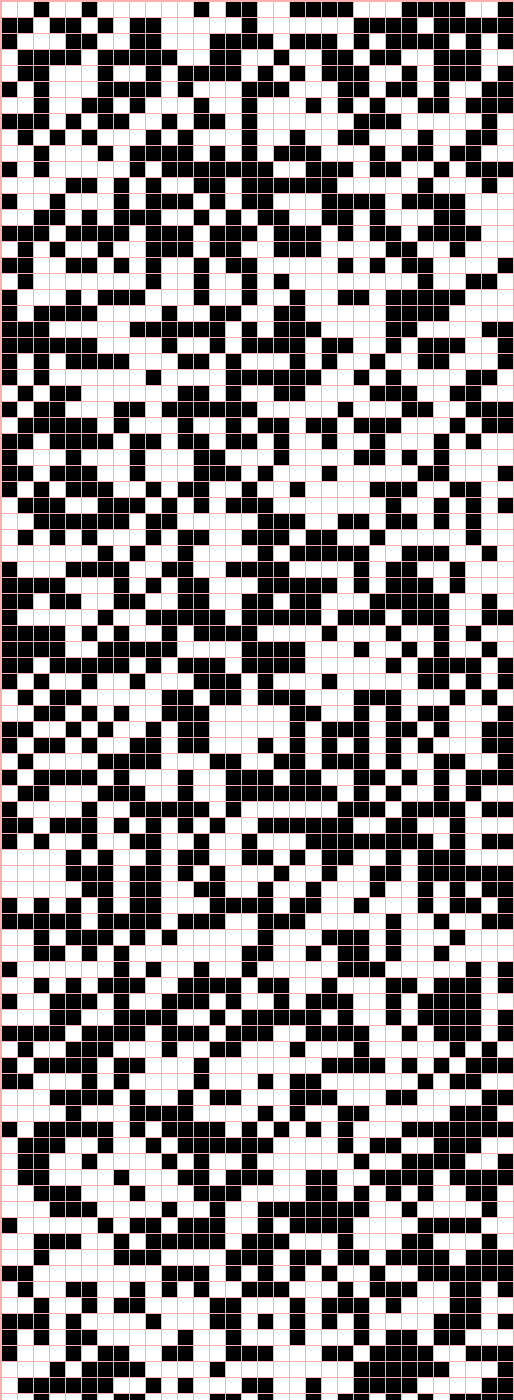}}
		\hfill
		\subfloat[$s_4$\label{taus_32_9650218_space}]{%
			\includegraphics[width=0.1\linewidth, height=5.0cm]{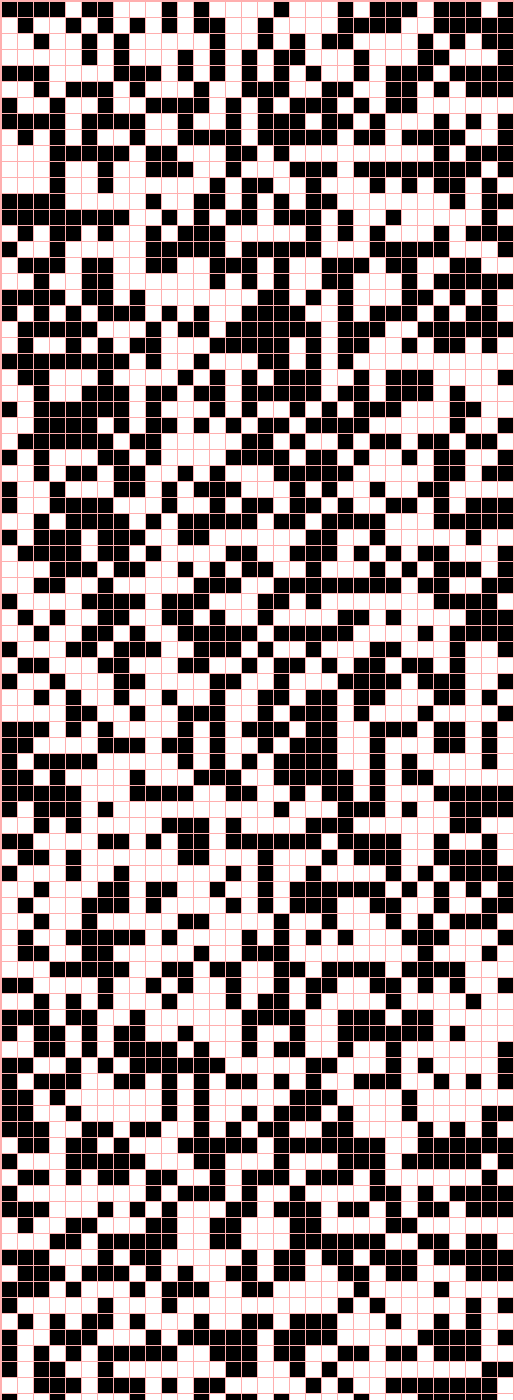}}
		\hfill
		\subfloat[$s_5$\label{taus_32_123456789123456789_space}]{%
			\includegraphics[width=0.1\linewidth, height=5.0cm]{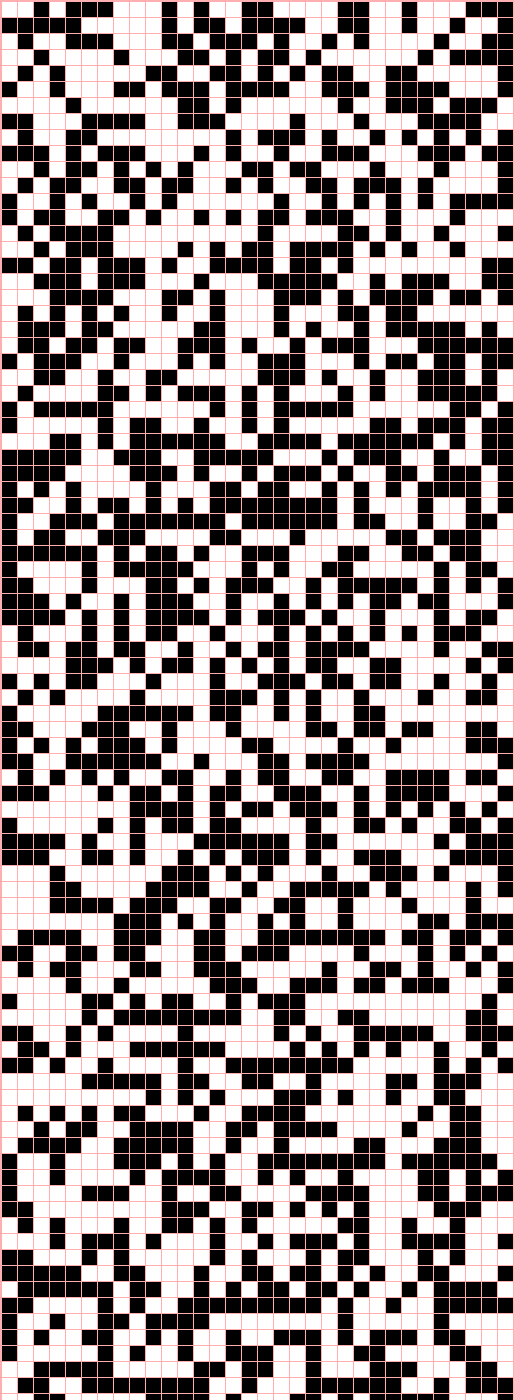}}
		%
		%
		\hfill \\
		\subfloat[$s_1$\label{dsfmt_52_7_space}]{%
			\includegraphics[width=0.2\linewidth, height=5.0cm]{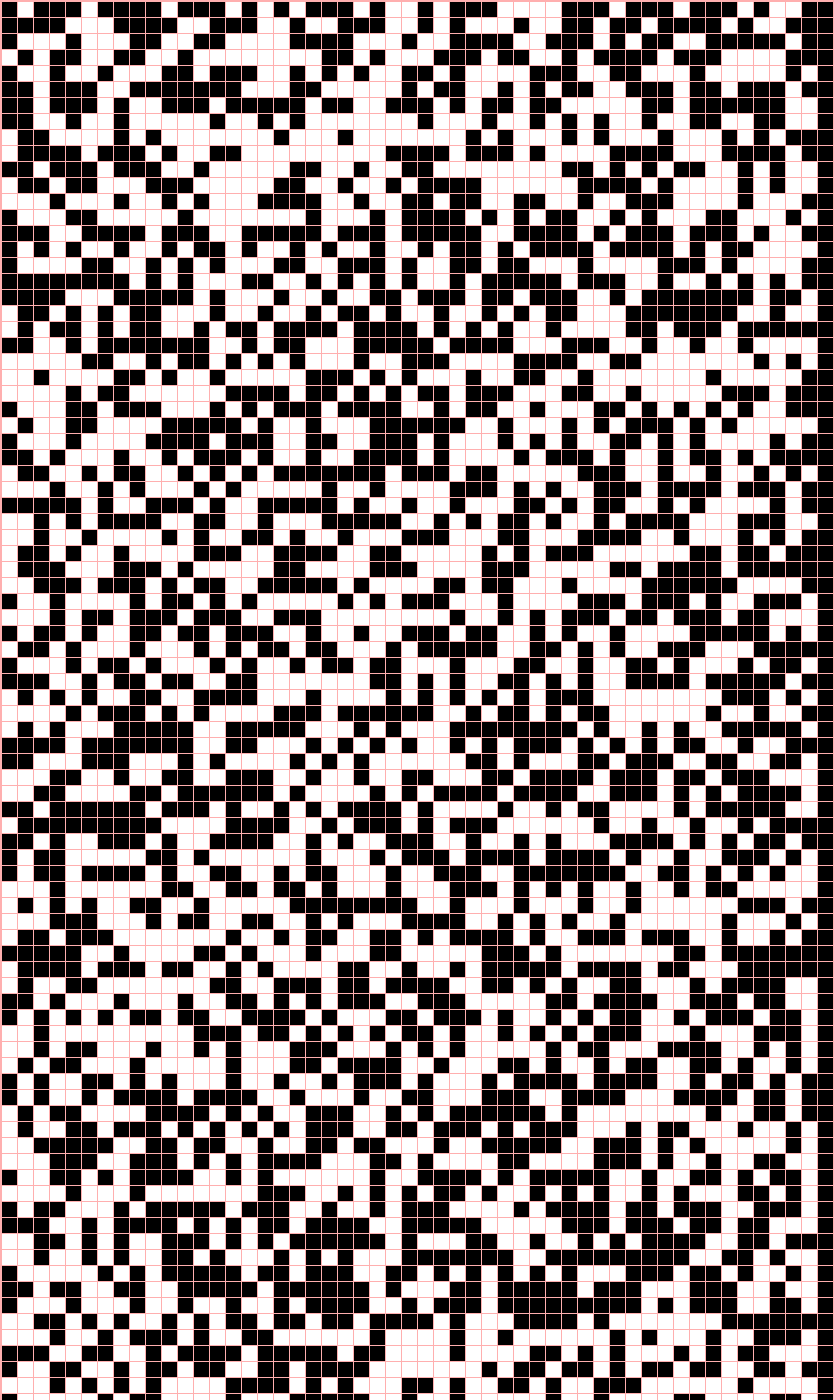}}
		\hfill
		\subfloat[$s_3$\label{dsfmt_52_12345_space}]{%
			\includegraphics[width=0.2\linewidth, height=5.0cm]{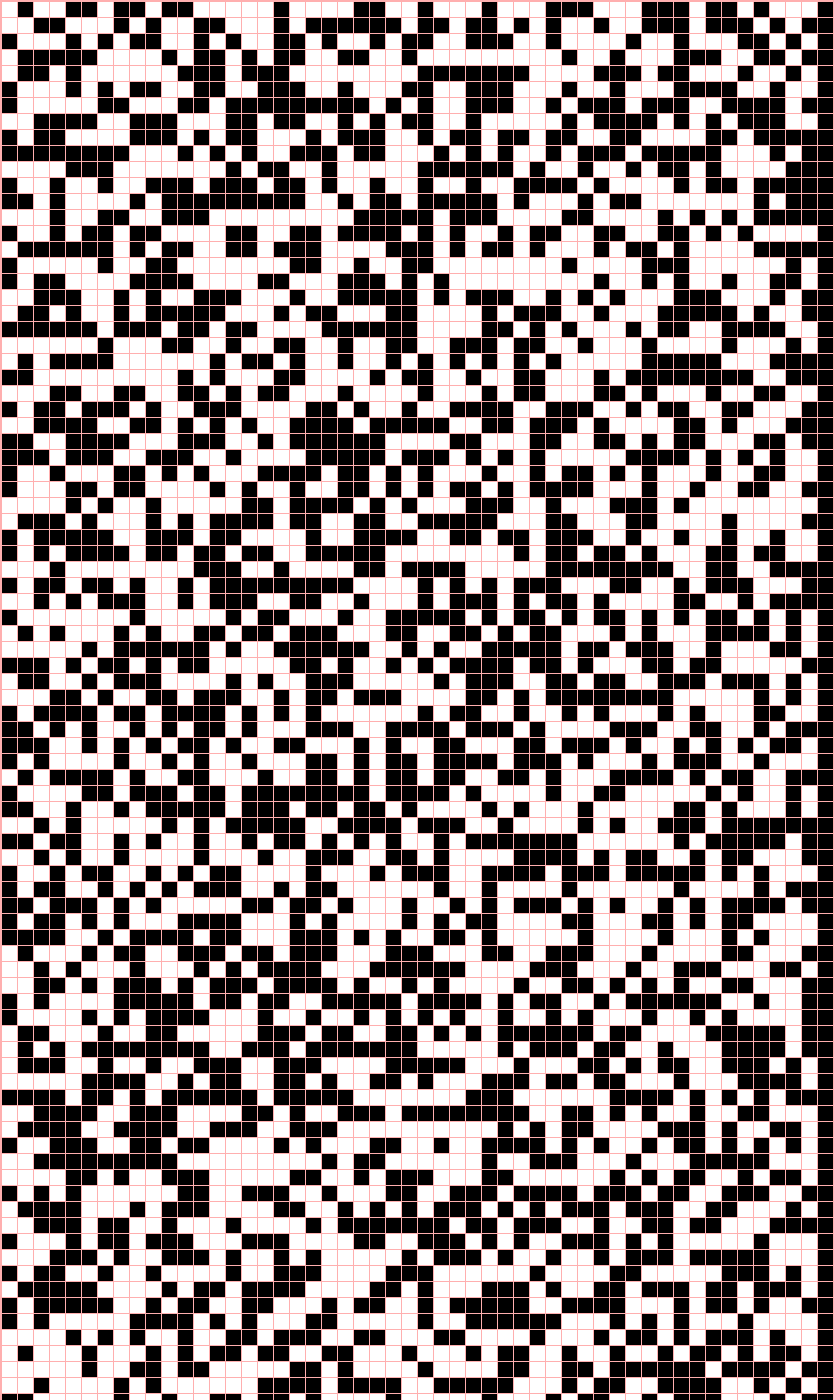}}
		\hfill
		\subfloat[$s_4$\label{dsfmt_52_9650218_space}]{%
			\includegraphics[width=0.2\linewidth, height=5.0cm]{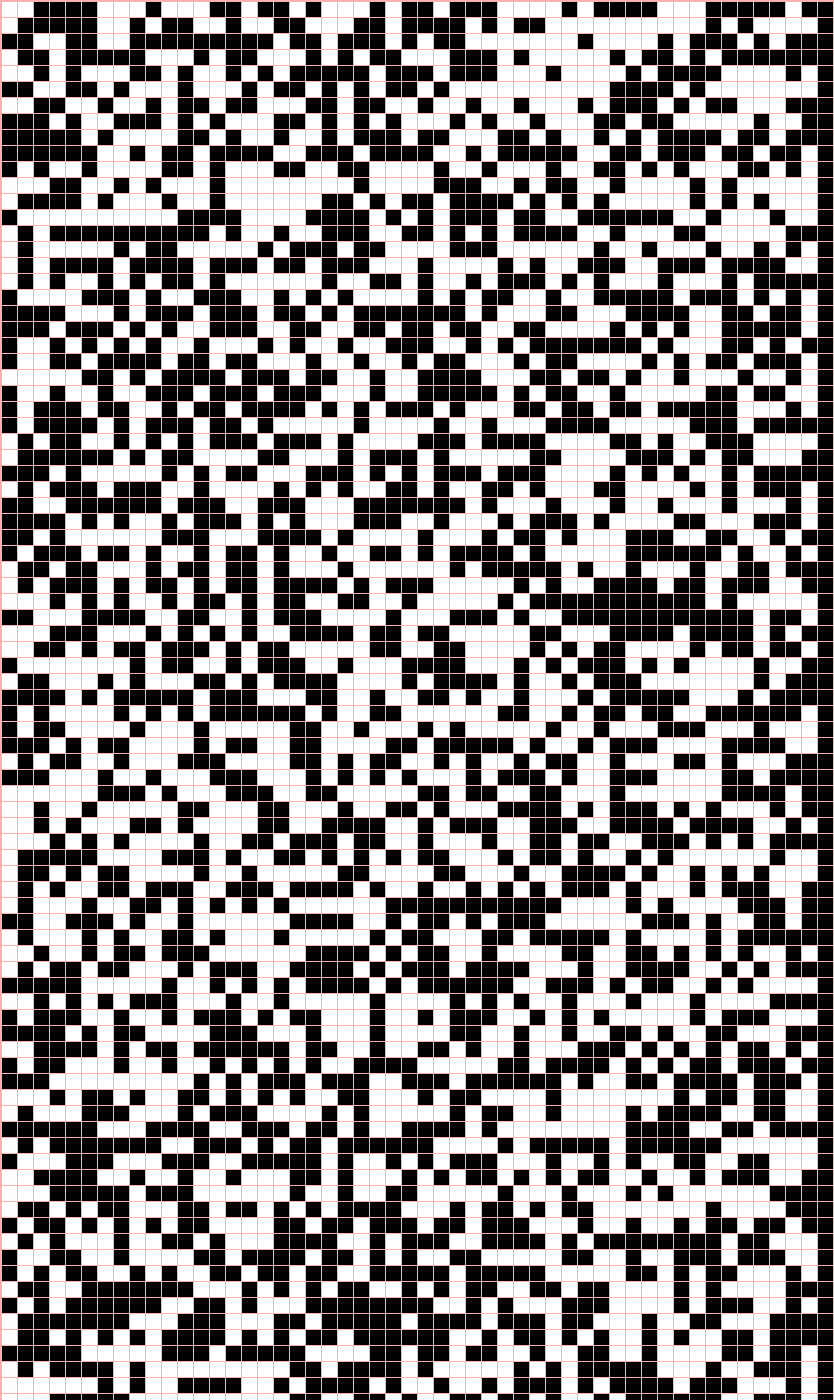}}
		\hfill
		\subfloat[$s_5$\label{dsfmt_52_123456789123456789_space}]{%
			\includegraphics[width=0.2\linewidth, height=5.0cm]{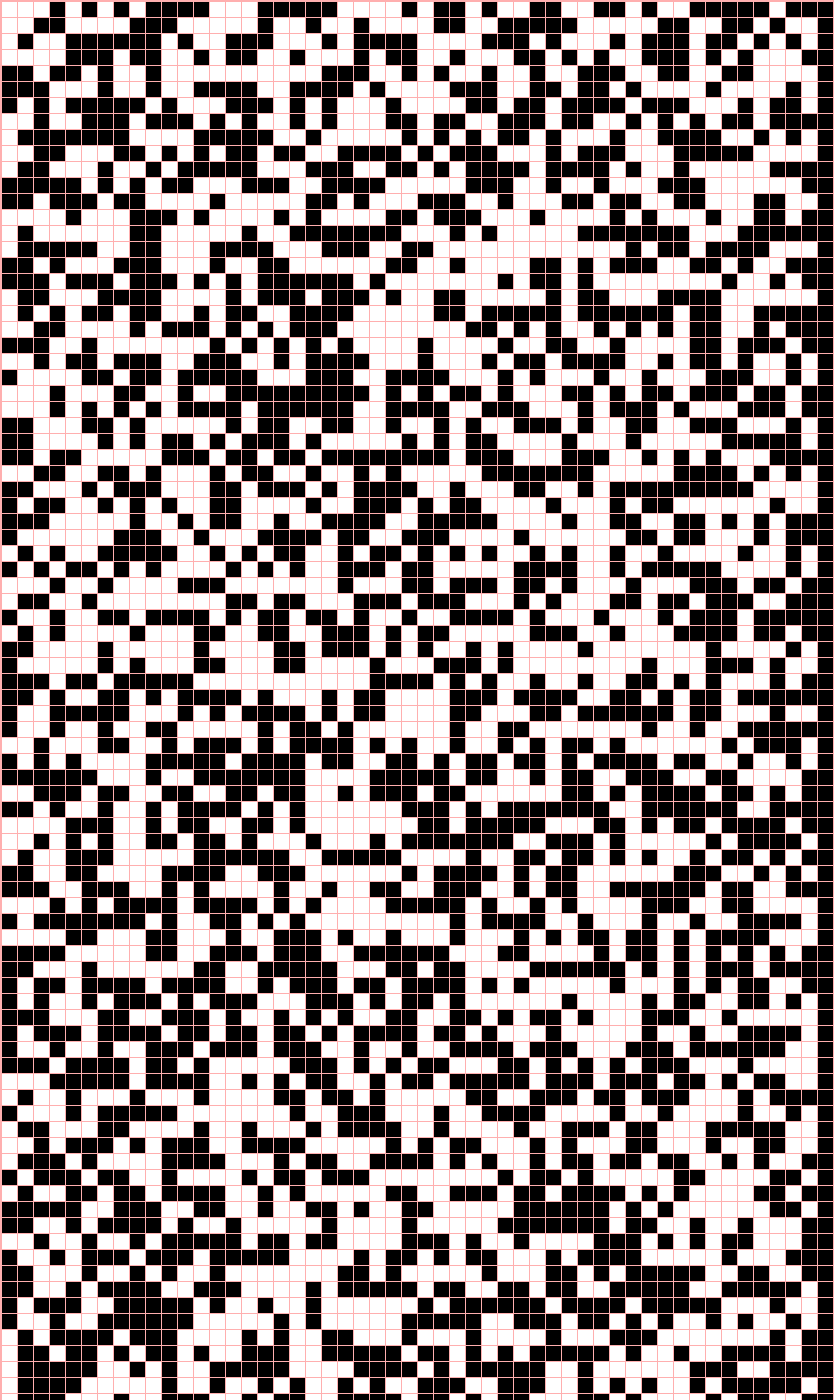}}
		\caption{Space-time diagram for  MRG31k3p (\ref{mrg_7_space} to \ref{mrg_123456789123456789_spaceo}) and PCG $32$-bit (\ref{pcg_32_7_space} to \ref{pcg_32_123456789123456789_space}), random (\ref{random_32_7_space} to \ref{random_32_123456789123456789_spaceo}), Tauss88 (\ref{taus_32_7_space} to \ref{taus_32_123456789123456789_space}) and dSFMT19937 $64$ bit (\ref{dsfmt_52_7_space} to \ref{dsfmt_52_123456789123456789_space})}
		\label{fig:dsfmt32_space-time}
	\end{figure}

	\begin{figure}[!h]
		\centering
		\vspace{-2.0em}
		\subfloat[$s_1$\label{lfsr258_7_space}]{%
			\includegraphics[width=0.2\linewidth, height=5.0cm]{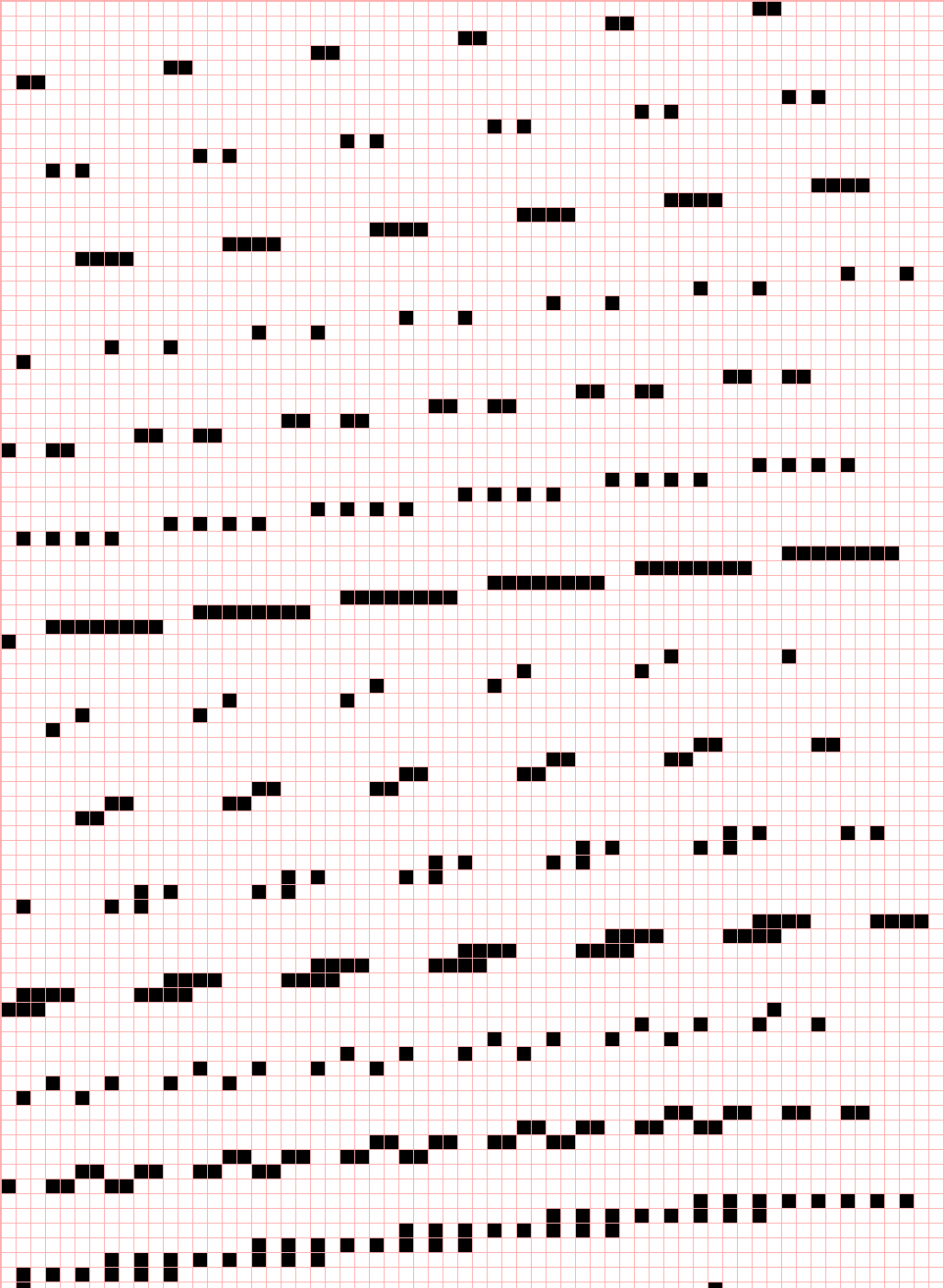}}
		\hfill
		\subfloat[$s_3$\label{lfsr258_12345_space}]{%
			\includegraphics[width=0.2\linewidth, height=5.0cm]{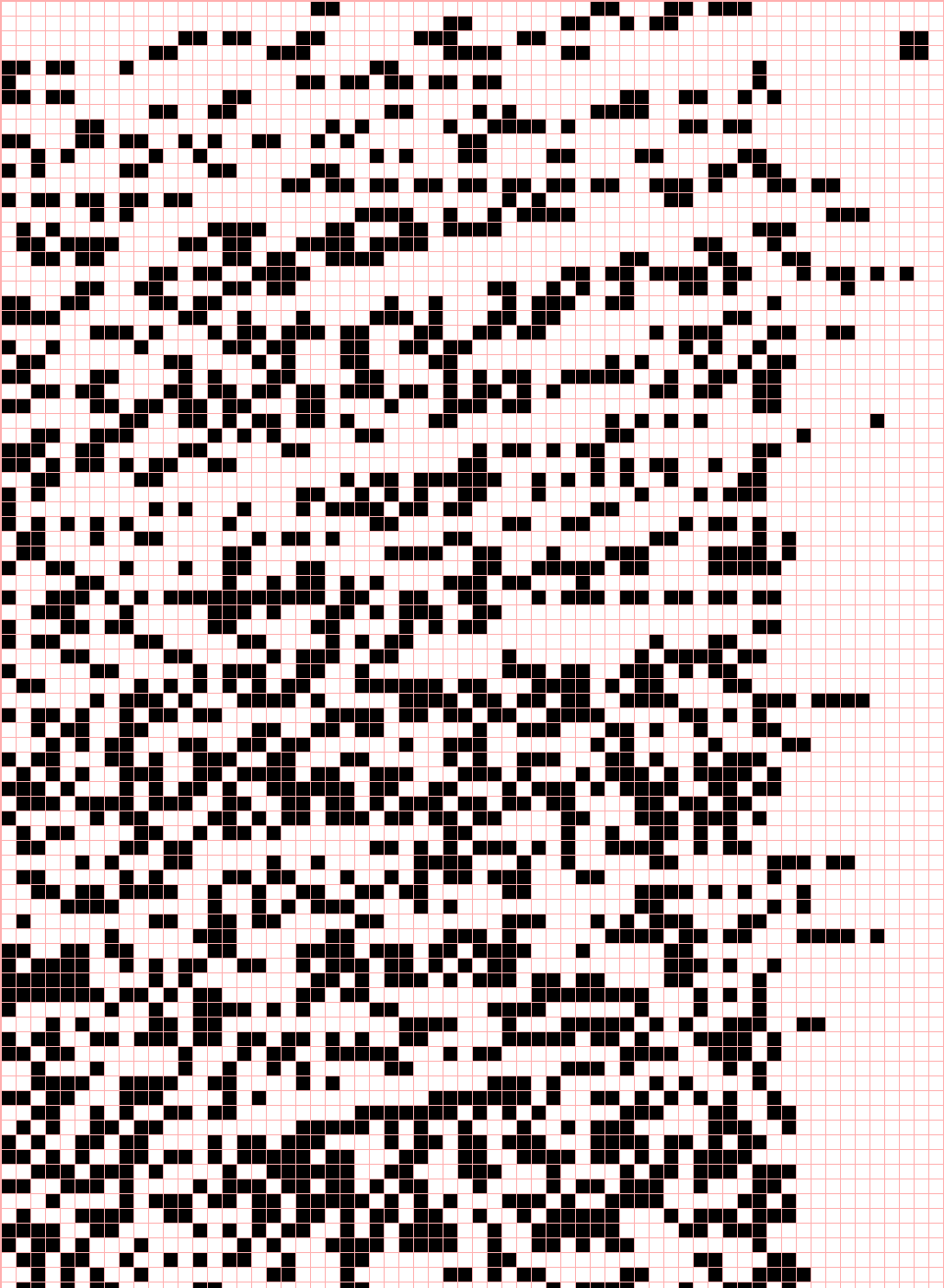}}
		\hfill
		\subfloat[$s_4$\label{lfsr258_9650218_space}]{%
			\includegraphics[width=0.2\linewidth, height=5.0cm]{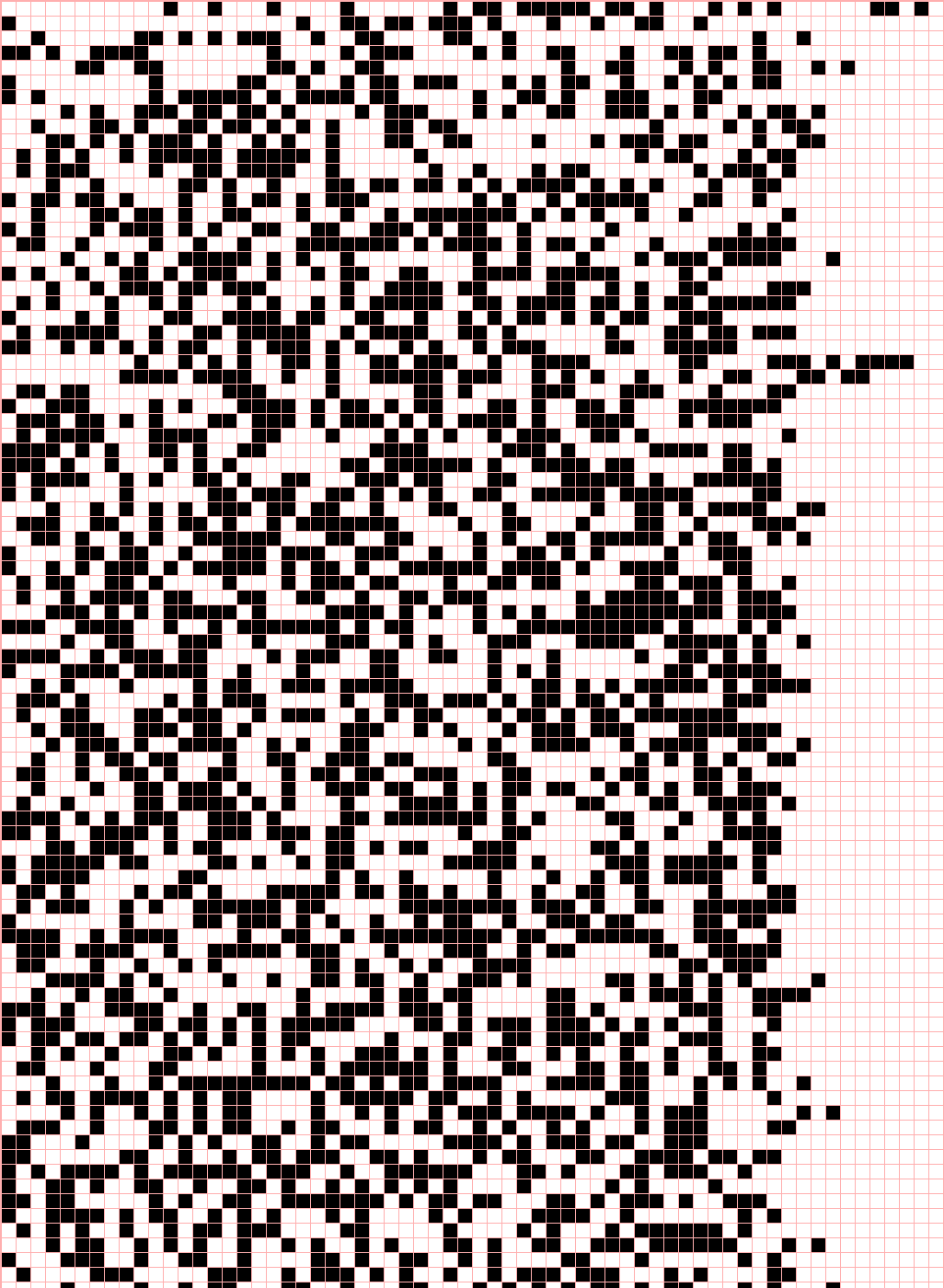}}
		\hfill
		\subfloat[$s_5$\label{lfsr258_123456789123456789_space}]{%
			\includegraphics[width=0.2\linewidth, height=5.0cm]{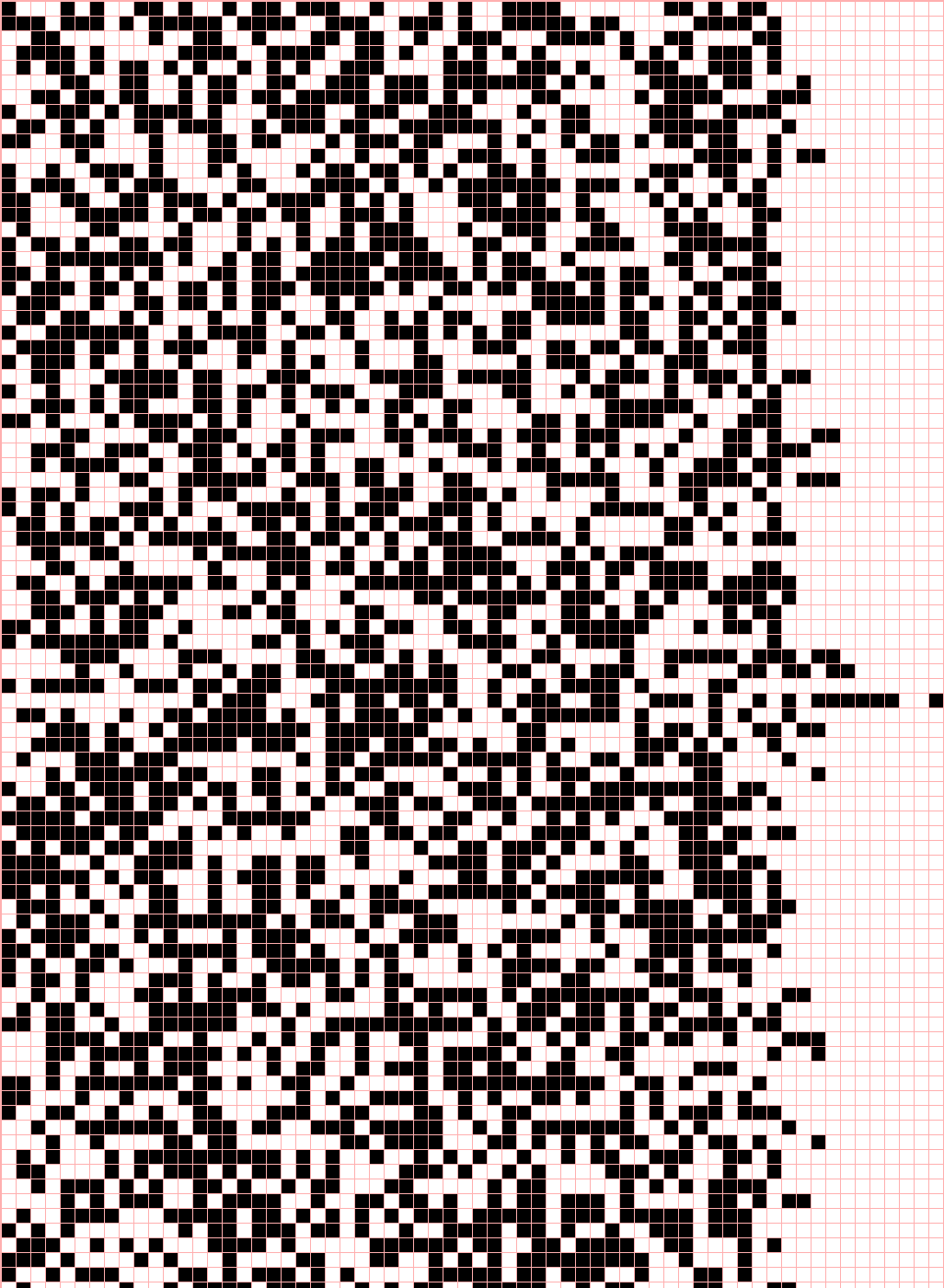}}
		%
		%
		\hfill\\
		\subfloat[$s_1$\label{lfsr_double_7_space}]{%
			\includegraphics[width=0.1\linewidth, height=5.0cm]{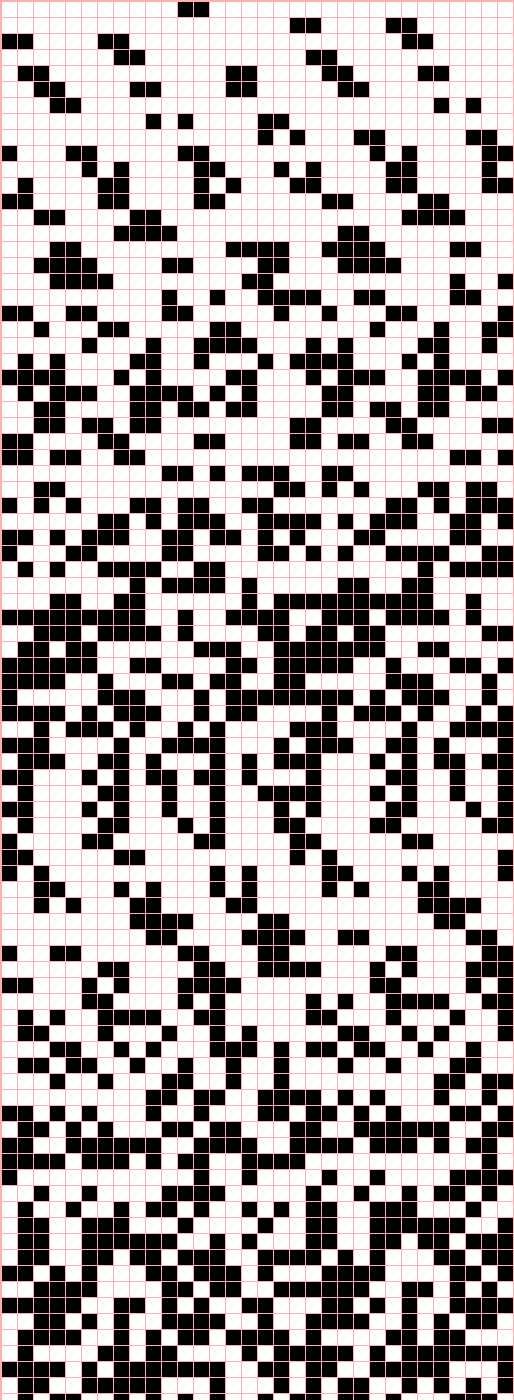}}
		\hfill
		\subfloat[$s_3$\label{lfsr_double_12345_space}]{%
			\includegraphics[width=0.1\linewidth, height=5.0cm]{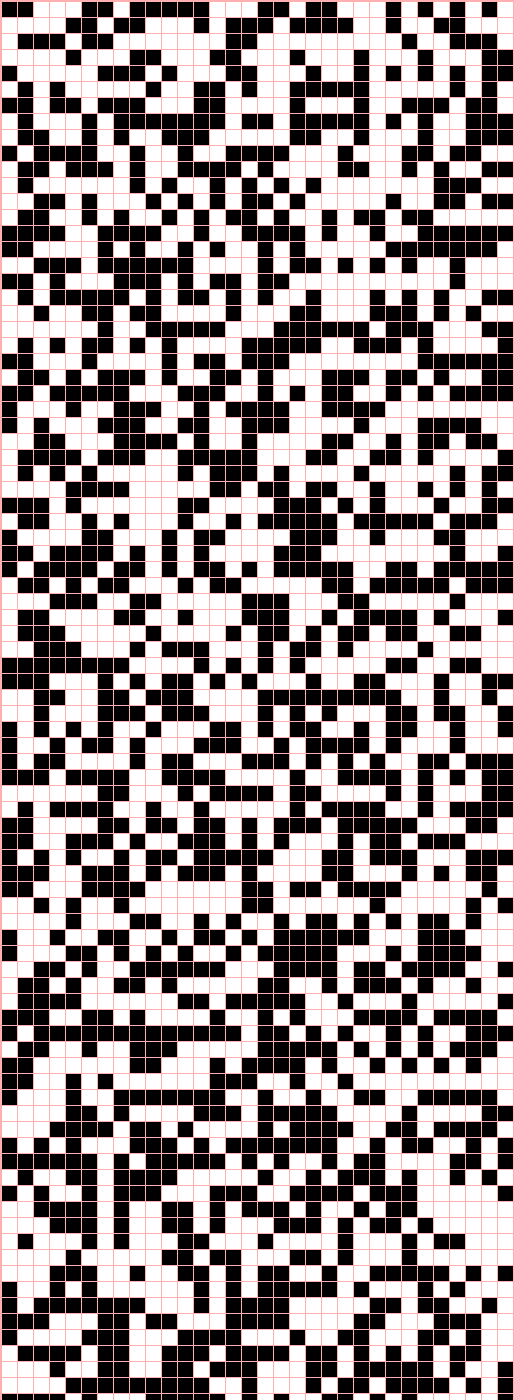}}
		\hfill
		\subfloat[$s_4$\label{lfsr_double_9650218_space}]{%
			\includegraphics[width=0.1\linewidth, height=5.0cm]{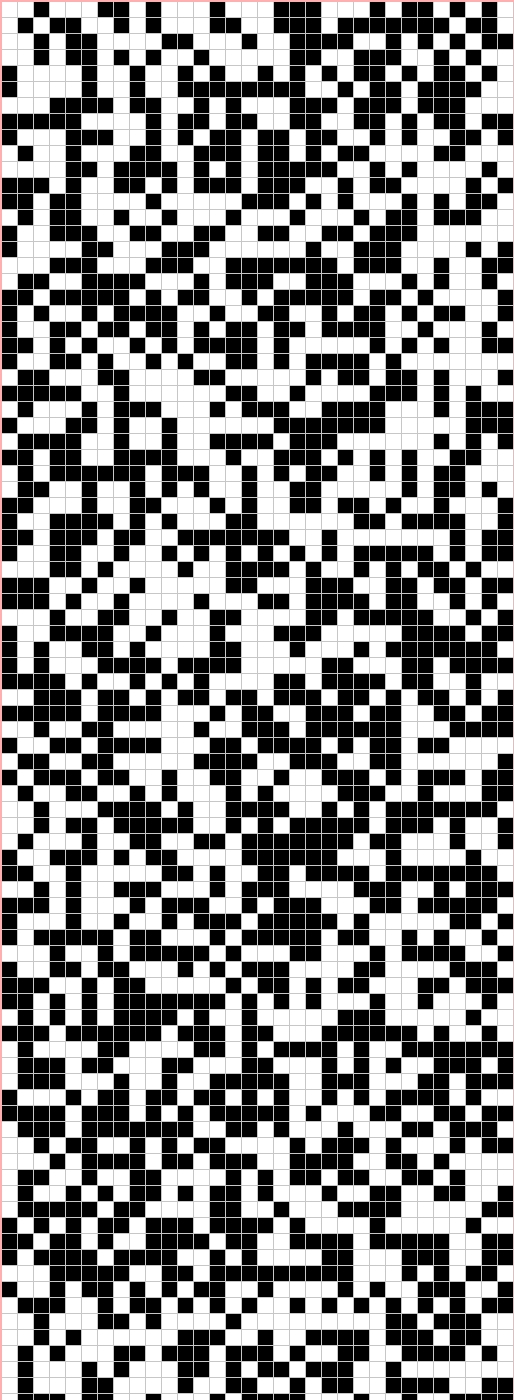}}
		\hfill
		\subfloat[$s_5$\label{lfsr_double_123456789123456789_spaceo}]{%
			\includegraphics[width=0.1\linewidth, height=5.0cm]{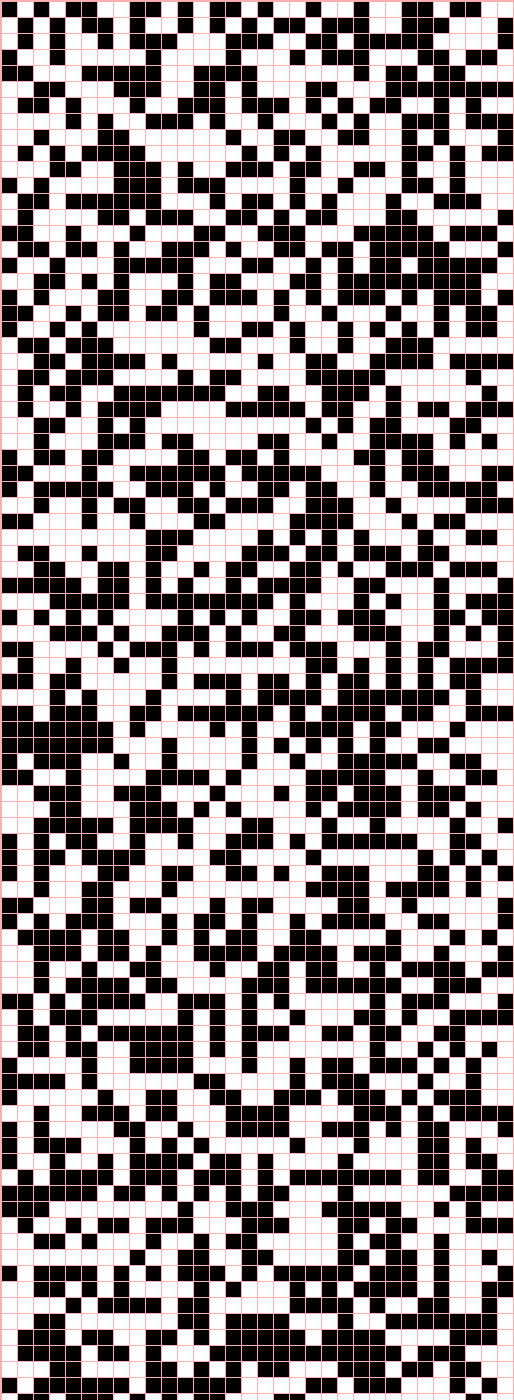}}
		\hfill
		\subfloat[$s_1$\label{xor32_7_space}]{%
			\includegraphics[width=0.1\linewidth, height=5.0cm]{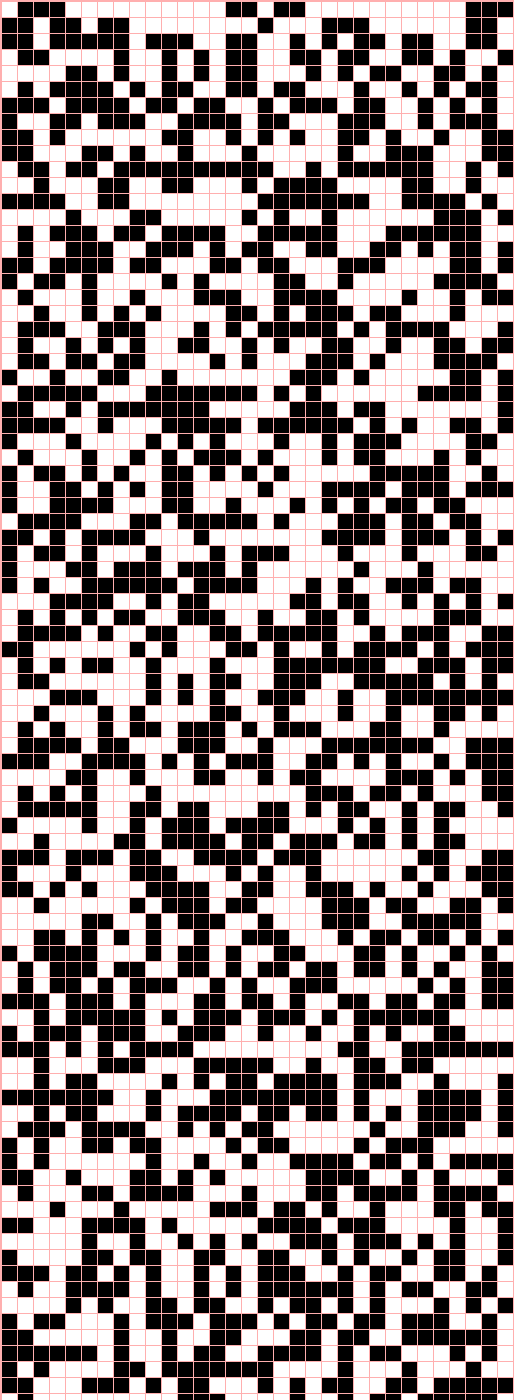}}
		\hfill
		\subfloat[$s_3$\label{xor32_12345_space}]{%
			\includegraphics[width=0.1\linewidth, height=5.0cm]{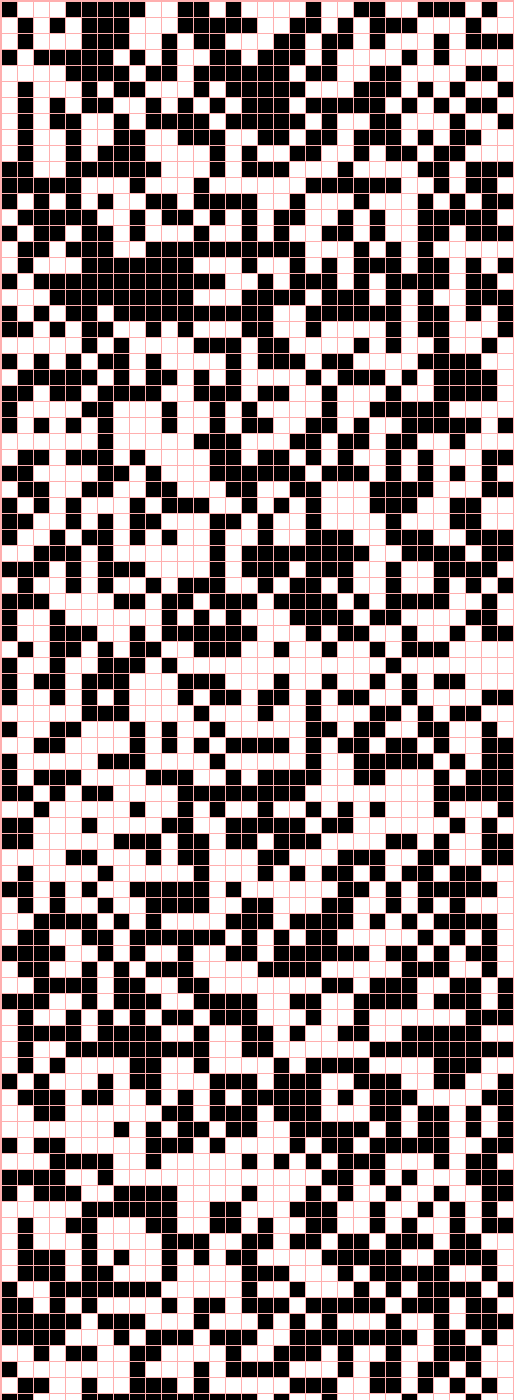}}
		\hfill
		\subfloat[$s_4$\label{xor32_9650218_space}]{%
			\includegraphics[width=0.1\linewidth, height=5.0cm]{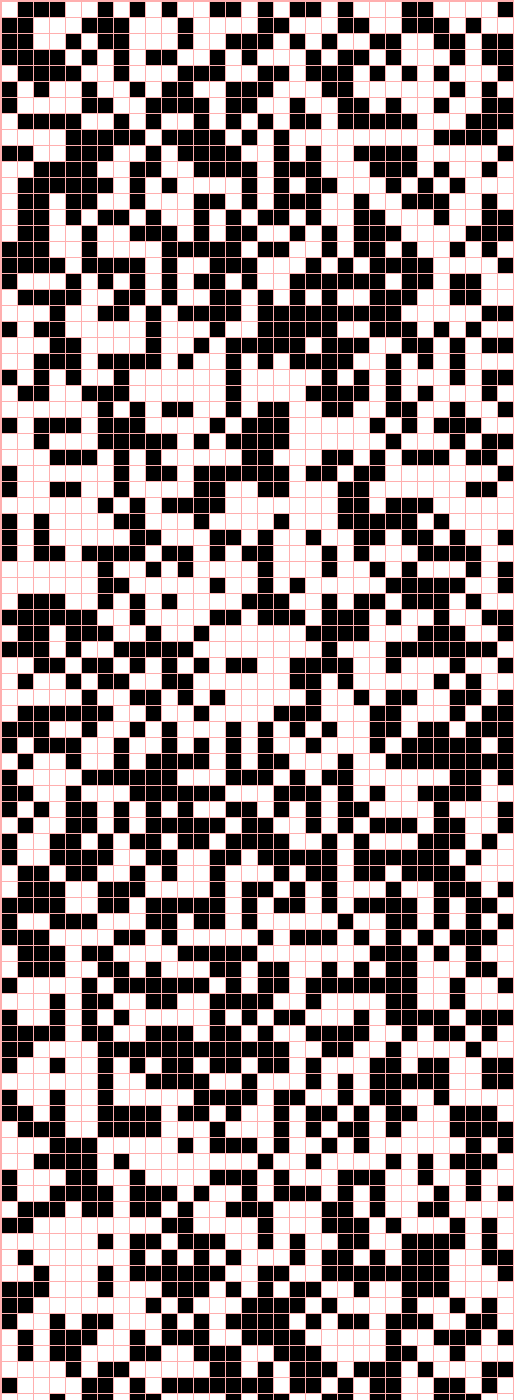}}
		\hfill
		\subfloat[$s_5$\label{xor32_123456789123456789_spaceo}]{%
			\includegraphics[width=0.1\linewidth, height=5.0cm]{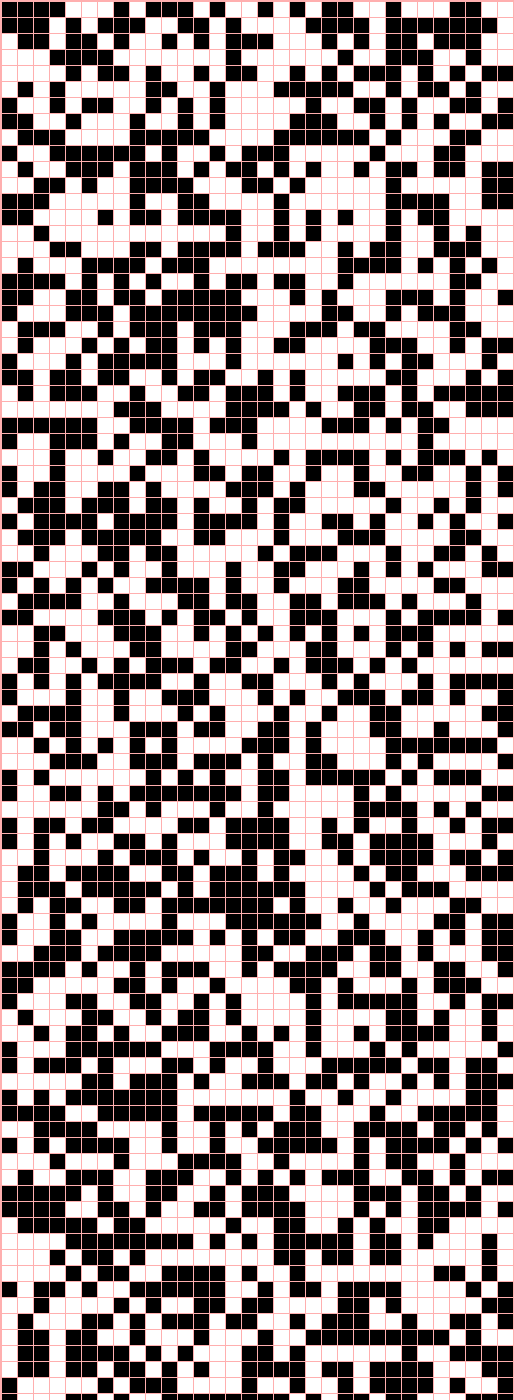}}
		%
		\hfill\\
		\subfloat[$s_1$\label{well512_7_space}]{%
			\includegraphics[width=0.1\linewidth, height=5.0cm]{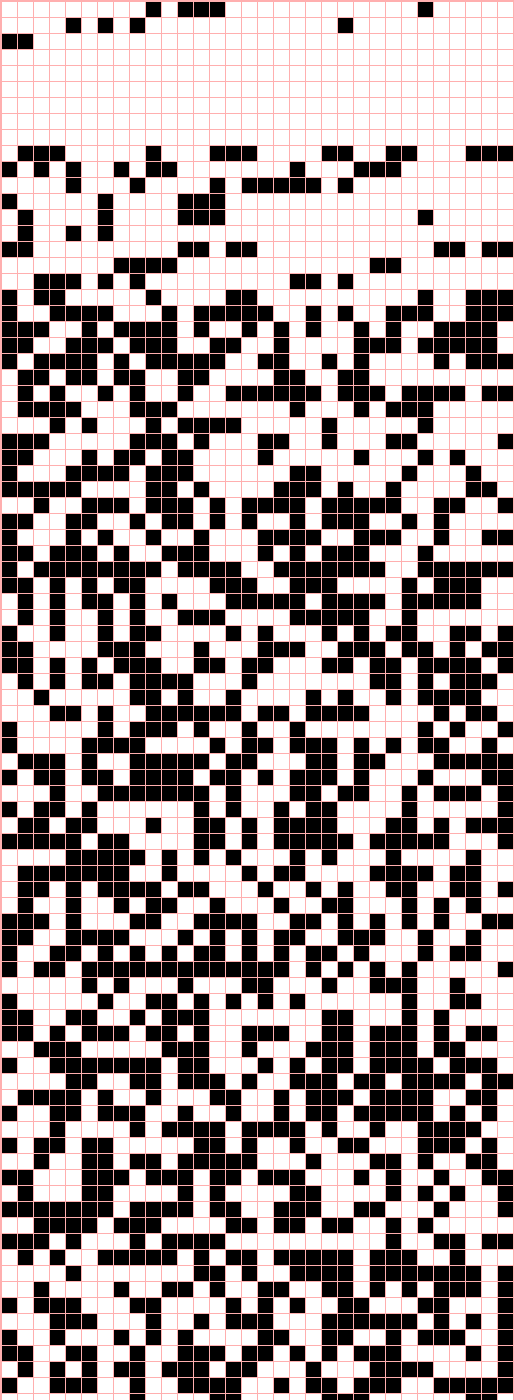}}
		\hfill
		\subfloat[$s_3$\label{well512_12345_space}]{%
			\includegraphics[width=0.1\linewidth, height=5.0cm]{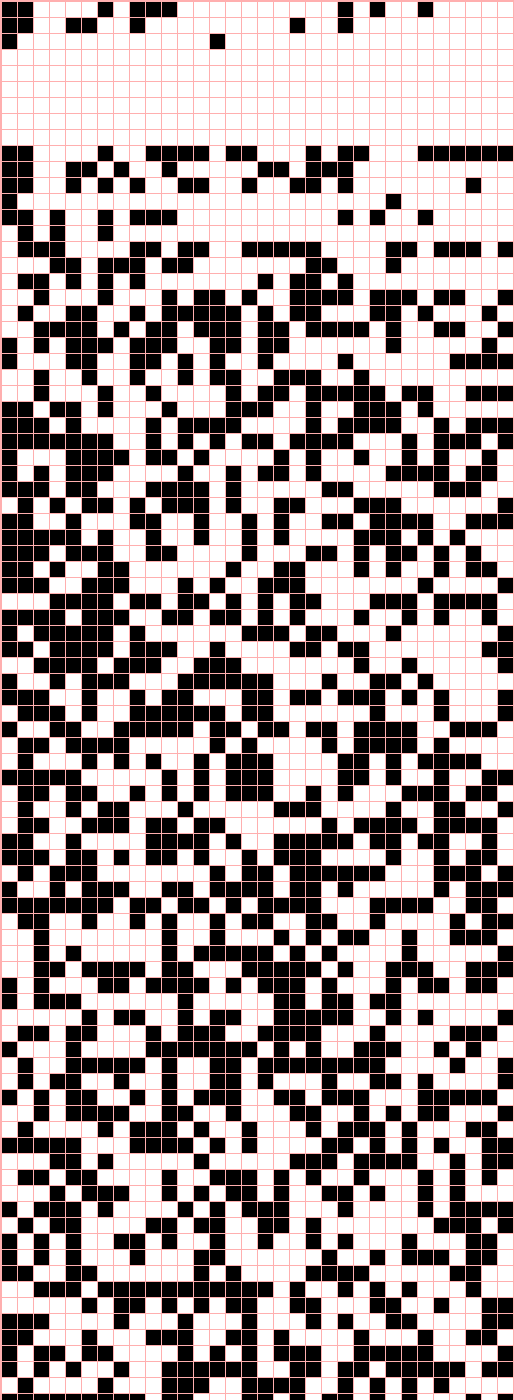}}
		\hfill
		\subfloat[$s_4$\label{well512_9650218_space}]{%
			\includegraphics[width=0.1\linewidth, height=5.0cm]{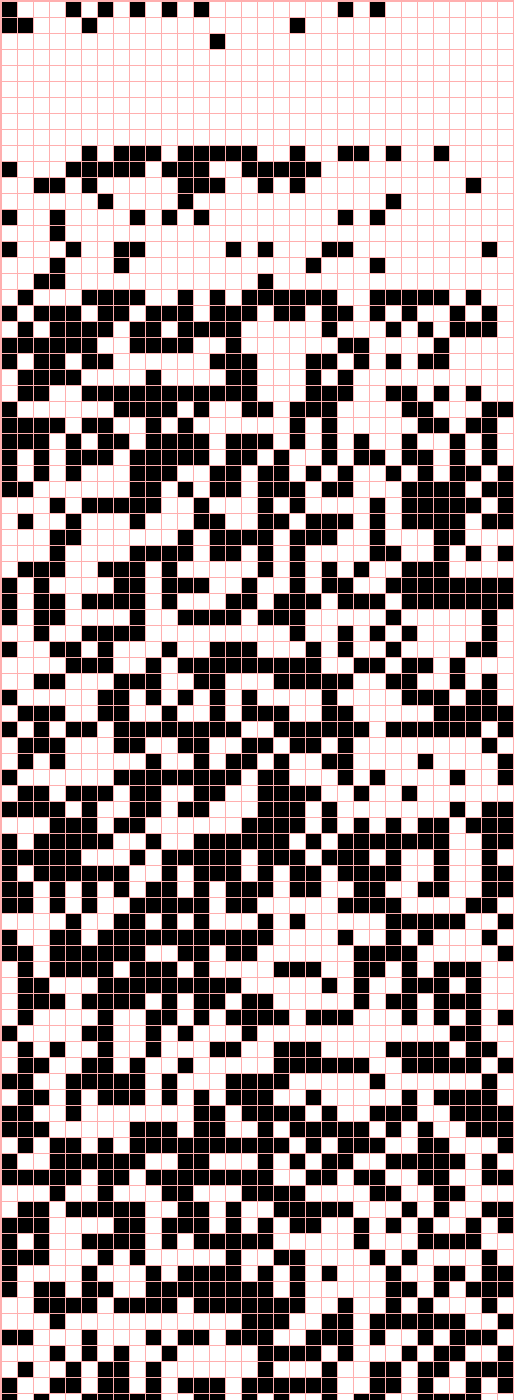}}
		\hfill
		\subfloat[$s_5$\label{well512_123456789123456789_spaceo}]{%
			\includegraphics[width=0.1\linewidth, height=5.0cm]{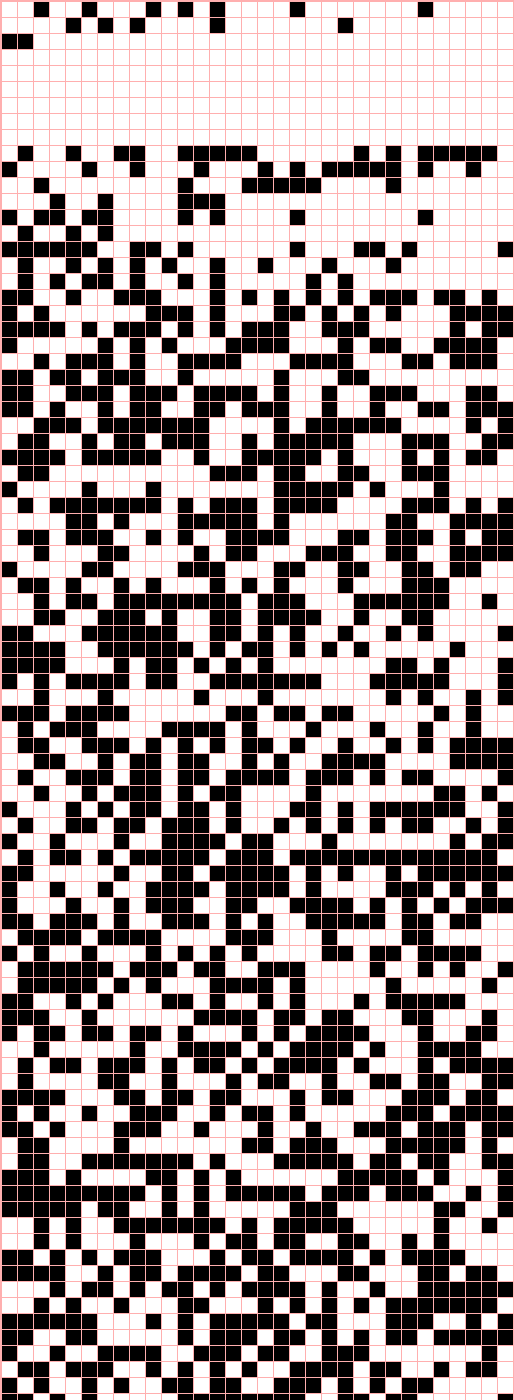}}
		\hfill
		\subfloat[$s_1$\label{well1024_7_space}]{%
			\includegraphics[width=0.1\linewidth, height=5.0cm]{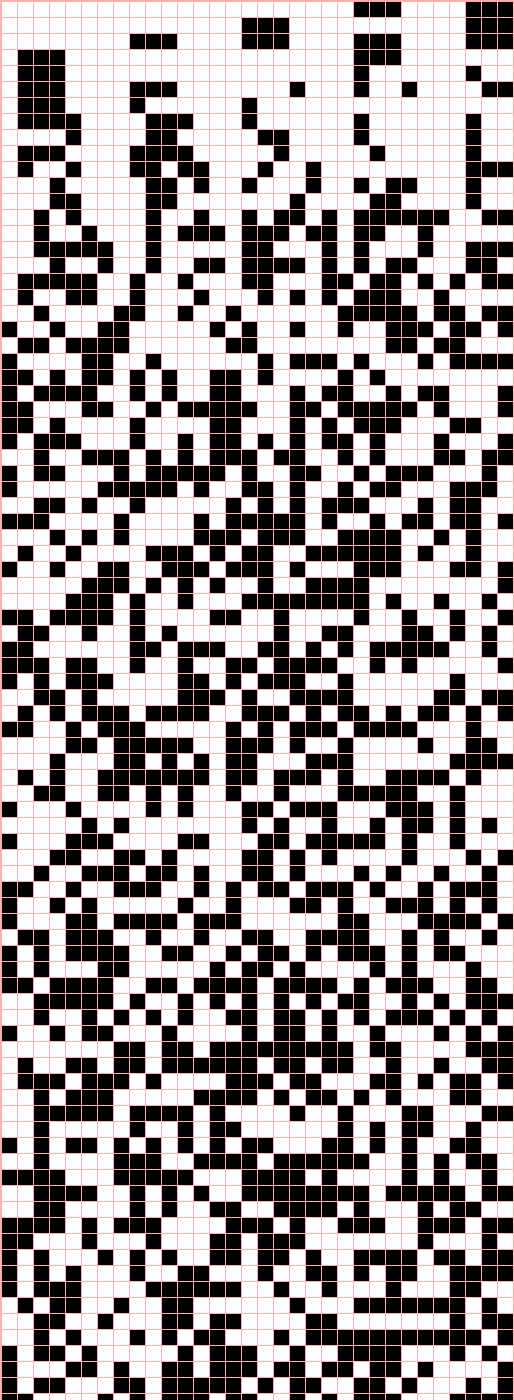}}
		\hfill
		\subfloat[$s_3$\label{well1024_12345_space}]{%
			\includegraphics[width=0.1\linewidth, height=5.0cm]{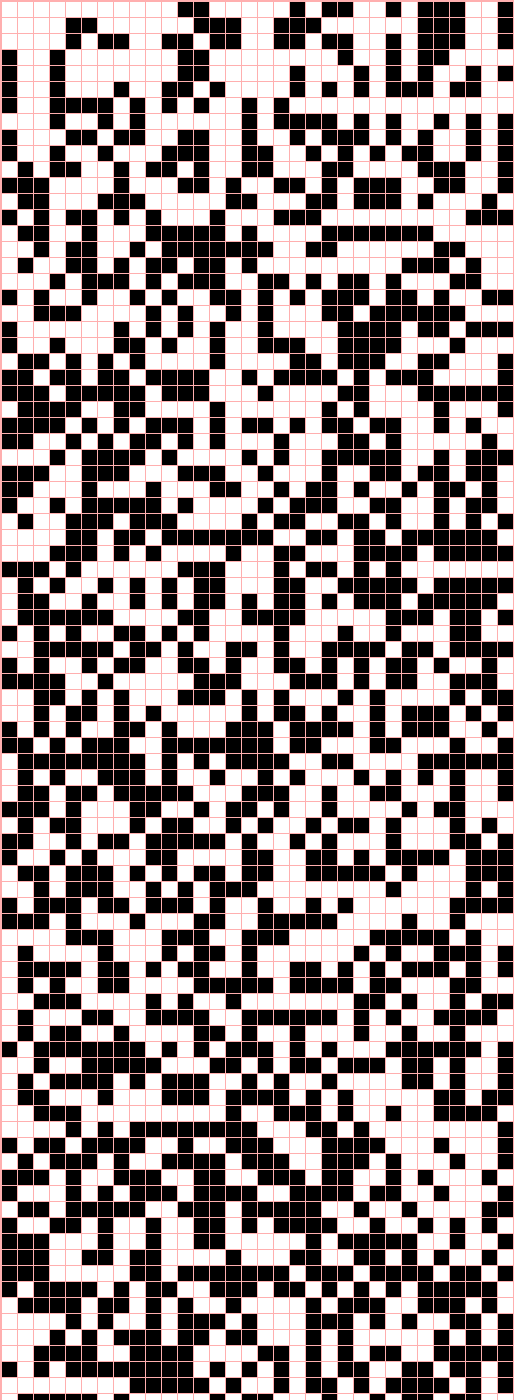}}
		\hfill
		\subfloat[$s_4$\label{well1024_9650218_space}]{%
			\includegraphics[width=0.1\linewidth, height=5.0cm]{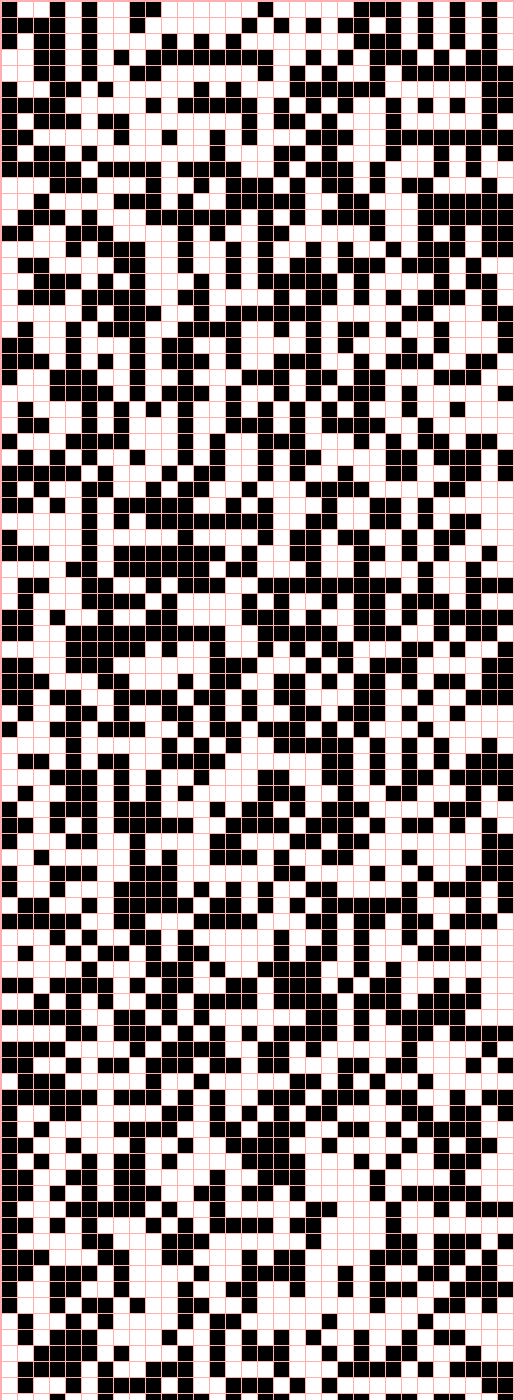}}
		\hfill
		\subfloat[$s_5$\label{well1024_123456789123456789_space}]{%
			\includegraphics[width=0.1\linewidth, height=5.0cm]{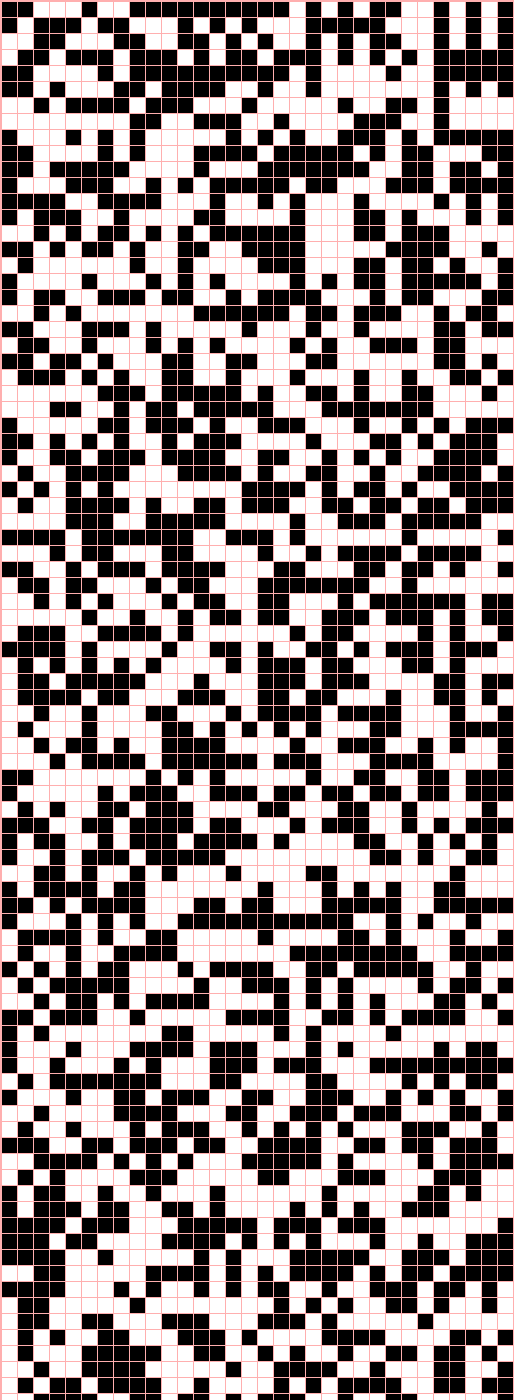}}
		\caption{Space-time diagram for LFSR258 (\ref{lfsr258_7_space} to \ref{lfsr258_123456789123456789_space}), LFSR113 (\ref{lfsr_double_7_space} to \ref{lfsr_double_123456789123456789_spaceo}) and xorshift (\ref{xor32_7_space} to \ref{xor32_123456789123456789_spaceo}), WELL512 (\ref{well512_7_space} to \ref{well512_123456789123456789_spaceo}) and WELL1024a (\ref{well1024_7_space} to \ref{well1024_123456789123456789_space})}
		\label{fig:well_space-time}
	\end{figure} 
	
	\begin{figure}[!h]
		\centering
		\vspace{-2.0em}
		\subfloat[$s_1$\label{xorshift64_7_space}]{%
			\includegraphics[width=0.2\linewidth, height=5.0cm]{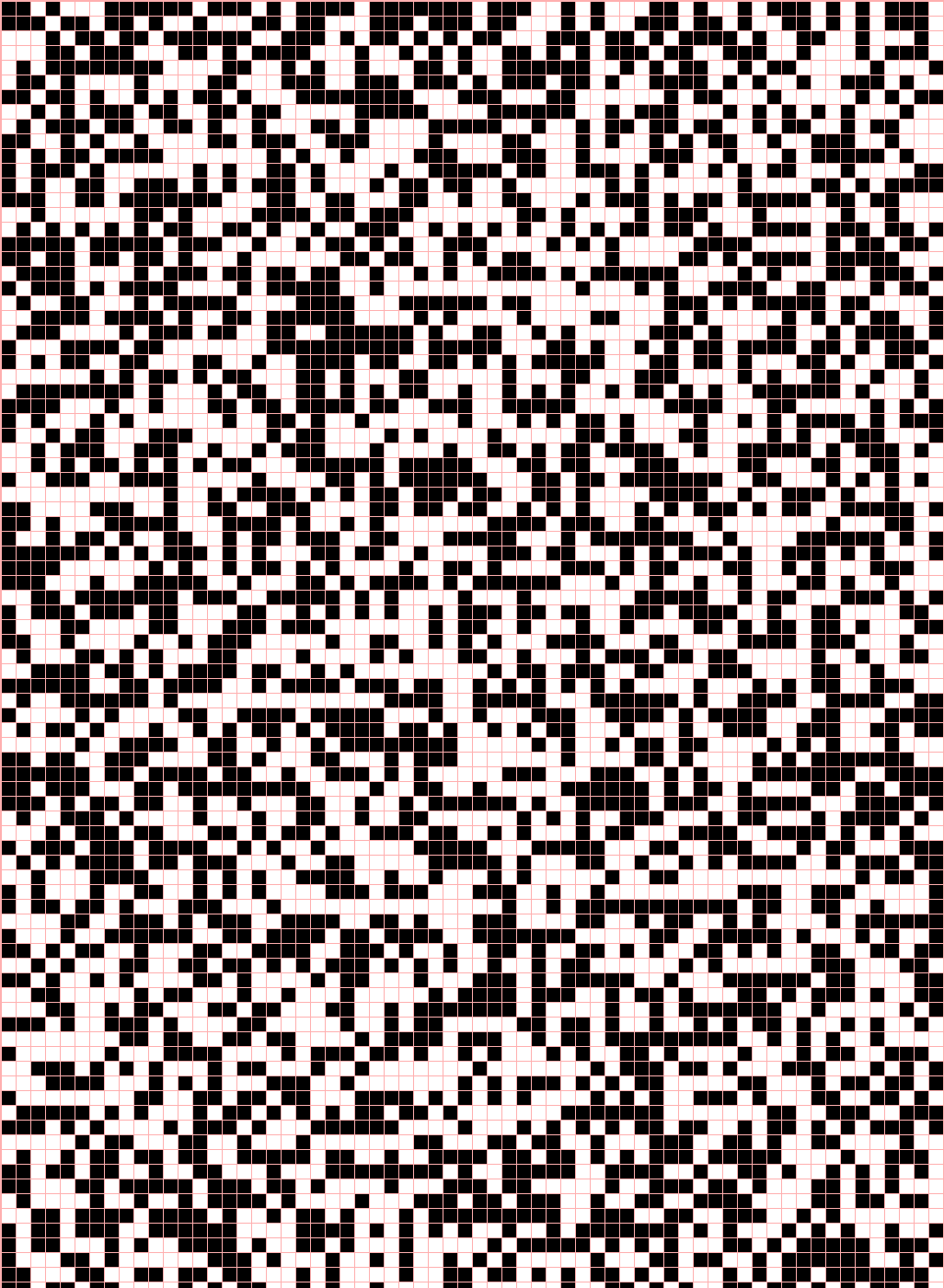}}
		\hfill
		\subfloat[$s_3$\label{xorshift64_12345_space}]{%
			\includegraphics[width=0.2\linewidth, height=5.0cm]{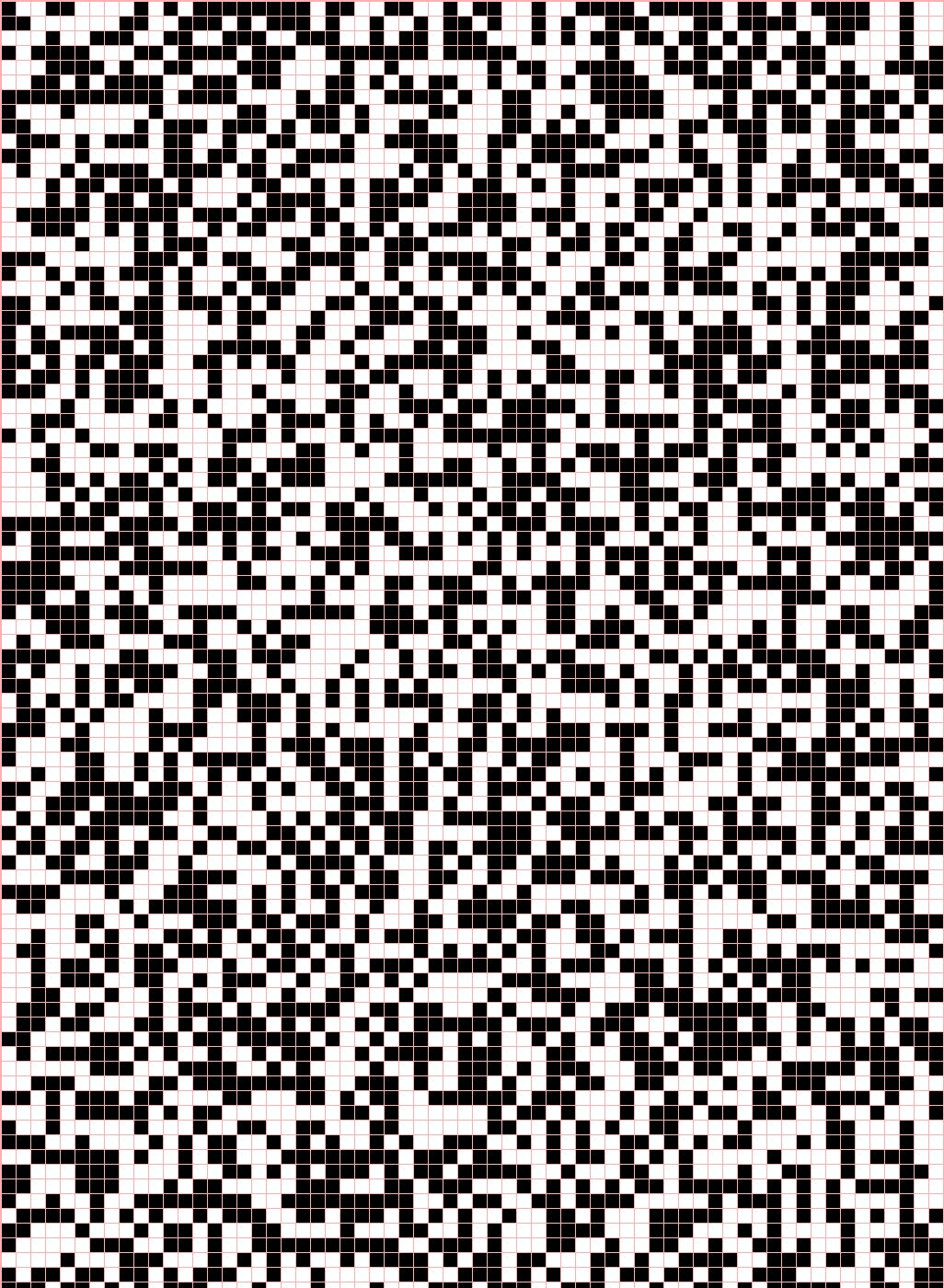}}
		\hfill
		\subfloat[$s_4$\label{xorshift64_9650218_space}]{%
			\includegraphics[width=0.2\linewidth, height=5.0cm]{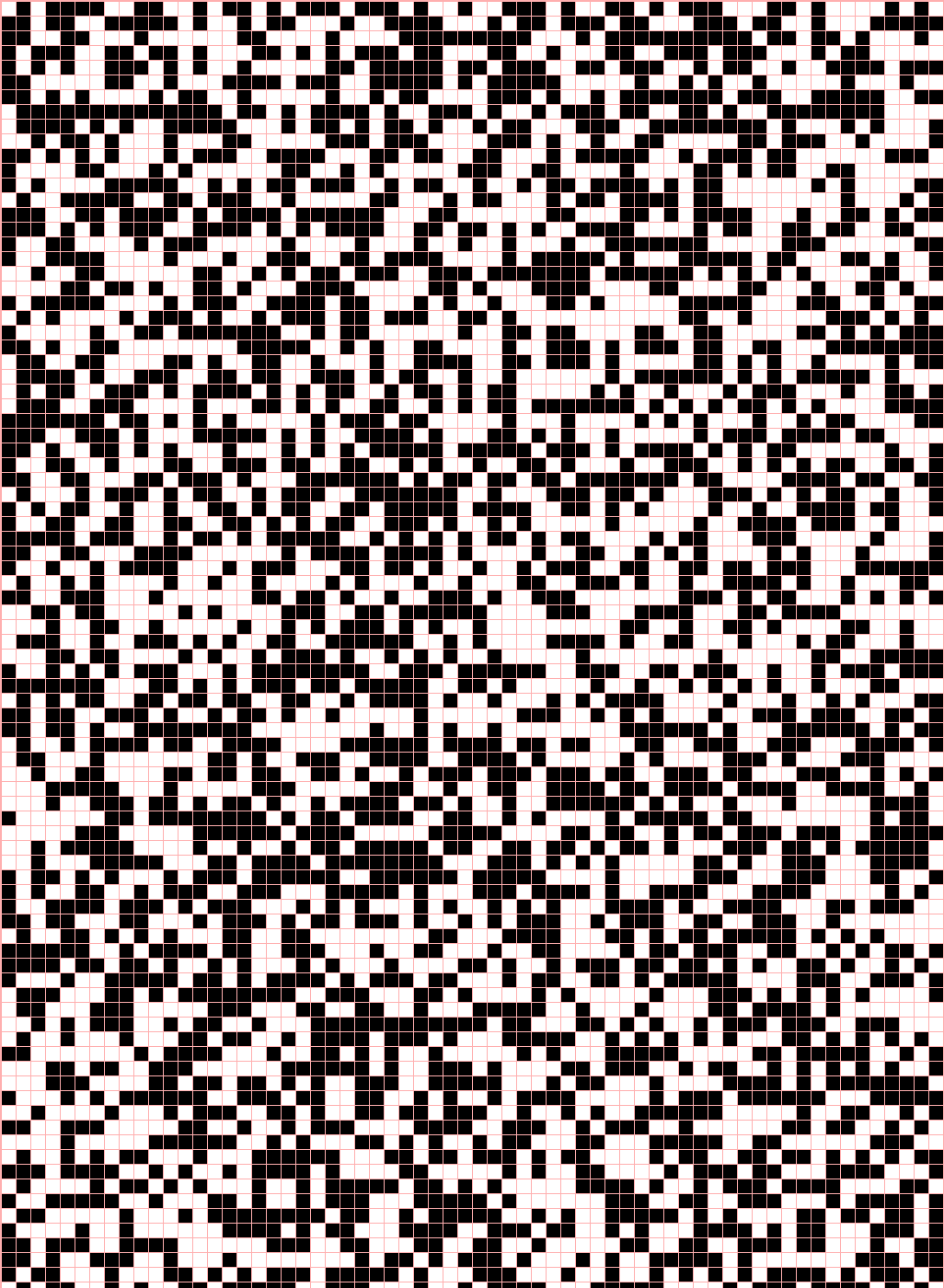}}
		\hfill
		\subfloat[$s_5$\label{xorshift64_123456789123456789_spaceo}]{%
			\includegraphics[width=0.2\linewidth, height=5.0cm]{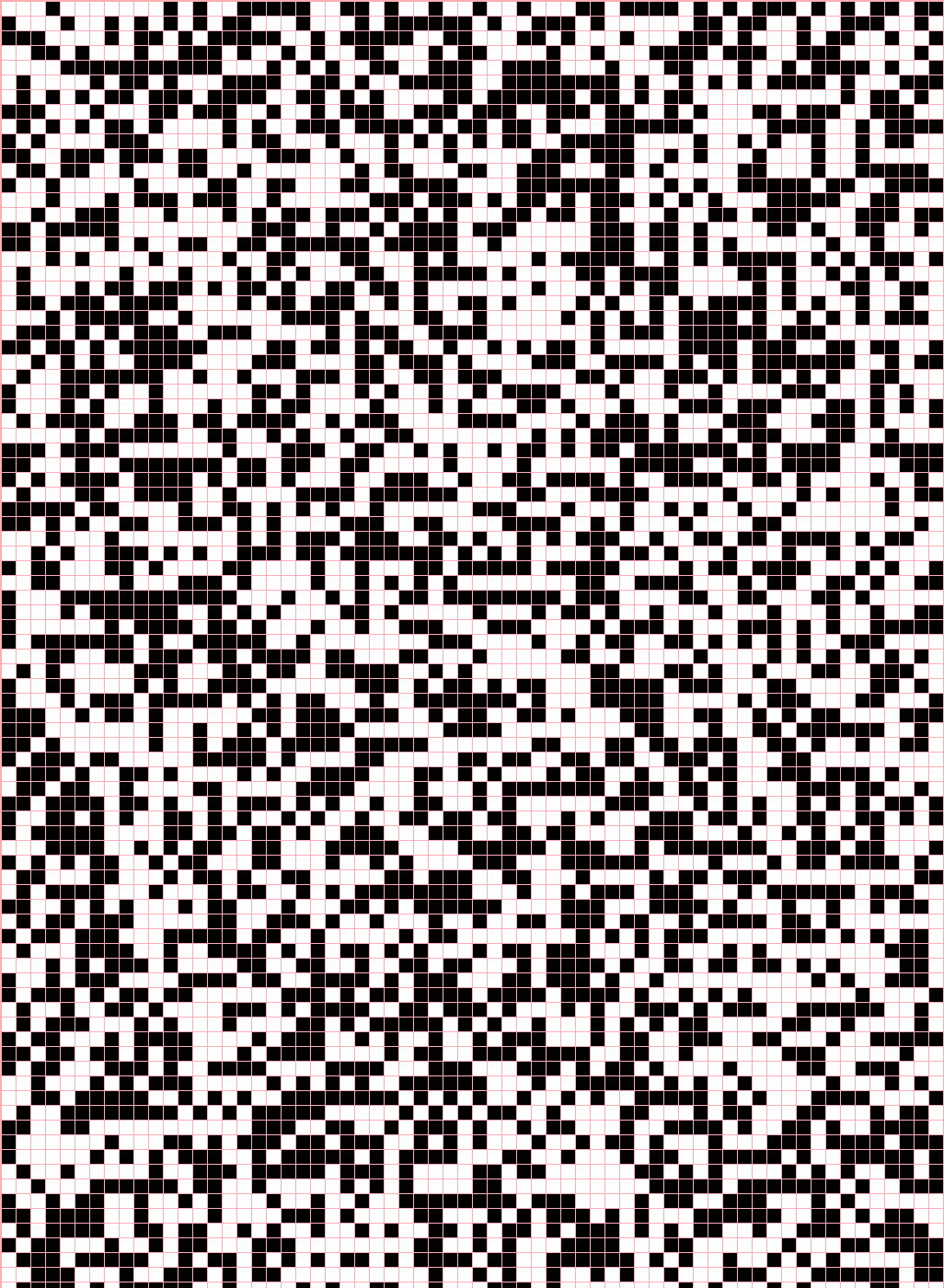}}
		%
		\hfill\\
		\subfloat[$s_1$\label{xorshift1024_7_space}]{%
			\includegraphics[width=0.2\linewidth, height=5.0cm]{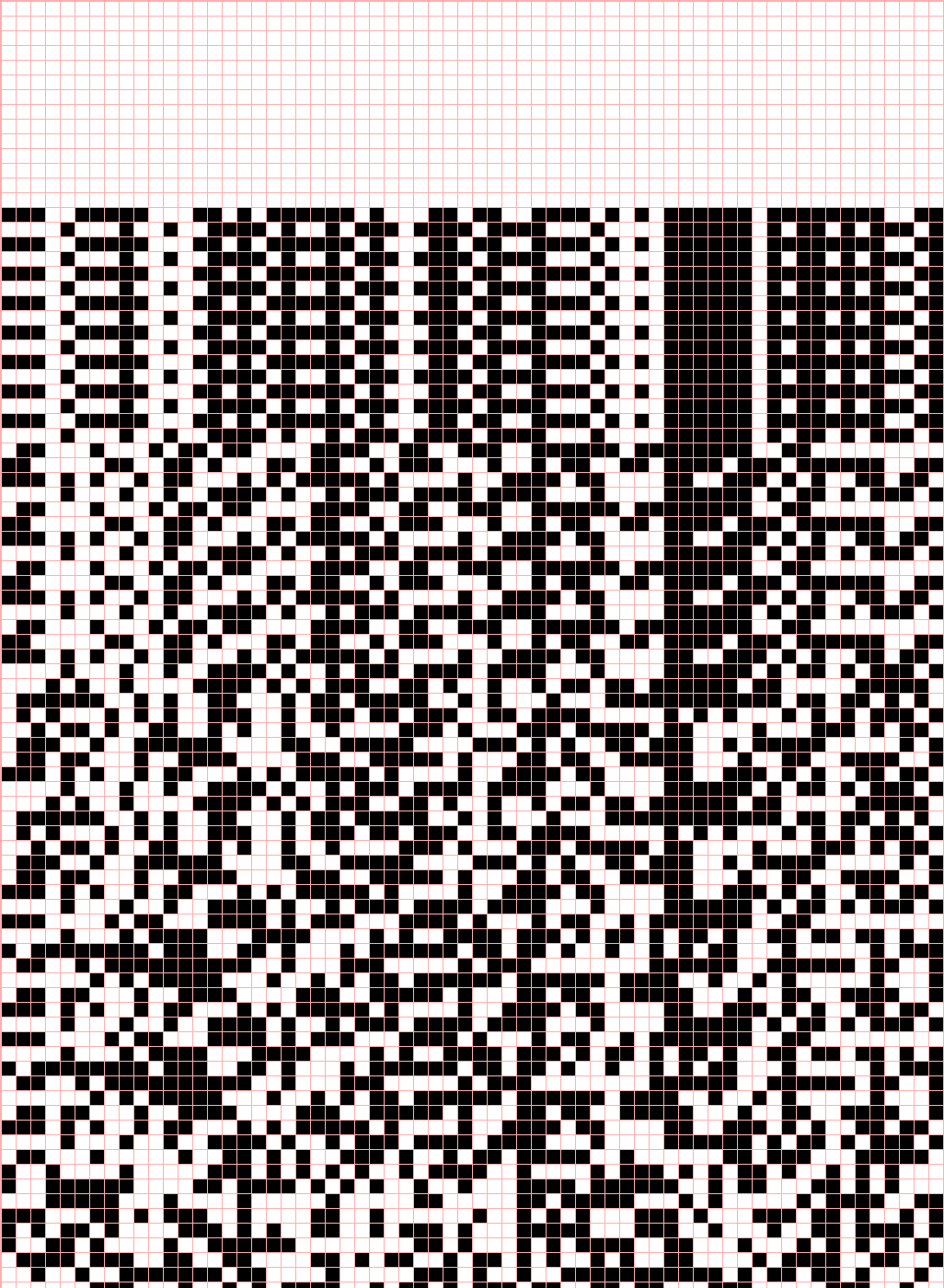}}
		\hfill
		\subfloat[$s_3$\label{xorshift1024_12345_space}]{%
			\includegraphics[width=0.2\linewidth, height=5.0cm]{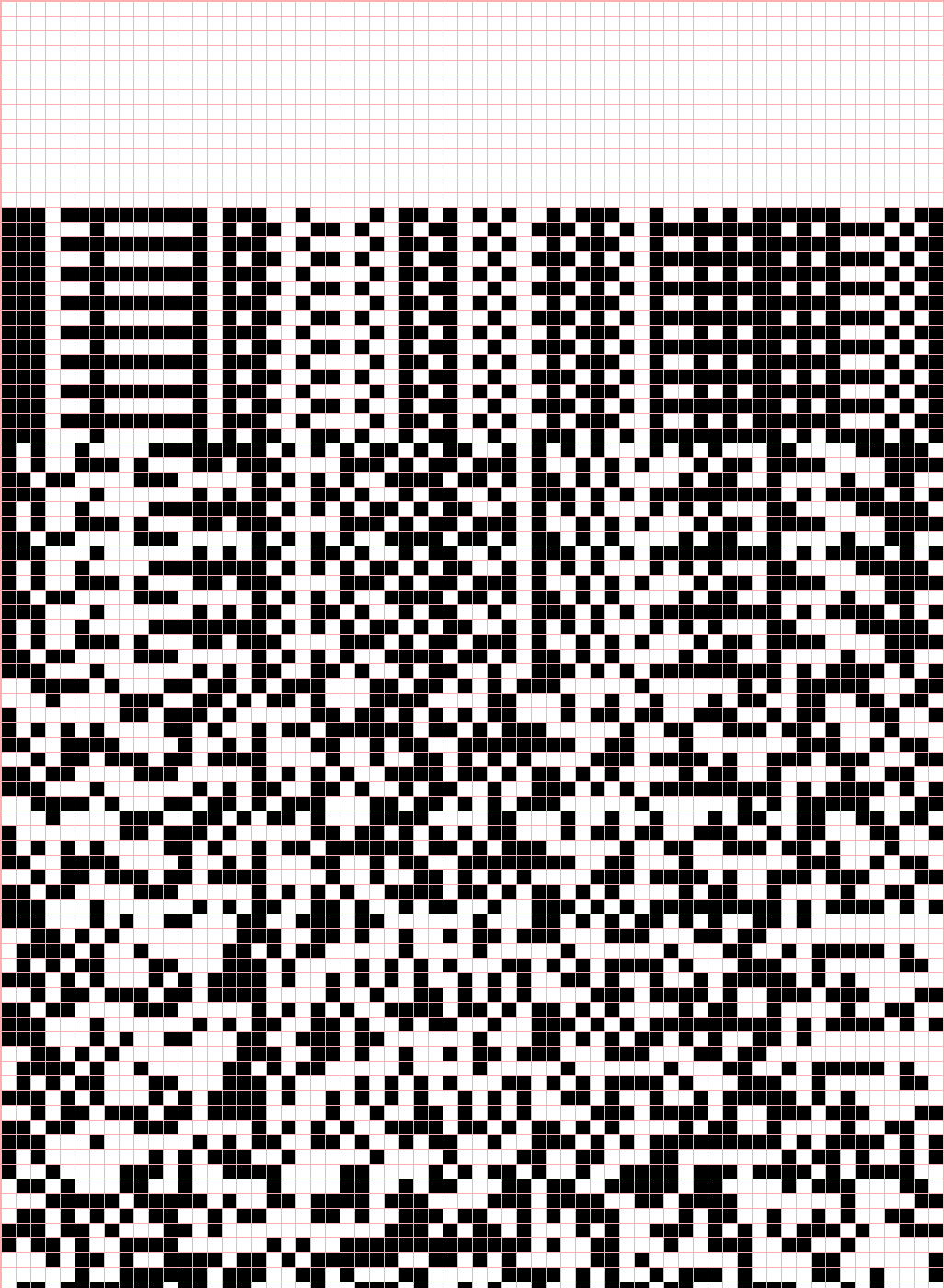}}
		\hfill
		\subfloat[$s_4$\label{xorshift1024_9650218_space}]{%
			\includegraphics[width=0.2\linewidth, height=5.0cm]{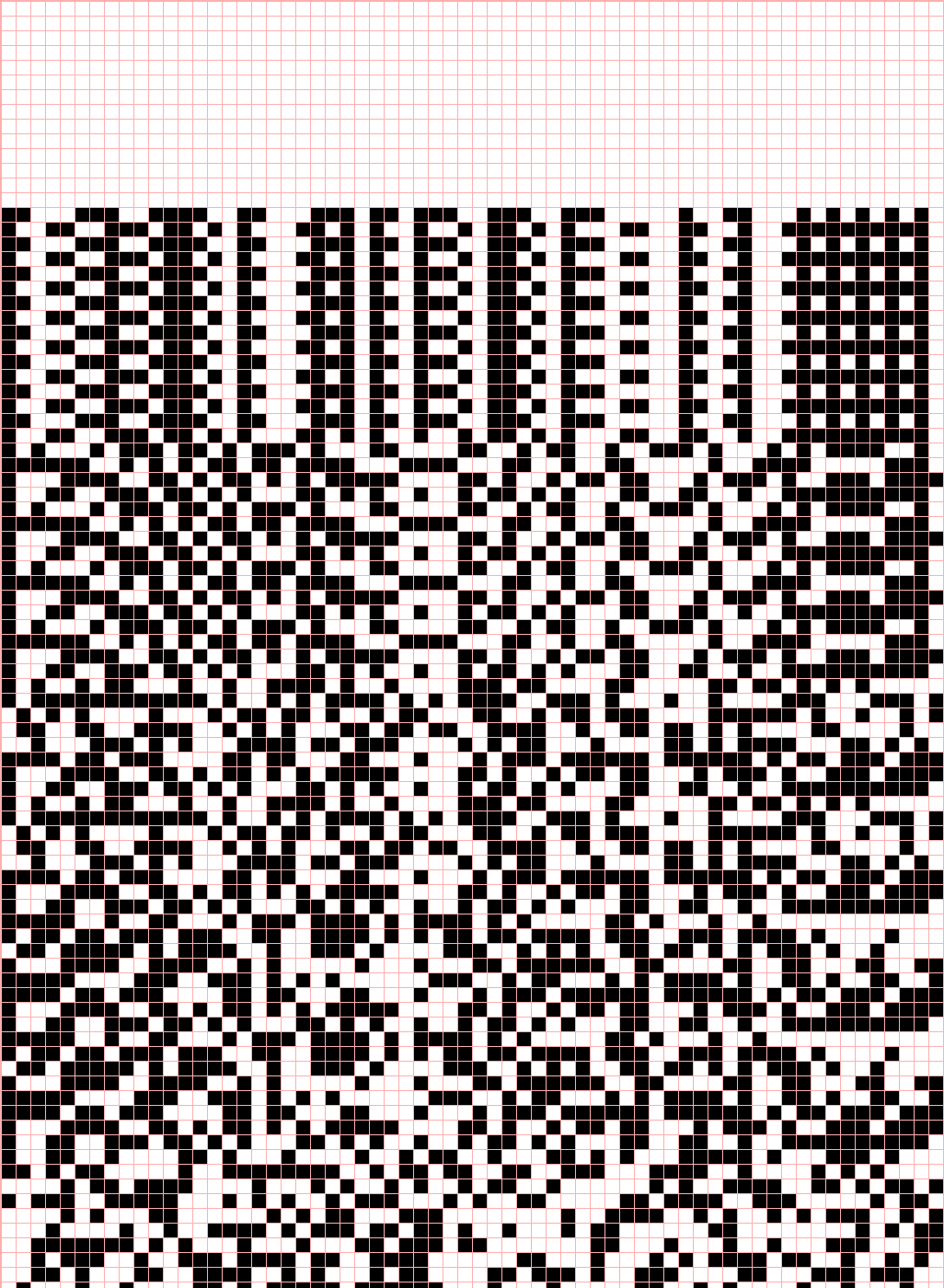}}
		\hfill
		\subfloat[$s_5$\label{xorshift1024_123456789123456789_space}]{%
			\includegraphics[width=0.2\linewidth, height=5.0cm]{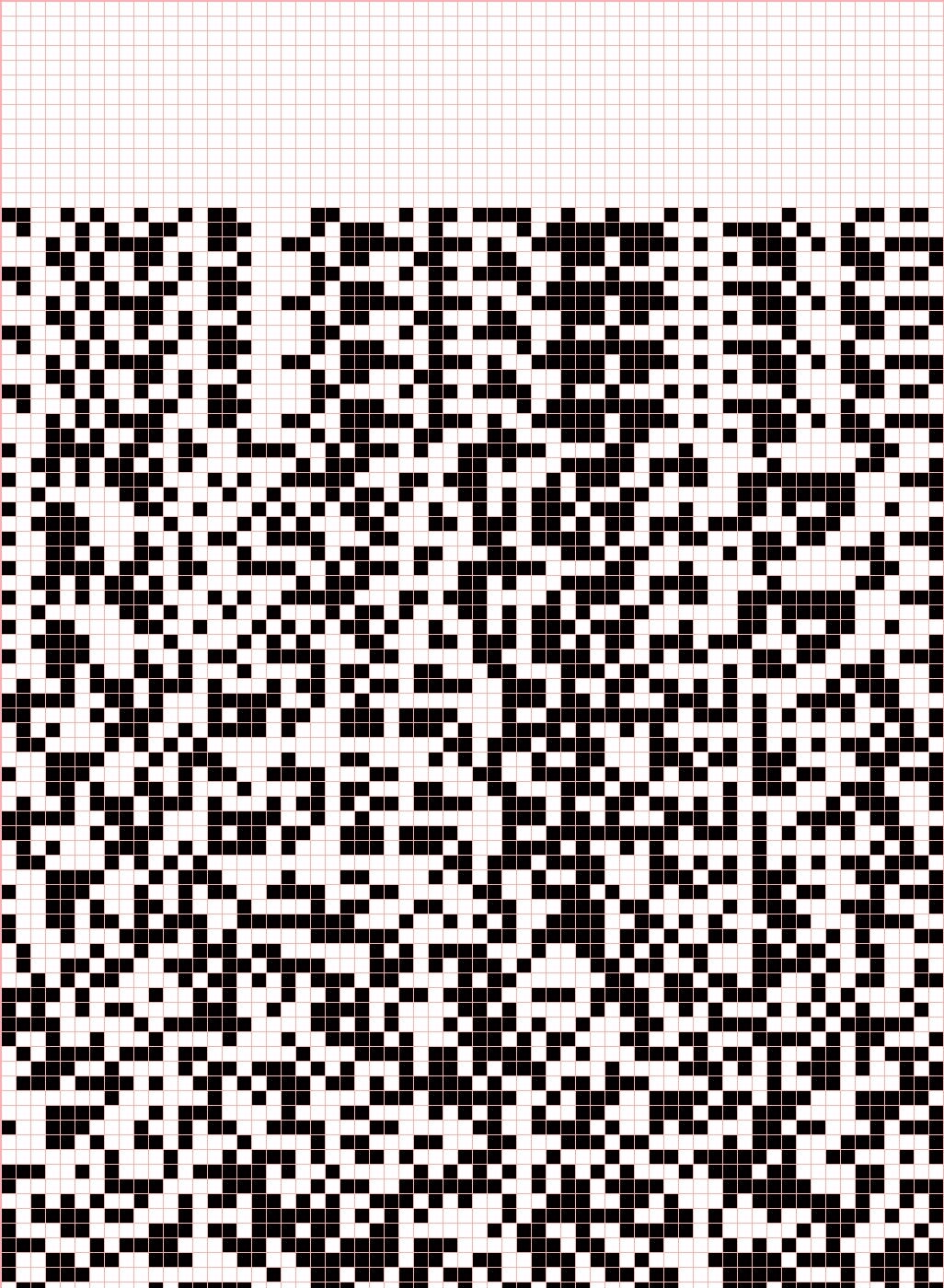}}
		%
		\hfill\\
		\subfloat[$s_1$\label{xorshift_7_space}]{%
			\includegraphics[width=0.2\linewidth, height=5.0cm]{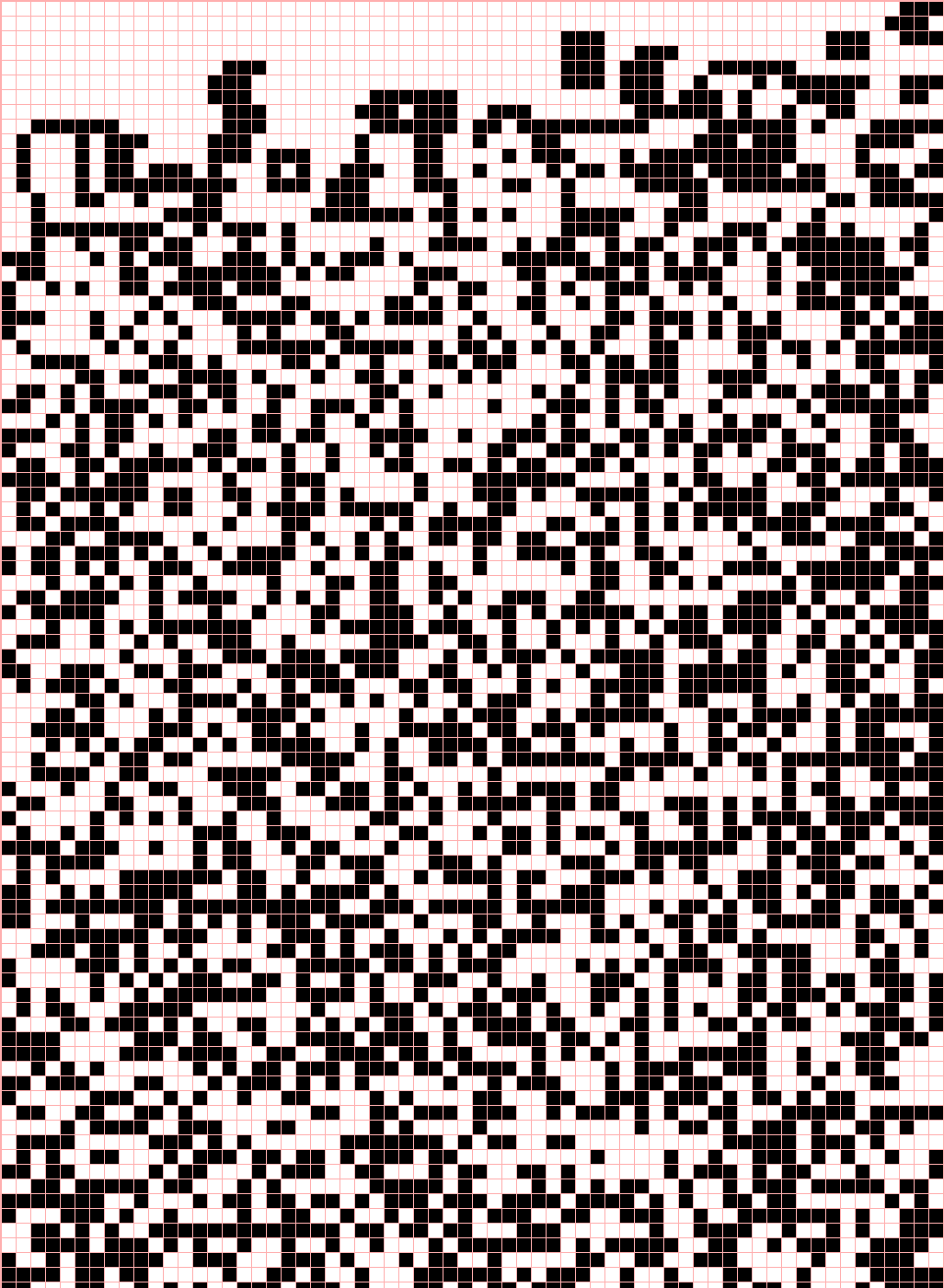}}
		\hfill
		\subfloat[$s_3$\label{xorshift_12345_space}]{%
			\includegraphics[width=0.2\linewidth, height=5.0cm]{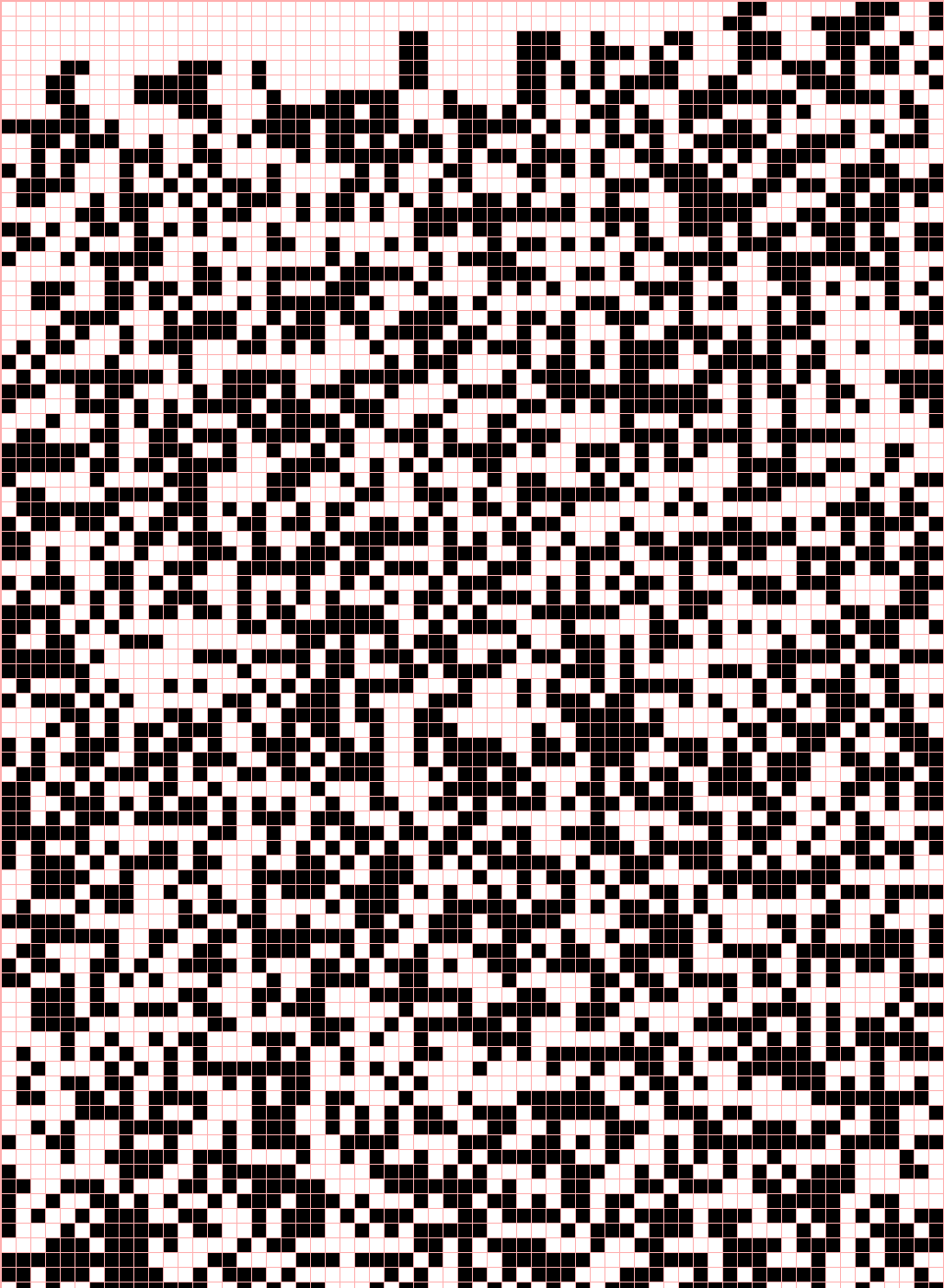}}
		\hfill
		\subfloat[$s_4$\label{xorshift_9650218_space}]{%
			\includegraphics[width=0.2\linewidth, height=5.0cm]{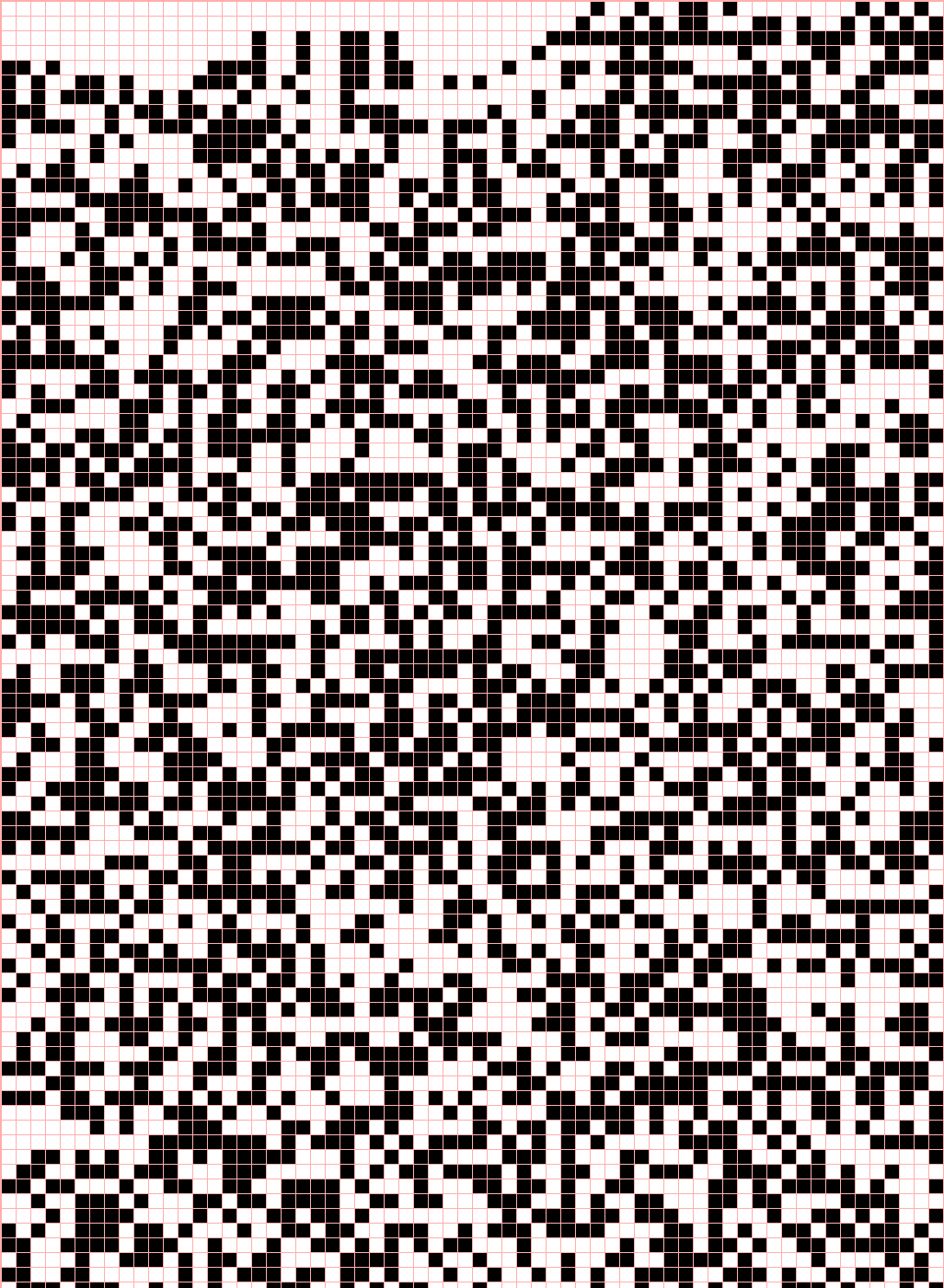}}
		\hfill
		\subfloat[$s_5$\label{xorshift_123456789123456789_space}]{%
			\includegraphics[width=0.2\linewidth, height=5.0cm]{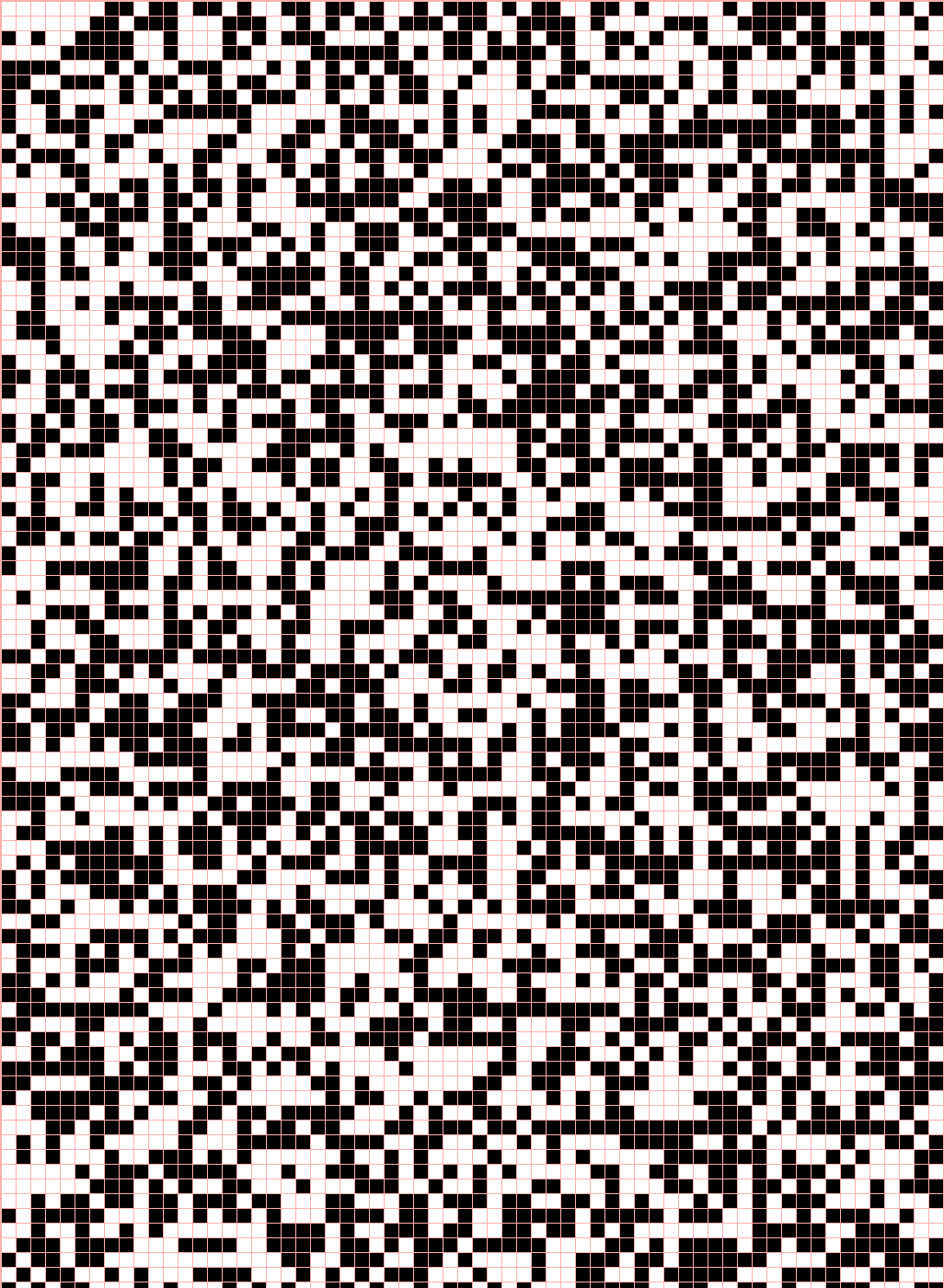}}
		\caption{Space-time diagram for xorshift64* (\ref{xorshift64_7_space} to \ref{xorshift64_123456789123456789_spaceo}), xorshift1024* (\ref{xorshift1024_7_space} to \ref{xorshift1024_123456789123456789_space}) and xorshift128+ (\ref{xorshift_7_space} to \ref{xorshift_123456789123456789_space})}
		\label{fig:xor_space-time}
	\end{figure} 
	
	\begin{figure}[!h]
		\centering
		\vspace{-2.0em}
		\subfloat[$s_1$\label{mt_32_7_space}]{%
			\includegraphics[width=0.1\linewidth, height=5.0cm]{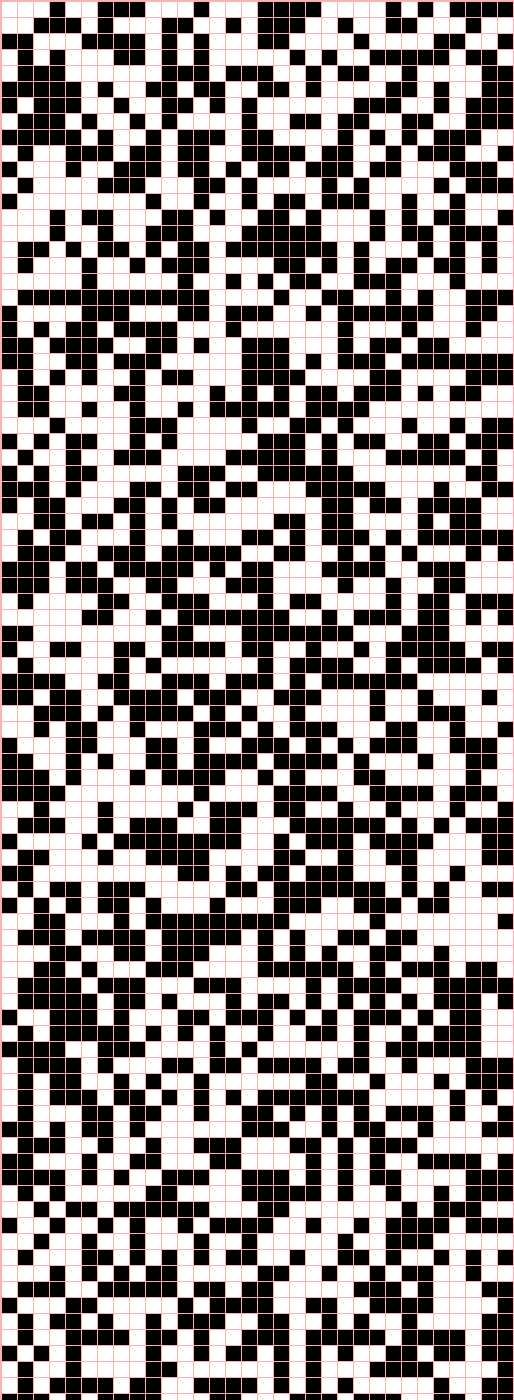}}
		\hfill
		\subfloat[$s_3$\label{mt_32_12345_space}]{%
			\includegraphics[width=0.1\linewidth, height=5.0cm]{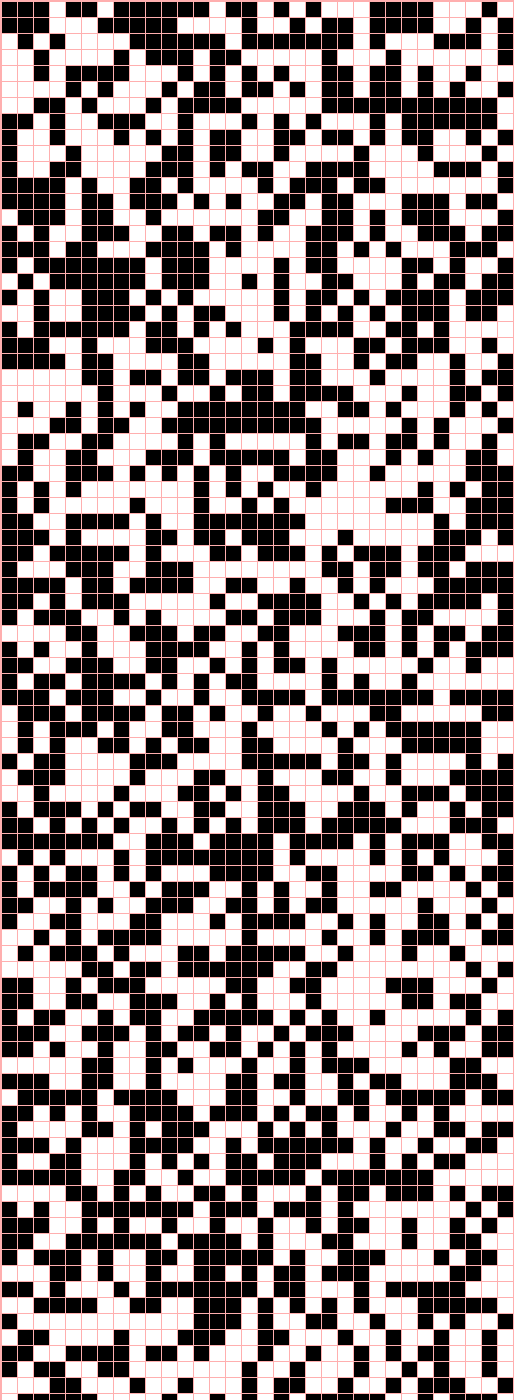}}
		\hfill
		\subfloat[$s_4$\label{mt_32_9650218_space}]{%
			\includegraphics[width=0.1\linewidth, height=5.0cm]{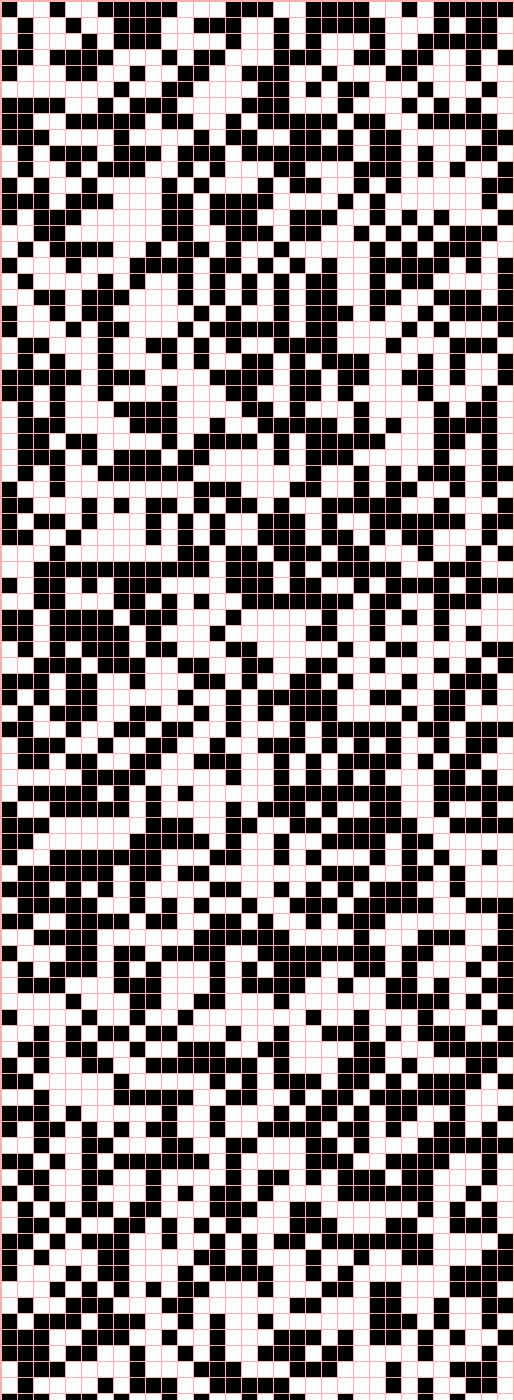}}
		\hfill
		\subfloat[$s_5$\label{mt_32_123456789123456789_spaceo}]{%
			\includegraphics[width=0.1\linewidth, height=5.0cm]{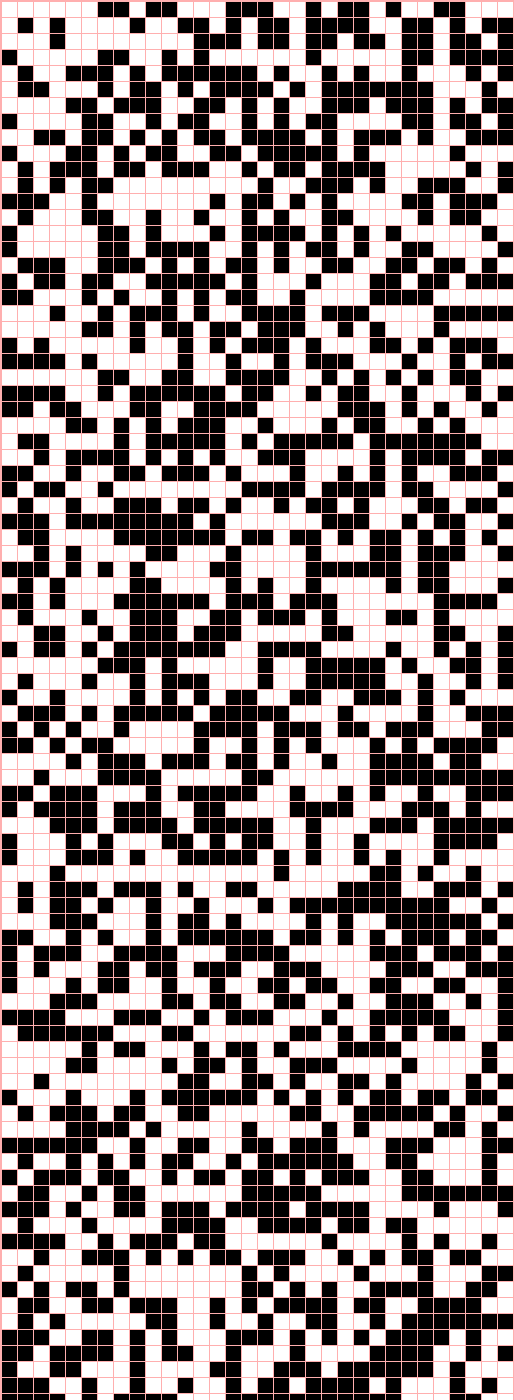}}
		\hfill
		\subfloat[$s_1$\label{sfmt_32_7_space}]{%
			\includegraphics[width=0.1\linewidth, height=5.0cm]{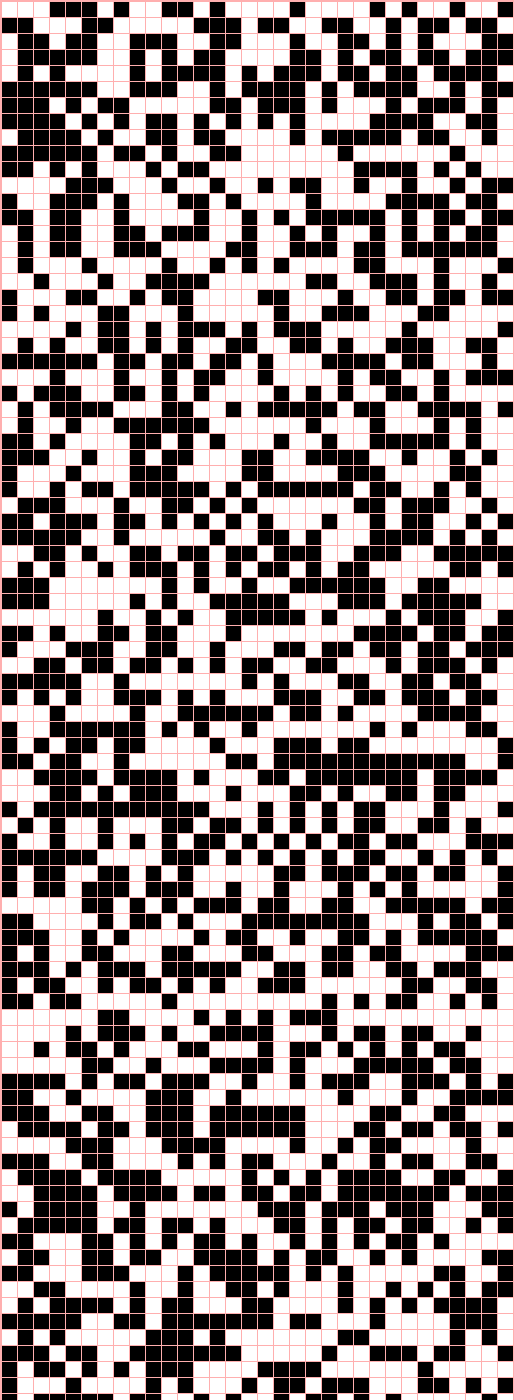}}
		\hfill
		\subfloat[$s_3$\label{sfmt_32_12345_space}]{%
			\includegraphics[width=0.1\linewidth, height=5.0cm]{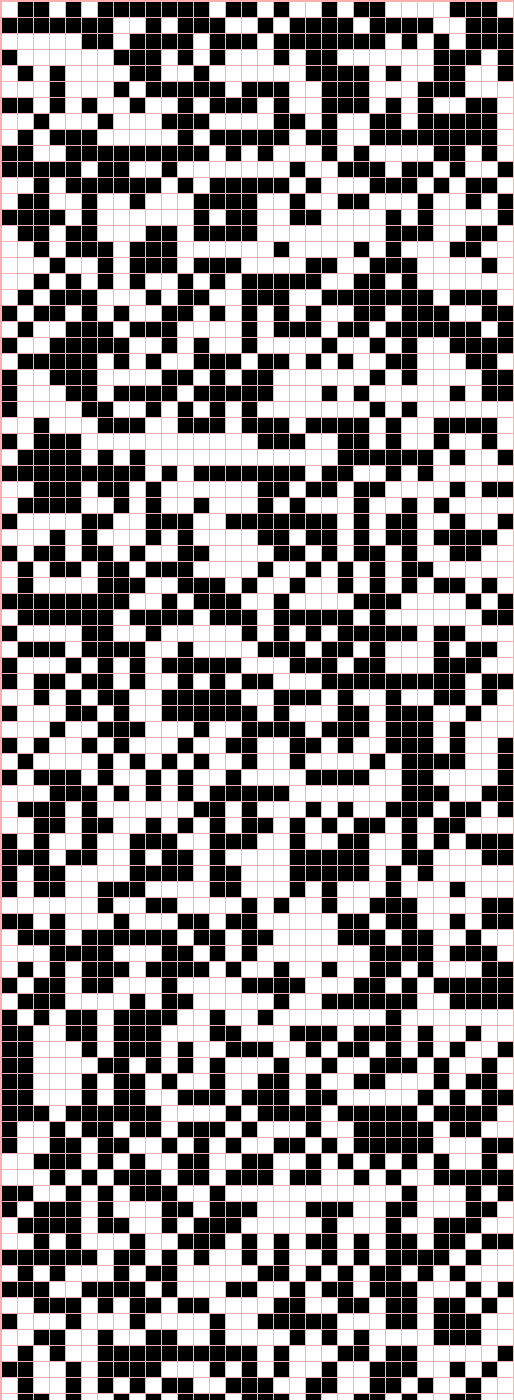}}
		\hfill
		\subfloat[$s_4$\label{sfmt_32_9650218_space}]{%
			\includegraphics[width=0.1\linewidth, height=5.0cm]{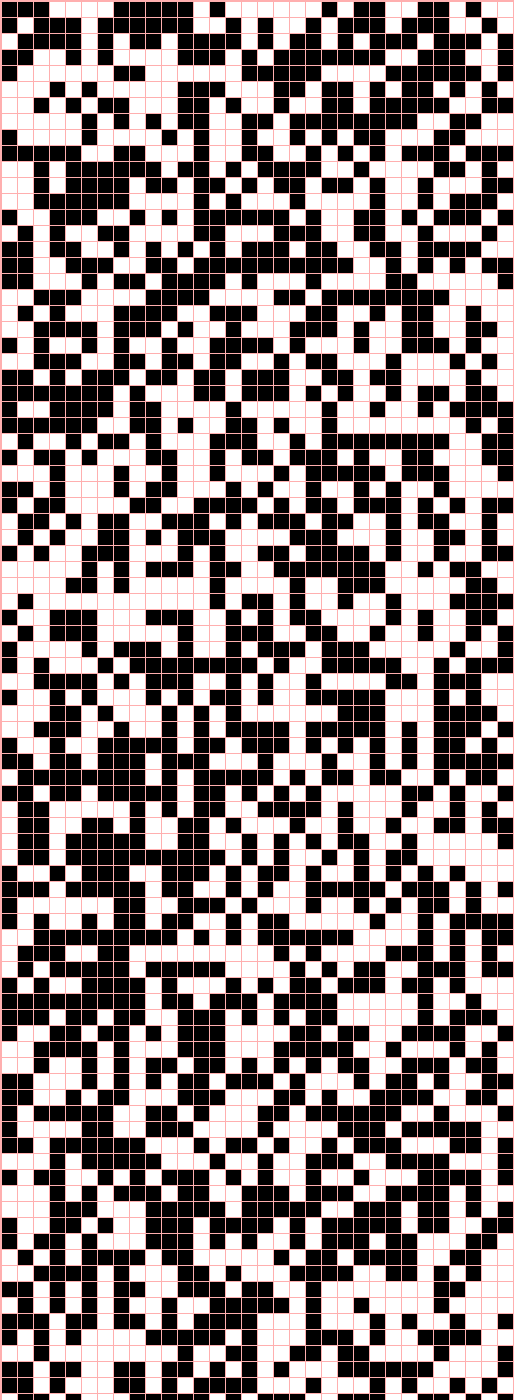}}
		\hfill
		\subfloat[$s_5$\label{sfmt_32_123456789123456789_spaceo}]{%
			\includegraphics[width=0.1\linewidth, height=5.0cm]{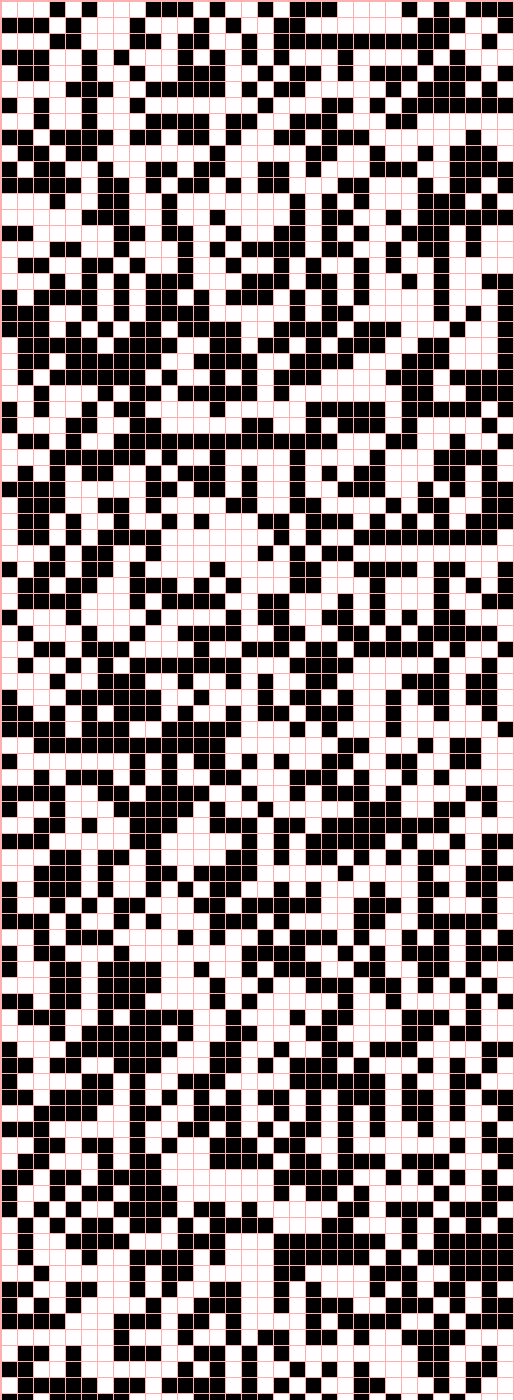}}
		\hfill\\
		\subfloat[$s_1$\label{mt_64_7_space}]{%
			\includegraphics[width=0.2\linewidth, height=5.0cm]{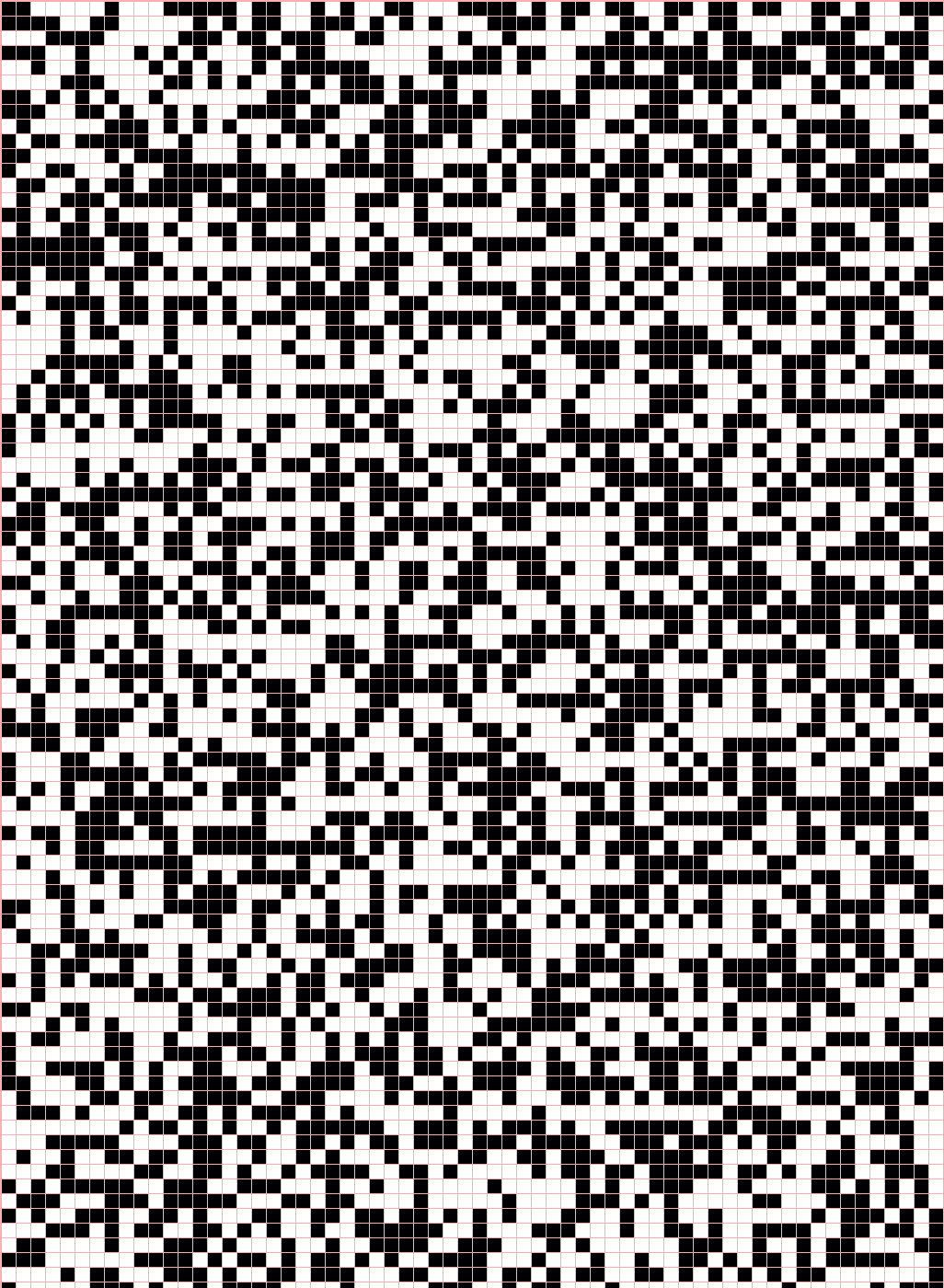}}
		\hfill
		\subfloat[$s_3$\label{mt_64_12345_space}]{%
			\includegraphics[width=0.2\linewidth, height=5.0cm]{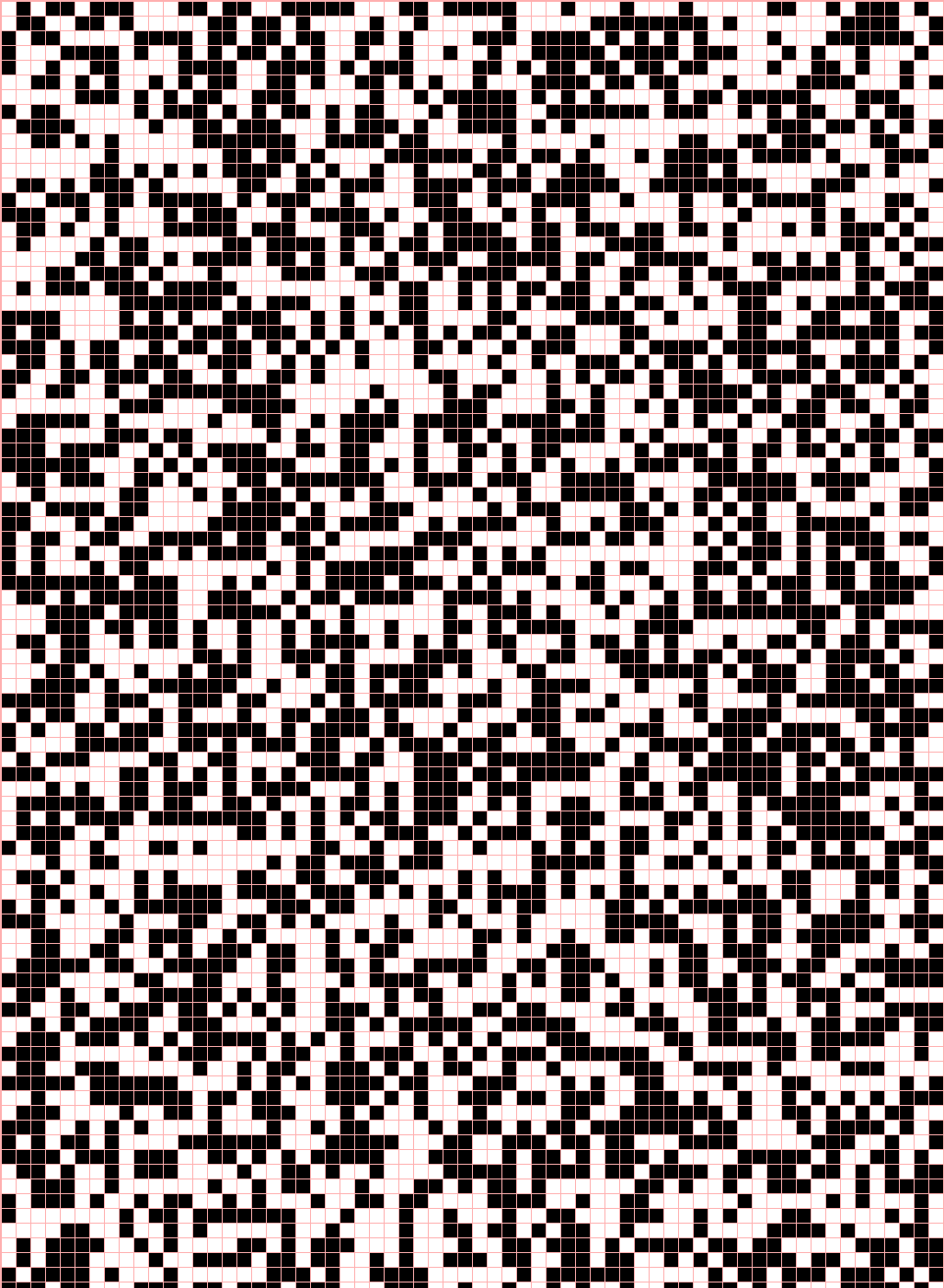}}
		\hfill
		\subfloat[$s_4$\label{mt_64_9650218_space}]{%
			\includegraphics[width=0.2\linewidth, height=5.0cm]{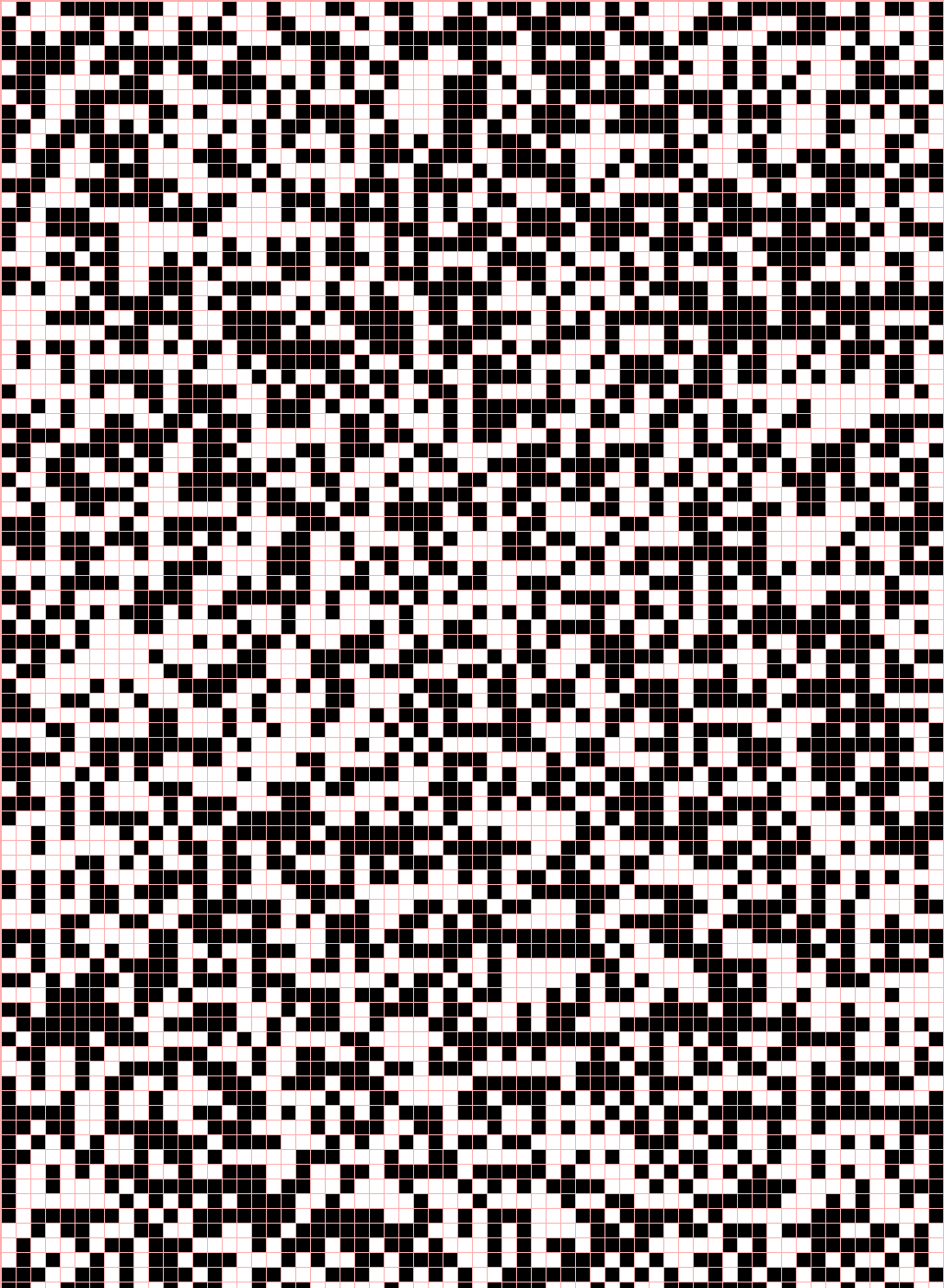}}
		\hfill
		\subfloat[$s_5$\label{mt_64_123456789123456789_space}]{%
			\includegraphics[width=0.2\linewidth, height=5.0cm]{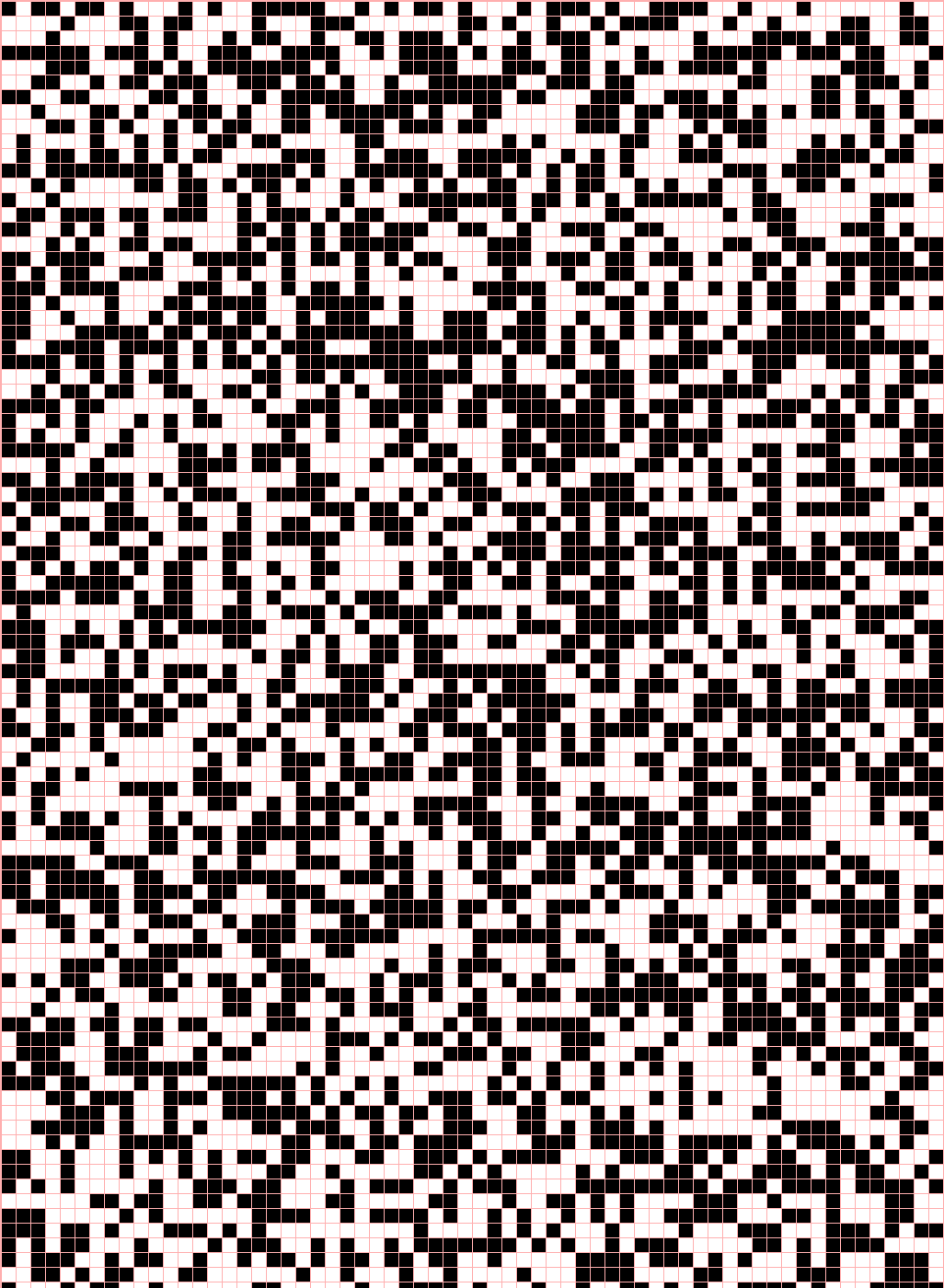}}
		%
		%
		\hfill\\
		\subfloat[$s_1$\label{sfmt_64_7_space}]{%
			\includegraphics[width=0.2\linewidth, height=5.0cm]{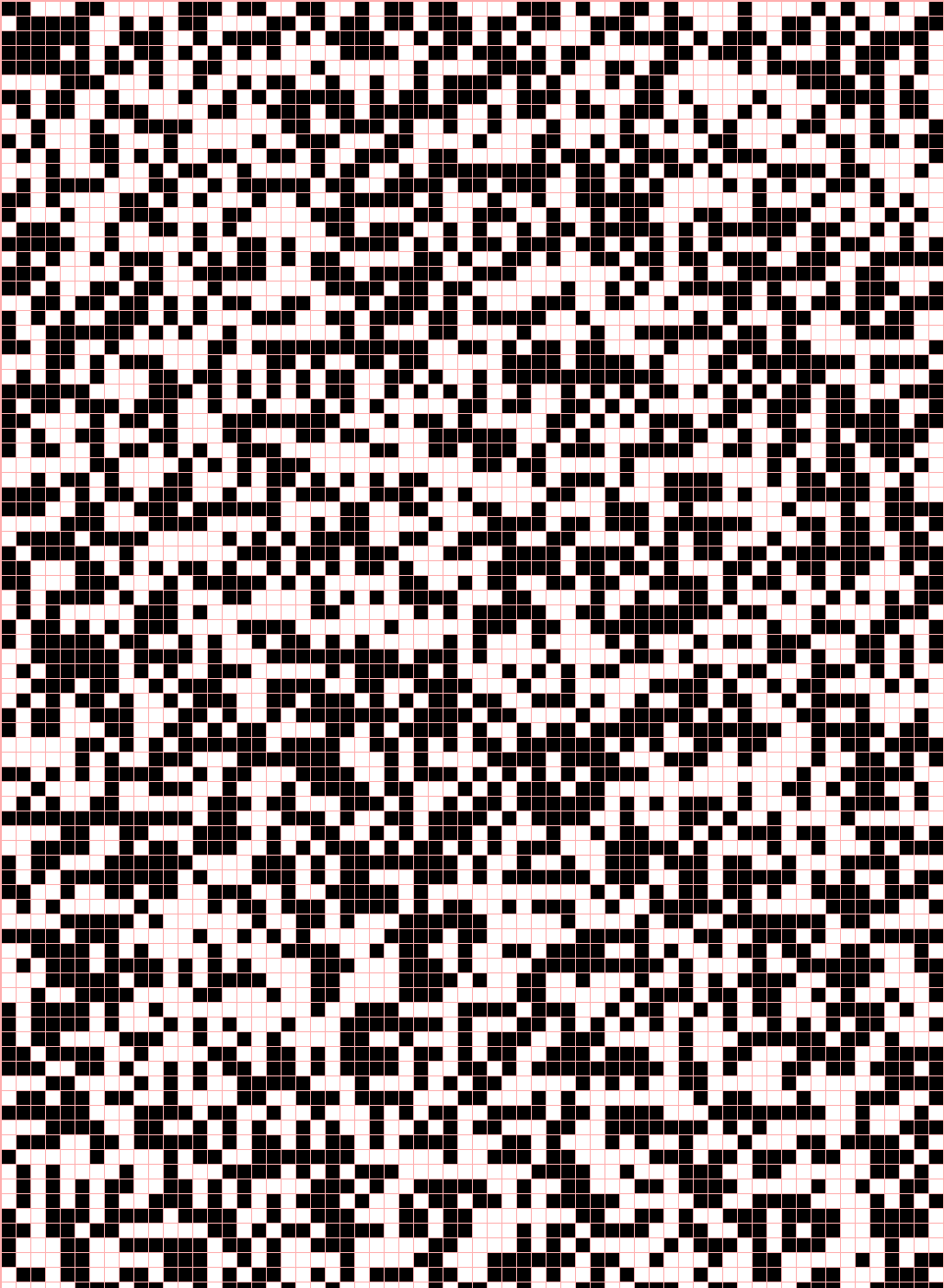}}
		\hfill
		\subfloat[$s_3$\label{sfmt_64_12345_space}]{%
			\includegraphics[width=0.2\linewidth, height=5.0cm]{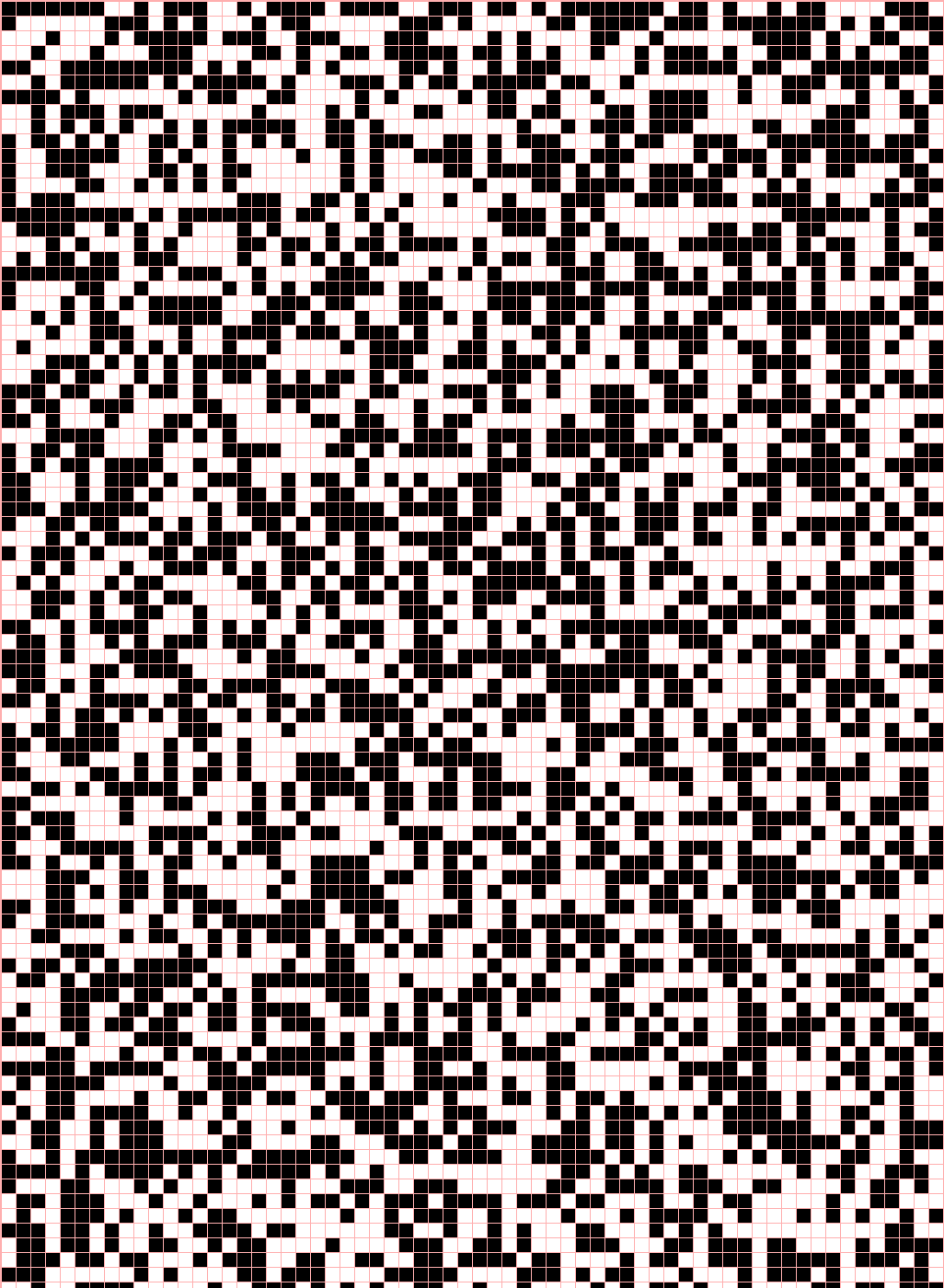}}
		\hfill
		\subfloat[$s_4$\label{sfmt_64_9650218_space}]{%
			\includegraphics[width=0.2\linewidth, height=5.0cm]{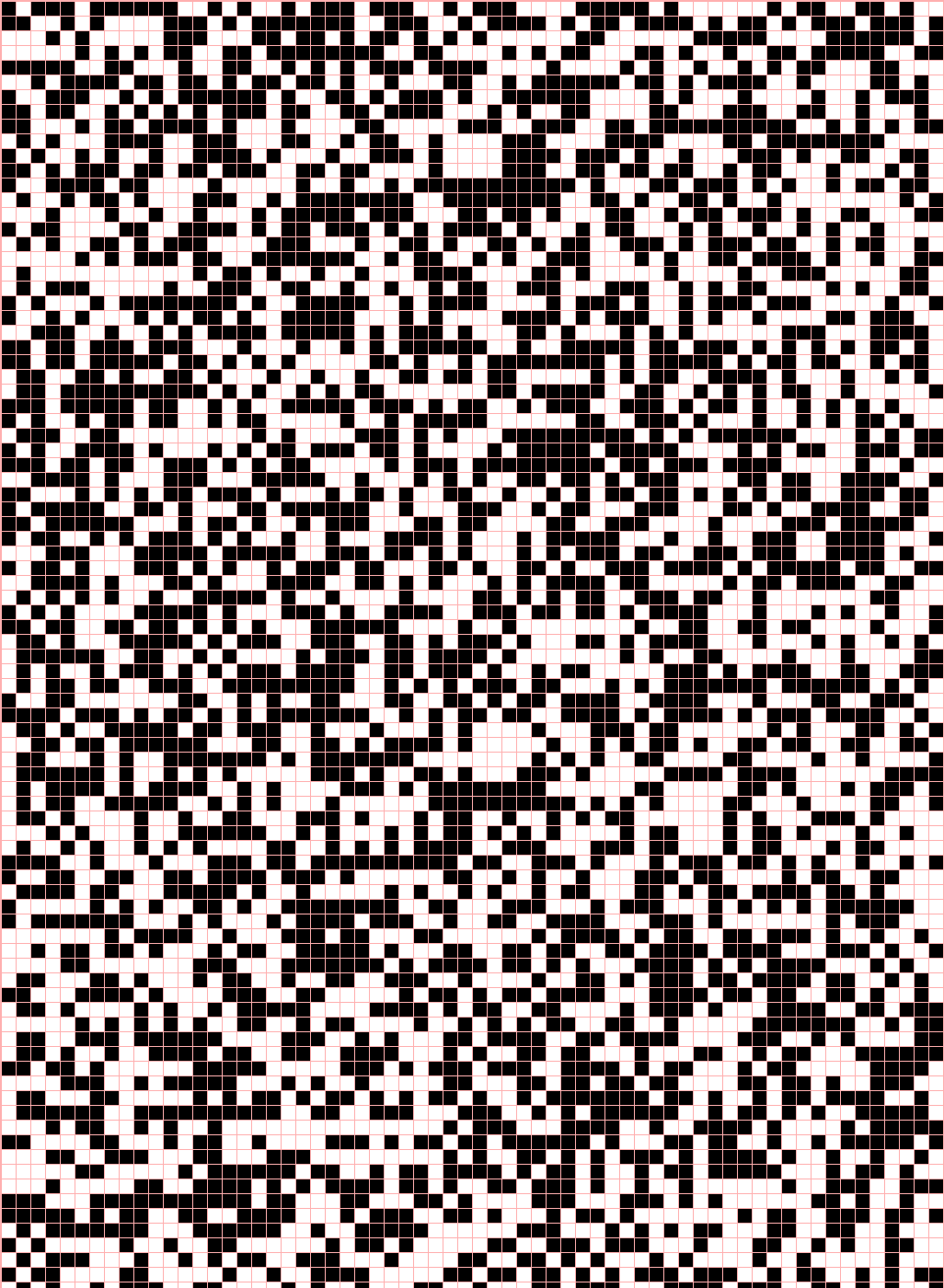}}
		\hfill
		\subfloat[$s_5$\label{sfmt_64_123456789123456789_space}]{%
			\includegraphics[width=0.2\linewidth, height=5.0cm]{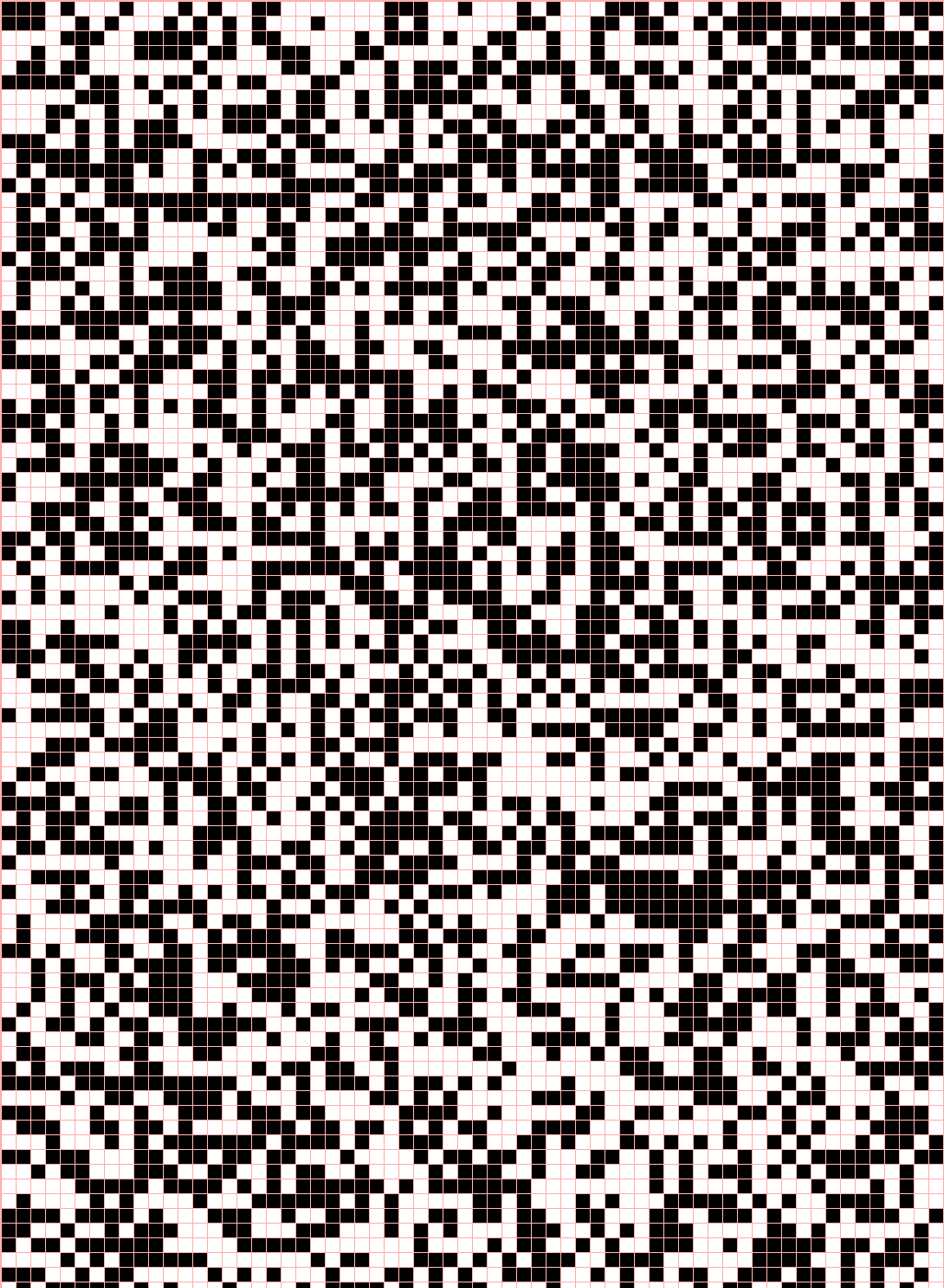}}
		\caption{Space-time diagram for MT19937 $32$ bit (\ref{mt_32_7_space} to \ref{mt_32_123456789123456789_spaceo}), SFMT19937 $32$ bit (\ref{sfmt_32_7_space} to \ref{sfmt_32_123456789123456789_spaceo}), MT19937 $64$ bit (\ref{mt_64_7_space} to \ref{mt_64_123456789123456789_space}) and SFMT19937 $64$ bit (\ref{sfmt_64_7_space} to \ref{sfmt_64_123456789123456789_space})}
		\label{fig:sfmt64_space-time}
	\end{figure} 
	
	\begin{figure}[!h]
		\centering
		\subfloat[$s_1$\label{dsfmt_7_space}]{%
			\includegraphics[width=0.1\linewidth, height=5.0cm]{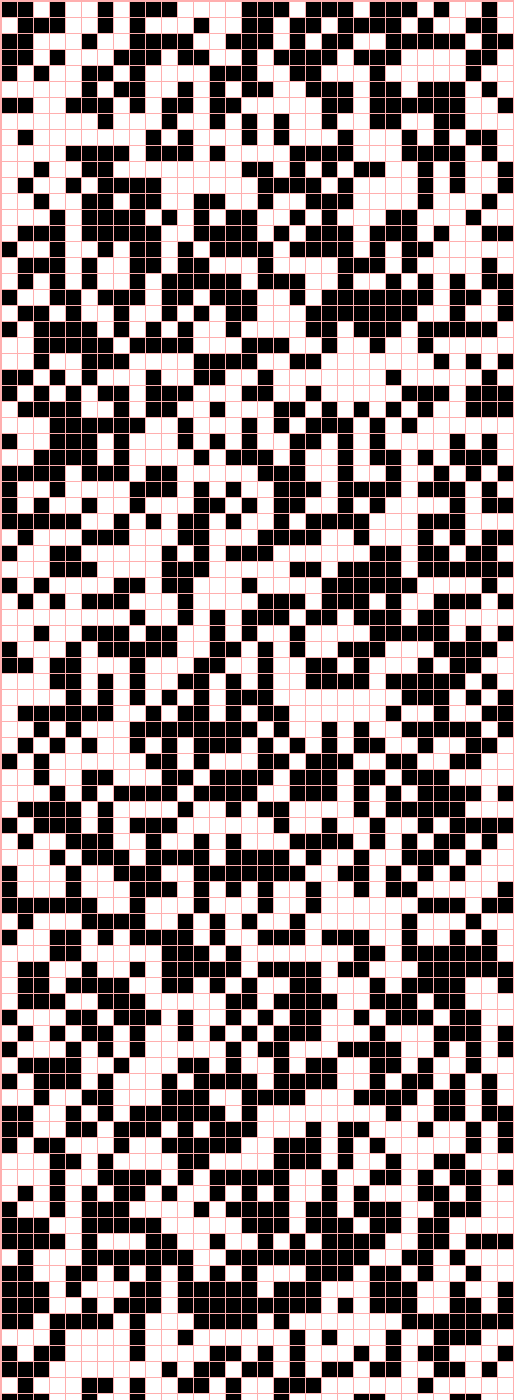}}
		\hfill
		\subfloat[$s_3$\label{dsfmt_12345_space}]{%
			\includegraphics[width=0.1\linewidth, height=5.0cm]{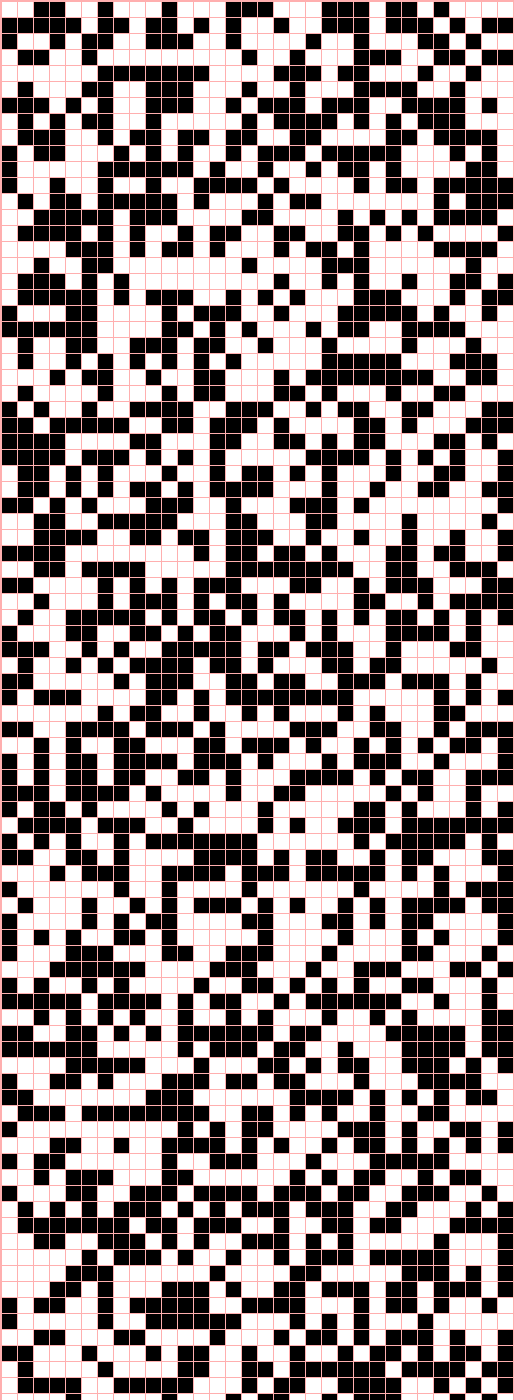}}
		\hfill
		\subfloat[$s_4$\label{dsfmt_9650218_space}]{%
			\includegraphics[width=0.1\linewidth, height=5.0cm]{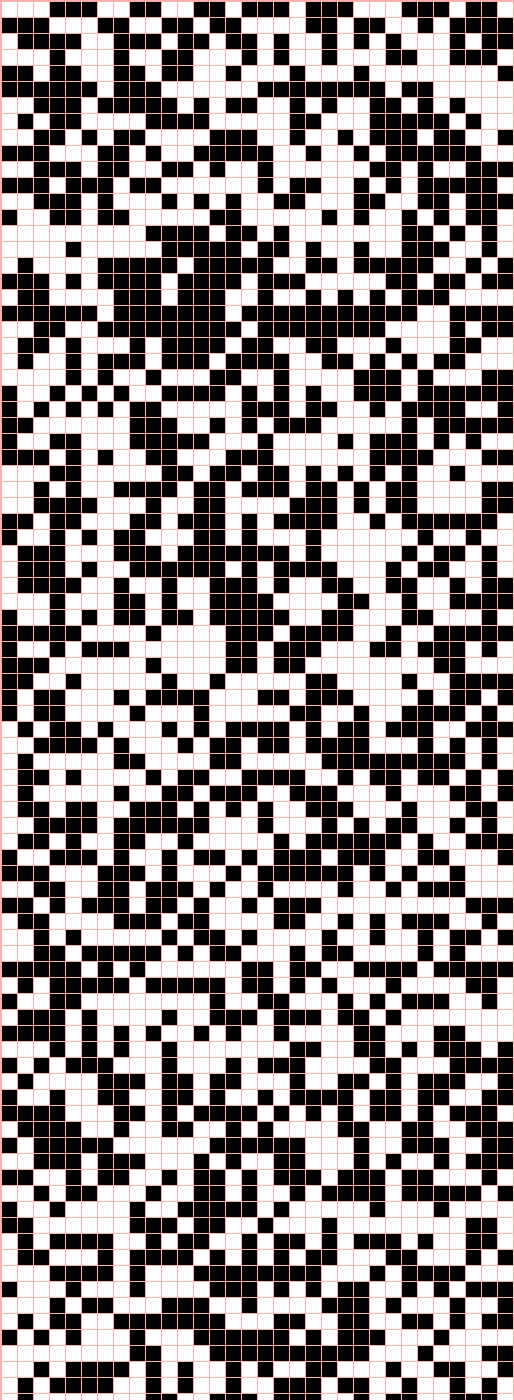}}
		\hfill
		\subfloat[$s_5$\label{dsfmt_123456789123456789_spaceo}]{%
			\includegraphics[width=0.1\linewidth, height=5.0cm]{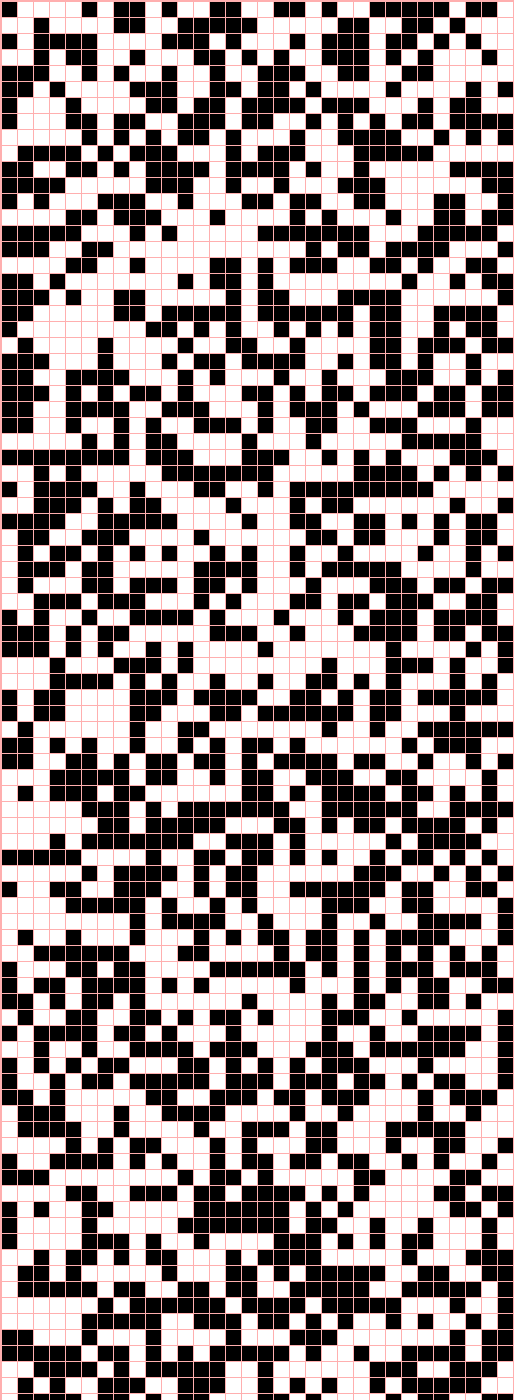}}
		\hfill
		\subfloat[$s_1$\label{rule30_7_space}]{%
			\includegraphics[width=0.1\linewidth, height=5.0cm]{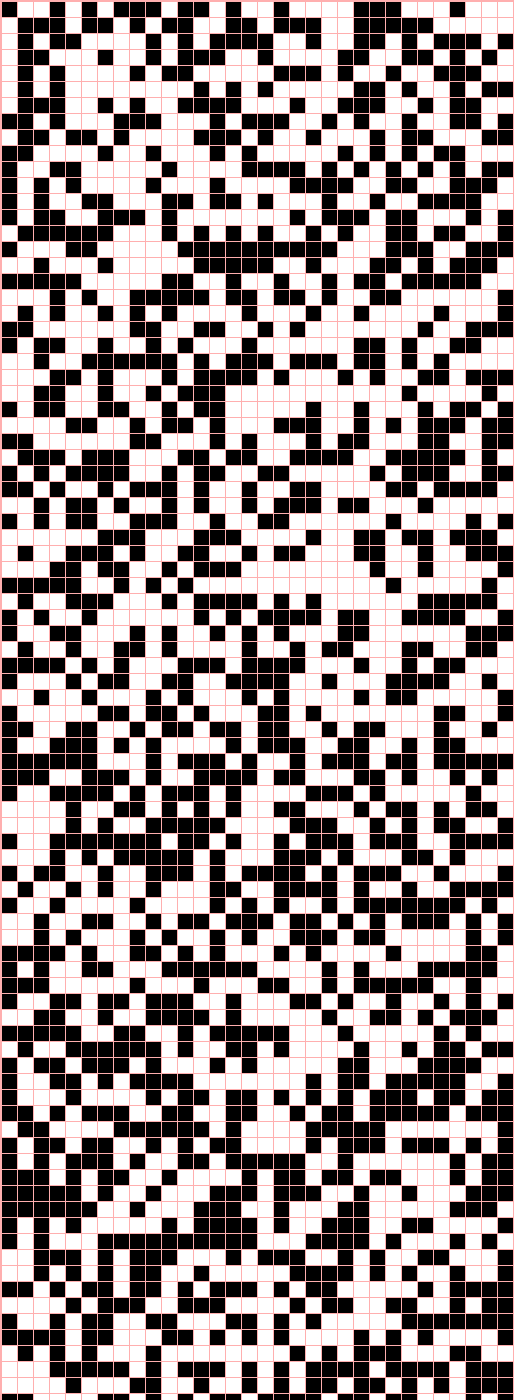}}
		\hfill
		\subfloat[$s_3$\label{rul30_12345_space}]{%
			\includegraphics[width=0.1\linewidth, height=5.0cm]{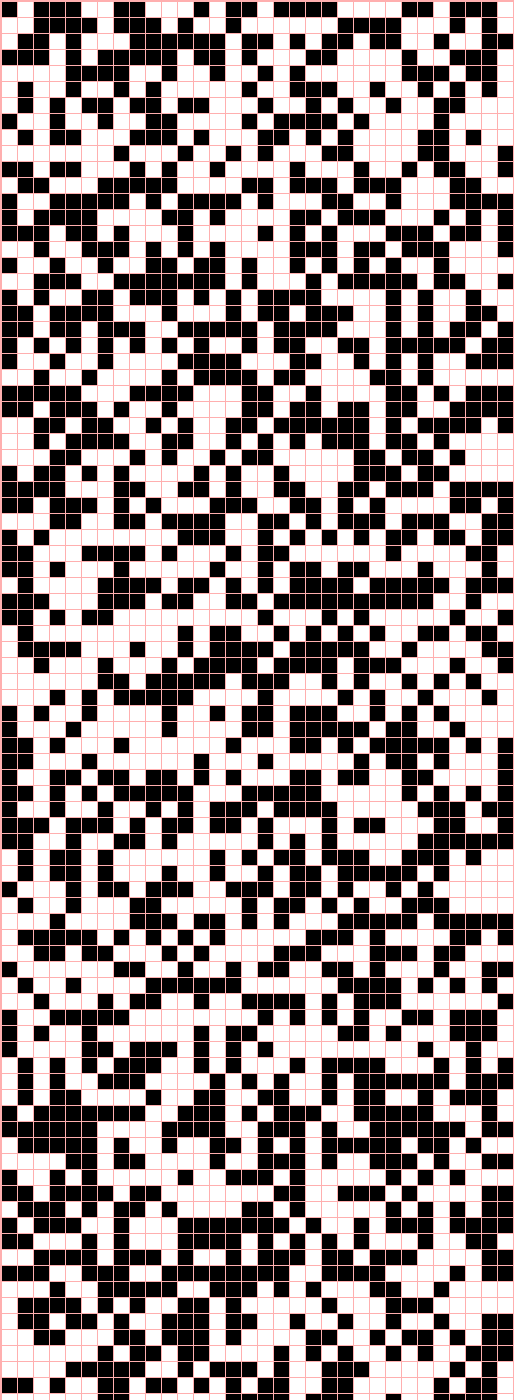}}
		\hfill
		\subfloat[$s_4$\label{rule30_9650218_space}]{%
			\includegraphics[width=0.1\linewidth, height=5.0cm]{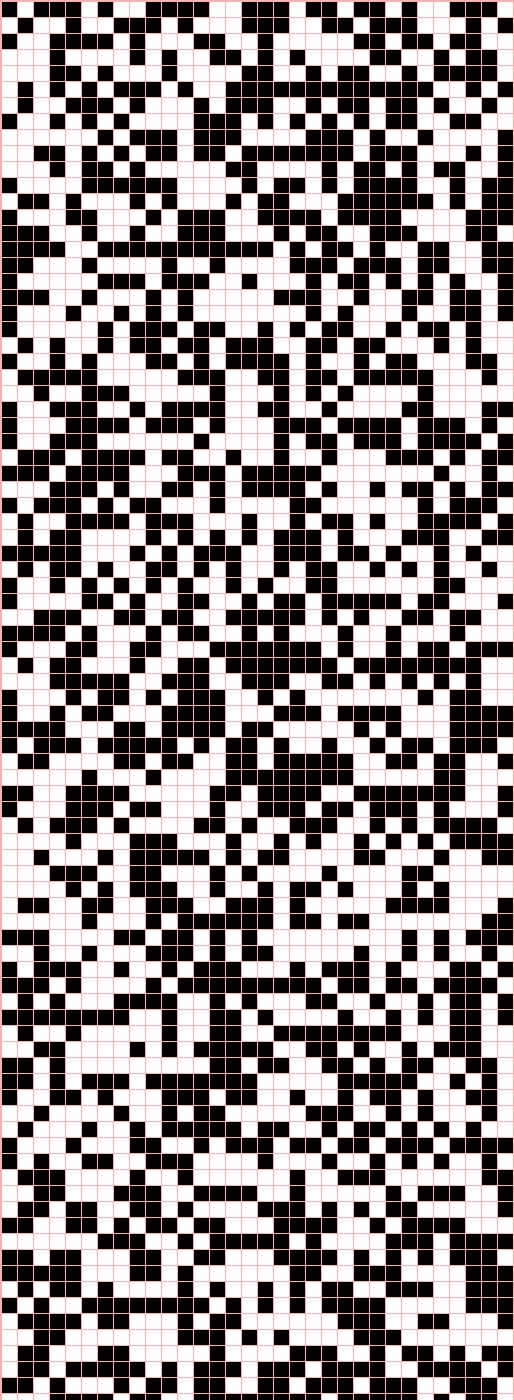}}
		\hfill
		\subfloat[$s_5$\label{rule30_123456789123456789_spaceo}]{%
			\includegraphics[width=0.1\linewidth, height=5.0cm]{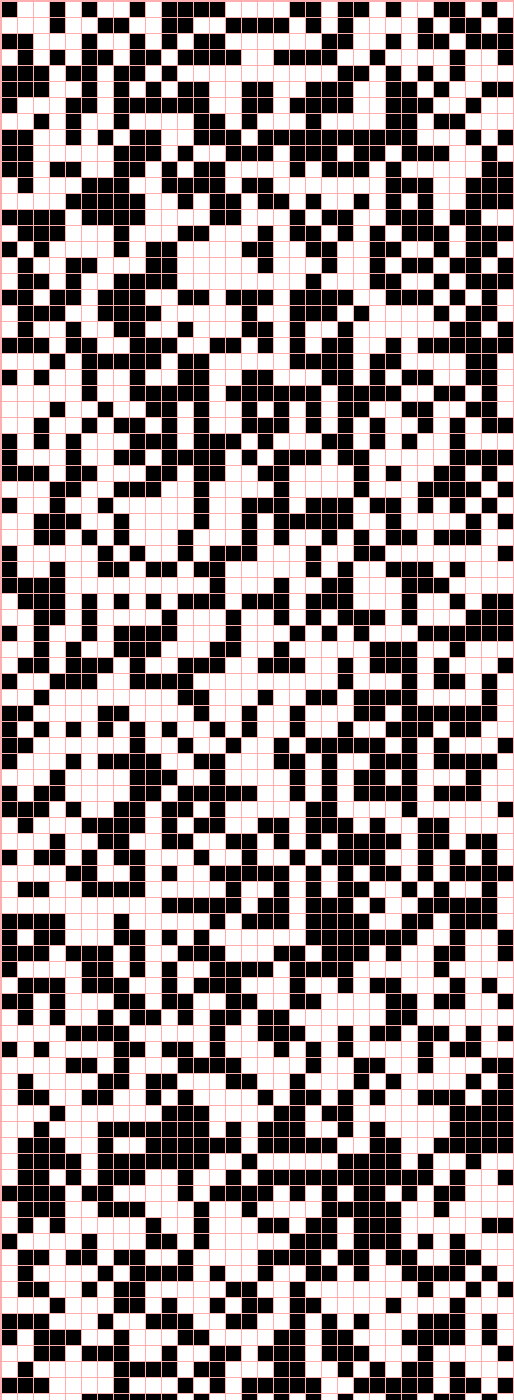}}
		\hfill\\
		\subfloat[$s_1$\label{rule30-45_7_space}]{%
			\includegraphics[width=0.1\linewidth, height=5.0cm]{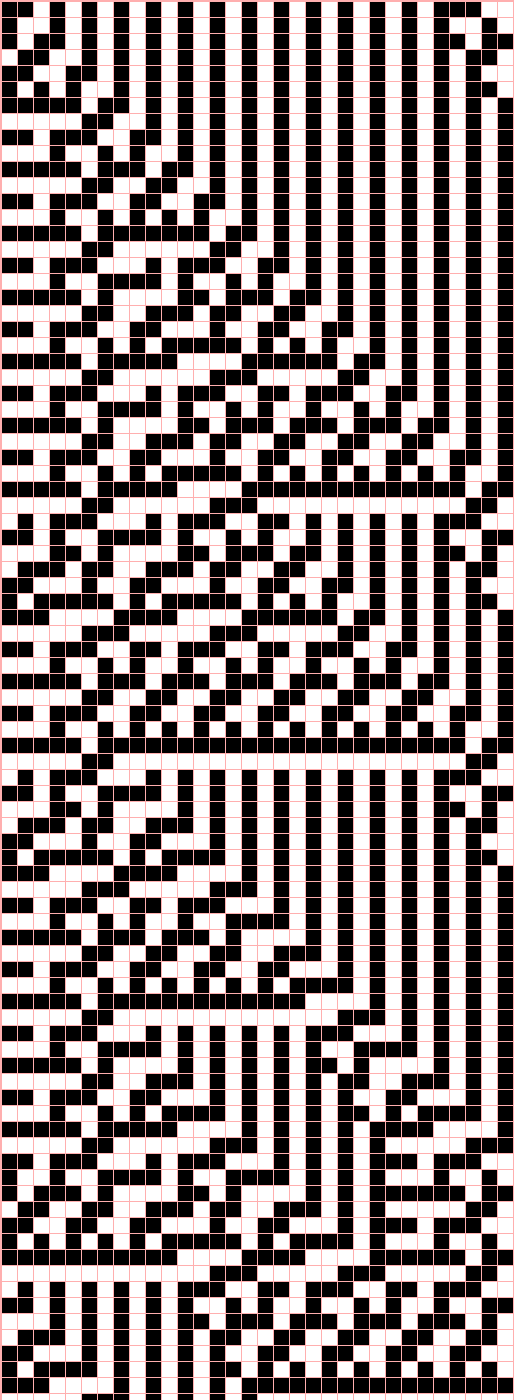}}
		\hfill
		\subfloat[$s_3$\label{rule30-45_12345_space}]{%
			\includegraphics[width=0.1\linewidth, height=5.0cm]{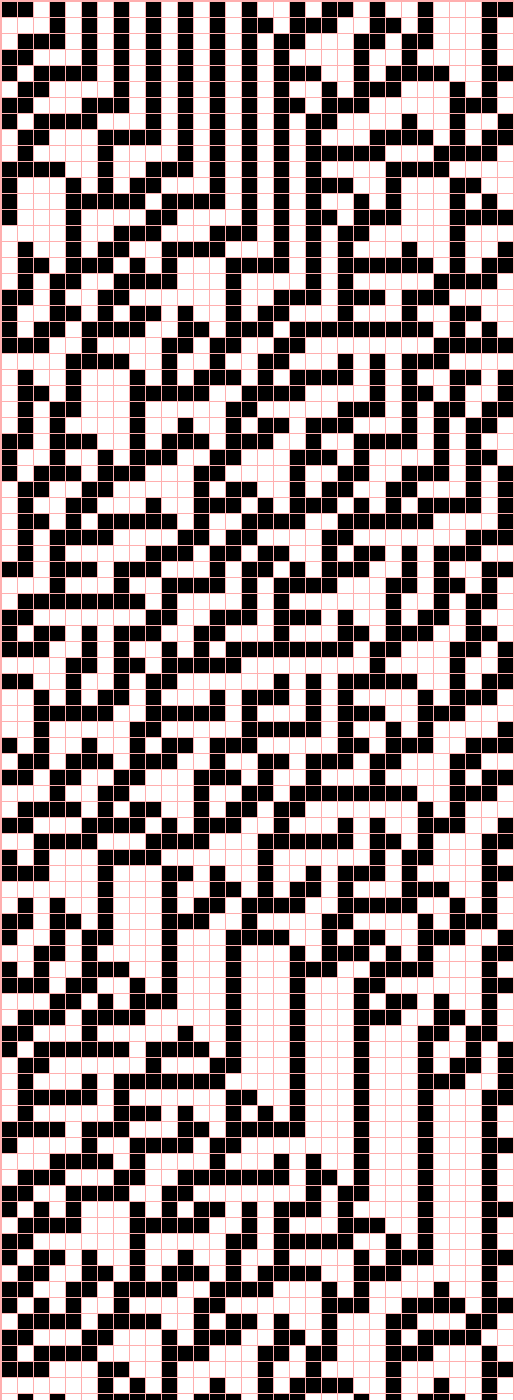}}
		\hfill
		\subfloat[$s_4$\label{rule30-45_9650218_space}]{%
			\includegraphics[width=0.1\linewidth, height=5.0cm]{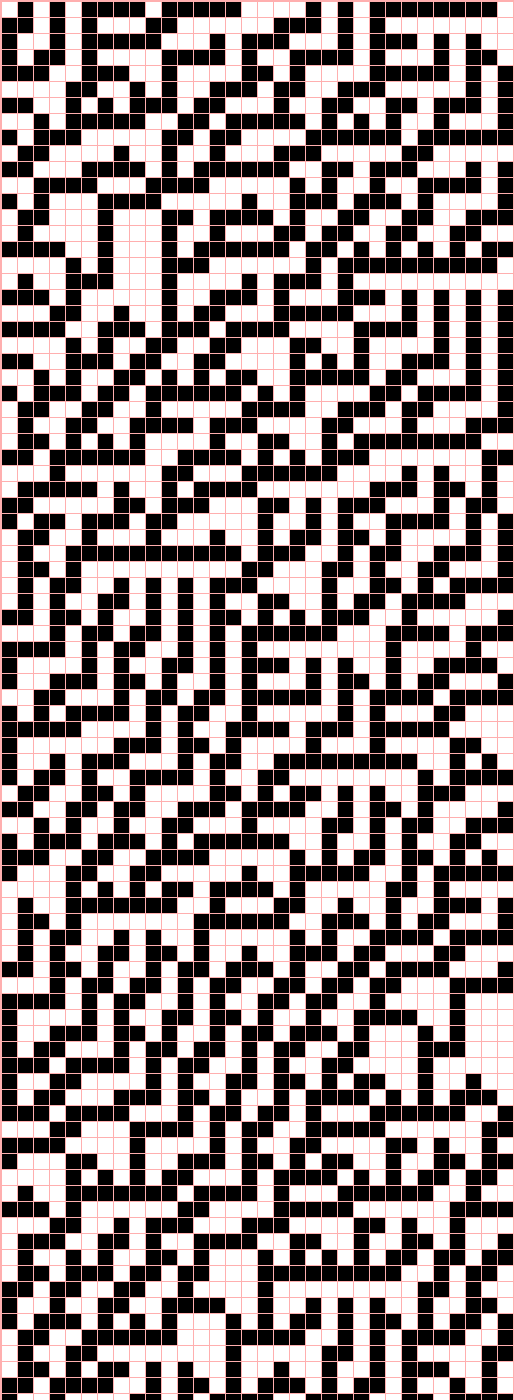}}
		\hfill
		\subfloat[$s_5$\label{rule30-45_123456789123456789_space}]{%
			\includegraphics[width=0.1\linewidth, height=5.0cm]{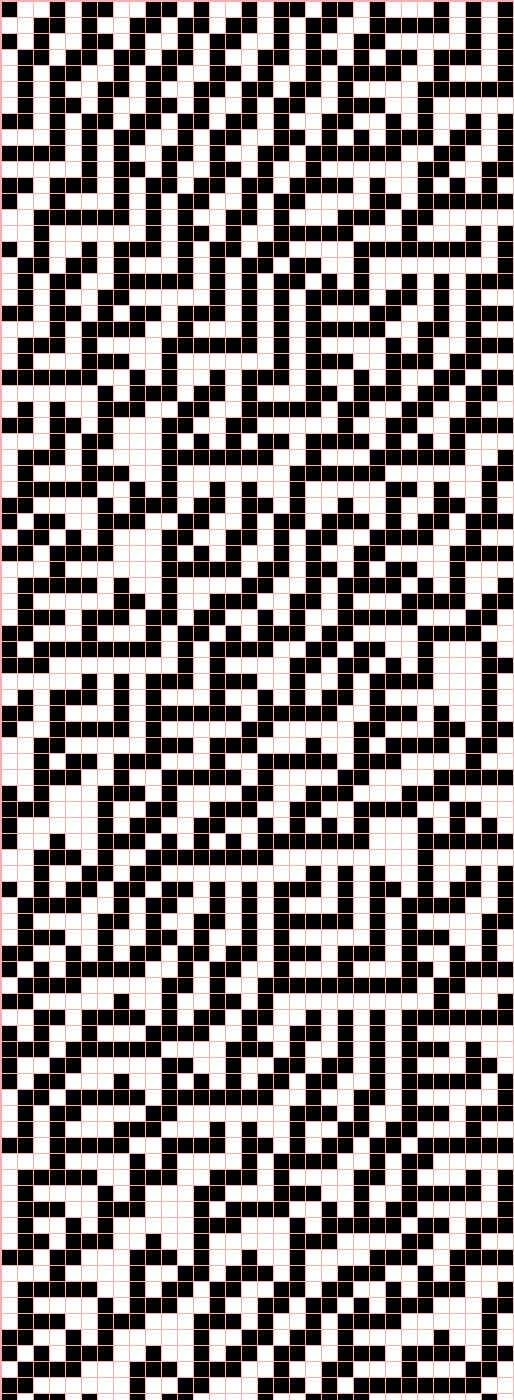}}
		\hfill
		%
		\subfloat[$s_1$\label{maxlength0_7_space}]{%
			\includegraphics[width=0.1\linewidth, height=5.0cm]{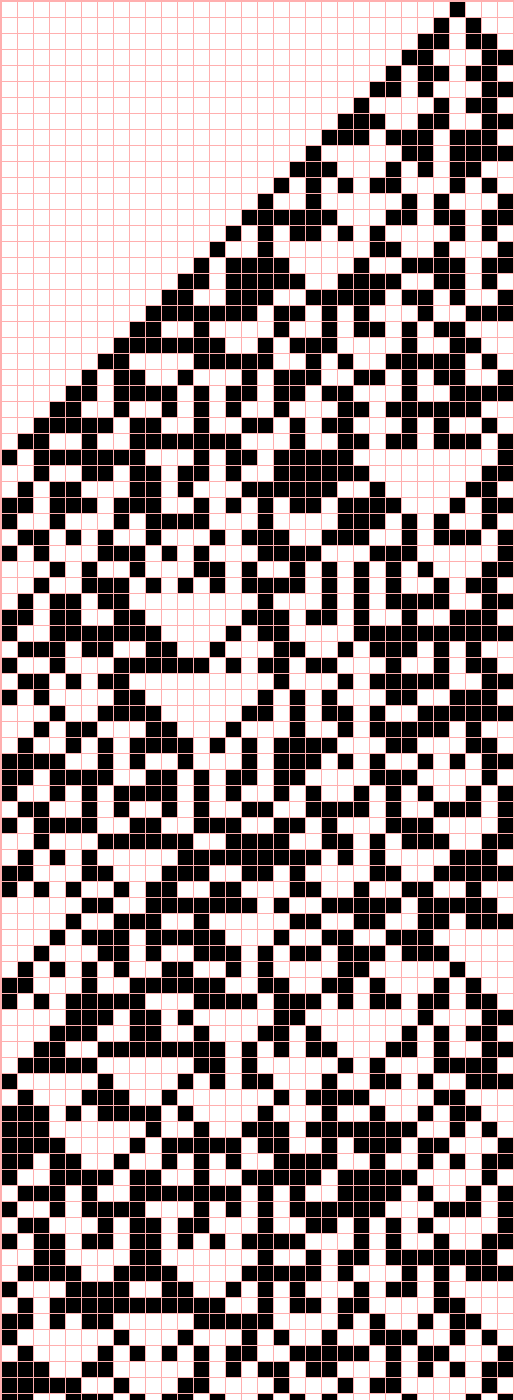}}
		\hfill
		\subfloat[$s_3$\label{maxlength0_12345_space}]{%
			\includegraphics[width=0.1\linewidth, height=5.0cm]{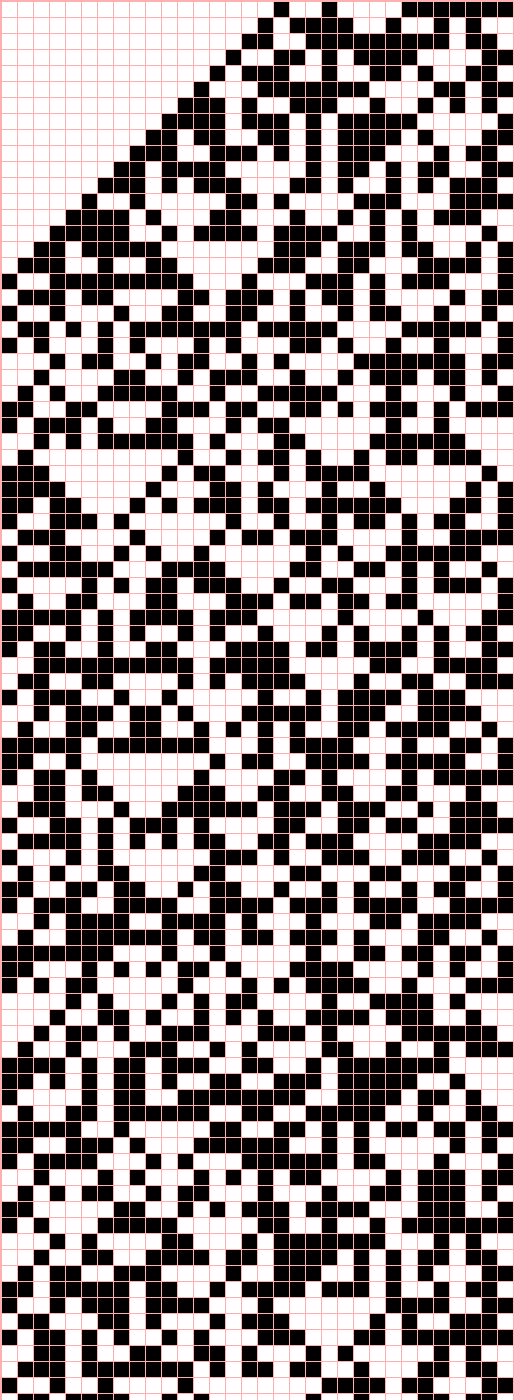}}
		\hfill
		\subfloat[$s_4$\label{maxlength0_9650218_space}]{%
			\includegraphics[width=0.1\linewidth, height=5.0cm]{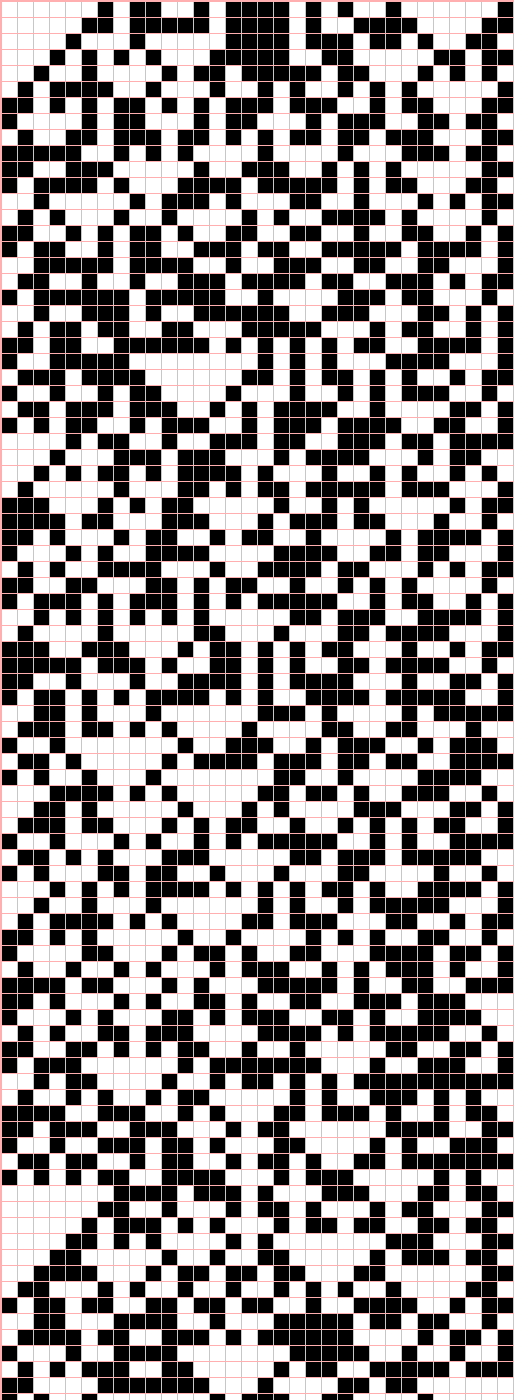}}
		\hfill
		\subfloat[$s_5$\label{maxlength0_123456789123456789_spaceo}]{%
			\includegraphics[width=0.1\linewidth, height=5.0cm]{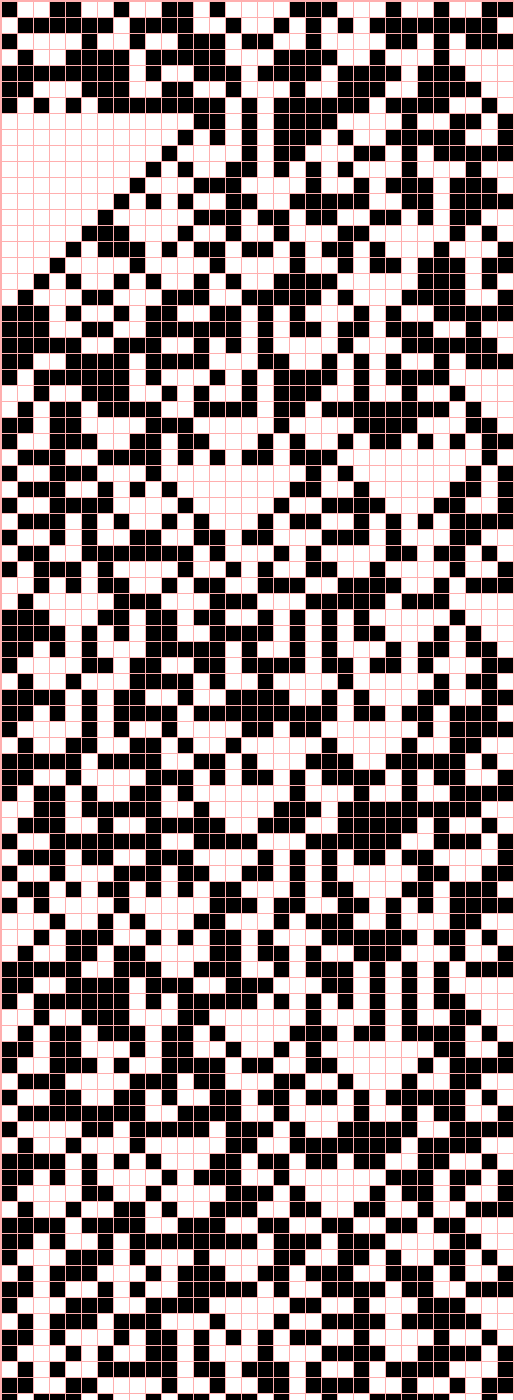}}
		\hfill\\
		\subfloat[$s_1$\label{maxlength1_7_space}]{%
			\includegraphics[width=0.1\linewidth, height=5.0cm]{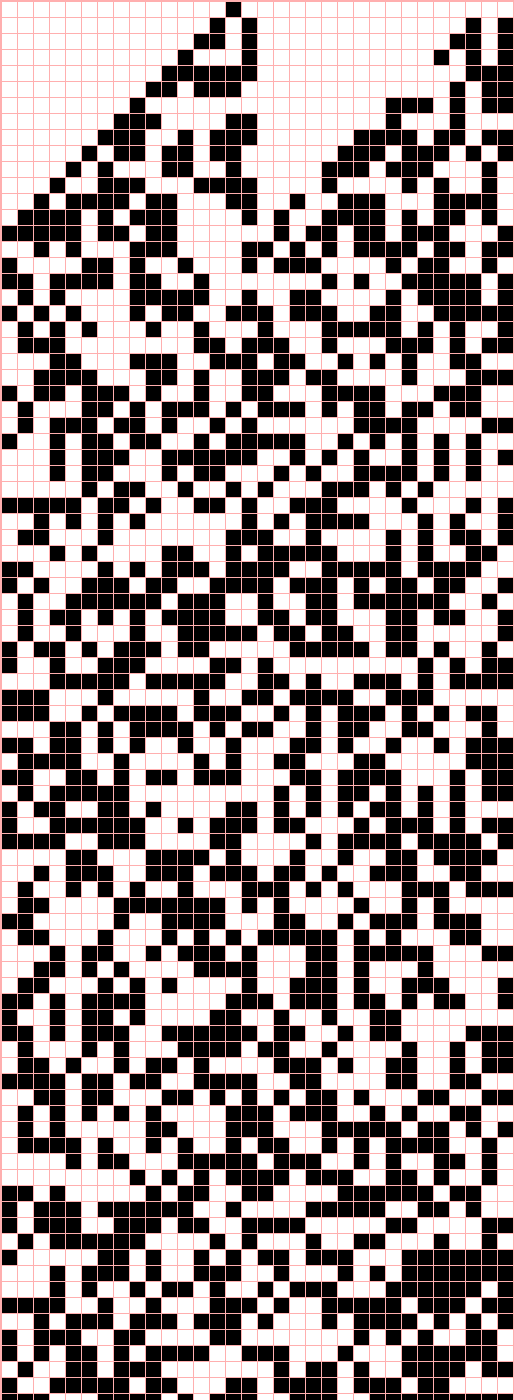}}
		\hfill
		\subfloat[$s_3$\label{maxlength1_12345_space}]{%
			\includegraphics[width=0.1\linewidth, height=5.0cm]{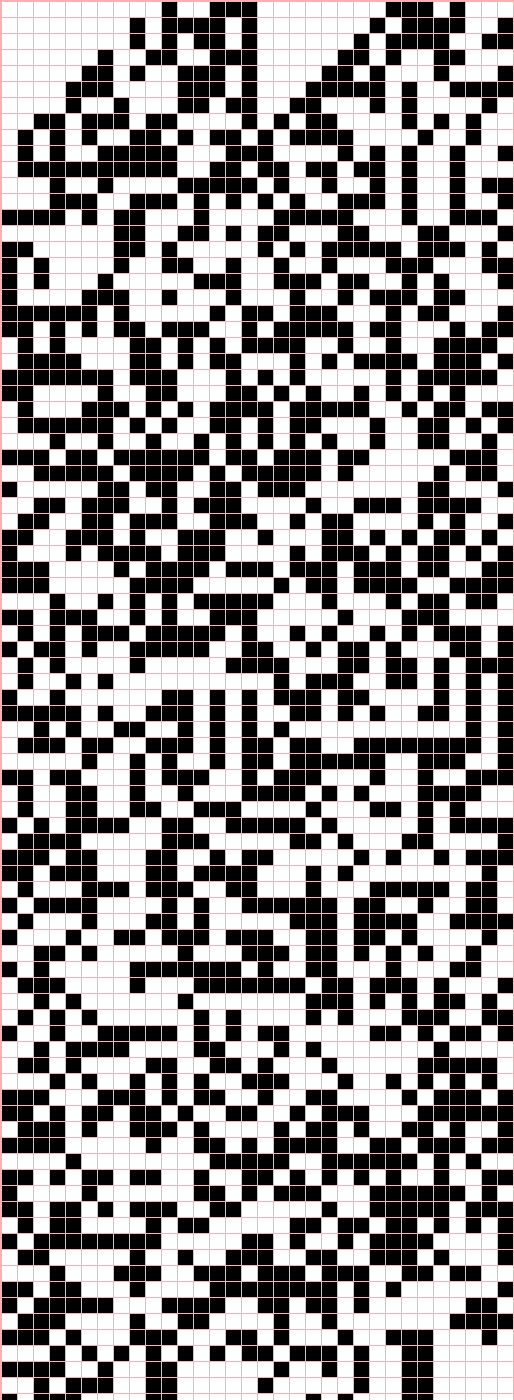}}
		\hfill
		\subfloat[$s_4$\label{maxlength1_9650218_space}]{%
			\includegraphics[width=0.1\linewidth, height=5.0cm]{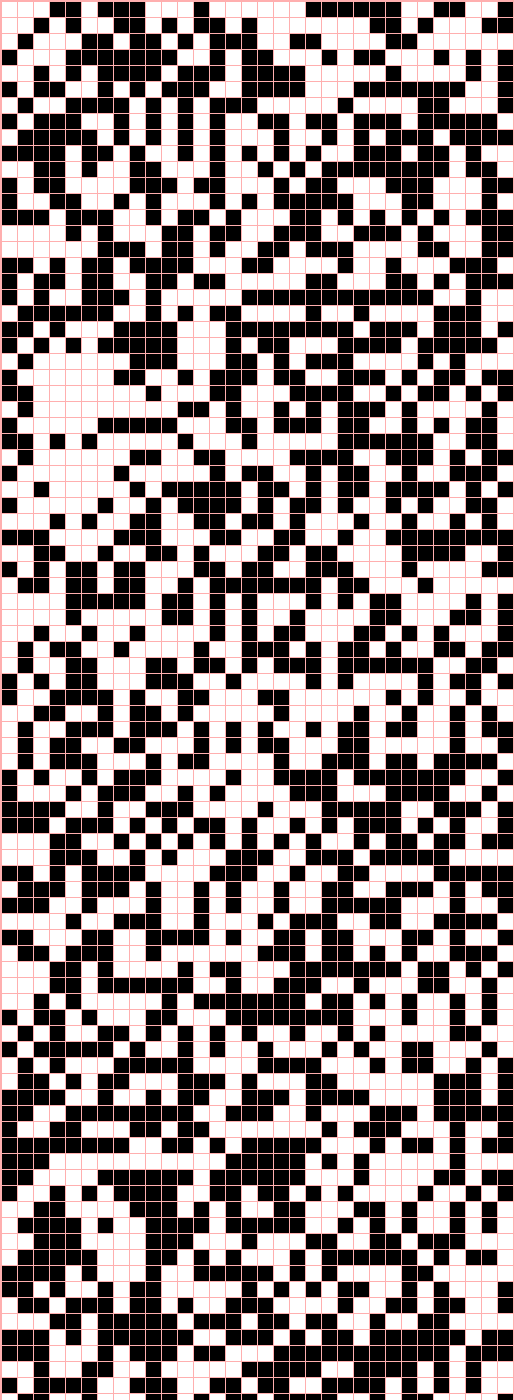}}
		\hfill
		\subfloat[$s_5$\label{maxlength1_123456789123456789_space}]{%
			\includegraphics[width=0.1\linewidth, height=5.0cm]{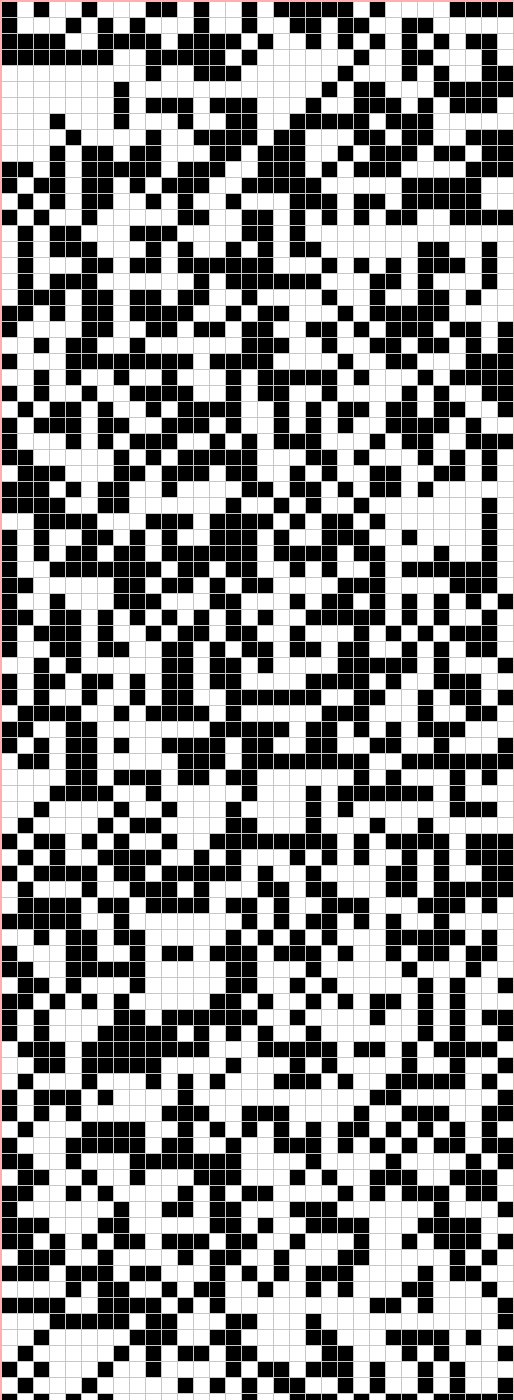}}
		\hfill
			\subfloat[$s_1$\label{nonlinear_7_space}]{%
			\includegraphics[width=0.12\linewidth, height=5.0cm]{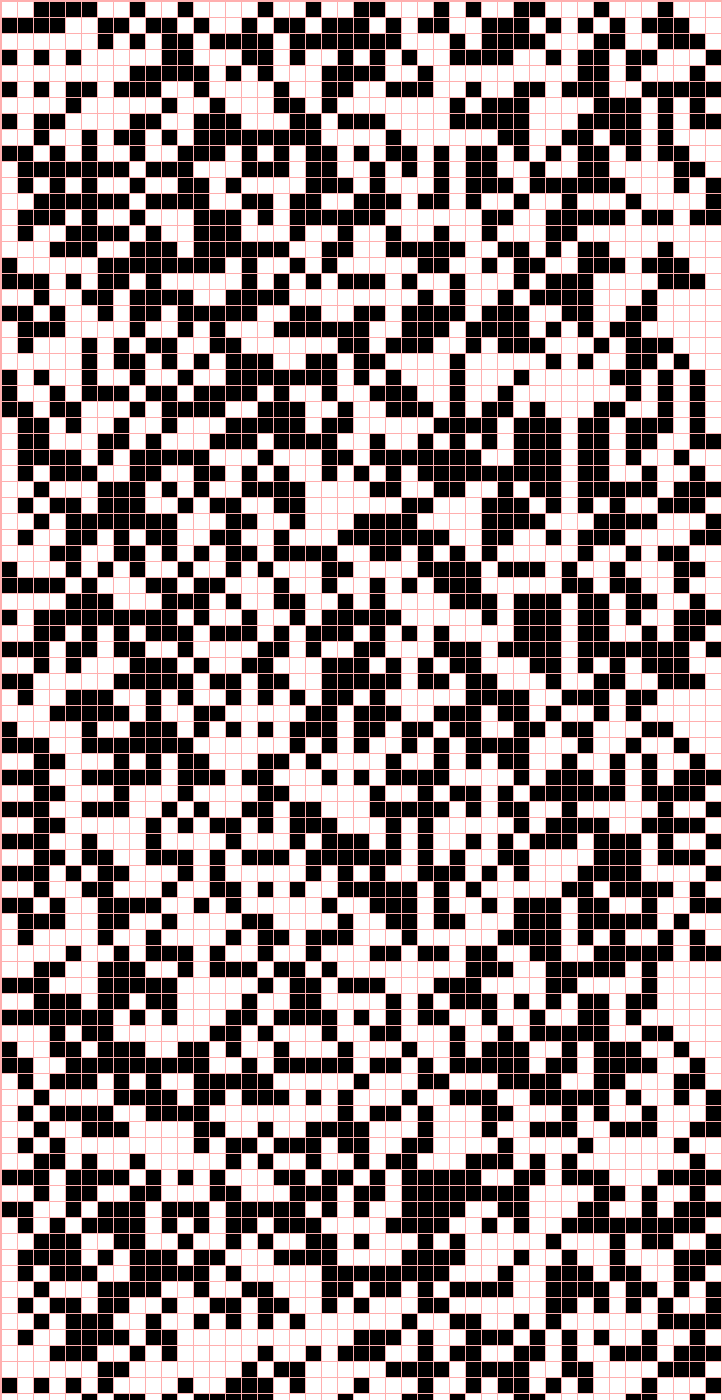}}
		\hfill
		\subfloat[$s_3$\label{nonlinear_12345_space}]{%
			\includegraphics[width=0.12\linewidth, height=5.0cm]{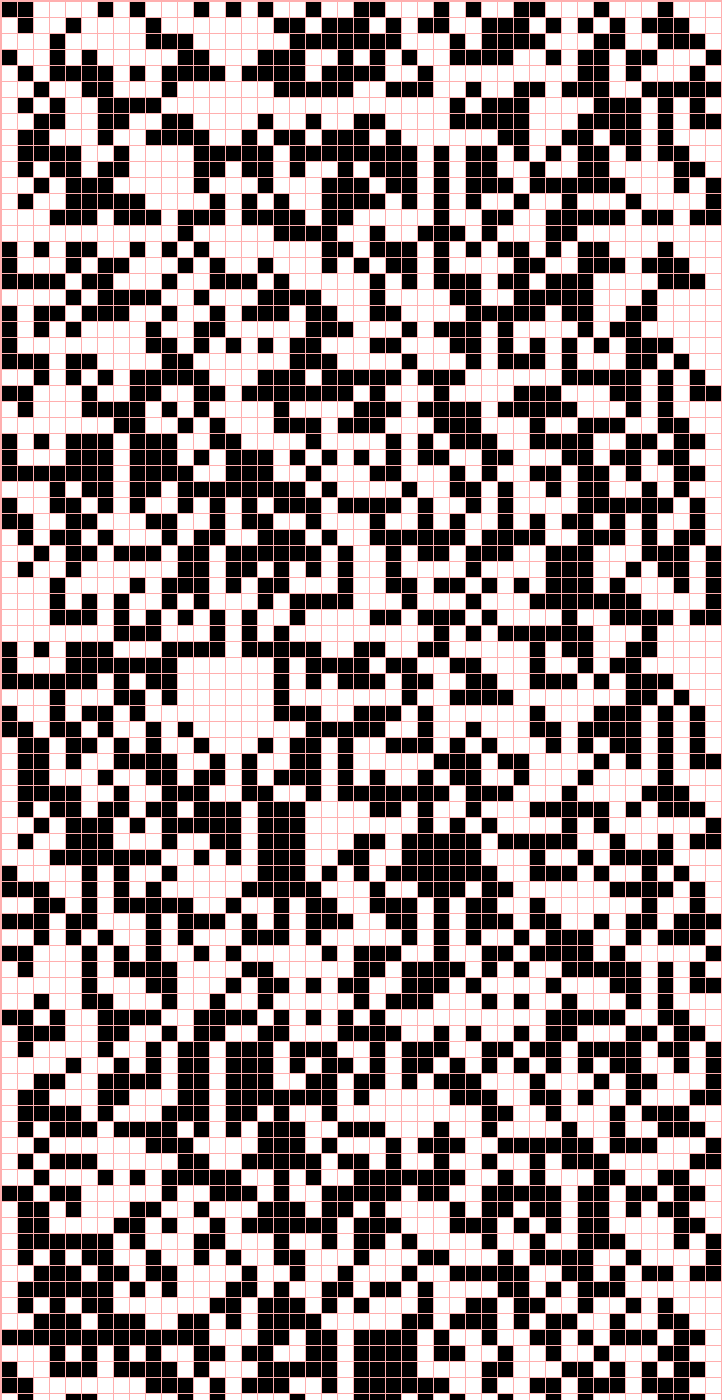}}
		\hfill
		\subfloat[$s_4$\label{nonlinear_9650218_space}]{%
			\includegraphics[width=0.12\linewidth, height=5.0cm]{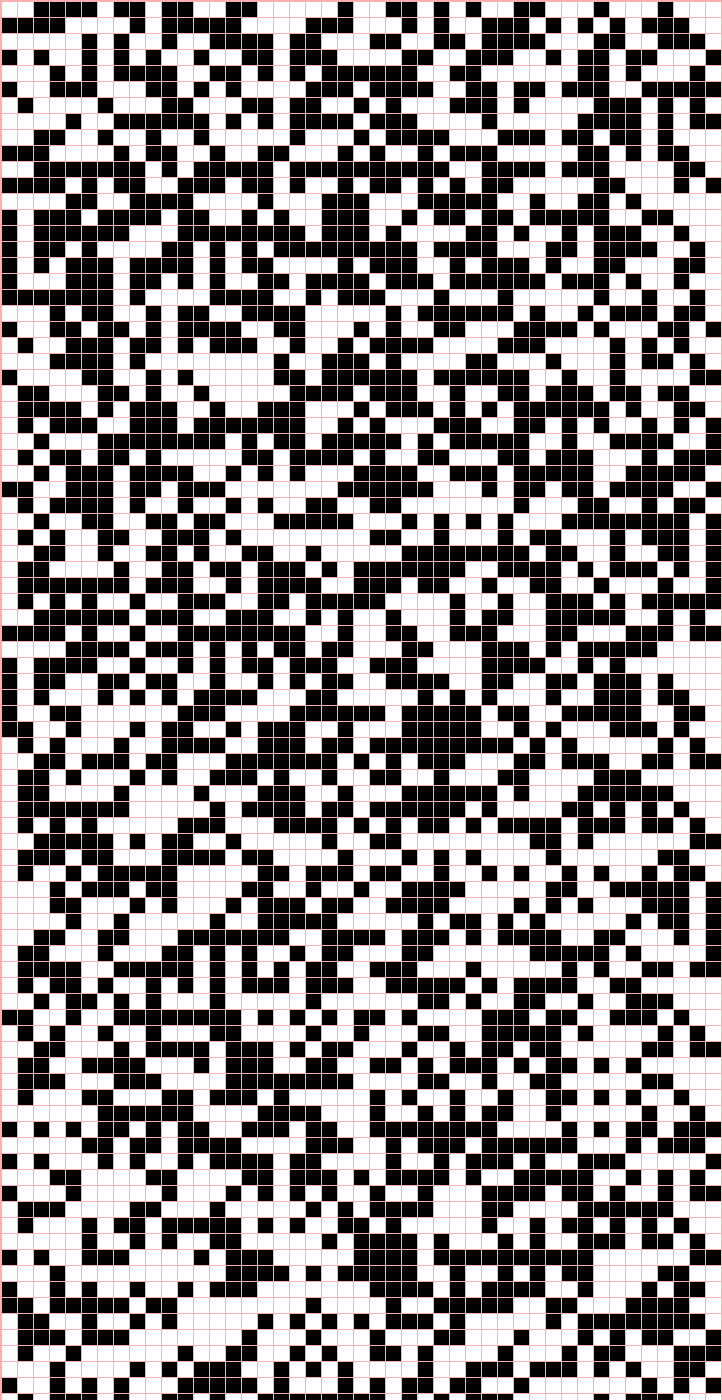}}
		\hfill
		\subfloat[$s_5$\label{nonlinear_123456789123456789_spaceo}]{%
			\includegraphics[width=0.12\linewidth, height=5.0cm]{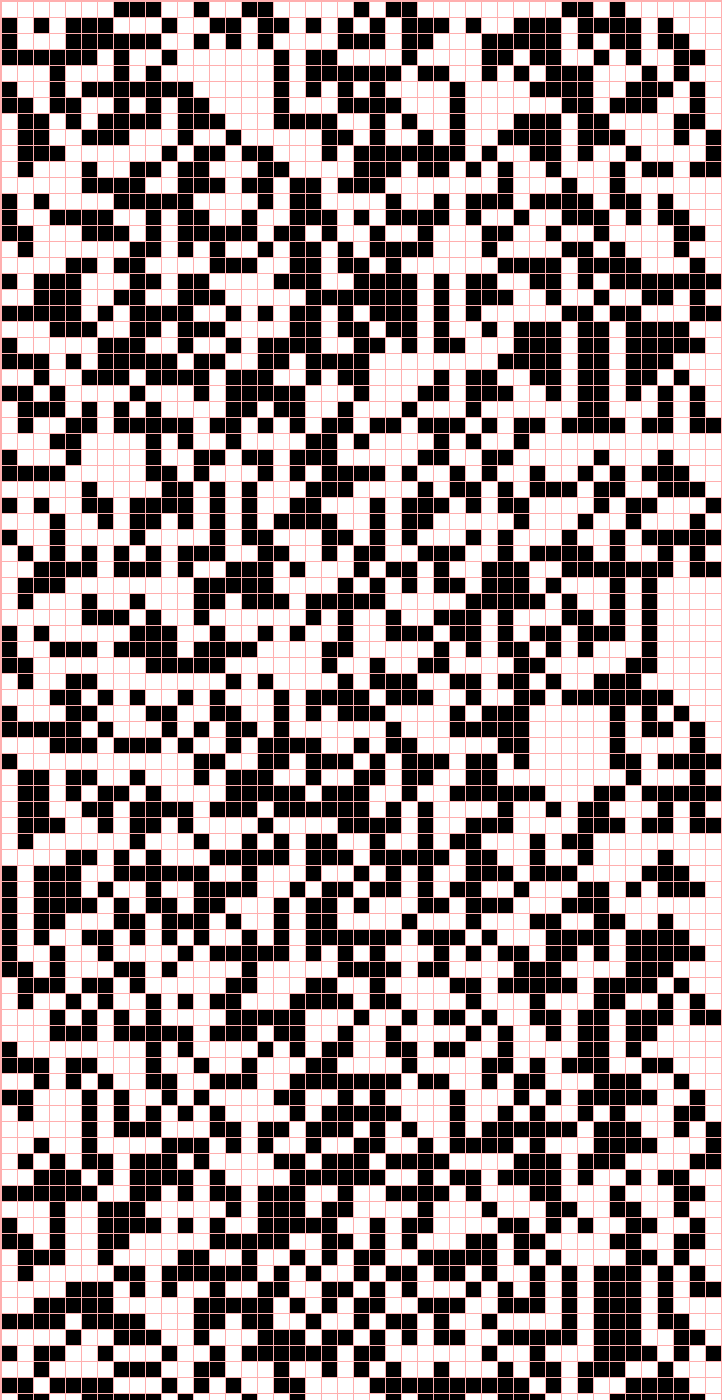}}
		\hfill\\
		\caption{Space-time diagram for dSFMT19937 $32$ bit (\ref{dsfmt_7_space} to \ref{dsfmt_123456789123456789_spaceo}), rule 30 (\ref{rule30_7_space} to \ref{rule30_123456789123456789_spaceo}), rule 30-45 (\ref{rule30-45_7_space} to \ref{rule30-45_123456789123456789_space}), max-length CA with $\gamma=0$ (\ref{maxlength0_7_space} to \ref{maxlength0_123456789123456789_spaceo}), max-length CA with $\gamma =1$ (\ref{maxlength1_7_space} to \ref{maxlength1_123456789123456789_space}) and non-linear $2$-state CA (\ref{nonlinear_7_space} to \ref{nonlinear_123456789123456789_spaceo})
			}
			\label{fig:ca_space-time}
		\end{figure} 
		%
	
	\begin{itemize}[leftmargin=1pt]
		\item For \verb|minstd_rand|, the last $6$ bits of the generated numbers are fixed and for Knuth's \verb|MMIX|, last $2$ bits of four consecutive numbers form a pattern.
		\item For \verb|rand|, \verb|lrand|, Borland's LCG, \verb|MRG31k3p| and \verb|random|, the percentage of black and white boxes representing $1$s and $0$s are not same. Even for \verb|PCG-32|, there is pattern visible in the diagrams.
		\item \verb|LFSR113| forms pattern for some seeds. For WELL and Xorshift generators, dependency on seed is visible for the initial numbers.
		\item For MTs and SFMTs, the dependency on seed is visible for very few levels. For dSFMTs, there is pattern visible in the diagrams.
		
		\item Among the CA-based generators, rule $30-45$ has visible patterns whereas max-length CAs have dependency on seed up to some initial configurations. For non-linear $2$-state CA, there are some minute cluster of black and white boxes. However, the figures for Decimal CA, ECA rule $30$ and $3$-state CA appear relatively random with no dependency on seed.
		\item For the LCGs, the dependency on seed is less visible than the LFSRs.
		\item If observed closely, every PRNG has some kind of clubbing of white boxes and black boxes, that is, none of the figures is actually free of pattern. However, for the good PRNGs, these patterns are non-repeating.
	\end{itemize}

	\subsection{Final Ranking and Remark}\label{sec:final_rank}
{Many time the blind test results and space-time diagram results do not accord with each other. For example, one can see \verb|xorshift1024*| (Figure~\ref{fig:xor_space-time} and Table~\ref{tab:blind_test}). In that case, although this PRNG passes many blind tests but it clearly shows pattern in space-time diagram as initial couple of numbers generated by it are fixed. This shows an evidence of flaw in the blind tests which can be detected by graphical tests that allow human intervention in the decision. So, in such cases, we change the ranking of the PRNG depending on its behavior in space-time diagram. Further, we can also distinguish between intra-class PRNGs using this tool, for example, \verb|WELL1024a| and \verb|MT19937-32| had same rank in Table~\ref{tab:blind_test_avg}. But, as \verb|WELL1024a| as visible patterns (see Figure~\ref{fig:well_space-time}), it loses its status to be included in the same group as \verb|MT19937-32|.}
Hence, using the space-time diagrams along with the statistical tests, we can further improve the rankings of the PRNGs --
	\begin{itemize}
		\item \verb|SFMT19937-64| holds the first position as it appears more random than \verb|SFMT19937-32.| Decimal CA gives a tough competition to \verb|SFMT19937-64|, sometimes even outperforming it which proves that it is at least at par with it (if not better!). So both of them are ranked $1$.
		
		\item Rule $30$ holds the $3^{rd}$ rank, whereas \verb|MT19937-64| holds rank $4$. \verb|PCG-32| is better than \verb|MT19937-32| and \verb|dSFMT-32.| So, it holds rank $5$. The next rank holder is \verb|MT19937-32.| 
		
		\item \verb|dSFMT-32| has less dependency on seed than \verb|WELL1024a| and \verb|xorshift128+.| So, it is ranked on the $7^{th}$ position.
		
		\item \verb|xorshift64*| has no dependency on seed, so it is ranked higher than \verb|WELL1024a| and \verb|xorshift128+.|  
		
		\item \verb|WELL512a| is ranked lower than \verb|WELL1024a| and \verb|xorshift128+|, as it has more dependency on seed. As \verb|Tauss88| (rank $11$) sometimes cannot pass any tests, so it is ranked lower than \verb|WELL512a| (rank $10$).
		
		\item \verb|dSFMT-52| has less dependency on seeds than non-linear $2$-state CA based PRNG and max-length CA with $\gamma=1$. So, it holds rank $12$.
		
		\item Non-linear $2$-state CA based PRNG, max-length CA with $\gamma=1$ based PRNG and $3$-state CA form the group of $13$ rank holders.
		
		\item Although \verb|LFSR113| and {\verb|xorshift1024*|} can perform well for some seeds, but because of its dependency on seeds and visible patterns in the space-time diagram, these are ranked lower than max-length CA with $\gamma=1$. Therefore, \verb|LFSR113| and {\verb|xorshift1024*|} downgrade to rank $14$.
		
		
		\item Knuth's \verb|MMIX| and \verb|xorshift32| generator are in the same group (rank $15$).
		
		\item Max-length CA with $\gamma=0$ has better rank (rank $16$) than rule $30-45$ CA (rank $17$).
		
		\item Among \verb|rand|, \verb|lrand|, \verb|minstd_rand|, Borland's LCG, \verb|MRGk13p| and \verb|random|, the ranking is \verb|minstd_rand| (rank $23$) < \verb|MRGk13p| (rank $22$) < Borland's LCG (rank $21$)< \verb|random| (rank $20$) < \verb|rand| (rank $19$) < \verb|lrand| (rank $18$), where `<' indicates left PRNG has poorer performance than the right one.
		
		\item \verb|LFSR258| is the worst generator among the selected PRNGs.
	\end{itemize}
\end{enumerate}

\begin{table}[!h]
		\vspace{-2.0em}
		\setlength{\tabcolsep}{1.9pt}
		\scriptsize
		\renewcommand{\arraystretch}{1.30}
	\centering
	\small
	\caption{Summary of all empirical test results and final ranking}
	\label{tab:final_rank_comparison}
	\resizebox{1.00\textwidth}{!}{
		\begin{tabular}{|c|c|c|c|c|c|c|c|p{10.2em}|c|c|c|}
			\hline
			\multicolumn{2}{|c|}{\multirow{2}{*}{\theadfont{Name of the PRNGs}}} & \multicolumn{3}{c|}{\theadfont{Fixed Seeds}} &  \multicolumn{2}{c|}{\theadfont{Random Seeds}} & \multirow{2}{*}{\theadfont{Lattice Test}} & \multirow{2}{*}{\theadfont{Space-time Diagram}} & \multicolumn{3}{c|}{\theadfont{Ranking}} \\
			\cline{3-7}\cline{10-12}
			\multicolumn{2}{|c|}{ } & Diehard & TestU01 & NIST & Estimate & Range &  & & {\theadfont{$1^{st}$ Level}} & \theadfont{$2^{nd}$ Level} &{\theadfont{Final Rank}}\\
			\hline
			\multirow{7}{*}{\rotatebox{90}{LCGs}}& MMIX & 4-6 & 16-19 & 7-8 & 6.5 & 2-9  & Not Filled & Last $2$ bits fixed & 8 & 9 & 15\\
			\cline{2-12}
			& minstd\_rand & 0 & 1 & 1-2 &0.38 & 0-1 &  Not Filled & last $6$ bits fixed & 12 & 14 & 23\\
			\cline{2-12}
			& Borland LCG & 1 & 3 & 4-5 & 1.9 & 1-2 & Not Filled &  Last $2$ bits fixed & 11 & 12 & 21\\
			\cline{2-12}
			& rand & 1 & 1-3 & 2-3 &  &  &  Not Filled & More 0s than 1s & 11 & 13 & 19\\
			\cline{2-12}
			& lrand48 & 1 & 2-3 & 2 & 1 & 1 & Not Filled & More 0s than 1s & 11 & 13 & 18\\
			\cline{2-12}
			& MRG31k3p & 0-1 & 1-2 & 1-2 & 0.9 & 0-1 & Scattered & LSB is 0, block of 0s, dependency on seed & 12 & 14 & 22\\
			\cline{2-12}
			& PCG-32 & 9-11 & 24-25 & 14-15 & 9.3 & 6-12 & Relatively Filled & Independent of seed & 2 & 4 & 5\\
			\cline{2-12}
			\hline
			\multirow{16}{*}{\rotatebox{90}{LFSRs}}& random & 1 & 1-3 & 1 & 1 & 1 & Not Filled & MSB is 0, blocks of 0s & 11 & 13 & 20\\
			\cline{2-12}
			& Tauss88 & 9-11 & 21-23 & 14-15 & 9.0 & 0-12 & Relatively Filled & Independent of seed, block of $0$s & 4 & 7 & 11\\
			\cline{2-12}
			& LFSR113 & 5-11 & 6-23 & 1-15 & 9.3 & 6-12 & Relatively Filled & Dependency on seed, Block of 0s & 7 & 7 & 14\\
			\cline{2-12}
			& LFSR258 & 0-1 & 0-5 & 0-2 &  1.8 & 1-2 & Scattered & Pattern & 12 & 14 & 24\\
			\cline{2-12}
			& WELL512a & 7-10 & 23 & 14-15 & 8.5 & 5-11 & Relatively filled & First few numbers are fixed with seed dependency & 5 & 6 & 10\\
			\cline{2-12}
			& WELL1024a & 9-10 & 24-25 & 14-15 & 9.2 & 6-11 & Relatively Filled & Dependency on seed & 3 & 4 & 9\\
			\cline{2-12}
			& MT19937-32 & 9-10 & 25 & 13-15 & 9.3 & 6-12 & Relatively Filled & Independent of seed & 3 & 4 & 6\\
			\cline{2-12}
			& MT19937-64 & 8-11 & 24-25 & 15 & 9.4 & 6-11 & Relatively Filled & Independent of seed & 2 & 3 & 4\\
			\cline{2-12}
			& SFMT19937-32 & 9-10 & 25 & 15 & 9.5 & 5-12 & Relatively Filled & Independent of seed & 1 & 1 & 2\\
			\cline{2-12}
			& SFMT19937-64 & 9-11 & 25 & 15 & 9.52 & 6-12 & Relatively Filled & Independent of seed & 1 & 1 & 1\\
			\cline{2-12}
			& dSFMT-32 & 7-11 & 24-25 & 13-15 & 9.3 & 5-11 & Relatively Filled & Independent of seed & 5 & 5 & 7\\
			\cline{2-12}
			& dSFMT-52 & 5-7 & 9-11 & 3 & 5.97 & 3-7 & Relatively Filled & Less dependency on seed & 9 & 10 & 12\\
			\cline{2-12}
			&  xorshift32 & 2-4 & 17 & 2-13 & 5.5 & 3-7 & Not Filled & Blocks of 0s & 9 & 10 & 15\\
			\cline{2-12}
			&  xorshift64* & 7-10 & 25 & 14-15 & 8.0 & 6-11 & Relatively Filled & Independent of seed & 5 & 6 & 8\\
			\cline{2-12}
			&  xorshift1024* & 6-9 & 20-21 & 6-15 & 7.0 & 4-9 & Not Filled & Dependency on seed, Pattern & 6 & 8 & 14\\
			\cline{2-12}
			&  xorshift128+ & 8-10 & 24-25 & 14-15 & 9.4 & 6-12 & Relatively Filled & Dependency on seed for first few numbers & 4 & 4 & 9\\
			\hline
			\multirow{7}{*}{\rotatebox{90}{CAs}}& Rule $30$ & 8-11 & 24-25 & 15 & 10.2 & 7-12 & Relatively Filled & Independent of seed & 2 & 2 & 3\\
			\cline{2-12}
			& Hybrid CA with Rules $30$ \& $45$ & 0-3 & 1-8 & 0-3 & 2.0 & 0-3 & Not Filled & Pattern & 11 & 12 & 17\\
			\cline{2-12}
			& Maximal Length CA with $\gamma=0$ & 0-2 & 12 & 10-11 & 1.6 & 1-2 & Not Filled & Pattern & 10 & 11 & 16\\
			\cline{2-12}
			& Maximal Length CA with $\gamma=1$ & 3-4 & 15-17 & 14 & 1.8 & 1-4 & Relatively Filled & Dependency on seed for first few numbers & 8 & 11 & 13\\
			\cline{2-12}
			& Non-linear $2$-state CA & 5-7 & 11 & 3-4 & 5.85 & 2-8 & Relatively Filled &  Less dependency on seed & 9 & 9 & 13\\
			\cline{2-12}
			& $3$-state CA & ${2-3}$ & ${11-12}$ & ${4-6}$ & ${2.7}$ & ${1-4}$ & {Relatively Filled} & {Independent of seed} & ${10}$ & ${11}$ & ${13}$\\
			\cline{2-12}
			& Decimal CA & ${9-11}$ & ${25}$ & ${15}$ & ${9.59}$ & ${6-12}$ & {Relatively Filled} & {Independent of seed} & ${1}$ & ${1}$ & ${1}$\\
			\hline
	\end{tabular}}
\end{table}

Based on the empirical tests, we finally rank the selected $30$ PRNGs into $24$ groups. This final ranking is shown in Table~\ref{tab:final_rank_comparison}.
			
\section{Conclusion}\label{conclusion}
\noindent {In this paper, we have surveyed the evolution of PRNGs over several technologies -- LCGS, LFSRs and CA-based. Our target has been to test the well-known PRNGs which are currently in use using conventional testbeds and check how they actually perform in similar platform with the same seeds. We have used three empirical test-beds -- Diehard, battery \emph{rabbit} of TestU01 and NIST for blind statistical tests with some fixed seeds. Using these results, a first level ranking of the PRNGs is done. Then, to enhance this ranking, we have used results of the rough estimate behavior of the PRNGs on Diehard battery of tests for $1000$ seeds. However, as the underlying approximations about distributions in the testbeds and the thresholds are not always reliable, all blind tests have inherent incompleteness. To address this, we further use two graphical tests -- lattice tests and space-time diagram test that incorporate human intervention to decide further. We have observed that many times the visual behavior of a PRNG do not agree with the blind test results. Therefore, giving preference to the visualization given by the space-time diagrams, a final ranking has been done in Table~\ref{tab:final_rank_comparison}. According to our tests, Decimal CA based PRNGs and SFMT-64 bit generator are at par and the best pseudo-random number generators among all our selected PRNGs. Nevertheless, this ranking is not absolute and can be changed based on different metric. Further, the disagreement between blind test and space-time diagram results indicates the requirement of developing better metric that can detect non-randomness and rank the PRNGs by theoretical analysis.}

\section*{Acknowledgments}
\noindent 
The authors are grateful to Mr. Krishnendu Maity for his contribution in testing the generators and studying the results. We also thank the anonymous reviewers for their comments and suggestions which have improved the work.
  
 \bibliographystyle{elsarticle-num} 
\bibliography{References_survey}

\end{document}